\documentclass[11pt]{article}  

\usepackage{a41}
\usepackage{cite}
\usepackage{graphicx}
\usepackage{epsfig}
\usepackage{amsmath}
\usepackage{amssymb}
\usepackage{ulem}
\usepackage{mdwlist}
\usepackage{color}
\usepackage{float}
\usepackage[rflt]{floatflt}
\usepackage{slashed}
\usepackage{axodraw}

\usepackage{array}

\usepackage[english]{babel}
\usepackage[latin1]{inputenc}
\usepackage[T1]{fontenc}
\usepackage{ae}

\usepackage{url}

\usepackage{amsmath, amsthm, amssymb}

\newtheorem{thm}{Theorem}[section]

\usepackage{array}

\usepackage[english]{babel}
\usepackage[latin1]{inputenc}
\usepackage[T1]{fontenc}
\usepackage{ae}

\newtheorem{definition}[thm]{Definition}

\newcommand{\gsim}{\raisebox{-0.07cm   }
{$\, \stackrel{>}{{\scriptstyle\sim}}\, $}}
\newcommand{\GeV}{\rm GeV}
\newcommand{\MS}{\overline{{\rm MS}}}

\newcommand{\Li}{{\rm Li}}

\newcommand{\ep}{\varepsilon}

\usepackage{rotating}

\usepackage{graphicx}

\newcounter{mmacnt}
\def\restartmma{\setcounter{mmacnt}{0}}
\restartmma \catcode`|=\active
\def|#1|{\mathrm{#1}}
\catcode`|=12
\newenvironment{mma}{
 \par\smallskip
 \catcode`|=\active
 \parskip=0pt\parindent=0pt 
 \small
 \def\In##1\\{%
   \def\linebreak{\hfill\break\null\qquad}%
   \refstepcounter{mmacnt}
   \hangindent=2.5em\hangafter=0
   \leavevmode
   \llap{\tiny\sffamily In[\arabic{mmacnt}]:=\kern.5em}%
   \mathversion{bold}\footnotesize$\displaystyle##1$\normalsize
   \mathversion{normal}\par
 }%
 \def\Print##1\\{%
   \def\linebreak{\hfill\break}%
   \hangindent=2.5em\hangafter=0
   \leavevmode ##1\par}%
 \def\Out##1\\{%
   \def\linebreak{$\hfill\break\null\hfill$}%
   \kern\abovedisplayskip\par
   \hangindent=2.5em\hangafter=0
   \leavevmode
   \llap{\tiny\sffamily Out[\arabic{mmacnt}]=\kern.5em}
   \footnotesize$\displaystyle##1$\normalsize\hfill\null\par
   \kern\belowdisplayskip
 }%
 \def\Warning##1##2\\{%
   \def\linebreak{\hfill\break}%
   \hangindent=2.5em\hangafter=0
   \leavevmode
   {\scriptsize##1 : ##2}\par}%
}{%
 \par\smallskip
}

\usepackage{color}

\newenvironment{fshaded}{%
\MakeFramed {\FrameRestore}
}%
{\endMakeFramed}

\usepackage{tikz}
\usetikzlibrary{matrix}

\allowdisplaybreaks[4]

\begin{document}
\setlength{\baselineskip}{0.515cm}
\sloppy
\thispagestyle{empty}
\begin{flushleft}
DESY 13--210
\hfill 
\\
DO--TH 13/11\\
MITP/14-028\\
SFB/CPP-14-26\\
LPN 14-074\\
June 2014\\
\end{flushleft}

\mbox{}
\vspace*{\fill}
\begin{center}

{\LARGE\bf The 3-Loop Non-Singlet Heavy Flavor}

\vspace*{2mm}
{\LARGE\bf Contributions and Anomalous Dimensions for the}

\vspace*{3mm}
{\LARGE\bf Structure Function \boldmath $F_2(x,Q^2)$ and Transversity}

\vspace{4cm}
\large
J.~Ablinger$^a$,
A.~Behring$^b$,
J.~Bl\"umlein$^b$,
A.~De Freitas$^b$,
A.~Hasselhuhn$^a$,

\vspace*{1mm}
A.~von Manteuffel$^c$,
M.~Round$^{a,b}$,
C.~Schneider$^a$ and
F.~Wi\ss{}brock$^{a,b}$\footnote{Present address: IHES, 35 Route de Chartres, 91440 
Bures-sur-Yvette, France.}

\vspace{1.5cm}
\normalsize   
{\it $^a$~Research Institute for Symbolic Computation (RISC),\\
                          Johannes Kepler University, Altenbergerstra\ss{}e 69,
                          A--4040, Linz, Austria}\\

\vspace*{3mm}
{\it  $^b$ Deutsches Elektronen--Synchrotron, DESY,}\\
{\it  Platanenallee 6, D-15738 Zeuthen, Germany}

\vspace*{3mm}
{\it  $^c$ PRISMA Cluster of Excellence and Institute of Physics, J. Gutenberg University,}\\
{\it D-55099 Mainz, Germany.}

\end{center}
\normalsize
\vspace{\fill}
\begin{abstract}
\noindent 
We calculate the massive flavor non-singlet Wilson coefficient for the heavy flavor contributions to 
the structure function $F_2(x,Q^2)$ in the asymptotic region $Q^2 \gg m^2$ and the associated operator matrix 
element $A_{qq,Q}^{(3), \rm NS}(N)$ to 3-loop order in Quantum Chromodynamics at general values of the Mellin 
variable $N$. This matrix element is associated to the vector current and axial vector current for the even 
and the odd moments $N$, respectively. We also calculate the corresponding operator matrix elements for 
transversity, compute the contributions to the 3-loop anomalous dimensions to $O(N_F)$ and compare to results 
in the literature. The 3-loop matching of the flavor non-singlet distribution in the variable flavor number scheme
is derived. All results can be expressed in terms of nested harmonic sums in $N$ space and harmonic 
polylogarithms in $x$-space. Numerical results are presented for the non-singlet charm quark contribution to 
$F_2(x,Q^2)$.
\end{abstract}

\vspace*{\fill}
\noindent
\numberwithin{equation}{section}

\newpage
\section{Introduction}
\label{sec:1}

\vspace*{1mm}
\noindent
The heavy flavor corrections to the structure functions in unpolarized deep-inelastic scattering yield 
large contributions in particular in the range of small values of the Bjorken variable $x$. Due to the current
precision of the world deep-inelastic data which amounts for the structure function $F_2(x,Q^2)$ to $O(1\%)$ in a wide 
kinematic range, for the precision determination of the parton 
distributions \cite{PDF}, the strong coupling constant $\alpha_s(M_Z^2)$ \cite{Bethke:2011tr} and of the
mass of the charm quark $m_c$ \cite{Alekhin:2012vu}, the heavy flavor corrections have to be known at 3-loop 
order. The heavy flavor corrections are known to 2-loop order in semi-analytic form \cite{CHTL}\footnote{For a precise
implementation in Mellin-space, see \cite{Alekhin:2003ev}.}. Due to a factorization of the heavy flavor Wilson 
coefficients into massive operator matrix elements (OMEs) and massless Wilson coefficients in the asymptotic region 
$Q^2 \gg m^2$ \cite{Buza:1995ie}, one may calculate the Wilson coefficients analytically in this region. In case of the 
structure function $F_2(x,Q^2)$ the asymptotic representation holds for scales $Q^2~\gsim~10~m^2$ \cite{Buza:1995ie}, 
while much larger scales are required for the structure function $F_L(x,Q^2)$. Analytic expressions for the heavy
flavor Wilson coefficients in the asymptotic region and for the corresponding massive operator matrix elements were 
calculated in Refs.~\cite{Buza:1995ie,Bierenbaum:2007qe,Buza:1996xr,Bierenbaum:2007pn,Buza:1996wv,Bierenbaum:2009zt,Buza:1997mg,
Blumlein:2014fqa} at 2-loop order. The asymptotic 3-loop heavy flavor corrections to $F_L(x,Q^2)$ were computed 
in \cite{Blumlein:2006mh,Behring:2014eya}. At 3-loop order, a series of Mellin moments has been calculated for the Wilson 
coefficients contributing to the structure function $F_2(x,Q^2)$ in \cite{Bierenbaum:2009mv}
and for transversity in \cite{Blumlein:2009rg}. The logarithmic corrections and the parts of the constant contribution
being determined by renormalization have been given in \cite{Behring:2014eya}, cf. also  \cite{Bierenbaum:2008yu}.
For general values of the Mellin variable $N$, all OMEs corresponding to the color factors $\propto N_F T_F^2 C_{F,A}$
were calculated in \cite{Ablinger:2010ty,Blumlein:2012vq} and for the gluonic OME $A_{gg,Q}$ also for the terms 
$\propto T_F^2 C_{F,A}$ in \cite{Ablinger:2014uka}. Recently, also the OME $A_{gq}^{(3)}(N)$ has been computed
\cite{Ablinger:2014lka}.

In the present paper, we calculate the massive flavor  non-singlet Wilson coefficient contributing to the structure 
function
$F_2(x,Q^2)$ in the asymptotic region to 3-loop order and the associated massive operator matrix element 
$A_{qq,Q}^{(3),\rm NS}(N)$.
The latter quantity is given also for odd moments, which applies to the non-singlet contribution of the structure function 
$g_1(x,Q^2)$ as well as the charged current structure functions.
We repeat the calculation for the tensor and pseudo-tensor operators which yield the corresponding OMEs for transversity 
\cite{Barone:2001sp,Blumlein:2009rg}. As a by-product of the calculation, we obtain the contributions
$\propto T_F$ to the 3-loop anomalous dimensions in the vector and tensor case, in an independent calculation. In the case of 
transversity, this is the first calculation ab initio. We also present the heavy-to-light transition relations 
for the non-singlet distributions 
in the variable flavor number scheme (VFNS) to 3-loop order.

The paper is organized as follows. In Section~2, we give an outline on the flavor non-singlet OMEs and the massive 
Wilson coefficient to 3-loop order. In Section~3, we summarize technical aspects of the present calculation.
The contributions to the 3-loop anomalous dimensions in the vector and transversity case are derived in Section~4.
In Section~5, the OME in the vector case are given in Mellin space and in Section~6 the non-singlet Wilson coefficient 
is presented. Here we give also numerical results on its contribution to the structure function $F_2(x,Q^2)$. 
The non-singlet transition functions in the VFNS are discussed in Section~7. 
The OME for transversity is calculated in Section~8. Section~9  contains the conclusions. 

In the Appendices we present a series of master 
integrals used in the present calculation and
summarize the corresponding representations in $x$-space. 
\section{Basic Formalism}
\label{sec:2}

\vspace*{1mm}
\noindent
The massive flavor non-singlet operator matrix element is given by \cite{Bierenbaum:2009mv,Blumlein:2009rg} 
\begin{eqnarray}
A_{qq,Q}^{\rm NS, \MS} (\Delta.p)^N = \langle q|O^{\sf NS}_{F,a;\mu_1, \ldots, \mu_n}|q \rangle \Delta_{\mu_1} ... 
\Delta_{\mu_N} 
= 1+ \sum_{k=2}^\infty a_s^{(k)} A_{qq,Q}^{(k),\rm NS, \MS},
\end{eqnarray}
with $a_s(\mu^2) = \alpha_s(\mu^2)/(4\pi)$ and where the local operators in the vector and transversity case read
\begin{eqnarray}
O^{\rm NS}_{q,r;\mu_1, \ldots, \mu_n} &=& i^{n-1} {\bf S} [\overline{\psi}
\gamma_{\mu_1} (\gamma_5)D_{\mu_2} \ldots D_{\mu_n} \frac{\lambda_r}{2}
\psi] - {\rm trace~terms}
\label{eq:OP1}
\\
O^{\rm NS, TR}_{q,r;\mu, \mu_1, \ldots, \mu_n} &=& i^{n-1} {\bf S} [\overline{\psi}
\sigma_{\mu, \mu_1} (\gamma_5)D_{\mu_2} \ldots D_{\mu_n} \frac{\lambda_r}{2}
\psi] - {\rm trace~terms}~.
\label{eq:OP2}
\end{eqnarray}
Here $\Delta$ denotes a light-like vector.
The operators are contracted between massless quark states $|q\rangle$. ${\bf S}$ denotes
the symmetrization operator for the tensor indices, 
$\sigma_{\mu\nu} = (i/2)[\gamma_\mu \gamma_\nu - \gamma_\nu \gamma_\mu]$, $D_\mu$ is the covariant derivative, 
and $\lambda_r$ are the $N_F^2-1$
matrices generating the $SU(N_F)$ flavor group of $N_F$ massless quarks.

The expansion coefficients at 2- and 3-loop order of the OME
$A_{qq,Q}^{\rm NS, \MS}$ in the on-shell scheme with charge renormalization in the                     $\overline{\rm MS}$
scheme read \cite{Bierenbaum:2009mv} 
\begin{eqnarray}
\label{eq:Aqq2} 
A_{qq,Q}^{(2),\rm NS, \MS}&=&
                  \frac{\beta_{0,Q}\gamma_{qq}^{(0)}}{4}   
                    \ln^2 \Bigl(\frac{m^2}{\mu^2}\Bigr)
                 +\frac{\hat{\gamma}_{qq}^{(1), {\rm NS}}}{2}
                    \ln \Bigl(\frac{m^2}{\mu^2}\Bigr)
                 +a_{qq,Q}^{(2),{\rm NS}}
                 -\frac{\beta_{0,Q}\gamma_{qq}^{(0)}}{4}\zeta_2~,
\\
\label{eq:Aqq3}
A_{qq,Q}^{(3),{\rm NS}, \MS}&=&
     -\frac{\gamma_{qq}^{(0)}\beta_{0,Q}}{6}
          \Bigl(
                 \beta_0
                +2\beta_{0,Q}
          \Bigr)   
             \ln^3 \Bigl(\frac{m^2}{\mu^2}\Bigr)
         +\frac{1}{4}
          \Biggl\{
                   2\gamma_{qq}^{(1),\rm NS}\beta_{0,Q}
                  -2\hat{\gamma}_{qq}^{(1),{\rm NS}} 
                             \Bigl(
                                    \beta_0
                                   +\beta_{0,Q}
                             \Bigr)
\nonumber\\ &&
                  +\beta_{1,Q}\gamma_{qq}^{(0)}
          \Biggr\}
             \ln^2 \Bigl(\frac{m^2}{\mu^2}\Bigr)
         +\frac{1}{2}
          \Biggl\{
                   \hat{\gamma}_{qq}^{(2),{\rm NS}}
                  -\Bigl(
                           4a_{qq,Q}^{(2),{\rm NS}}
                          -\zeta_2\beta_{0,Q}\gamma_{qq}^{(0)}
                                    \Bigr)(\beta_0+\beta_{0,Q})
\nonumber\\ &&
                  +\gamma_{qq}^{(0)}\beta_{1,Q}^{(1)}
          \Biggr\}
             \ln \Bigl(\frac{m^2}{\mu^2}\Bigr)
         +4\overline{a}_{qq,Q}^{(2),{\rm NS}}(\beta_0+\beta_{0,Q})
         -\gamma_{qq}^{(0)}\beta_{1,Q}^{(2)}
         -\frac{\gamma_{qq}^{(0)}\beta_0\beta_{0,Q}\zeta_3}{6}
\nonumber\\ &&
         -\frac{\gamma_{qq}^{(1),{\rm NS}}\beta_{0,Q}\zeta_2}{4}
         +2 \delta m_1^{(1)} \beta_{0,Q} \gamma_{qq}^{(0)}
         +\delta m_1^{(0)} \hat{\gamma}_{qq}^{(1),{\rm NS}}
         +2 \delta m_1^{(-1)} a_{qq,Q}^{(2),{\rm NS}}
\nonumber\\ &&
         +a_{qq,Q}^{(3),{\rm NS}},
\end{eqnarray}

\noindent
where $\mu$ denotes the factorization scale. Here we consider the case of a single heavy quark only\footnote{
For the more general case of two heavy quarks, contributing at 3-loop order, see \cite{BW1}.}. The expressions
depend on the expansion coefficients of the anomalous dimensions $\gamma_{ij}^{(k)}$, the massless and massive 
QCD $\beta$-function and heavy quark mass, as well as the parts of the OME
$a_{ij}, \bar{a}_{ij}$ up to 3-loop order, and $\zeta_k = \sum_{l=1}^\infty (1/l^k), k \in \mathbb{N}, k\geq 2$ 
denote the values of the Riemann $\zeta$-function at integer values. For details see 
Refs.~\cite{Bierenbaum:2009mv,Blumlein:2009rg}. We also use the notation
\begin{eqnarray}
\hat{f}(N_F) =  f(N_F+1) - f(N_F)~.
\end{eqnarray}
The contribution $\propto N_F$ of the non-singlet anomalous dimension is obtained form the term $\propto \ln(m^2/\mu^2)$ in
Eq.~(\ref{eq:Aqq3}) as a by-product of the present calculation.

The massive flavor non-singlet Wilson coefficient in the asymptotic region $Q^2 \gg m^2$ can be expressed
in terms of the massive OMEs and the massless Wilson coefficients. The latter are known to 3-loop order 
\cite{Vermaseren:2005qc}. The flavor non-singlet Wilson coefficient reads \cite{Buza:1995ie,Bierenbaum:2009mv}
\begin{eqnarray}
L_{q,2}^{\rm NS}(N_F+1) &=& a_s^2\Biggl[A_{qq,Q}^{(2),\rm NS}(N_F+1) + \hat{C}_{q,2}^{(2),\rm NS}(N_F)\Biggr]
\nonumber \\
&+& a_s^3\Biggl[
  A_{qq,Q}^{(3),\rm NS}(N_F+1) 
+ A_{qq,Q}^{(2),\rm NS}(N_F+1) {C}_{q,2}^{(1),\rm NS}(N_F+1)
+ \hat{C}_{q,2}^{(3),\rm NS}(N_F) \Biggr].
\nonumber\\
\label{eq:LQ2}
\end{eqnarray}
Here the notation `$N_F+1$' in the OMEs $A_{qq,Q}$ symbolically denotes that these are calculated at 
$N_F$ massless and one massive flavor. 

The 2-loop result has been calculated in Refs.~\cite{Buza:1995ie,Bierenbaum:2007qe} and the $O(\varepsilon)$
term $\bar{a}_{qq,Q}^{(2),\rm NS}$ needed for the renormalization in \cite{Bierenbaum:2008yu}. The new contribution
computed here is $a_{qq,Q}^{(3),\rm NS}$ both in the vector and transversity cases for the even moments.
The odd moments refer to the corresponding polarized cases. The $\gamma_5$--problem in the flavor non-singlet case
is solved trivially since the operator is placed on the external line and the Ward-Takahashi identity \cite{WT} 
allows to map the vertex function into self-energy corrections, off the vertex. 

The heavy flavor non-singlet contribution to the unpolarized deep-inelastic structure function $F_2(x,Q^2)$ for pure photon 
exchange is given by
\begin{eqnarray}
\left. F_2^{Q\bar{Q},\rm NS}(x,Q^2)\right| = \sum_{k = 1}^{N_F} e_k^2 x L_{q,2}^{\rm NS}\left(x, 
N_F+1,\frac{Q^2}{\mu^2},\frac{m^2}{\mu^2}\right) \otimes \left[f_k(x,\mu^2,N_F) + \bar{f}(x,\mu^2,N_F)\right],
\label{eq:SF}
\end{eqnarray}
where  $N_F$ denotes the number of light quarks, $e_k$ their electric charge in units of the elementary charge and 
$f_k (\bar{f}_k)$ are the quark (antiquark) parton distribution functions. The operator $\otimes$ denotes the Mellin
convolution 
\begin{eqnarray}
A(x) \otimes B(x) = \int_0^1 dx_1 \int_0^1 dx_2 \delta(x - x_1 x_2) A(x_1) B(x_2)~.
\label{eq:MELC}
\end{eqnarray}
Here the diction `flavor non-singlet term' for (\ref{eq:SF}) was introduced in Ref.~\cite{Buza:1996wv}, rather meaning the 
individual flavor contribution. The non-singlet contribution in the sense of an associated operator would be $f_k + 
\bar{f}_k - \Sigma/N_F$, with $\Sigma = \sum_{l=1}^{N_F} (f_l + \bar{f}_l)$. For historical reasons we follow the former 
notion. 

Before we present the results of the calculation, let us turn to its technical details first.
\section{Calculation of the Diagrams and Feynman Integrals}
\label{sec:3}

\subsection{Diagrams and operator insertions}

\begin{figure}
\centering
\includegraphics[width=0.8\textwidth]{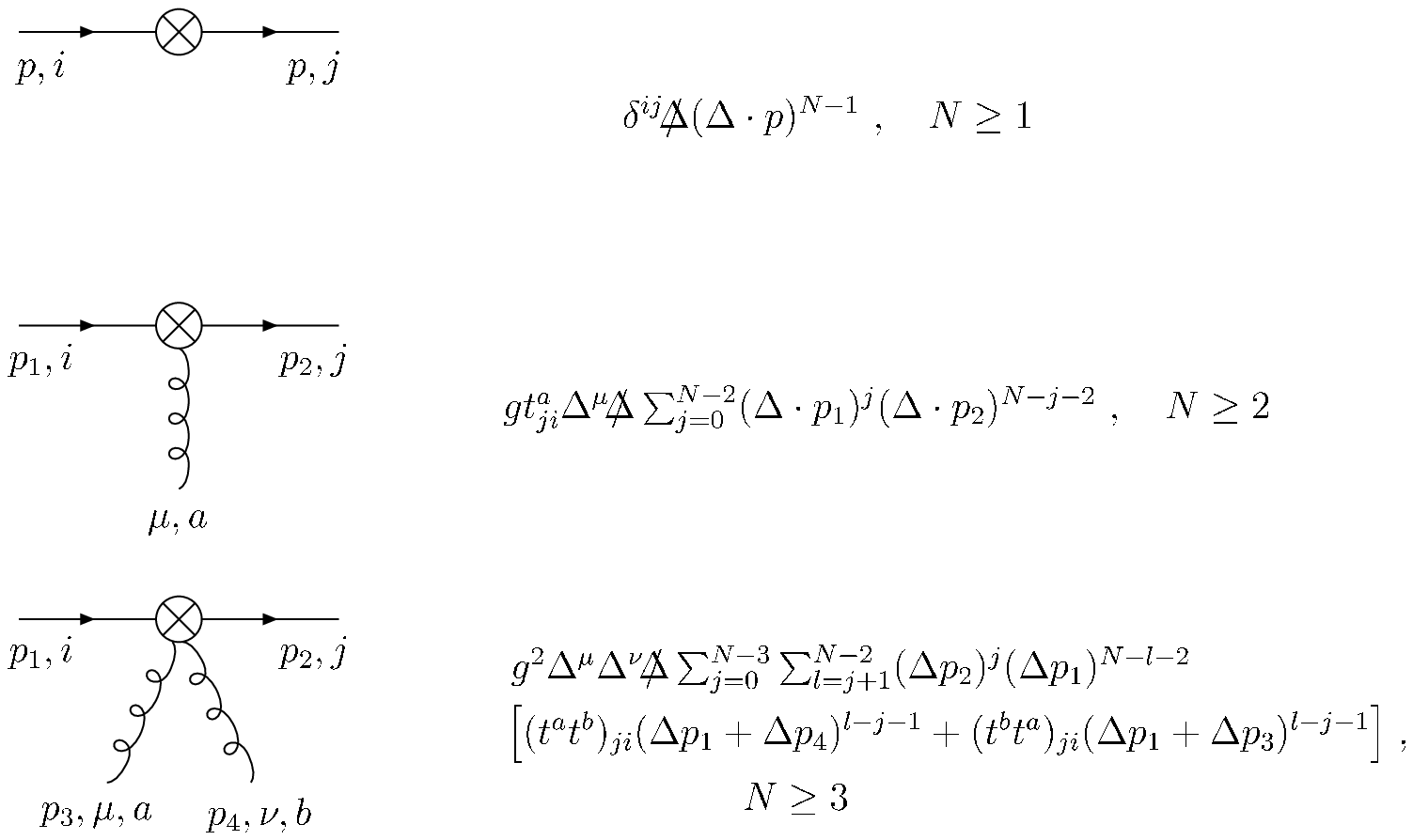}
\caption{\sf \small Feynman rules for non-singlet operator insertions. The insertions in the case of transversity are obtained 
replacing $\slashed \Delta$ by $\sigma^{\mu \nu} \Delta_{\nu}$.}
\label{feynrules}
\end{figure}

\vspace*{1mm}
\noindent
In order to calculate the operator matrix elements $A_{qq,Q}^{(3), \rm NS}$ and $A_{qq,Q}^{(3), \rm NS, TR}$, we make use of the
Feynman rules for operator insertions shown in Figure~\ref{feynrules}, together with the standard Feynman rules of QCD, from 
which we construct and later evaluate the corresponding Feynman diagrams. There is a total of 112 diagrams needed for 
$A_{qq,Q}^{(3), \rm NS}$, as well as for $A_{qq,Q}^{(3), \rm NS, TR}$, which we generated using an extension of {\tt QGRAF} 
\cite{Nogueira:1991ex} including also local operators \cite{Bierenbaum:2009mv}. The diagrams for $A_{qq,Q}^{(3), \rm NS}$ look 
exactly the same as the diagrams for  $A_{qq,Q}^{(3), \rm NS, TR}$, the only difference being in the explicit expressions for 
the 
operator insertions, where $\slashed \Delta$ is replaced by $\sigma^{\mu \nu} \Delta_{\nu}$ in the case of transversity.
As we perform the renormalization for the reducible set of Feynman diagrams, cf.~\cite{Bierenbaum:2009mv}, the self-energy 
contribution
\begin{eqnarray}
\Sigma_3 &=&
C_F C_A T_F \frac{8}{3} \frac{1}{\varepsilon^3}
+ \left[C_F T_F^2 \left(\frac{64}{9} + \frac{32}{9} N_F \right)
- C_A C_F T_F \frac{40}{9}
- C_F^2 T_F \frac{8}{3} \right] 
\frac{1}{\varepsilon^2}
\nonumber\\ &&
+ \Bigg[
C_F C_A T_F \left(\frac{454}{27} + \zeta_2 \right)
+ C_F T_F^2 \left(\frac{80}{27} + \frac{40}{27} N_F\right)
-26 C_F^2 T_F \Biggr]
\frac{1}{\varepsilon}
\nonumber\\ &&
+ C_F C_A T_F
\left(\frac{1879}{162}
-\frac{5}{3} \zeta_2 +\frac{17}{3} \zeta_3\right)
+ C_F T_F^2 \left[
\frac{604}{81}
+\frac{674}{81} N_F
+ \left(\frac{8}{3} + \frac{4}{3} N_F\right) \zeta_2\right]
\nonumber\\ &&
- C_F^2 T_F \left(
  \frac{335}{18}
+ \zeta_2 + 8 \zeta_3
\right),
\end{eqnarray}
has to be accounted for both the vector and transversity cases. Here $\ep = D-4$ is the parameter emerging in dimensional 
regularization.
For the vector flavor non-singlet OME, it will lead
to a vanishing 1$^{\rm{st}}$ moment due to fermion-number conservation.

\begin{figure}
\begin{minipage}[c]{0.23\linewidth}
     \includegraphics[width=1\textwidth]{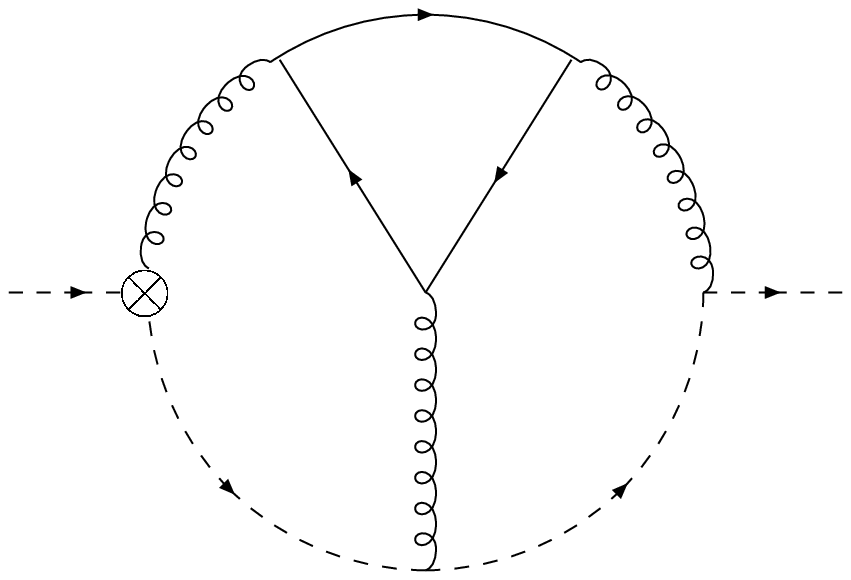}
\vspace*{-8mm}
\begin{center}
{\footnotesize (a)}
\end{center}
\end{minipage}
\hspace*{1mm}
\begin{minipage}[c]{0.23\linewidth}
     \includegraphics[width=1\textwidth]{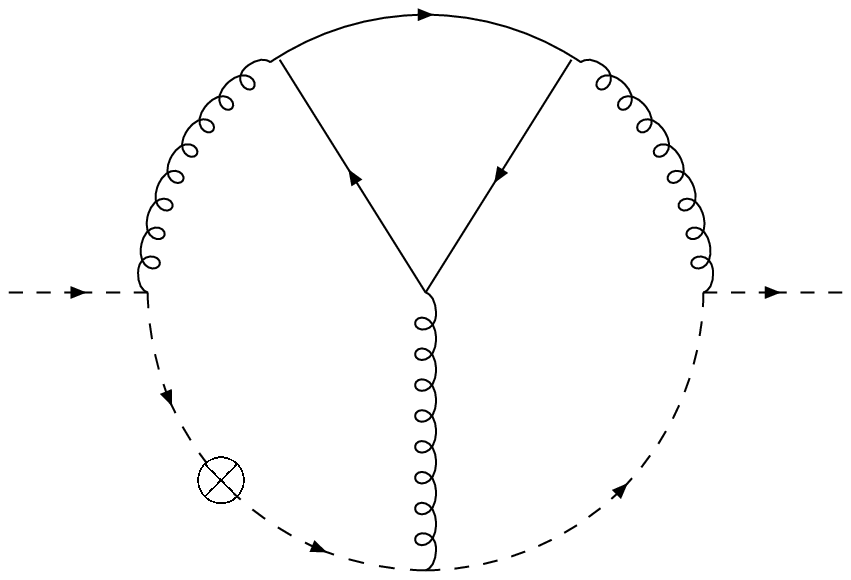}
\vspace*{-8mm}
\begin{center}
{\footnotesize (b)}
\end{center}
\end{minipage}
\hspace*{1mm}
\begin{minipage}[c]{0.23\linewidth}
     \includegraphics[width=1\textwidth]{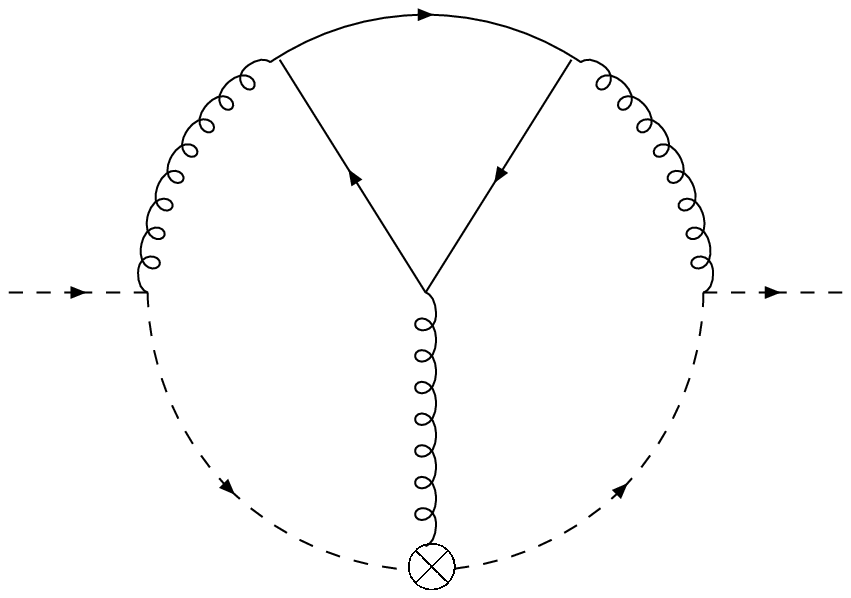}
\vspace*{-8mm}
\begin{center}
{\footnotesize (c)}
\end{center}
\end{minipage}
\hspace*{1mm}
\begin{minipage}[c]{0.23\linewidth}
     \includegraphics[width=1\textwidth]{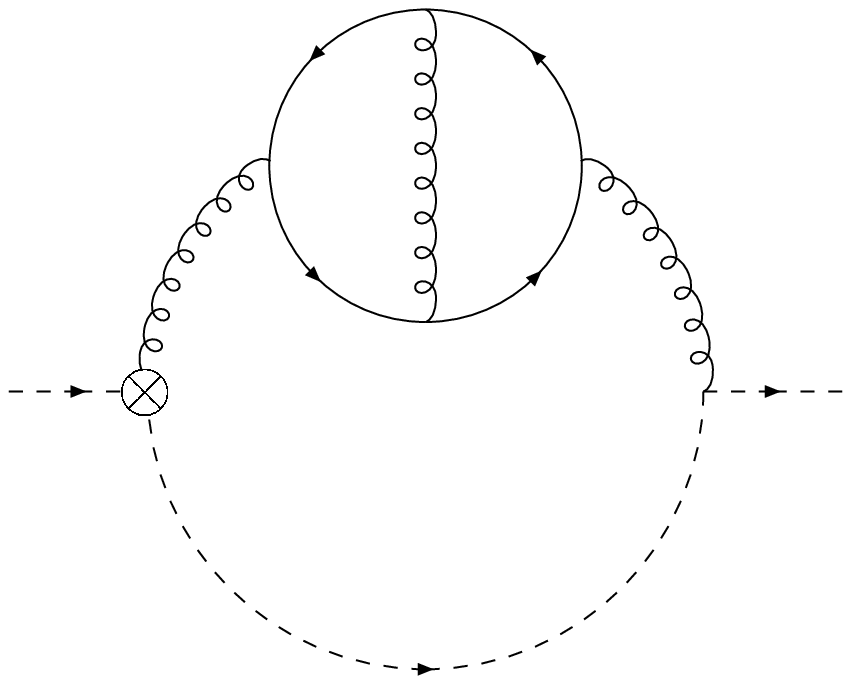}
\vspace*{-8mm}
\begin{center}
{\footnotesize (d)}
\end{center}
\end{minipage}

\vspace*{4mm}
\begin{minipage}[c]{0.23\linewidth}
     \includegraphics[width=1\textwidth]{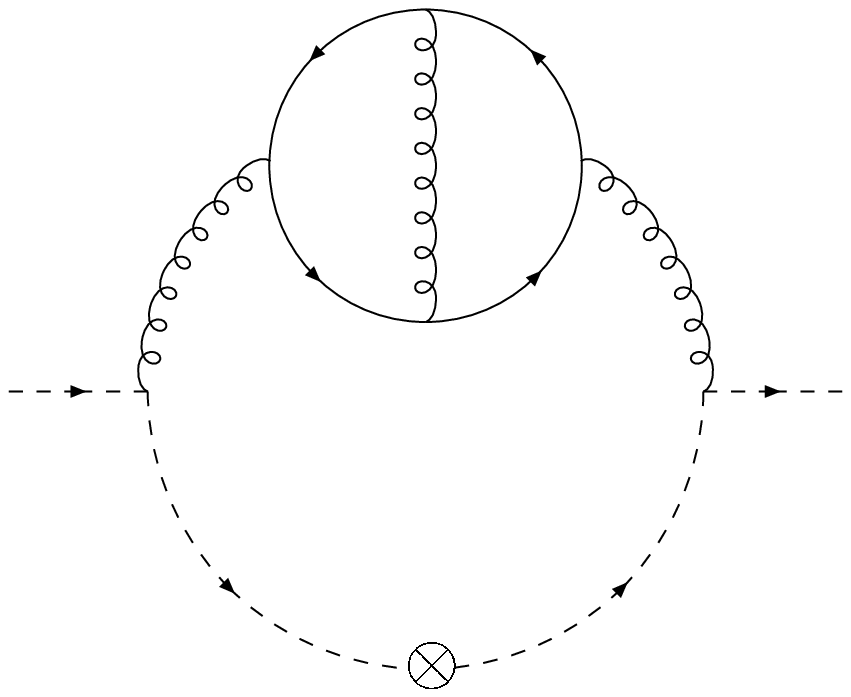}
\vspace*{-8mm}
\begin{center}
{\footnotesize (e)}
\end{center}
\end{minipage}
\hspace*{1mm}
\begin{minipage}[c]{0.23\linewidth}
     \includegraphics[width=1\textwidth]{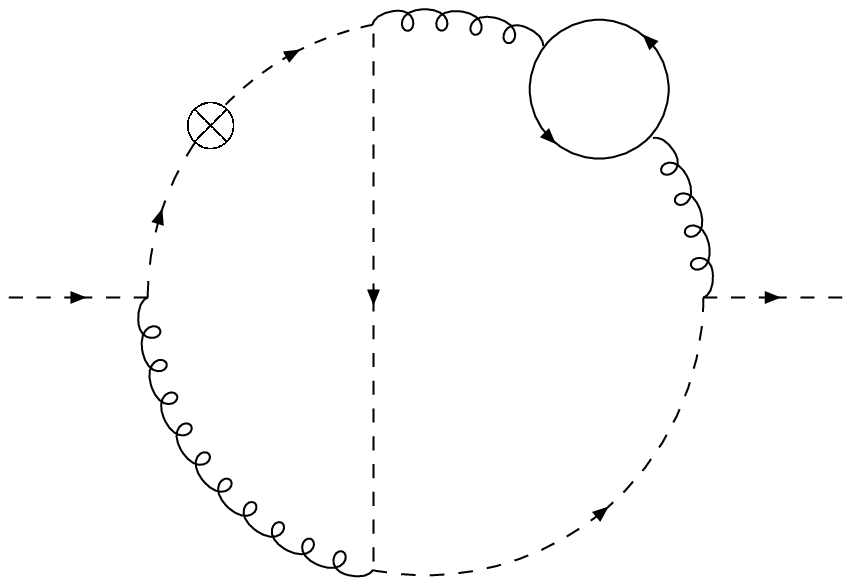}
\vspace*{-8mm}
\begin{center}
{\footnotesize (f)}
\end{center}
\end{minipage}
\hspace*{1mm}
\begin{minipage}[c]{0.23\linewidth}
     \includegraphics[width=1\textwidth]{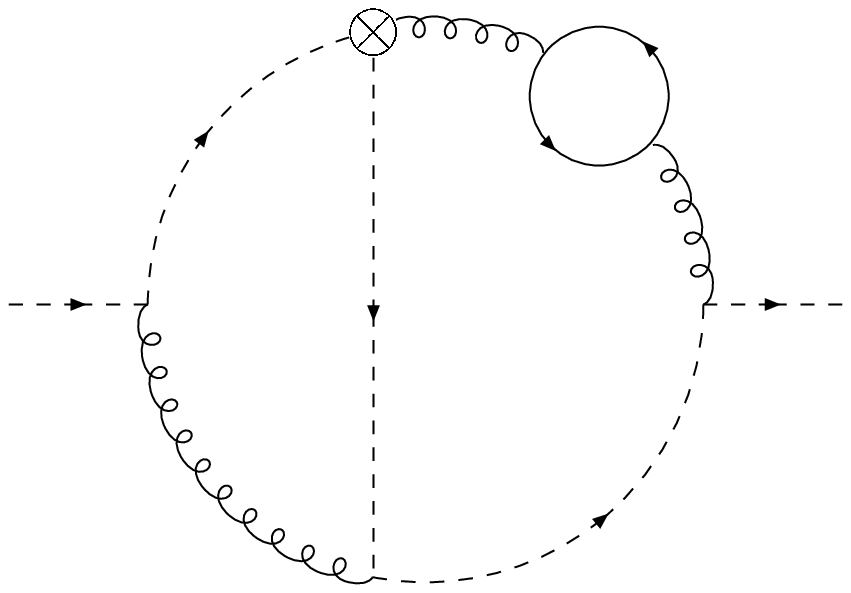}
\vspace*{-8mm}
\begin{center}
{\footnotesize (g)}
\end{center}
\end{minipage}
\hspace*{1mm}
\begin{minipage}[c]{0.23\linewidth}
     \includegraphics[width=1\textwidth]{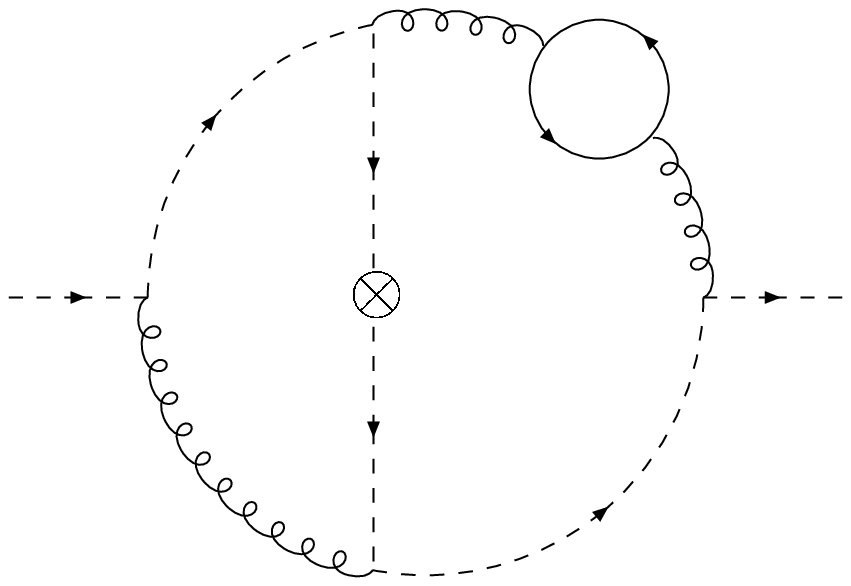}
\vspace*{-8mm}
\begin{center}
{\footnotesize (h)}
\end{center}
\end{minipage}

\vspace*{4mm}
\begin{minipage}[c]{0.23\linewidth}
     \includegraphics[width=1\textwidth]{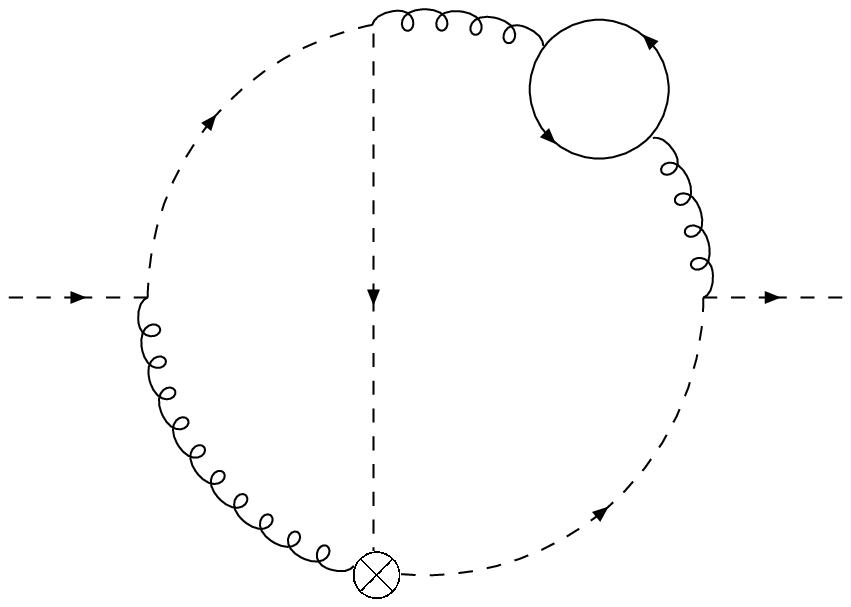}
\vspace*{-8mm}
\begin{center}
{\footnotesize (i)}
\end{center}
\end{minipage}
\hspace*{1mm}
\begin{minipage}[c]{0.23\linewidth}
     \includegraphics[width=1\textwidth]{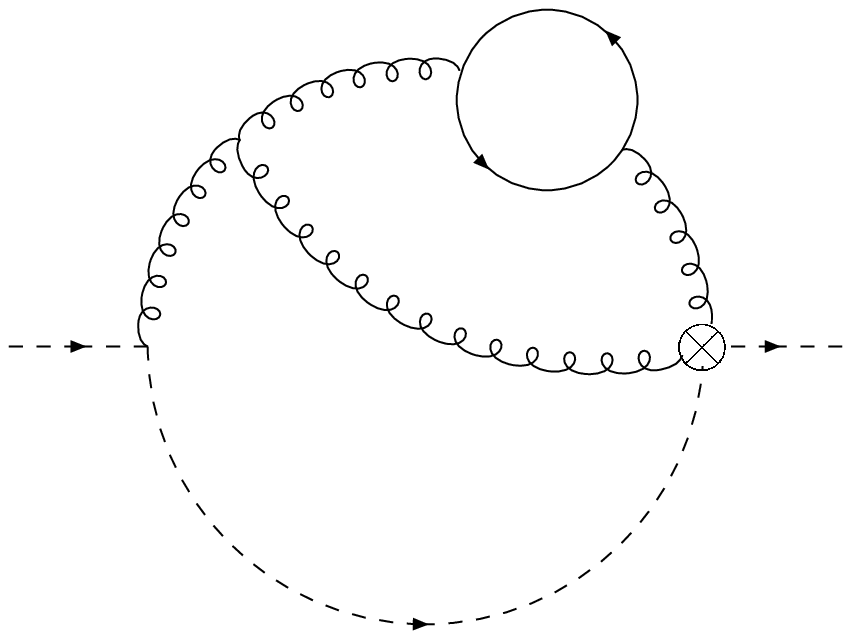}
\vspace*{-8mm}
\begin{center}
{\footnotesize (j)}
\end{center}
\end{minipage}
\hspace*{1mm}
\begin{minipage}[c]{0.23\linewidth}
     \includegraphics[width=1\textwidth]{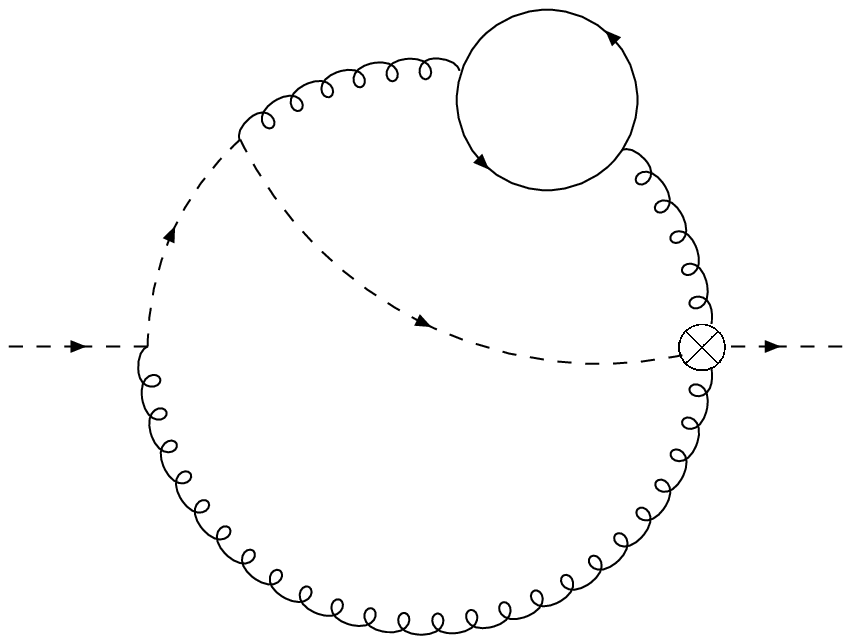}
\vspace*{-8mm}
\begin{center}
{\footnotesize (k)}
\end{center}
\end{minipage}
\hspace*{1mm}
\begin{minipage}[c]{0.23\linewidth}
     \includegraphics[width=1\textwidth]{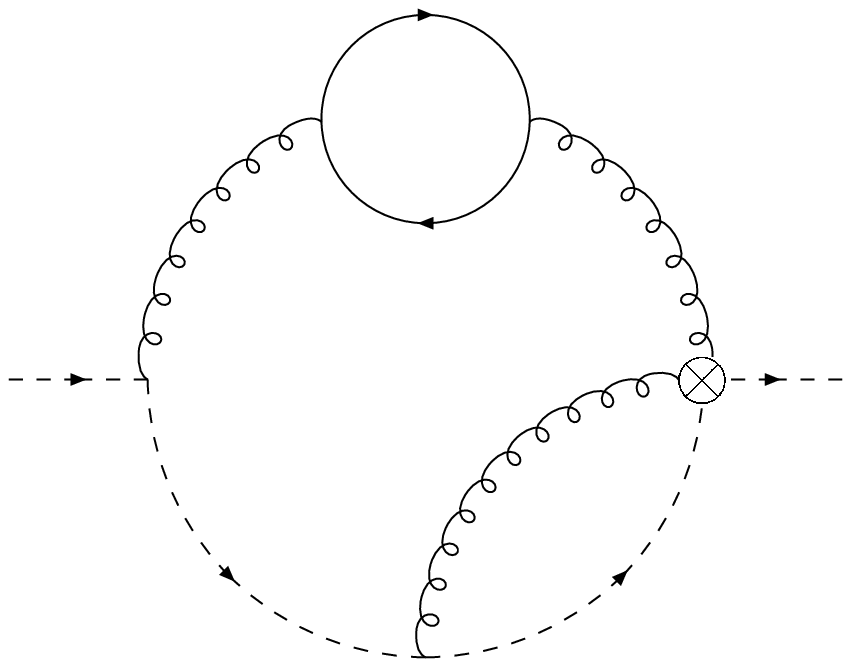}
\vspace*{-8mm}
\begin{center}
{\footnotesize (l)}
\end{center}
\end{minipage}

\vspace*{4mm}
\begin{minipage}[c]{0.23\linewidth}
     \includegraphics[width=1\textwidth]{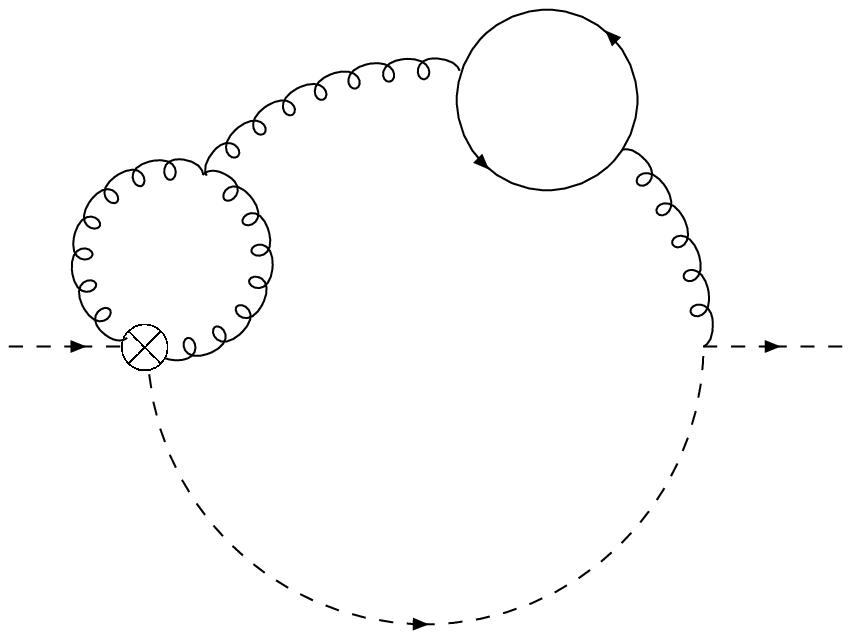}
\vspace*{-8mm}
\begin{center}
{\footnotesize (m)}
\end{center}
\end{minipage}
\hspace*{1mm}
\begin{minipage}[c]{0.23\linewidth}
     \includegraphics[width=1\textwidth]{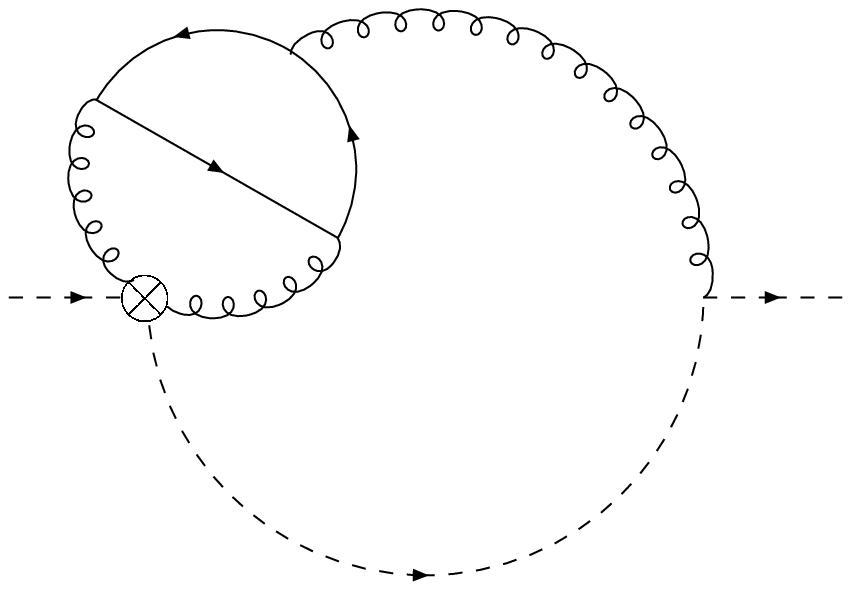}
\vspace*{-8mm}
\begin{center}
{\footnotesize (n)}
\end{center}
\end{minipage}
\hspace*{1mm}
\begin{minipage}[c]{0.23\linewidth}
     \includegraphics[width=1\textwidth]{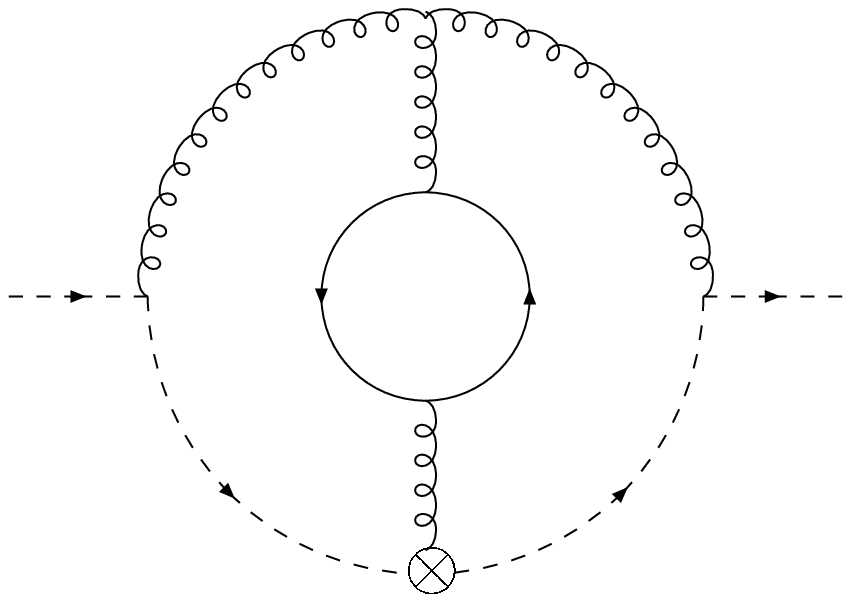}
\vspace*{-8mm}
\begin{center}
{\footnotesize (o)}
\end{center}
\end{minipage}
\hspace*{1mm}
\begin{minipage}[c]{0.23\linewidth}
     \includegraphics[width=1\textwidth]{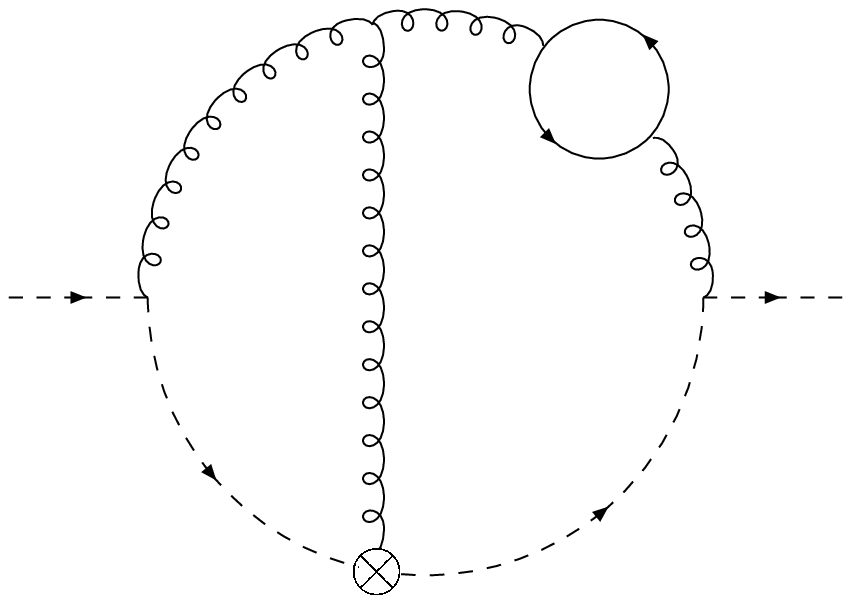}
\vspace*{-8mm}
\begin{center}
{\footnotesize (p)}
\end{center}
\end{minipage}

\vspace*{4mm}
\begin{minipage}[c]{0.23\linewidth}
     \includegraphics[width=1\textwidth]{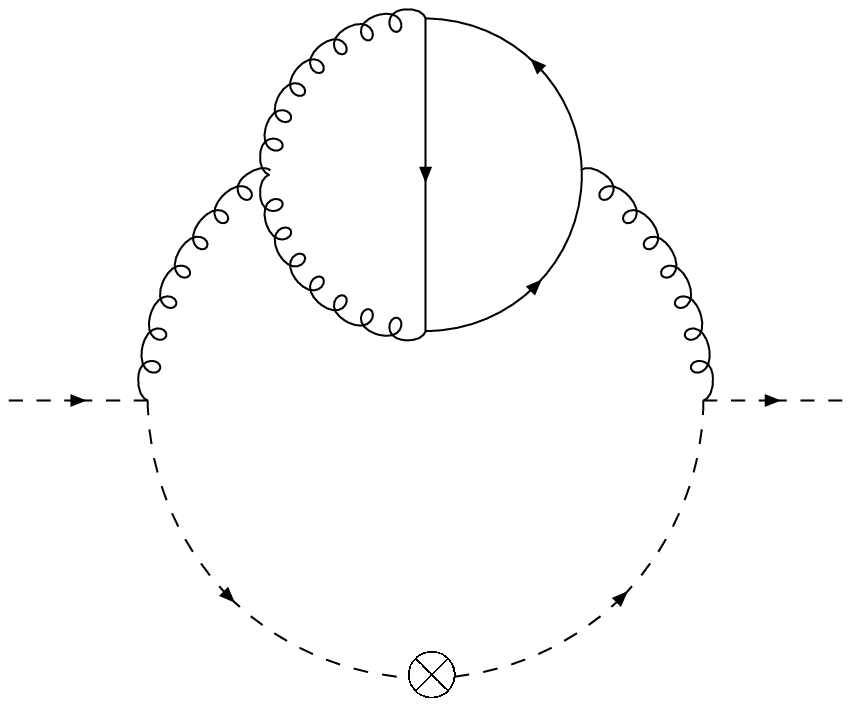}
\vspace*{-8mm}
\begin{center}
{\footnotesize (q)}
\end{center}
\end{minipage}
\hspace*{1mm}
\begin{minipage}[c]{0.23\linewidth}
     \includegraphics[width=1\textwidth]{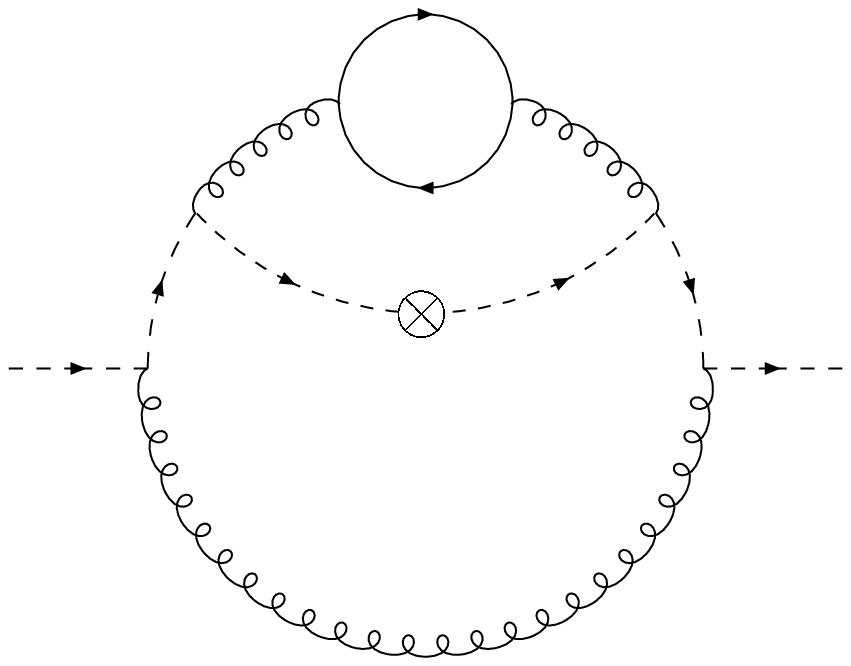}
\vspace*{-8mm}
\begin{center}
{\footnotesize (r)}
\end{center}
\end{minipage}
\hspace*{1mm}
\begin{minipage}[c]{0.23\linewidth}
     \includegraphics[width=1\textwidth]{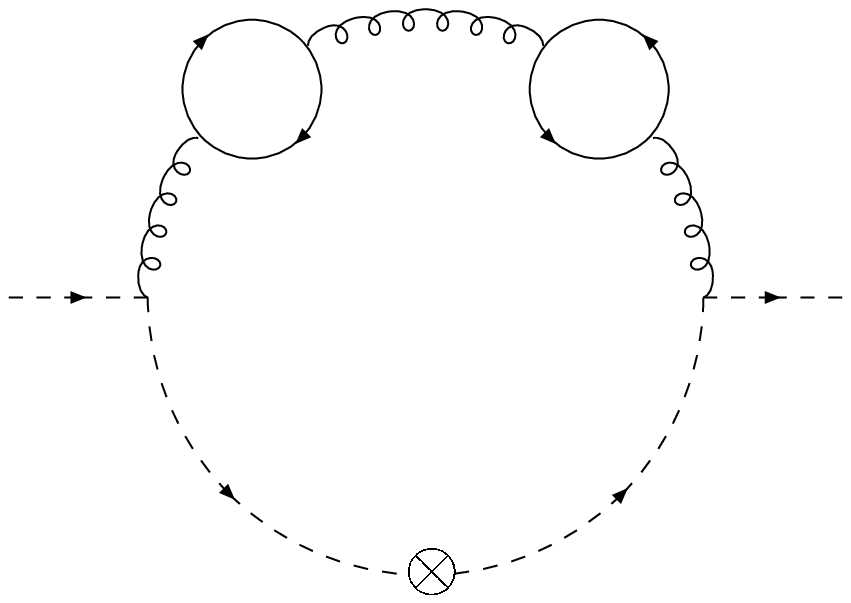}
\vspace*{-8mm}
\begin{center}
{\footnotesize (s)}
\end{center}
\end{minipage}
\hspace*{1mm}
\begin{minipage}[c]{0.23\linewidth}
     \includegraphics[width=1\textwidth]{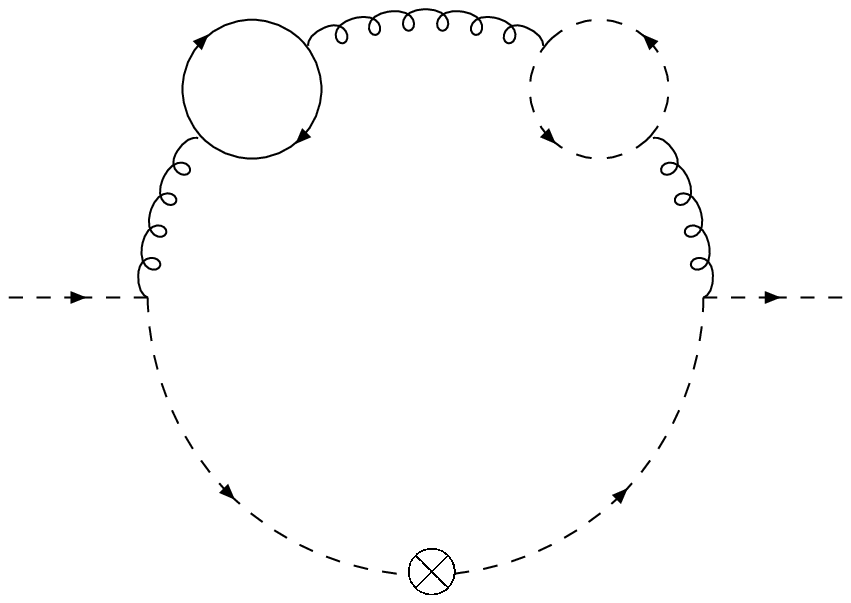}
\vspace*{-8mm}
\begin{center}
{\footnotesize (t)}
\end{center}
\end{minipage}
\caption{\sf \small Sample of diagrams for $A_{qq,Q}^{(3), \rm NS}$ and $A_{qq,Q}^{(3), \rm NS, TR}$. The dashed arrow lines 
represent
massless quarks, while the solid arrow lines represent massive quarks, and curly lines are gluons.
The operator insertion Feynman rules for $A_{qq,Q}^{(3), \rm NS, TR}$ are obtained from those of $A_{qq,Q}^{(3), \rm NS}$ 
by making the replacement $\slashed \Delta \rightarrow \sigma^{\mu \nu} \Delta_{\nu}$. 
}
\label{samplediagrams}
\end{figure}

A sample of the diagrams is shown in Figure~\ref{samplediagrams}. The most complicated topologies are the Benz-like 
ones\footnote{Finite Benz-type graphs can be completely calculated using the method of hyperlogarithms 
\cite{Ablinger:2014yaa}.}
of Figures~\ref{samplediagrams}a--\ref{samplediagrams}c, with a massive triangle in the loop, and those with four massive lines
as in the diagrams in Figures~\ref{samplediagrams}d and \ref{samplediagrams}e. No ladder or non-planar diagrams are involved in the 
calculation of these operator matrix elements, which simplifies matters considerably.

A {\tt FORM} \cite{Tentyukov:2007mu} program was written, cf.~\cite{Bierenbaum:2009mv}, in order to replace the propagators, 
vertices and operator insertions appearing in the output of {\tt QGRAF} by the corresponding Feynman rules, and also to introduce the 
corresponding projectors and perform the Dirac matrix algebra in the numerator of the diagrams. After this, the diagrams end up 
being 
expressed as linear combinations of scalar integrals. For example, the diagram
(also shown without labels in Figure~\ref{samplediagrams}b) 
\begin{equation}
\includegraphics[width=0.38\textwidth]{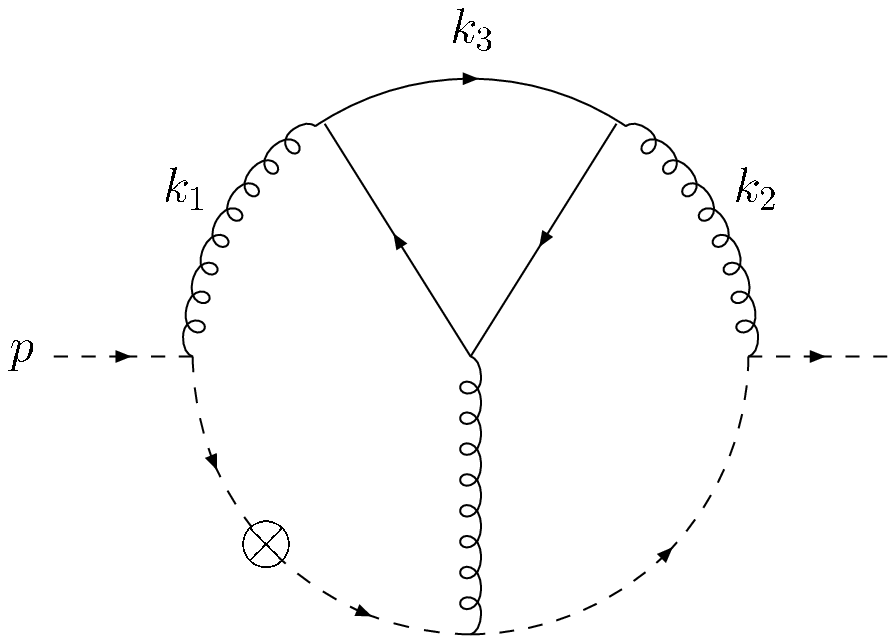}
\label{benzdiag}
\end{equation}

is written as a 
linear combination of integrals of the form
\begin{equation}
I_{\nu_1, \ldots , \nu_9}^{a,b,c,N} =
\int \frac{d^D k_1}{(2\pi)^D} \frac{d^D k_2}{(2\pi)^D} \frac{d^D k_3}{(2\pi)^D}
\frac{(\Delta.k_1)^a (\Delta.k_2)^b (\Delta.k_3)^c \left( \Delta.p - \Delta.k_1 
\right)^{N-1}}{D_1^{\nu_1} D_2^{\nu_2} D_3^{\nu_3} D_4^{\nu_4} D_5^{\nu_5} D_6^{\nu_6} D_7^{\nu_7} D_8^{\nu_8} D_9^{\nu_9}} \, ,
\label{scalint1}
\end{equation}
where $p$ is the momentum of the external massless quark, which is taken on-shell ($p^2=0$), $a$, $b$, $c$ and $\nu_1 \ldots \nu_9$ are
integers, and

\begin{eqnarray}
& D_1 = k_1^2, \quad D_2 = (k_1-p)^2, \quad D_3 = k_2^2, \quad D_4 = (k_2-p)^2,  \nonumber \\
& D_5 = k_3^2-m^2, \quad D_6 = (k_1-k_3)^2-m^2, \quad D_7 = (k_2-k_3)^2-m^2, \nonumber \\
& D_8 = (k_1-k_2)^2 \quad {\rm and} \quad D_9 = (k_3-p)^2-m^2\, . & 
\end{eqnarray}
Here $m$ is the mass of the heavy quark and $a.b$ denotes the Minkowski product. The denominators $D_1$ to $D_8$ are the 
inverse of
the propagators appearing in the diagram.
An additional auxiliary inverse propagator $D_9$ has been introduced in such a
way that all possible scalar products of momenta $k_i.k_j$ and $k_i.p$ ($i,j=1,2,3$) can be uniquely expressed as linear
combinations of all inverse propagators $D_1$ to $D_9$. A scalar integral with irreducible numerators will be represented in
this way by one or more negative integers among the exponents $\nu_1$ to $\nu_9$ in $I_{\nu_1, \ldots , \nu_9}^{a,b,c,N}$.

In Eq.~(\ref{scalint1}) there are also powers of the type $\Delta.k_i$ in the numerator. These come from the contraction of 
the
$\slashed \Delta$ in the Feynman rule for the operator insertion in the diagram (\ref{benzdiag}) with an internal momentum 
after the
Dirac matrix algebra is performed in the numerator. At most one such power can appear in the case of this diagram, i.e., $a,b,c 
\ge 0$ and
$a+b+c \leq 1$, but in the case of diagrams with 3-point and 4-point vertex operator insertions, higher powers can also appear.
We will see in the next Section that all operator insertions can be written in terms of artificial propagators, and introducing 
enough auxiliary propagators of this new type, the scalar products of internal momenta with $\Delta$ will be expressed as 
linear combinations of artificial propagators.

\subsection{Reduction to master integrals using integration by parts identities}

\vspace{1mm}
\noindent
We apply integration by parts identities \cite{IBP} to the Feynman diagrams with operator insertions to reduce
the problem to the calculation of a set of master integrals.
The reductions were realized extending the {\tt C++}-code 
{\tt Reduze\;2}~\cite{vonManteuffel:2012np}\footnote{The package {\tt Reduze\;2} uses the packages {\tt Fermat} \cite{FERMAT}
and {\tt Ginac} \cite{Bauer:2000cp}.}. 
In the following we describe the general implementation needed for the calculation of all the massive 3-loop 
operator matrix elements and focus later to the non-singlet case.
\def\CrossCirc(#1,#2)#3{
  \BCirc(#1,#2){#3}
  \Line(#1 #3 neg 2.0 sqrt div add, #2 #3 neg 2.0 sqrt div add)(#1 #3 2.0 sqrt div add, #2 #3 2.0 sqrt div add)
  \Line(#1 #3 neg 2.0 sqrt div add, #2 #3 2.0 sqrt div add)(#1 #3 2.0 sqrt div add, #2 #3 neg 2.0 sqrt div add)
}

\def\BTriRot(#1,#2)(#3,#4)#5{
\put(0,0){
\special{"
 0.5 #5 neg #1 #3 add add mul 0.5 #2 #4 add mul translate
 1.0 #4 #2 sub mul 1.0 #3 #1 sub mul atan rotate
 \axocolor  -0.5 #5 mul \axoxo add  0.5 #5 mul \axoyo add
            -0.5 #5 mul \axoxo add -0.5 #5 mul \axoyo add
             0.5 #5 mul \axoxo add  0.0 \axoyo add
 \axowidth \axoscale btriangle
}}}

\newcommand{\picbox}[1]{\vcenter{\hbox{#1}}}
\subsubsection{Mapping operator insertions to linear propagators}
\label{sec:opprops}

\vspace{1mm}
\noindent
In order to systematically reduce loop integrals with operator insertions 
with {\tt Reduze\;2}, resummed contributions are introduced which effectively lead to linear propagators.
Let us consider operator insertions involving fermion lines, where we suppress all factors independent of potential loop 
momenta. An insertion on a fermion line (FF) is mapped to scalar propagators introducing a new variable $x$ and 
summing over $N$  \cite{Ablinger:2012qm} according to\footnote{Here and in the following corresponding
shifts in $N$ have to be considered.}
\begin{equation}
\picbox{\begin{picture}(55,10)(-27,-5)
 \ArrowLine(-20,0)(-5,0)
 \ArrowLine(5,0)(20,0)
 \CrossCirc(0,0){5}
 \Text(-25,0)[]{$p$}
 \Text(25,0)[]{$p$}
\end{picture}}
\propto
(\Delta.p)^{N-1}~(N\ge 1)
\longrightarrow
\sum_{N=0}^{\infty} x^N (\Delta.p)^N = \frac{1}{1-x \Delta.p} =~
\picbox{\begin{picture}(55,10)(-27,-5)
 \Line(-20,0)(20,0)
 \BTriRot(-20,0)(20,0){5}
 \Text(-25,0)[]{$p$}
 \Text(25,0)[]{$p$}
\end{picture}}
\label{eq:PIC1}
\end{equation}
The last equality defines a graphical representation for linear propagators.
Similarly, we find for the second type of operator insertion (FFV) shown
in Figure~\ref{feynrules} that the corresponding terms can be resummed into
a generating function in the following way
\begin{equation}
\picbox{\begin{picture}(60,25)(-30,-20)
 \ArrowLine(-20,0)(-5,0)
 \ArrowLine(5,0)(20,0)
 \Gluon(0,-20)(0,0){2}{3}
 \CrossCirc(0,0){5}
 \Text(-25,0)[]{$p_1$}
 \Text(25,0)[]{$p_2$}
\end{picture}}
\longrightarrow
\sum_{N=0}^\infty x^N \sum_{k=0}^N (\Delta.p_1)^k (\Delta.p_2)^{N-k}
= \frac{1}{(1-x \Delta.p_1)(1-x \Delta.p_2)} =
\picbox{\begin{picture}(60,25)(-30,-20)
 \Line(-20,0)(0,0)
 \BTriRot(-20,0)(0,0){5}
 \Line(0,0)(20,0)
 \BTriRot(0,0)(20,0){5}
 \Line(0,-20)(0,0)
 \Vertex(0,0){1}
 \Text(-25,0)[]{$p_1$}
 \Text(25,0)[]{$p_2$}
\end{picture}}
\end{equation}
after an appropriate shift in $N$ and a partial fractioning.
Note that the relative orientation of $p_1$ and $p_2$ in the linear 
propagators is important.
We observe that it is possible to effectively decompose vertices with $n$ edges into a combination of $n-2$-vertices and $n-1$
auxiliary edges. For $n=4,5$ we have (FFVV):
\begin{equation}
\picbox{\begin{picture}(60,25)(-30,-20)
 \ArrowLine(-20,0)(-5,0)
 \ArrowLine(5,0)(20,0)
 \Gluon(-15,-20)(0,0){2}{3}
 \LongArrow(-22,-20)(-16,-13)
 \LongArrow(22,-20)(16,-13)
 \Gluon(15,-20)(0,0){2}{3}
 \CrossCirc(0,0){5}
 \Text(-25,0)[]{$p_1$}
 \Text(25,0)[]{$p_2$}
 \Text(-30,-20)[]{$p_3$}
 \Text(30,-20)[]{$p_4$}
\end{picture}}
\longrightarrow
\picbox{\begin{picture}(100,25)(-50,-20)
 \Line(-40,0)(-20,0)
 \BTriRot(-40,0)(-20,0){5}
 \Line(-20,0)(20,0)
 \BTriRot(-20,0)(20,0){5}
 \Line(20,0)(40,0)
 \BTriRot(20,0)(40,0){5}
 \Line(-35,-20)(-20,0)
 \Line(35,-20)(20,0)
 \Vertex(-20,0){1}
 \Vertex(20,0){1}
 \Text(-45,0)[]{$p_1$}
 \Text(45,0)[]{$p_2$}
 \Text(0,10)[]{$p_1+p_3$}
\end{picture}}
\quad
\picbox{\begin{picture}(100,25)(-50,-20)
 \Line(-40,0)(-20,0)
 \BTriRot(-40,0)(-20,0){5}
 \Line(-20,0)(20,0)
 \BTriRot(-20,0)(20,0){5}
 \Line(20,0)(40,0)
 \BTriRot(20,0)(40,0){5}
 \Line(-35,-20)(20,0)
 \Line(35,-20)(-20,0)
 \Vertex(-20,0){1}
 \Vertex(20,0){1}
 \Text(-45,0)[]{$p_1$}
 \Text(45,0)[]{$p_2$}
 \Text(0,10)[]{$p_1+p_4$}
\end{picture}}
\end{equation}
and (FFVVV):
\begin{align}
\label{eq:PIC2}
\picbox{\begin{picture}(60,25)(-30,-20)
 \ArrowLine(-20,0)(-5,0)
 \ArrowLine(5,0)(20,0)
 \Gluon(-20,-20)(0,0){2}{4}
 \Gluon(0,-20)(0,0){2}{3}
 \Gluon(20,-20)(0,0){2}{4}
 \CrossCirc(0,0){5}
\end{picture}}
&\longrightarrow
\picbox{\begin{picture}(64,25)(-32,-20)
 \Line(-32,0)(-16,0)
 \BTriRot(-32,0)(-16,0){5}
 \Line(-16,0)(0,0)
 \BTriRot(-16,0)(0,0){5}
 \Line(0,0)(16,0)
 \BTriRot(0,0)(16,0){5}
 \Line(16,0)(32,0)
 \BTriRot(16,0)(32,0){5}
 \Line(-32,-20)(-16,0)
 \Line(0,-20)(0,0)
 \Line(32,-20)(16,0)
 \Vertex(-16,0){1}
 \Vertex(0,0){1}
 \Vertex(16,0){1}
\end{picture}}
\quad
\picbox{\begin{picture}(64,25)(-32,-20)
 \Line(-32,0)(-16,0)
 \BTriRot(-32,0)(-16,0){5}
 \Line(-16,0)(0,0)
 \BTriRot(-16,0)(0,0){5}
 \Line(0,0)(16,0)
 \BTriRot(0,0)(16,0){5}
 \Line(16,0)(32,0)
 \BTriRot(16,0)(32,0){5}
 \Line(-32,-20)(-16,0)
 \Line(0,-20)(16,0)
 \Line(32,-20)(0,0)
 \Vertex(-16,0){1}
 \Vertex(0,0){1}
 \Vertex(16,0){1}
\end{picture}}
\quad
\picbox{\begin{picture}(64,25)(-32,-20)
 \Line(-32,0)(-16,0)
 \BTriRot(-32,0)(-16,0){5}
 \Line(-16,0)(0,0)
 \BTriRot(-16,0)(0,0){5}
 \Line(0,0)(16,0)
 \BTriRot(0,0)(16,0){5}
 \Line(16,0)(32,0)
 \BTriRot(16,0)(32,0){5}
 \Line(-32,-20)(0,0)
 \Line(0,-20)(-16,0)
 \Line(32,-20)(16,0)
 \Vertex(-16,0){1}
 \Vertex(0,0){1}
 \Vertex(16,0){1}
\end{picture}}
\notag\\
&\phantom{\longrightarrow}
\picbox{\begin{picture}(64,25)(-32,-20)
 \Line(-32,0)(-16,0)
 \BTriRot(-32,0)(-16,0){5}
 \Line(-16,0)(0,0)
 \BTriRot(-16,0)(0,0){5}
 \Line(0,0)(16,0)
 \BTriRot(0,0)(16,0){5}
 \Line(16,0)(32,0)
 \BTriRot(16,0)(32,0){5}
 \Line(-32,-20)(0,0)
 \Line(0,-20)(16,0)
 \Line(32,-20)(-16,0)
 \Vertex(-16,0){1}
 \Vertex(0,0){1}
 \Vertex(16,0){1}
\end{picture}}
\quad
\picbox{\begin{picture}(64,25)(-32,-20)
 \Line(-32,0)(-16,0)
 \BTriRot(-32,0)(-16,0){5}
 \Line(-16,0)(0,0)
 \BTriRot(-16,0)(0,0){5}
 \Line(0,0)(16,0)
 \BTriRot(0,0)(16,0){5}
 \Line(16,0)(32,0)
 \BTriRot(16,0)(32,0){5}
 \Line(-32,-20)(16,0)
 \Line(0,-20)(-16,0)
 \Line(32,-20)(0,0)
 \Vertex(-16,0){1}
 \Vertex(0,0){1}
 \Vertex(16,0){1}
\end{picture}}
\quad
\picbox{\begin{picture}(64,25)(-32,-20)
 \Line(-32,0)(-16,0)
 \BTriRot(-32,0)(-16,0){5}
 \Line(-16,0)(0,0)
 \BTriRot(-16,0)(0,0){5}
 \Line(0,0)(16,0)
 \BTriRot(0,0)(16,0){5}
 \Line(16,0)(32,0)
 \BTriRot(16,0)(32,0){5}
 \Line(-32,-20)(16,0)
 \Line(0,-20)(0,0)
 \Line(32,-20)(-16,0)
 \Vertex(-16,0){1}
 \Vertex(0,0){1}
 \Vertex(16,0){1}
\end{picture}}
\end{align}
Since we consider 3-loop diagrams, one of the four linear propagators
in each term on the r.h.s.\ of the last mapping is redundant.
In the corresponding diagrams, one of the legs will be external such
that a combination of two linear propagators attached to the external leg
can be split via partial fractioning.

For operator insertions at purely gluonic edges (VV) and vertices (VVV,VVVV, VVVVV)\footnote{These terms will be required 
in future calculations of gluonic operators, see also 
Ref.~\cite{Ablinger:2014uka}.} we obtain similar patterns, although there are more
combinatorial possibilities.
In all cases, we find the relevant combinations of linear propagators
represented by up to three connected edges in scalar diagrams with three-vertices
only.
The orientation of the linear momenta is such that they correspond
to the direction of a transversal of these edges.
If two consecutive edges differ only by an external momentum, one
of them is ignored in the mapping to linear propagators.
\subsubsection{Construction of integral families}
\label{sec:families}

\vspace{1mm}
\noindent
In order to systematically handle our loop integrals, we index them
using integral families.
An integral family defines a set of scalar propagators which is complete
and minimal in the sense that any scalar product of a loop momentum with
a loop or external momentum can be expressed uniquely as a linear
combination of inverse propagators, where the coefficients are kinematic
invariants.
As pointed out above already, this may require the introduction of auxiliary
propagators, which are not fixed by the individual diagrams.
In the following we describe how we construct a set of integral families
such that all terms emerging from the diagrams can be indexed by
powers of propagators from an integral family.
Different routings of massive lines and the various combinations for
the linear propagators require the introduction of many families.
Note that in general one diagram will lead to a sum of terms which need
to be matched to different integral families.

Our guiding principle is to cover all terms by a minimal number
of families, which preferably have a large number of permutation
symmetries.
We start by considering diagrams without operator insertions.
Possible 3-loop self-energy diagrams are described by Benz,
ladder or crossed topologies (8 propagators) and their sub-topologies.
They come in different variants, depending on the routing of the massive
lines.
Let us consider the following diagrams
\begin{eqnarray}
&&\picbox{\begin{picture}(120,80)(-60,-40)
\SetWidth{1.3}
\ArrowArcn(0,0)(40,135,45)
\ArrowLine(-28.284271247461900976,28.284271247461900976)(0,0)
\ArrowLine(0,0)(28.284271247461900976,28.284271247461900976)
\SetWidth{0.5}
\ArrowArcn(0,0)(40,180,135)
\ArrowArcn(0,0)(40,45,0)
\ArrowArcn(0,0)(40,0,270)
\ArrowArcn(0,0)(40,270,180)
\ArrowLine(-60,0)(-40,0)
\ArrowLine(40,0)(60,0)
\ArrowLine(0,0)(0,-40)
\Text(-52,-8)[]{$p$}
\Text(-30,10)[]{$k_1$}
\Text(0,30)[]{$k_3$}
\Text(32,10)[]{$k_2$}
\end{picture}}
~~
\picbox{\begin{picture}(148,72)(-74,-36) 
\SetWidth{1.3}
\ArrowLine(-18,36)(-18,-36)
\ArrowLine(18,-36)(18,36)
\ArrowLine(-18,36)(18,36)
\ArrowLine(18,-36)(-18,-36)
\SetWidth{0.5}
\ArrowArcn(-18,0)(36,180,90)
\ArrowArcn(-18,0)(36,270,180)
\ArrowArcn(18,0)(36,90,0)
\ArrowArcn(18,0)(36,0,270)
\ArrowLine(-74,0)(-54,0)
\ArrowLine(54,0)(74,0)
\Text(-66,-8)[]{$p$}
\Text(-38,20)[]{$k_1$}
\Text(0,27)[]{$k_3$}
\Text(38,20)[]{$k_2$}
\end{picture}}
~~
\picbox{\begin{picture}(120,80)(-60,-40)
\SetWidth{1.3}
\ArrowArcn(0,0)(40,135,45)
\ArrowLine(-28.284271247461900976,28.284271247461900976)(0,0)
\Line(0,0)(28.284271247461900976,-28.284271247461900976)
\ArrowLine(-28.284271247461900976,-28.284271247461900976)(-3,-3)
\Line(3,3)(28.284271247461900976,28.284271247461900976)
\ArrowArcn(0,0)(40,315,225)
\SetWidth{0.5}
\ArrowArcn(0,0)(40,180,135)
\ArrowArcn(0,0)(40,45,0)
\ArrowArcn(0,0)(40,0,315)
\ArrowArcn(0,0)(40,225,180)
\ArrowLine(-60,0)(-40,0)
\ArrowLine(40,0)(60,0)
\Text(-52,-8)[]{$p$}
\Text(-30,10)[]{$k_1$}
\Text(0,30)[]{$k_3$}
\Text(32,10)[]{$k_2$}
\end{picture}}
\nonumber\\
&&
\picbox{\begin{picture}(120,80)(-60,-40)
\SetWidth{1.3}
\ArrowArcn(0,0)(40,180,135)
\ArrowArcn(0,0)(40,270,180)
\ArrowLine(-28.284271247461900976,28.284271247461900976)(0,0)
\ArrowLine(0,0)(0,-40)
\SetWidth{0.5}
\ArrowArcn(0,0)(40,135,45)
\ArrowArcn(0,0)(40,45,0)
\ArrowArcn(0,0)(40,0,270)
\ArrowLine(0,0)(28.284271247461900976,28.284271247461900976)
\ArrowLine(-60,0)(-40,0)
\ArrowLine(40,0)(60,0)
\Text(-52,-8)[]{$p$}
\Text(-30,10)[]{$k_3$}
\Text(0,30)[]{$k_1$}
\Text(32,10)[]{$k_2$}
\end{picture}}
~~
\picbox{\begin{picture}(148,72)(-74,-36) 
\SetWidth{1.3}
\ArrowLine(-18,36)(-18,-36)
\ArrowArcn(-18,0)(36,180,90)
\ArrowArcn(-18,0)(36,270,180)
\SetWidth{0.5}
\ArrowArcn(18,0)(36,90,0)
\ArrowArcn(18,0)(36,0,270)
\ArrowLine(18,-36)(18,36)
\ArrowLine(-18,36)(18,36)
\ArrowLine(18,-36)(-18,-36)
\ArrowLine(-74,0)(-54,0)
\ArrowLine(54,0)(74,0)
\Text(-66,-8)[]{$p$}
\Text(-38,20)[]{$k_3$}
\Text(0,27)[]{$k_1$}
\Text(38,20)[]{$k_2$}
\end{picture}}
\end{eqnarray}
The kinematics requires 9 propagators per family, that
is, one auxiliary propagator.
Requiring a large number of permutation symmetries and a small number
of families, leads to the following choice of two families\footnote{The names given to the families ($B1$, $C1$, etc.)
are arbitrary. The rationale behind the names chosen will be explained elsewhere.}  -- one
which covers the depicted four planar topologies and one for the non-planar topology.
\begin{equation}
\begin{tabular}{c|l}
\multicolumn{2}{l}{\it planar family B1:}\\[2mm]
index & denominator\\
\hline
1 & $k_1^2$\\
2 & $(k_1-p)^2$\\
3 & $k_2^2$\\
4 & $(k_2-p)^2$\\
5 & $k_3^2$\\
6 & $(k_1-k_3)^2 - m^2$\\
7 & $(k_2-k_3)^2 - m^2$\\
8 & $(k_1-k_2)^2$\\
9 & $(k_3-p)^2 - m^2$
\end{tabular}
\qquad
\begin{tabular}{c|l}
\multicolumn{2}{l}{\it non--planar family C1:}\\[2mm]
index & denominator\\
\hline
1 & $k_1^2$\\
2 & $(k_1-p)^2$\\
3 & $k_2^2$\\
4 & $(k_2-p)^2$\\
5 & $k_3^2-m^2$\\
6 & $(k_1-k_3)^2 - m^2$\\
7 & $(k_2-k_3)^2 - m^2$\\
8 & $(k_1+k_2-k_3-p)^2 -m^2$\\
9 & $(k_3-p)^2 - m^2$
\end{tabular}
\nonumber
\end{equation}

We now consider diagrams with operator insertions.
From the kinematical point of view, we deal with a 4-point function
with degenerate kinematics:

\vspace*{1.5mm}
\begin{equation}
\label{degenkin}
\begin{picture}(1,1)
\LongArrow(-15,15)(-5,5)
\LongArrow(-15,-15)(-5,-5)
\LongArrow(4,4)(14,14)
\LongArrow(4,-4)(14,-14)
\CCirc(0,0){6.5}{Black}{Gray}
\Text(-20,12)[]{$\Delta$}
\Text(19,12)[]{$\Delta$}
\Text(-20,-12)[]{$p$}
\Text(19,-12)[]{$p$}
\end{picture}
\end{equation}
\vspace*{1.5mm}

\noindent
In addition to the 9 standard propagators we provide 3 additional
linear propagators within each family to be able to express
$\Delta.k_1$, $\Delta.k_2$, $\Delta.k_3$ via
inverse propagators, where the $k_i$ are the loop momenta.
Different operator insertions require different
forms and combinations of the linear propagators.
We construct the relevant combinations using graphs, where we
substitute the operator insertions according to the rules
explained in the previous section.
It is advantageous for the construction to consider all
diagrams which will eventually contribute, even if they are not entering
the quantities discussed in this work.
The decomposition of 4- and 5-vertices typically fix all three linear
propagators up to an overall sign in the resulting families.
Requiring for each linear propagator edge also the presence of a standard
propagator with that momentum, together with the conventions for our
operator--free families discussed above, fixes the integral families.
Consequently, we start with diagrams with the most involved vertices and
work our way towards diagrams with simpler vertices but more ambiguities
in the family construction.
An example for a constructed graph and the corresponding family is:
\begin{equation}
\picbox{\begin{picture}(120,80)(-60,-40)
\SetWidth{1.3}
\CArc(0,0)(40,45,135)
\Line(-28.284271247461900976,28.284271247461900976)(0,0)
\Line(0,0)(28.284271247461900976,28.284271247461900976)
\SetWidth{0.5}
\CArc(0,0)(40,135,45)
\Line(-60,0)(-40,0)
\Line(40,0)(60,0)
\Line(0,0)(0,-40)
\BTriRot(0,0)(-28.284271247461900976,28.284271247461900976){7}
\BTriRot(28.284271247461900976,28.284271247461900976)(0,0){7}
\BTriRot(-10,40)(10,40){7}
\end{picture}}
\quad
\begin{tabular}{c|l}
\multicolumn{2}{l}{\it family B1a:}\\[2mm]
index & denominator\\
\hline
1 & $k_1^2$\\
2 & $(k_1-p)^2$\\
3 & $k_2^2$\\
4 & $(k_2-p)^2$\\
5 & $k_3^2$\\
6 & $(k_1-k_3)^2 - m^2$\\
7 & $(k_2-k_3)^2 - m^2$\\
8 & $(k_1-k_2)^2$\\
9 & $(k_3-p)^2 - m^2$\\
10 & $ m^2(1 - x \Delta.(k_3-k_1))$\\
11 & $ m^2(1 - x \Delta.k_3)$\\
12 & $ m^2(1 - x \Delta.(k_3-k_2))$
\end{tabular}
\nonumber
\end{equation}

We construct families for diagrams with insertions of type
$FFVVV$, $FFVV$, $FFV$ and $FF$ for the massive fermion (in that order)
and then for insertions of type $VVVVV$, $VVVV$, $VVV$, $VV$, where the
latter scalar families also cover insertions involving massless fermions
and ghosts.

Finally, we arrange the integral families such that families based on planar
rather than non-planar diagrams, with fewer massive propagators, or
with higher permutation symmetries (in that order) are preferred.
In this way we construct a total number of 24 integral families for the whole present 3-loop project
(14 based on planar diagrams, 10 based on non-planar diagrams).
It is curious to note that the above construction is quite stringent
and leads to a relatively small set of integral families.
As we will see, in the case of the OMEs $A_{qq,Q}^{(3), \rm NS, TR}$ only three families contribute.
\subsubsection{Reduze\;2: application and new features}
\label{sec:reduze}

\vspace{1mm}
\noindent
We employ {\tt Reduze\;2}~\cite{vonManteuffel:2012np}\footnote{
For other public reduction programs see \cite{Anastasiou:2004vj, Smirnov:2013dia, Lee:2012cn}.
}
for the reduction of loop integrals with integration--by--part identities and
shift relations.
{\tt Reduze\;2} implements a distributed variant of Laporta's algorithm
\cite{Laporta:1996mq,Laporta:2001dd,Tkachov:1981wb,Chetyrkin:1981qh}.
For this work, various aspects of the program were improved and extended.
These improvements were implemented in a generic way
and will, along with further new features, become publicly available
with the upcoming {\tt Reduze} release.

For this work, we extended various features of the program to
support the linear propagators needed here.
The previous version, {\tt Reduze\;2.0}, contained rudimentary support
for bilinear propagators of the type $1/(q_1.q_2 -m^2)$
where $q_1$ and $q_2$ are linear combinations of loop and external
momenta.
The new version of {\tt Reduze} significantly extends support for this
propagator type.
Permutation symmetries are supported also if bilinear propagators are
involved and explicitely verified to avoid user errors.
In our example family {\it B1a} there is a symmetry under the following
permutation of the propagators
\begin{equation}
  1 \leftrightarrow 3, \quad 2 \leftrightarrow 4,\quad
  6 \leftrightarrow 7, \quad 10 \leftrightarrow 12,
\end{equation}
corresponding to the shift of loop momenta $k_1 \leftrightarrow k_2$.
Restricting to permutations of subsets of propagators one can consider
more general shifts.
We extended the combinatorial shift finder
(job \verb+setup_sector_mappings_alt+) to determine such shifts
of loop momenta also if bilinear denominators are permuted and/or
a general crossing of external legs is involved.\footnote{
Currently, we restrict the algorithms to permutations without additional
minus factors, which is sufficient for the type of propagators
considered in this work.
Bilinear propagators with $m^2=0$, as present in different effective
theories, can allow for mappings between propagators which involve
additional minus signs:
$P^{-1}_a = q_1.q_2 \to q'_1.q'_2 = -P^{-1}_b$.
}
These shifts are used to determine relations between sectors,
possibly also between different families, and relations between integrals
of specific sectors.
In the present work, the only non-trivial crossing we need to consider is
$p \rightarrow -p$, which translates to $x \Delta.p \rightarrow -x \Delta.p$
at the level of the invariants.
Furthermore, we improved the zero sector recognition for sectors with bilinear
propagators.
Applied to the families used in this work, these features reduce
drastically the number of sectors for which a reduction needs to be performed.

In the presence of operator insertions, we find it more convenient to
calculate integrals with higher denominator powers in comparison to integrals
with additional numerators.
We therefore choose an appropriate integral ordering to reflect this
preference in the choice of the unreduced integrals in the reduction identities,
that is, for the basis of master integrals.
We added a new run-time option to the program which allows the user to select
between various integral orderings.

The new {\tt Reduze} version provides a family finding algorithm, which
systematically {\it constructs} families with a maximal number of permutation
symmetries for a given set of propagators.
Despite its current restriction to standard massless propagators, we found
it useful for the construction of our families.

In the setup of the input files for {\tt Reduze}, we implement the degenerate kinematics
(\ref{degenkin}) via three incoming momenta $p$, $-p$, $\Delta$ and appropriate
rules for the scalar products.
Some algorithms of {\tt Reduze} assume non-degenerate kinematics.
Certain modifications were required to allow {\tt Reduze} to be used more flexibly for
degenerate kinematics or other special setups.
Most notably, this includes control over crossings of external legs to be used
by the program via an explicit list.

Finally, the performance for long job queues and different usage aspects
of the program were improved.
We find it convenient to determine the master integrals needed in the
calculation before the actual reduction is performed.
This can be done more easily now using a new option we implemented in {\tt Reduze}.

Let us consider the following example as an illustration for our IBP reduction
procedure.
In the case of the diagram (\ref{benzdiag}), we can rewrite the integrals given in Eq.~(\ref{scalint1}), which 
belong to family $B1c$, as
\begin{equation}
I_{\nu_1, \ldots , \nu_{10}}^{a,b,c} =
\int \frac{d^D k_1}{(2\pi)^D} \frac{d^D k_2}{(2\pi)^D} \frac{d^D k_3}{(2\pi)^D}
\frac{(\Delta.k_1)^a (\Delta.k_2)^b (\Delta.k_3)^c}{D_1^{\nu_1} D_2^{\nu_2} 
\cdots D_9^{\nu_9} D_{10}^{\nu_{10}}}\, .
\label{scalint2}
\end{equation}
The operator insertion is resummed as
\begin{equation}
D_{10} = 1 - x(\Delta.p - \Delta.k_1)\, .
\end{equation}
If we appropriately introduce two additional linearly independent artificial denominators of this type, for example, 
\begin{equation}
D_{11} = 1 - x (\Delta.p - \Delta.k_2) \quad {\rm and} \quad D_{12} = 1 - x (\Delta.k_3)\, ,
\end{equation}
then also the scalar products of $\Delta$ with internal momenta in the numerator of Eq.~(\ref{scalint2}) can be expressed as linear 
combinations
of $D_{10}$, $D_{11}$ and $D_{12}$. In this way, all scalar integrals can be written as
\begin{equation}
I_{\nu_1, \ldots , \nu_{12}} =
\int \frac{d^D k_1}{(2\pi)^D} \frac{d^D k_2}{(2\pi)^D} \frac{d^D k_3}{(2\pi)^D}
\frac{1}{D_1^{\nu_1} D_2^{\nu_2} \cdots D_{12}^{\nu_{12}}}\, .
\label{scalint3}
\end{equation}
A given scalar integral will be completely identified by the set of indices $\nu_1$ to $\nu_{12}$.
Of course, the specific form of the set of inverse propagators $D_1, \ldots, D_{12}$ will depend on the
diagram we are considering. As outlined before,
the auxiliary propagators in a given set should always be chosen so that the set is complete and minimal, which means 
that any scalar product of a loop momentum with $\Delta$, $p$ or loop momenta can be expressed uniquely as a linear combination
of the inverse propagators  $D_1, \ldots, D_{12}$ defining a given integral family.
We have found that all the diagrams needed for
$A_{qq,Q}^{(3), \rm NS}$ and $A_{qq,Q}^{(3), \rm NS, TR}$ can be described by just three integral families, which
are shown in Table~\ref{families}.

\begin{table}
\begin{center}
\begin{tabular}{|l|l|l|l|}
\hline
          &  Family $B1b$              &  Family $B1c$                                   &  Family $B5a$             \\
\hline
$D_1$     &  $k_1^2$                 &  $k_1^2$                                      &  $k_1^2-m^2$            \\
$D_2$     &  $(k_1-p)^2$             &  $(k_1-p)^2$                                  &  $(k_1-p)^2-m^2$        \\
$D_3$     &  $k_2^2$                 &  $k_2^2$                                      &  $k_2^2-m^2$            \\
$D_4$     &  $(k_2-p)^2$             &  $(k_2-p)^2$                                  &  $(k_2-p)^2-m^2$        \\
$D_5$     &  $k_3^2-m^2$             &  $k_3^2-m^2$                                  &  $k_3^2$                \\
$D_6$     &  $(k_3-k_1)^2-m^2$       &  $(k_3-k_1)^2-m^2$                            &  $(k_3-k_1)^2-m^2$      \\
$D_7$     &  $(k_3-k_2)^2-m^2$       &  $(k_3-k_2)^2-m^2$                            &  $(k_3-k_2)^2-m^2$      \\
$D_8$     &  $(k_1-k_2)^2$           &  $(k_1-k_2)^2$                                &  $(k_1-k_2)^2$          \\
$D_9$     &  $(k_3-p)^2-m^2$         &  $(k_3-p)^2-m^2$                              &  $(k_3-p)^2$            \\
$D_{10}$  &  $1-x \Delta.k_1$  &  $1-x (\Delta.p - \Delta.k_1)$    &  $1-x \Delta.k_1$ \\
$D_{11}$  &  $1-x \Delta.k_3$  &  $1-x (\Delta.k_2 - \Delta.k_1)$  &  $1-x \Delta.k_3$ \\
$D_{12}$  &  $1-x \Delta.k_2$  &  $1-x \Delta.k_3$                       &  $1-x \Delta.k_2$ \\
\hline
\end{tabular}
\end{center}
\caption{\sf \small Propagators of the integral families used for the calculation of $A_{qq,Q}^{(3), \rm NS}$ and $A_{qq,Q}^{(3), 
\rm NS, TR}$.}
\label{families}
\end{table}

In this representation, scalar integrals will be functions of $x$. A given integral $I(x)$ will be written after reductions as a linear
combination of master integrals $J_i(x)$
\begin{equation}
I(x) = \sum_i c_i(x) J_i(x) \, ,
\label{Ix}
\end{equation}
where the coefficients $c_i(x)$ are rational functions of $x \Delta.p$, the mass $m$ and the dimension $D$. Once an integral
is obtained as a function of $x$, we can obtain the original integral we wanted to compute as a function of $N$
by extracting the $N^{\rm{th}}$ coefficient of the corresponding Taylor expansion in $x$, and doing the corresponding shift in 
$N$ 
implied by Eqs.~(\ref{eq:PIC1})-(\ref{eq:PIC2}). It must be remarked that the
coefficients $c_i(x)$ may contain poles in $\varepsilon = D-4$, which will require the calculation of the corresponding 
master integrals beyond order $\varepsilon^0$.

Since each diagram is written as a linear combination of scalar integrals $I(x)$, the diagrams themselves have
the structure of Eq.~(\ref{Ix}) in terms of master integrals, i.e., they are linear combinations of master integrals with
rational coefficients in $x$. We can summarize the steps needed to calculate a diagram as a function of $N$ as follows

\begin{enumerate}
\item Calculate the master integrals as functions of $N$, i.e., find the expressions $\hat{J}_i(N)$ in $J_i(x) = 
\sum_N^{\infty}
x^N \hat{J}_i(N)$; if it is clear from the context, we will write $J_i(N)$ instead of $\hat{J}_i(N)$.
\item Simplify the $J_i(N)$ in terms of special functions.
\item Assemble the diagram as a linear combination of master integrals\footnote{In principle, one could calculate the master
integrals
directly as functions of $x$ and then simplify
them as functions of $x$. But so far we are not aware of algorithms that are general and efficient enough to carry out this 
job.}
$\sum_i c'_i(x) J_i(x)$.
\item Extract the $N^{\rm{th}}$ term of the Taylor expansion in $x$ to obtain the diagram as a function of $N$. Shift $N$ as
needed according to the corresponding operator insertion of the diagram.
\end{enumerate}

\noindent In the next Section, we will present some examples illustrating how the first step is performed. The remaining three
steps
are done by applying up-to-date computer algebra technologies. More precisely, after carrying out step 1, the master integrals
$I_i(N)$ are given in terms of definite multi-sums. Then in step 2 the obtained sums are simplified to expressions in terms of
indefinite nested sums and products using the {\tt Mathematica} package {\tt EvaluateMultiSums}~\cite{EMSSP}. The backbone of 
this
machinery relies on the package {\tt Sigma}~\cite{SIG1,SIG2} that encodes advanced symbolic summation algorithms in the setting
of difference fields\cite{Karr:81,Schneider:01,Schneider:05a,Schneider:07d,Schneider:08c,
Schneider:10a,Schneider:10b,Schneider:10c,Schneider:13b}. In order to treat infinite summations these algorithms are 
supplemented
by the package {\tt HarmonicSums} \cite{HARMONICSUMS,Ablinger:2010kw} that can deal, e.g., with asymptotic expansions of the
arising special functions.\\
After carrying out step 3, large expressions in terms of power series are given and a non-trivial task is to derive the
$N^{\rm{th}}$ coefficient in step 4. Here the package {\tt SumProduction} \cite{EMSSP} crunches the large amount of power 
series
to a manageable number of basis sums and \texttt{HarmonicSums} calculates the $N^{\rm{th}}$ coefficient of this compactified
expression.
The result is given in terms of definite sums which originate from the application of Cauchy-products. Therefore the packages
\texttt{Sigma} and \texttt{EvaluateMultiSums} together with \texttt{HarmonicSums} are applied once more. As already mentioned
earlier, we end up at expressions of the desired diagrams in terms of harmonic sums.

In Table~\ref{masterslist}, we show the list of all master integrals used for the reductions, where
we identify each integral by giving the corresponding values of the indices $\nu_1 , \ldots, \nu_{12}$, see Eq.~(\ref{scalint3}).
We also indicate in the list the corresponding integral family and the order in $\varepsilon$ to which these integrals 
need to be expanded. 
Of course, this list is not unique, and one is free to choose a different basis of master integrals. For convenience, we have chosen a basis 
of master integrals with no negative powers of inverse propagators.

\begin{table}
\begin{center}
\begin{tabular}{|c|c|cccccccccccc|c|}
\hline
Integral  &  Family              & $\nu_1$ & $\nu_2$ & $\nu_3$ & $\nu_4$ & $\nu_5$ & $\nu_6$ &  $\nu_7$ & $\nu_8$ & $\nu_9$ & $\nu_{10}$ & $\nu_{11}$ & $\nu_{12}$     & order in $\varepsilon$  \\
\hline
$J_1$      &    $B1b$                &    0    &    0    &    2    &    0    &    0    &    1    &     1    &    1    &    1    
&     
1      &     0      &     0          &         1            \\ 
$J_2$      &    $B1b$                &    0    &    0    &    1    &    0    &    0    &    1    &     1    &    1    &    1    
&     
1      &     0      &     0          &         1            \\ 
$J_3$      &    $B1b$                &    1    &    0    &    1    &    0    &    0    &    0    &     1    &    1    &    1    
&     
1      &     0      &     0          &         1            \\ 
$J_4$      &    $B1b$                &    0    &    2    &    1    &    0    &    0    &    0    &     1    &    1    &    1    
&     
1      &     0      &     0          &         2            \\ 
$J_5$      &    $B1b$                &    0    &    1    &    1    &    0    &    0    &    0    &     1    &    1    &    1    
&     
1      &     0      &     0          &         0            \\ 
$J_6$      &    $B1b$                &    0    &    0    &    1    &    1    &    0    &    1    &     0    &    1    &    1    
&     
1      &     0      &     0          &         0            \\ 
$J_7$      &    $B1b$                &    1    &    0    &    0    &    1    &    0    &    1    &     0    &    1    &    1    
&     
1      &     0      &     0          &         1            \\ 
$J_8$      &    $B1b$                &    2    &    0    &    1    &    0    &    0    &    1    &     1    &    0    &    1    
&     
1      &     0      &     0          &         2            \\ 
$J_9$      &    $B1b$                &    1    &    0    &    1    &    0    &    0    &    1    &     1    &    0    &    1    
&     
1      &     0      &     0          &         1            \\ 
$J_{10}$   &    $B1b$                &    0    &    1    &    1    &    0    &    0    &    1    &     1    &    0    &    1    
&     
1      &     0      &     0          &         1            \\ 
$J_{11}$   &    $B1b$                &    1    &    0    &    1    &    1    &    0    &    1    &     1    &    0    &    0    
&     
1      &     0      &     0          &         0            \\ 
$J_{12}$   &    $B1b$                &    0    &    1    &    1    &    1    &    0    &    1    &     1    &    0    &    0    
&     
1      &     0      &     0          &         0            \\ 
$J_{13}$   &    $B1b$                &    0    &    0    &    1    &    0    &    0    &    1    &     0    &    1    &    1    
&     
1      &     0      &     0          &         2            \\ 
$J_{14}$   &    $B1b$                &    0    &    0    &    2    &    0    &    0    &    1    &     1    &    0    &    1    
&     
1      &     0      &     0          &         2            \\ 
$J_{15}$   &    $B1b$                &    0    &    0    &    1    &    0    &    0    &    1    &     1    &    0    &    1    
&     
1      &     0      &     0          &         2            \\ 
$J_{16}$   &    $B1b$                &    1    &    0    &    0    &    0    &    0    &    1    &     1    &    0    &    1    
&     
1      &     0      &     0          &         2            \\ 
$J_{17}$   &    $B1b$                &    0    &    1    &    1    &    0    &    0    &    1    &     1    &    0    &    0    
&     
1      &     0      &     0          &         2            \\ 
$J_{18}$   &    $B1b$                &    1    &    0    &    0    &    1    &    0    &    1    &     1    &    0    &    0    
&     
1      &     0      &     0          &         1            \\ 
$J_{19}$   &    $B1b$                &    0    &    1    &    0    &    1    &    1    &    0    &     1    &    1    &    0    
&     
1      &     0      &     0          &         2            \\ 
$J^*_{20}$ &    $B1b$                &    2    &    0    &    0    &    1    &    1    &    0    &     1    &    1    &    0    
&     
1      &     0      &     0          &         3            \\ 
$J^*_{21}$ &    $B1b$                &    1    &    0    &    0    &    1    &    1    &    0    &     1    &    1    &    0    
&     
1      &     0      &     0          &         1            \\ 
$J_{22}$   &    $B1b$                &    0    &    1    &    0    &    0    &    1    &    1    &     1    &    0    &    0    
&     
1      &     0      &     0          &         0            \\ 
$J_{23}$   &    $B1b$                &    1    &    0    &    1    &    1    &    0    &    1    &     0    &    1    &    1    
&     
1      &     0      &     1          &         0            \\ 
$J_{24}$   &    $B1b$                &    1    &    1    &    1    &    0    &    0    &    1    &     1    &    0    &    1    
&     
1      &     0      &     1          &         0            \\ 
$J_{25}$   &    $B1b$                &    1    &    1    &    1    &    1    &    0    &    1    &     1    &    0    &    0    
&     
1      &     0      &     1          &         0            \\ 
$J_{26}$   &    $B1b$                &    0    &    1    &    1    &    0    &    0    &    1    &     1    &    0    &    1    
&     
1      &     0      &     1          &         0            \\ 
$J_{27}$   &    $B1b$                &    2    &    0    &    1    &    0    &    0    &    1    &     1    &    0    &    1    
&     
1      &     0      &     1          &         0            \\ 
$J_{28}$   &    $B1b$                &    1    &    0    &    1    &    0    &    0    &    1    &     1    &    0    &    1    
&     
1      &     0      &     1          &         0            \\ 
$J_{29}$   &    $B1c$                &    0    &    1    &    1    &    0    &    1    &    1    &     0    &    1    &    0    
&     
1      &     1      &     0          &         0            \\ 
$J_{30}$   &    $B1c$                &    1    &    1    &    1    &    0    &    0    &    0    &     1    &    1    &    1    
&     
1      &     1      &     0          &         0            \\ 
$J_{31}$   &    $B5a$                &    0    &    1    &    0    &    1    &    1    &    1    &     1    &    0    &    0    
&     
0      &     1      &     0          &         1            \\ 
$J_{32}$   &    $B1b$                &    1    &    0    &    1    &    0    &    0    &    1    &     1    &    0    &    0    
&     
0      &     0      &     0          &         4            \\ 
$J_{33}$   &    $B1b$                &    0    &    0    &    0    &    0    &    1    &    1    &     1    &    0    &    0    
&     
0      &     0      &     0          &         2            \\ 
$J_{34}$   &    $B5a$                &    1    &    0    &    1    &    0    &    0    &    1    &     1    &    0    &    0    
&     
0      &     0      &     0          &         2            \\ 
\hline
\end{tabular}
\end{center}
\caption{\sf \small List of master integrals identified by the indices $\nu_1$ to $\nu_{12}$ according to Eq.~(\ref{scalint3}), and 
by the
integral families defined in Table~\ref{families}. In the last column, we indicate the order in $\varepsilon$ to which each
integral needs to be expanded. An asterisk means that the integral is not only needed in the form presented in Eq.~(\ref{scalint3}), 
but
also in the form obtained by performing $p \rightarrow -p$ in this equation, which translates to $x \Delta.p \rightarrow -x 
\Delta.p$ at the level of the invariants.}
\label{masterslist}
\end{table}

Although the presence of the last three indices, $\nu_{10}$, $\nu_{11}$ and $\nu_{12}$ indicate in Table~\ref{masterslist} 
that the integrals are functions of $x$, one may as well interpret them
as functions of $N$ by undoing the replacements as made in Eq.~(\ref{eq:PIC1})--(\ref{eq:PIC2}).
In Appendix~\ref{sec:MIres}  we give the results for the master integrals as functions of $N$. 
\subsection{Calculation of the master integrals}

Compared with the master integrals required for the calculation of other operator matrix elements, the master integrals
for $A_{qq,Q}^{(3), \rm NS}$ and $A_{qq,Q}^{(3), \rm NS, TR}$ listed in Table~\ref{masterslist} are relatively simple. At most
six propagators appear in these integrals (not counting the artificial propagators arising from operator insertions), and 
of these, at most three are massive, with the exception of the two integrals in family $B5a$, 
where four massive propagators appear. 
This simplifies the calculation of these integrals considerably and, in fact,
many can be calculated solely in terms of Euler Beta-functions or a single sum of Beta-functions, which can then be performed
using the package {\tt Sigma}. Other more complicated cases required the solution of the master integrals in terms of
hypergeometric functions \cite{GHYP,Slater,Appell}, and the use of the corresponding series representations, in order to 
obtain a result in terms of a multiple sum to be solved by {\tt Sigma, EvaluateMultiSums} and {\tt HarmonicSums.}
In this Section we will show some examples describing the way these calculations are
performed.
To begin with, let us consider the integral $J_2$. After introducing Feynman parameters and performing the momentum integrals, 
we obtain the following expression
\begin{eqnarray}
J_2 &=& \int_0^1 dx \int_0^1 dy \int_0^1 dz \int_0^1 dw \,\,\, \Gamma \left(-1-\frac{3}{2}\varepsilon\right)
x^{-1-\varepsilon} (1-x)^{\varepsilon/2} y^{\varepsilon} z^{\varepsilon/2} (1-z)^{-1-\varepsilon/2} \,\,\,\,,
\nonumber \\
&& \phantom{\int_0^1 dx \int_0^1 dy} \,\,\,\,\,\,\,\,\,\,\,
 \times w^{\varepsilon/2} \left[1-(x+z-x z)(1-w) \right]^N 
\left(1+\frac{1-y z}{y z} x \right)^{1+\frac{3}{2} \varepsilon} \, .
\end{eqnarray}
For simplicity, we have set the mass $m$ and $\Delta.p$ to 1, and an overall factor of $i$ has been omitted.
After doing a binomial expansion twice, integrating in $w$ and in $x$ in terms of a hypergeometric function, and using the following analytic 
continuation,
\begin{equation}
{}_2F_1(\alpha, \beta; \gamma; z) = (1-z)^{-\alpha} {}_2F_1\left(\alpha, \gamma-\beta; \gamma; \frac{z}{z-1}\right)
\end{equation}
we find,
\begin{eqnarray}
J_2 &=& \int_0^1 dy \int_0^1 dz \,\,\, \Gamma \left(-1-\frac{3}{2}\varepsilon\right) 
\sum_{j=0}^N \sum_{k=0}^j (-1)^j \binom{N}{j} \binom{j}{k} \frac{\Gamma(1+\varepsilon/2) \Gamma(j+1)}{\Gamma(j+2+\varepsilon/2)}
\nonumber \\ 
&& \phantom{\int_0^1 dy \int_0^1 dz}
\times y^{-1-\varepsilon/2} z^{j-k-1-\varepsilon} (1-z)^{-1-\varepsilon/2}
\frac{\Gamma(k-\varepsilon) \Gamma(j-k+1+\varepsilon/2)}{\Gamma(j+1-\varepsilon/2)}
\nonumber \\ 
&& \phantom{\int_0^1 dy \int_0^1 dz}
\times {}_2F_1\left(-1-\frac{3}{2} \varepsilon, j-k+1+\varepsilon/2; j+1-\varepsilon/2; 1-y z \right) \, .
\end{eqnarray}
At this point, we can use the series representation of the hypergeometric function since the argument is bounded between 0 and 1, and
the series converges. We obtain
\begin{eqnarray}
J_2 &=& \int_0^1 dy \int_0^1 dz \,\,\, 
\sum_{j=0}^N \sum_{k=0}^j (-1)^j \binom{N}{j} \binom{j}{k} \frac{\Gamma(1+\varepsilon/2) \Gamma(j+1)}{\Gamma(j+2+\varepsilon/2)}
\Gamma(k-\varepsilon) 
\nonumber \\ 
&& \phantom{\int_0^1 dy \int_0^1 dz}
\sum_{m=0}^{\infty} \frac{\Gamma\left(m-1-\frac{3}{2} \varepsilon\right) \Gamma(m+j-k+1+\varepsilon/2)}{m! \Gamma(m+j+1-\varepsilon/2)} (1-y z)^m
\nonumber \\ 
&& \phantom{\int_0^1 dy \int_0^1 dz}
\times y^{-1-\varepsilon/2} z^{j-k-1-\varepsilon} (1-z)^{-1-\varepsilon/2},
\nonumber \\
&=&
\sum_{j=0}^N \sum_{k=0}^j \sum_{m=0}^{\infty} \sum_{i=0}^m (-1)^{i+j}  
\binom{N}{j} \binom{j}{k} \binom{m}{i} \frac{\Gamma(1+\varepsilon/2) \Gamma(j+1)}{\Gamma(j+2+\varepsilon/2)} 
\frac{\Gamma(k-\varepsilon)}{i-\varepsilon/2}
\nonumber \\ 
&& \phantom{ \sum_{j=0}^N \sum_{k=0}^j }
\times 
\frac{\Gamma\left(m-1-\frac{3}{2} \varepsilon\right) \Gamma(m+j-k+1+\varepsilon/2)}{m! \Gamma(m+j+1-\varepsilon/2)}
\frac{\Gamma(i+j-k-\varepsilon) \Gamma(-\varepsilon/2)}{\Gamma\left(i+j-k-\frac{3}{2} \varepsilon\right)} \, .
\end{eqnarray}
This multiple sum is convergent and can be solved by our summation packages mentioned above
in terms of harmonic sums. The result is given in Eq.~(\ref{J2N}).


Other integrals required the introduction of Appell hypergeometric functions. For example, after Feynman parameterization, 
the integral $J_{10}$ turned out to be given by
\begin{eqnarray}
J_{10} &=& \int_0^1 dx \cdots \int_0^1 dz_2 \sum_{i=0}^N (-1)^i \binom{N}{i} \Gamma\left(-1-\frac{3}{2} \varepsilon\right)
x^{i+\varepsilon/2} (1-x)^{\varepsilon/2} y^{\varepsilon/2} (1-y)^{\varepsilon/2} 
\nonumber \\ &&
\phantom{\int_0^1 dx \cdots \int_0^1 dz_2}
\times z_1^{-1-\varepsilon/2} z_2^{i-1-\varepsilon/2}
\theta(1-z_1-z_2) \left(1-z_1 \frac{x}{x-1}-z_2 \frac{y}{y-1}\right)^{1+\frac{3}{2} \varepsilon},
\nonumber \\ 
&=& \int_0^1 dx \int_0^1 dy \sum_{i=0}^N (-1)^i \binom{N}{i} \Gamma\left(-1-\frac{3}{2} \varepsilon\right)
x^{i+\varepsilon/2} (1-x)^{\varepsilon/2} y^{\varepsilon/2} (1-y)^{\varepsilon/2}
\nonumber \\ &&
\,\,\,\,\,
\times \frac{\Gamma(-\varepsilon/2) \Gamma(i-\varepsilon/2)}{\Gamma(i+1-\varepsilon)}
F_1\left(-1-\frac{3}{2} \varepsilon; -\frac{\varepsilon}{2}, i-\frac{\varepsilon}{2}; i+1-\varepsilon; \frac{x}{x-1},\frac{y}{y-1}\right) \, ,
\end{eqnarray}
where the ellipsis in the first line of the expression given above stand for the missing integrals in $y$ and $z_1$. 
Now we use the following analytic continuation,
\begin{equation}
F_1\left(a;b,b';c;\frac{x}{x-1},\frac{y}{y-1}\right) = (1-x)^b (1-y)^{b'} F_1(c-a;b,b';c;x,y)
\end{equation}
and obtain,
\begin{eqnarray}
J_{10} &=& \int_0^1 dx \int_0^1 dy \sum_{i=0}^N (-1)^i \binom{N}{i} \Gamma\left(-1-\frac{3}{2} \varepsilon\right)
\frac{\Gamma(-\varepsilon/2) \Gamma(i-\varepsilon/2)}{\Gamma(i+1-\varepsilon)}
x^{i+\varepsilon/2} y^{\varepsilon/2} (1-y)^i
\nonumber \\ &&
\phantom{\int_0^1 dx \cdots \int_0^1 dy} \times
F_1\left(i+2+\frac{\varepsilon}{2}; -\frac{\varepsilon}{2}, i-\frac{\varepsilon}{2}; i+1-\varepsilon; x,y\right) \, .
\label{J10appell2}
\end{eqnarray}
It would be nice if we could use at this point the series expansion representation of the Appell hypergeometric function.
However, if we do so here, we will get divergent sums, since the parameters of the $F_1$ function do not satisfy the conditions
for convergence. We therefore rewrite our result using
\begin{equation}
F_1(a;b,b';c;x,y) = \frac{\Gamma(c)}{\Gamma(a) \Gamma(c-a)} \int_0^1 dz \,\,\, z^{a-1} (1-z)^{c-a-1} (1-zx)^{-b} (1-zy)^{-b'} \, ,
\end{equation}
and  get
\begin{eqnarray}
J_{10} &=& \int_0^1 dx \int_0^1 dy \int_0^1 dz \sum_{i=0}^N (-1)^i \binom{N}{i} 
\frac{\Gamma(-\varepsilon/2) \Gamma(i-\varepsilon/2)}{\Gamma(i+2+\varepsilon/2)}
x^{i+\varepsilon/2} y^{\varepsilon/2} (1-y)^i
\nonumber \\ &&
\phantom{\int_0^1 dx \cdots \int_0^1 dy \int+0^1 dz} \times
z^{i+1+\varepsilon/2} (1-z)^{-2-\frac{3}{2} \varepsilon} (1-z x)^{\varepsilon/2} (1-z y)^{-i+\varepsilon/2} \, .
\end{eqnarray}
It is clear that the divergence of the $F_1$ function in Eq.~(\ref{J10appell2}) is partially due to the high negative power in
the term $(1-z)^{-2-\frac{3}{2} \varepsilon}$ appearing above. We can lower this power applying integration by parts in $z$, which leads to
\begin{equation}
J_{10} = \frac{1}{1+\frac{3}{2} \varepsilon} \left(-K_1+\frac{\varepsilon}{2} K_2-K_3\right) \, ,
\end{equation}
where
\begin{eqnarray}
K_1 &=& \int_0^1 dx \int_0^1 dy \int_0^1 dz \sum_{i=0}^N (-1)^i \binom{N}{i} 
\frac{\Gamma(-\varepsilon/2) \Gamma(i-\varepsilon/2)}{\Gamma(i+1+\varepsilon/2)}
x^{i+\varepsilon/2} y^{\varepsilon/2} (1-y)^i
\nonumber \\ &&
\phantom{\int_0^1 dx \cdots \int_0^1 dy \int_0^1} \,\,\, \times
z^{i+\varepsilon/2} (1-z)^{-1-\frac{3}{2} \varepsilon} (1-z x)^{\varepsilon/2} (1-z y)^{-i+\varepsilon/2} \, ,
\\
K_2 &=& \int_0^1 dx \int_0^1 dy \int_0^1 dz \sum_{i=0}^N (-1)^i \binom{N}{i} 
\frac{\Gamma(-\varepsilon/2) \Gamma(i-\varepsilon/2)}{\Gamma(i+2+\varepsilon/2)}
x^{i+1+\varepsilon/2} y^{\varepsilon/2} (1-y)^i
\nonumber \\ &&
\phantom{\int_0^1 dx \cdots \int_0^1 dy \int_0^1} \,\,\, \times
z^{i+1+\varepsilon/2} (1-z)^{-1-\frac{3}{2} \varepsilon} (1-z x)^{-1+\varepsilon/2} (1-z y)^{-i+\varepsilon/2} \, ,
\\
K_3 &=& \int_0^1 dx \int_0^1 dy \int_0^1 dz \sum_{i=0}^N (-1)^i \binom{N}{i} 
\frac{\Gamma(-\varepsilon/2) \Gamma(i+1-\varepsilon/2)}{\Gamma(i+2+\varepsilon/2)}
x^{i+\varepsilon/2} y^{1+\varepsilon/2} (1-y)^i
\nonumber \\ &&
\phantom{\int_0^1 dx \cdots \int_0^1 dy \int_0^1} \,\,\, \times
z^{i+1+\varepsilon/2} (1-z)^{-1-\frac{3}{2} \varepsilon} (1-z x)^{\varepsilon/2} (1-z y)^{-i-1+\varepsilon/2} \, .
\end{eqnarray}

These integrals will still produce divergent sums if they are re-expressed in terms of hypergeometric functions. 
In order to deal with $K_2$, we now apply integration by parts in $x$, leading to
\begin{equation}
K_2 = \frac{2}{\varepsilon}\left(K_1-K_{2a}\right) \, ,
\end{equation}
where
\begin{equation}
K_{2a} = \int_0^1 dy \int_0^1 dz \sum_{i=0}^N (-1)^i \binom{N}{i} 
\frac{\Gamma(-\varepsilon/2) \Gamma(i-\varepsilon/2)}{\Gamma(i+2+\varepsilon/2)}
y^{\varepsilon/2} (1-y)^i z^{i+\varepsilon/2} (1-z)^{-1-\varepsilon} (1-z y)^{-i+\varepsilon/2} \, .
\end{equation}
Now our original integral becomes,
\begin{equation}
J_{10} = -\frac{1}{1+\frac{3}{2} \varepsilon}\left(K_{2a}+K_3\right) \, .
\label{J10K2aK3}
\end{equation}

We see that integral $K_1$ does not need to be calculated, since it disappeared from the expression above. 
The integral $K_{2a}$ can be done in terms of a ${}_3F_2$ hypergeometric function,
\begin{eqnarray}
K_{2a} &=& \sum_{i=0}^N (-1)^i \binom{N}{i} 
\frac{\Gamma(-\varepsilon/2) \Gamma(1+\varepsilon/2) \Gamma(i+1) \Gamma(-\varepsilon)}{(i+1+\varepsilon/2) (i-\varepsilon/2) \Gamma(i+2+\varepsilon/2)}
\nonumber \\ && \phantom{\sum_{i=0}^N} \times
{}_3F_2\left(i+1+\varepsilon/2, i-\varepsilon/2, 1+\varepsilon/2; i+1-\varepsilon/2, i+2+\varepsilon/2; 1\right), 
\nonumber \\
&=& \sum_{i=0}^N (-1)^i \binom{N}{i} \frac{\Gamma(-\varepsilon/2) \Gamma(-\varepsilon) \Gamma(i+1)}{\Gamma(i+2+\varepsilon/2)}
\nonumber \\ && \phantom{\sum_{i=0}^N} \times
\sum_{k=0}^{\infty} \frac{\Gamma(k+i+1+\varepsilon/2) \Gamma(k+i-\varepsilon/2) \Gamma(k+1+\varepsilon/2)}{k! \Gamma(k+i+1-\varepsilon/2) \Gamma(k+i+2+\varepsilon/2)} \, .
\label{K2afinal}
\end{eqnarray}

Integral $K_3$ is slightly more difficult. In order to do it, we first perform the following change of variables,
\begin{equation}
z = 1-z' y' \, , \quad y = \frac{1-z'}{1-z' y'} \, ,
\end{equation}
which leaves the limits of integration from 0 to 1 unchanged, see also \cite{Hamberg}. This leads to
\begin{eqnarray}
K_3 &=& \int_0^1 dx \int_0^1 dy \int_0^1 dz \sum_{i=0}^N (-1)^i \binom{N}{i} 
\frac{\Gamma(-\varepsilon/2) \Gamma(i+1-\varepsilon/2)}{\Gamma(i+2+\varepsilon/2)}
x^{i+\varepsilon/2} y^{-1-\frac{3}{2} \varepsilon} (1-y)^i 
\nonumber \\ && \phantom{\int_0^1 dx \int_0^1 dy \int_0^1 dz \sum_{i=0}^N} \times
z^{-1-\varepsilon} (1-z)^{1+\varepsilon/2} \frac{[1-(1-z y) x]^{\varepsilon/2}}{1-z y} \, ,
\end{eqnarray}
where we have dropped the primes in the variables $y$ and $z$.
The integral in $x$ can be done in terms of a ${}_2F_1$ hypergeometric function, which can then be expanded in terms of its series representation.
Integrating then in the remaining variables, we obtain
\begin{eqnarray}
K_3 &=&  \sum_{i=0}^N (-1)^i \binom{N}{i} \frac{\Gamma(i+1-\varepsilon/2) \Gamma(i+1) \Gamma(2+\varepsilon/2)}{\Gamma(i+2+\varepsilon/2)}
\nonumber \\ && \,\,\,\,\,\, \times
\left[ 
\sum_{k=1}^{\infty} \sum_{j=0}^{k-1} (-1)^j \binom{k-1}{j} 
\frac{\Gamma(k-\varepsilon/2) \Gamma\left(j-\frac{3}{2} \varepsilon\right) \Gamma(j-\varepsilon)}{k! (k+i+1+\varepsilon/2) \Gamma\left(j+i+1-\frac{3}{2} \varepsilon\right) \Gamma(j+2-\varepsilon/2)}
\right.
\nonumber \\ && \left. \phantom{\times \left[ \right.} \,\,\,\,\,\,
+\frac{\Gamma(-\varepsilon/2)}{i+1+\varepsilon/2}
\sum_{k=0}^{\infty} \frac{\Gamma(k-\varepsilon) \Gamma\left(k-\frac{3}{2} \varepsilon\right)}{\Gamma(k+2-\varepsilon/2) \Gamma\left(k+i+1-\frac{3}{2} \varepsilon\right)} 
\right] \, .
\label{K3final}
\end{eqnarray}
The sums in Eqs.~(\ref{K2afinal}) and (\ref{K3final}) are convergent and can be performed using our summation
toolbox, and after combining them according to Eq.~(\ref{J10K2aK3}) we obtain the result in Eq.~(\ref{J10N}).

Except for integrals $J_{31}$ and $J_{34}$, all of the other integrals were calculated in a similar way. 
The integral $J_{34}$ is, in fact, just a constant (independent of $N$). 
In spite of that, it is still not easy to compute\footnote{Integrals with fixed values of $N$ or with no operator insertions 
can be obtained up to order $\varepsilon^0$ using the program {\tt MATAD} \cite{Steinhauser:2000ry}. Unfortunately, $J_{34}$ is 
needed up to order 
$\varepsilon^2$.}.
After Feynman parameterization we obtain,
\begin{eqnarray}
J_{34} &=& -\int_0^1 dx \int_0^1 dy \int_0^1 dz \,\,\,\Gamma\left(-2-\frac{3}{2} \varepsilon\right) \frac{(x (1-x) y (1-y))^{\varepsilon/2}}{(z (1-z))^{1+\varepsilon/2}}
\nonumber \\ && \phantom{-\int_0^1 dx \int_0^1 dy \int_0^1 dz} \times
\left(\frac{z}{x (1-x)}+\frac{1-z}{y (1-y)}\right)^{2+\frac{3}{2} \varepsilon} \, .
\label{J34a}
\end{eqnarray}
From here, we can obtain a Mellin-Barnes integral representation \cite{MELB}, using
\begin{equation}
\frac{1}{(A+B)^{\nu}} =  \frac{1}{2 \pi i} \int_{-i \infty}^{+i \infty} d\sigma \,\,\, \frac{\Gamma(-\sigma) \Gamma(\sigma+\nu)}{\Gamma(\nu)} \frac{A^{\sigma}}{B^{\sigma+\nu}} \, ,
\end{equation}
in order to split the last term in Eq.~(\ref{J34a}). This allows to compute the Feynman parameter integrals in terms of 
Beta-functions. We get
\begin{eqnarray}
J_{34} &=& -\frac{1}{2 \pi i} \int_{-i \infty}^{+i \infty} d\sigma \,\,\, \Gamma(-\sigma) \Gamma\left(\sigma-2-\frac{3}{2} \varepsilon\right)
\frac{\Gamma(-\sigma+1+\varepsilon/2)^2 \Gamma(\sigma-1-\varepsilon)^2}{\Gamma(-2 \sigma+2+\varepsilon) \Gamma(2 \sigma-2-2 \varepsilon)}
\nonumber \\ && \phantom{-\frac{1}{2 i \pi} \int_{-i \infty}^{+i \infty} d\sigma} \times
\frac{\Gamma(\sigma-\varepsilon/2) \Gamma(-\sigma+2+\varepsilon)}{\Gamma(2+\varepsilon/2)} \, .
\end{eqnarray}
This integral can now be calculated with the help of the {\tt Mathematica} package {\tt MB} \cite{Czakon:2005rk,Smirnov:2009up}, 
which allows to analytically continue the integral for
$\varepsilon \rightarrow 0$ and afterwards perform the $\varepsilon$ expansion. We obtain
\begin{eqnarray}
J_{34} &=& \frac{16}{\varepsilon^3}-\frac{92}{3 \varepsilon^2}+\frac{6 \zeta_2+35}{\varepsilon}-\frac{23 \zeta_2}{2}+2 \zeta_3-\frac{127}{6}
+\varepsilon \left(\frac{57 \zeta_2^2}{40}+\frac{105 \zeta_2}{8}-\frac{67 \zeta_3}{6}+\frac{351}{16}\right)
\nonumber \\ &&
+\varepsilon^2 \left(-\frac{1493 \zeta_2^2}{160}+\left(\frac{3 \zeta_3}{4}-\frac{127}{16}\right) \zeta_2+\frac{103 \zeta_3}{8}+\frac{3 \zeta_5}{10}+\frac{5633}{96}\right) 
+b_0+\varepsilon b_1+\varepsilon^2 b_2 \, , \,\,\,\,\,\, \phantom{.}
\end{eqnarray}
where $b_0$, $b_1$ and $b_2$ are contour integrals, 
\begin{equation}
b_0 = -\frac{1}{2 \pi i} \int_{-i \infty}^{+i \infty} d\sigma \,\,\, \Gamma(-\sigma) \Gamma(\sigma-2)
\Gamma(\sigma) \Gamma(-\sigma+2) 
\frac{\Gamma(-\sigma+1)^2 \Gamma(\sigma-1)^2}{\Gamma(-2 \sigma+2) \Gamma(2 \sigma-2)} \, ,
\end{equation}
and
\begin{eqnarray}
b_1 &=& \frac{1}{2 \pi i} \int_{-i \infty}^{+i \infty} d\sigma \,\,\, \Gamma(-\sigma) \Gamma(\sigma-2)
\Gamma(\sigma) \Gamma(-\sigma+2) 
\frac{\Gamma(-\sigma+1)^2 \Gamma(\sigma-1)^2}{\Gamma(-2 \sigma+2) \Gamma(2 \sigma-2)}
\nonumber \\ && \phantom{-\frac{1}{2 i \pi} \int_{-i \infty}^{+i \infty} d\sigma}  \times
\left( \frac{1}{2}+\gamma_E+\psi(2-2 \sigma)-\psi(1-\sigma)-\psi(2-\sigma)+\frac{3}{2} \psi(\sigma-2)
\right.
\nonumber \\ && \phantom{-\frac{1}{2 i \pi} \int_{-i \infty}^{+i \infty} d\sigma \times \left( \right.}
\left.
+2 \psi(\sigma-1)+\frac{1}{2} \psi(\sigma)-2 \psi(2 \sigma-2) \right) \, .
\end{eqnarray}
Here $\bar{N} = N \exp(\gamma_E)$ and $\gamma_E$ denotes the Euler-Mascheroni constant.
The expression for $b_2$ is somewhat large, so it will be omitted here. These integrals can be computed
by closing the contour to the left (or the right) and summing the residues.
For example, in the case of $b_0$, closing the contour to the left we get
\begin{equation}
b_0 = \sum_{k=1}^{\infty} \left(\frac{4}{(k+1)^3}-\frac{2}{(k+2)^3}-\frac{2}{k^3}\right) = -\frac{7}{4} \, .
\end{equation}
For more complicated expressions, the sum of residues can be performed using the package {\tt Sigma}.
The result for $J_{34}$ is given in Eq.~(\ref{J31N}).
In the case of integral $J_{31}$, we obtain
\begin{eqnarray}
J_{31} &=& \frac{1}{N+1+\varepsilon/2} \int_0^1 dx \int_0^1 dy \int_0^1 dz \,\,\, \Gamma\left(-1-\frac{3}{2} \varepsilon\right) 
\frac{(x (1-x) y (1-y))^{\varepsilon/2}}{(z (1-z))^{1+\varepsilon/2}}
\nonumber \\ && \phantom{\frac{1}{N+1+\varepsilon/2} \int_0^1 dx \int_0^1 dy \int_0^1 dz} \times
\left(\frac{z}{x (1-x)}+\frac{1-z}{y (1-y)}\right)^{1+\frac{3}{2} \varepsilon} \, .
\end{eqnarray}
We see that the $N$ dependence factorizes outside of the integral, which is then just a constant. Except for this $N$-dependent factor, 
this integral is very similar to $J_{34}$, and can be computed in an analogous way. The result is shown in Eq.~(\ref{J34N}).
\section{The Contributions to the 3-Loop Anomalous Dimensions}
\label{sec:4}

\vspace*{1mm}
\noindent
In the calculation of the massive OME  Eq.~(\ref{eq:Aqq3}) one obtains from the contribution $\propto \ln^2(m^2/\mu^2)$
the 2-loop non-singlet anomalous dimension and from the term $\propto \ln(m^2/\mu^2)$ the part of
the 3-loop anomalous dimension $\propto N_F$. 
The anomalous dimensions $\gamma^{\pm \rm NS}$ have the representation
\begin{eqnarray}
\gamma^{\pm}_{\rm NS}(N) &=& \sum_{k=1}^\infty a_s^k \left[\gamma^{(k-1)}_{qq,\rm NS}(N) 
+ (-1)^N \gamma^{(k-1)}_{q\bar{q},\rm NS}(N)\right]~,
\end{eqnarray}
where $\gamma_{\rm NS}^+(N)$ is defined for even values of $N$ and $\gamma_{\rm NS}^-(N)$ for odd values of $N$.
These are also the sets of values from which the analytic continuation to $N \in \mathbb{C}$ is performed.
The anomalous dimensions are expressed in terms 
of harmonic sums \cite{HSUM}
\begin{eqnarray}
S_{b,\vec{a}}(N) = \sum_{k=1}^N \frac{({\rm sign}(b))^k}{k^{|b|}} S_{\vec{a}}(k),~~~S_\emptyset = 1,~b, a_i \in
\mathbb{Z}, b, a_i \neq 0, N > 0, N \in \mathbb{N}.
\end{eqnarray}
In the following, we drop the argument $N$ of the harmonic sums and use the short-hand 
notation $S_{\vec{a}}(N) \equiv S_{\vec{a}}$.
\subsection{The Vector Case}

We obtain the complete non-singlet anomalous dimensions at 2-loop order
\begin{eqnarray}
\gamma^{(1)}_{qq,\rm NS}(N) &=& 
\textcolor{blue}{C_F C_A } \Biggl[-32 S_{-2,1}
+\frac{536}{9}S_1+2 \left(16 
S_1-\frac{8}{N (N+1)}\right) S_{-2}
\nonumber\\ &&
-\frac{P_1
}{9 N^3 (N+1)^3}
-\frac{88}{3} S_2 + 16[S_3 + 
S_{-3}]\Biggr]
\nonumber\\ &&
+4 \textcolor{blue}{C_F T_F N_F}  \Biggl[\frac{3 N^4+6 N^3+47 N^2+20 N-12}{9 N^2 (N+1)^2}
-\frac{40}{9} S_1+\frac{8}{3} S_2\Biggr]
\nonumber\\ &&
+ \textcolor{blue}{C_F^2} \Biggl[64 S_{-2,1}+2
\Biggl(\frac{8 (2 N+1)}{N^2 (N+1)^2}-16 S_2\Biggr) S_1
+\frac{8 
\big(3 N^2+3 N+2\big)}{N (N+1)} S_2 
\nonumber\\ &&
-\frac{P_2}{N^3 (N+1)^3}
+2 \left(
\frac{16}{N (N+1)}-32 S_1\right) S_{-2}
-32[ S_3 + S_{-3}]\Biggr],
\\
\gamma^{(1)}_{q\bar{q},\rm NS}(N) &=& 16 \textcolor{blue}{C_F \left(\frac{C_A}{2} - C_F\right)}
\frac{2 N^3+2 N^2+2 N+1} {N^3 (N+1)^3},
\end{eqnarray}
with the polynomials
\begin{eqnarray}
P_1 &=& 51 N^6+153 N^5+757 N^4+144 (-1)^N N^3+995 N^3+352 N^2+12 N+72~,
\\
P_2 &=& 3 N^6+9 N^5+9 N^4-32 (-1)^N N^3+27 N^3+8 N^2-8~.
\end{eqnarray}
They agree with previous results given in Refs.~\cite{ANDI2}.

For the contributions $\propto N_F$ to the 3-loop non-singlet anomalous dimension the present calculation yields 
\begin{eqnarray}
\gamma^{(2)}_{qq,\rm NS}(N) &=& 
\textcolor{blue}{C_F T_F N_F}
\Biggl\{
4 \textcolor{blue}{T_F N_F}
\Biggl[
\frac{P_4}{27 N^3 (N+1)^3}
-\frac{16}{27} S_1
-\frac{80}{27} S_2
+\frac{16}{9} S_3
\Biggr]
\nonumber\\ &&
+\textcolor{blue}{C_A}
\Biggl\{
\frac{1}{2} (-1)^N 
\Biggl[
 \frac{256 (4 N+1)}{9 (N+1)^4}
-\frac{128}{3 (N+1)^3} S_1
\Biggr]
+\frac{1}{2} 
\Biggl[
-\frac{8 P_8}{27 N^4 (N+1)^4}
+
\Biggl[
 \frac{128}{3} S_1
\nonumber\\ &&
-\frac{64 \big(10 N^2+10 N+3\big)}{9 N (N+1)}
\Biggr] S_{-3}
+\frac{5344}{27} S_2
+ 
\Biggl[
\frac{64 \big(16 N^2+10 N-3\big)}{9 N^2 (N+1)^2}
-\frac{1280}{9} S_1
\nonumber\\ && 
+\frac{256}{3} S_2
\Biggr] 
S_{-2}
-\frac{32 \big(14 N^2+14 N+3\big)}{3 N (N+1)} S_3
+\frac{320}{3} S_4
+\frac{256}{3} S_{-4}
-\frac{128}{3} S_{2,-2}
\nonumber\\ && 
-\frac{256}{3} S_{3,1}
+\frac{128 \big(10 N^2+10 N-3\big)}{9 N (N+1)} S_{-2,1}
+
\Biggl[
-\frac{16 P_6}{27 N^3 (N+1)^3}
+64 S_3
\nonumber\\ && 
+\frac{256}{3} S_{-2,1}
\Biggr] S_1
-\frac{512}{3} S_{-2,1,1}
\Biggr]
+\frac{1}{2} 
\Biggl[
\frac{32 \big(3 N^2+3 N+2\big)}{N (N+1)}
-128 S_1
\Biggr]
\zeta_3
\Biggr\}
\nonumber\\ &&
+\textcolor{blue}{C_F}
\Biggl\{
\frac{1}{2} (-1)^N 
\Biggl[
 \frac{256}{3 (N+1)^3} S_1
-\frac{512 (4 N+1)}{9 (N+1)^4}
\Biggr]
+\frac{1}{2} 
\Biggl[
-\frac{128}{3} S_2^2
-\frac{16 P_3}{9 N^2 (N+1)^2} S_2
\nonumber\\ && 
+\frac{4 P_7}{9 N^4 (N+1)^4}
+
\Biggl[
\frac{128 \big(10 N^2+10 N+3\big)}{9 N (N+1)}
-\frac{256}{3} S_1
\Biggr] S_{-3}
\nonumber\\ && 
+
\Biggl[
-\frac{128 \big(16 N^2+10 N-3\big)}{9 N^2 (N+1)^2}
+\frac{2560}{9} S_1
-\frac{512}{3} S_2
\Biggr] 
S_{-2}
-\frac{512}{3} [S_4 + S_{-4}] 
\nonumber\\ && 
+\frac{64 \big(29 N^2+29 N+12\big)}{9 N (N+1)} S_3
+\frac{256}{3} [S_{2,-2} + S_{3,1}]
+
\Biggl[
-\frac{8 P_5}{9 N^3 (N+1)^3}
\nonumber\\ &&
+\frac{1280}{9} S_2
-\frac{512}{3} [S_3 + S_{-2,1}]
\Biggr] 
S_1
-\frac{256 \big(10 N^2+10 N-3\big)}{9 N (N+1)} S_{-2,1}
+\frac{1024}{3} S_{-2,1,1}
\Biggr]
\nonumber\\ &&
+\frac{1}{2} 
\Biggl[
128 S_1
-\frac{32 \big(3 N^2+3 N+2\big)}{N (N+1)}
\Biggr]
\zeta_3 
\Biggr\}
\Biggr\},
\\
\gamma^{(2)}_{q\bar{q},\rm NS}(N) &=& 
\textcolor{blue}{C_F T_F N_F \left(\frac{C_A}{2}-C_F\right)} 
\Biggl[
\frac{64 \big(2 N^3+2 N^2+2 N+1\big)}{3 N^3 (N+1)^3} S_1
-\frac{16 P_9}{9 N^4 (N+1)^4}\Biggr],
\end{eqnarray}
with the polynomials
\begin{eqnarray}
P_3 &=& 15 N^4+30 N^3+79 N^2+16 N-24~,
\\
P_4 &=& 51 N^6+153 N^5+57 N^4+35 N^3+96 N^2+16 N-24~,
\\
P_5 &=& 165 N^6+495 N^5+495 N^4+421 N^3+240 N^2-16 N-48~,
\\
P_6 &=& 209 N^6+627 N^5+627 N^4+209 N^3-36 N^2-36 N-18~,
\\
P_7 &=& 207 N^8+828 N^7+1467 N^6+1707 N^5+650 N^4-163 N^3-320 N^2-80 N+24~,
\nonumber\\
\\
P_8 &=& 270 N^8+1080 N^7+365 N^6-1417 N^5-1087 N^4+45 N^3-128 N^2-72 N+72~,
\nonumber\\
\\
P_9 &=& 3 N^6+73 N^5+86 N^4+77 N^3+39 N^2-10 N-12~.
\end{eqnarray}
The result agrees with the moments given in \cite{ANDI3} and the general $N$ result in \cite{Moch:2004pa}.

\subsection{Transversity}
The complete 2-loop anomalous dimensions are obtained by
\begin{eqnarray}
\gamma_{qq,\rm NS, TR}^{(1)} &=& -\textcolor{blue}{C_F \left(\frac{C_A}{2}-C_F\right)} 
\Biggl[64 S_{-2,1}
+\frac{2 \big(17 N^2+17 N-12\big)}{3 N (N+1)}
-64 S_{-2} S_1
-\frac{1072}{9} S_1
\nonumber\\ &&
+\frac{176}{3} S_2
-32 S_3
-32 S_{-3}
\Biggr]
+ \textcolor{blue}{C_F^2} 
\Biggl[S_1 
\Biggl[\frac{1072}{9}
-32 S_2
\Biggr]
-\frac{104}{3} S_2
-\frac{43}{3}
\Biggr]
\nonumber\\ &+&
\textcolor{blue}{C_F T_F N_F}
\Biggl[
-\frac{160}{9} S_1+\frac{32}{3} S_2+\frac{4}{3}
\Biggr],
\\
\gamma_{q\bar{q},\rm NS, TR}^{(1)} &=& -\textcolor{blue}{C_F \left(\frac{C_A}{2}-C_F\right)} \frac{8}{N (N+1)}~. 
\end{eqnarray}
They agree with the results obtained in Refs.~\cite{TR2LOOP}. For the contribution $\propto N_F$ of the 3-loop anomalous 
dimensions we obtain
\begin{eqnarray}
\gamma_{qq, \rm NS, TR}^{(2)} 
&=& 
\textcolor{blue}{C_F^2 T_F N_F}
\Biggl\{
\frac{256}{3} S_{3,1}
+ 
\Biggl[
-\frac{512}{3} S_{-2,1}
+\frac{1280}{9} S_2
-\frac{512}{3} S_3
-\frac{440}{3}
\Biggr] S_1
-\frac{2560}{9} S_{-2,1}
\nonumber\\ && 
-\frac{256}{3} S_{-2,2}
+\frac{1024}{3} S_{-2,1,1}
+\frac{4 \big(207 N^3+414 N^2+311 N+56\big)}{9 N (N+1)^2}
-\frac{128}{3} S_2^2
-\frac{80}{3} S_2
\nonumber\\ && 
+
\Biggl[\frac{1280}{9}
-\frac{256}{3} S_1
\Biggr] S_{-3}
+
\Biggl[
\frac{2560}{9} S_1
-\frac{256}{3} S_2
\Biggr] S_{-2}
+\frac{1856}{9} S_3
-\frac{512}{3} S_4
\nonumber\\ && 
-\frac{256}{3} S_{-4}
+ \Biggl[128 S_1 - 96\Biggr] \zeta_3
\Biggr\}
\nonumber\\ &&
+\textcolor{blue}{C_F C_A T_F N_F}
\Biggl\{
\Biggl[
\frac{256}{3} S_{-2,1}
-\frac{16 \big(209 N^2+209 N-9\big)}{27 N (N+1)}
+64 S_3
\Biggr] S_1
-\frac{256}{3} S_{3,1}
\nonumber\\ && 
+\frac{1280}{9} S_{-2,1}
+\frac{128}{3} S_{-2,2}
-\frac{512}{3} S_{-2,1,1}
-\frac{16 \big(15 N^3+30 N^2+12 N-5\big)}{3 N (N+1)^2}
+ 
\Biggl[
\frac{128 }{3} S_1
\nonumber\\ && 
-\frac{640}{9}
\Biggr] S_{-3}
+\frac{5344}{27} S_2
+
\Biggl[
\frac{128}{3} S_2-\frac{1280}{9} S_1
\Biggr] S_{-2}
-\frac{448}{3} S_3
+\frac{320}{3} S_4
\nonumber\\ && 
+\frac{128}{3} S_{-4}
+
\Biggl[96-128 S_1\Biggr] \zeta_3
\Biggr\}
\nonumber\\ &&
+\textcolor{blue}{C_F T_F^2 N_F^2}
\Biggl\{
\frac{8 \big(17 N^2+17 N-8\big)}{9 N (N+1)}
-\frac{128}{27} S_1
-\frac{640}{27} S_2
+\frac{128}{9} S_3
\Biggr\},
\\
\gamma_{q\bar{q}, \rm NS, TR}^{(2)} &=& 
\frac{32}{3}\textcolor{blue}{C_F T_F N_F \left(\frac{C_A}{2}-C_F\right)} \Biggl[\frac{13 N+7}{3 N (N+1)^2}-\frac{2}{N (N+1)} 
S_1 
\Biggr].
\end{eqnarray}
We agree with the moments given in \cite{GRAC,Bagaev:2012bw} and note a typo in the 15$^{\rm th}$ moment of 
\cite{Velizhanin:2012nm}.
In Ref.~\cite{Velizhanin:2012nm} the complete result for the 3-loop anomalous dimension was guessed studying the
difference to the vector case, while the above result has been obtained in a direct calculation.
\section{The Flavor Non-Singlet  Massive Operator Matrix Element: Vector Case}
\label{sec:5}

\vspace*{1mm}
\noindent
The new contribution beyond the terms determined by renormalization and factorization at 3-loop order 
is $a_{qq,Q}^{(3), \rm NS}$, in Eq.~(\ref{eq:Aqq3}). It is obtained as the constant part in the dimensional parameter
$\ep$ of the unrenormalized 3-loop OME and given by
\begin{eqnarray}
\label{eq:aqq3v}
\lefteqn{a_{qq,Q}^{(3), \rm NS} =} \nonumber\\ && \textcolor{blue}{C_F^2 T_F} 
\Biggl\{
\Biggl[
 \frac{128}{27} S_2
-\frac{16 (2 N+1)}{27 N^2 (N+1)^2}\Biggr]
S_1^3
+\Biggl[
-\frac{8 P_{18}}{9 N^3 (N+1)^3}
-\frac{64}{9 N (N+1)} S_2
+\frac{64}{3} S_3
\nonumber\\ &&
-\frac{128}{9} S_{2,1}
-\frac{256}{9} S_{-2,1}
\Biggr] S_1^2
+\Biggl[
-\frac{64}{9} S_2^2
+\frac{16 P_{15}}{81 N^2 (N+1)^2} S_2
+\frac{8 P_{30}}{81 N^4 (N+1)^4}
+\frac{704}{9} S_4
\nonumber\\ &&
-\frac{64 \big(40 N^2+40 N+9\big)}{27 N (N+1)} S_3
+\frac{128}{9 N (N+1)} S_{2,1}
-\frac{320}{9} S_{3,1}
-\frac{256 \big(10 N^2+10 N-3\big)}{27 N (N+1)} S_{-2,1}
\nonumber\\ &&
-\frac{256}{9} S_{-2,2}
+\frac{1024}{9} S_{-2,1,1}
\Biggr]
S_1
-\frac{16 \big(31 N^2+31 N-6\big)}{27 N (N+1)} S_2^2
-\frac{8}{3} B_4 \gamma_{qq}^{0}
+\frac{P_{34}}{162 N^5 (N+1)^5}
\nonumber\\ &&
+
\Biggl[
\frac{256}{9} S_1
-\frac{128 \big(10 N^2+10 N+3\big)}{27 N (N+1)}
\Biggr] 
S_{-4}
+
\Biggl[
\frac{128}{9} S_1^2
-\frac{128 \big(10 N^2+10 N+3\big)}{27 N (N+1)} S_1
\nonumber\\ &&
+\frac{64 \big(112 N^3+224 N^2+169 N+39\big)}{81 N (N+1)^2}
+\frac{128}{9} S_2
\Biggr]
S_{-3}
+\frac{8 P_{17}}{81 N^2 (N+1)^2} S_3
\nonumber\\ &&
-\frac{176 \big(17 N^2+17 N+6\big)}{27 N (N+1)} S_4
+\frac{512}{9} S_5
+\frac{256}{9} S_{-5}
+
\Biggl[
\frac{256}{27} S_1^3
-\frac{128}{9 N (N+1)} S_1^2
\nonumber\\ &&
+\frac{128 P_{13}}{81 N^2 (N+1)^2} S_1
-\frac{64 P_{14}}{81 N^3 (N+1)^3}
-\frac{1280}{27} S_2
+\frac{512}{27} S_3
-\frac{512}{9} S_{2,1}
\Biggr]
S_{-2}
\nonumber\\ &&
+\frac{16 P_{11}}{9 N^2 (N+1)^2} S_{2,1}
+\frac{256}{9} S_{2,3}
-\frac{512}{9} S_{2,-3}
+\frac{16 \big(89 N^2+89 N+30\big)}{27 N (N+1)} S_{3,1}
-\frac{512}{9} S_{4,1}
\nonumber\\ &&
-\frac{128 \big(112 N^3+112 N^2-39 N+18\big)}{81 N^2 (N+1)} S_{-2,1}
+
\Biggl[
-\frac{8 P_{27}}{81 N^3 (N+1)^3}
+\frac{256}{27} S_3
+\frac{256}{3} S_{-2,1}
\Biggr] S_2
\nonumber\\ &&
-\frac{128 \big(10 N^2+10 N-3\big)}{27 N (N+1)} S_{-2,2}
+\frac{512}{9} S_{-2,3}
+\frac{512}{9} S_{2,1,-2}
+\frac{256}{9} S_{3,1,1}
\nonumber\\ &&
+\frac{512 \big(10 N^2+10 N-3\big)}{27 N (N+1)} S_{-2,1,1}
+\frac{512}{9} S_{-2,2,1}
-\frac{2048}{9} S_{-2,1,1,1}
+\Biggl[
\frac{P_{23}}{3 N^3 (N+1)^3}
\nonumber\\ &&
+\frac{8 P_{12}}{3 N^2 (N+1)^2} S_1
+
\Biggl[
\frac{64}{3} S_1
-\frac{32}{3 N (N+1)}
\Biggr] S_{-2}
+\frac{32}{3} S_3
+\frac{32}{3} S_{-3}
-\frac{64}{3} S_{-2,1}
\Biggr]
\zeta_2
\nonumber\\ &&
+(-1)^N 
\Biggl[
 \frac{64 \big(2 N^2+2 N+1\big)}{9 N^3 (N+1)^3} S_1^2
-\frac{32 P_{24}}{27 N^4 (N+1)^4} S_1
+\frac{16 P_{28}}{81 N^5 (N+1)^5}
\nonumber\\ &&
+\frac{64 \big(2 N^2+2 N+1\big)}{9 N^3 (N+1)^3} S_2
+\frac{16 \big(2 N^2+2 N+1\big)}{3 N^3 (N+1)^3} \zeta_2
\Biggr]
\nonumber\\ &&
+\gamma_{qq}^{0} 
\Biggl[
12 \zeta_4
+\frac{8}{3} S_{2,1,1}
+\frac{4}{3} S_2 \zeta_2
\Biggr]
+
\Biggl[
\frac{2 P_{16}}{9 N^2 (N+1)^2}
-\frac{1208}{9} S_1
-\frac{64}{3} S_2
\Biggr]
\zeta_3
\Biggr\}
\nonumber\\ && 
+ \textcolor{blue}{C_F T_F^2}
\Biggl\{
-\frac{4 P_{29}}{729 N^4 (N+1)^4}
-\frac{19424}{729} S_1
+\frac{1856}{81} S_2
-\frac{640}{81} S_3
+\frac{128}{27} S_4
\nonumber\\ &&
+\Biggl[
\frac{8 P_{10}}{27 N^2 (N+1)^2}
-\frac{320}{27} S_1
+\frac{64}{9} S_2
\Biggr]
\zeta_2
-\frac{128}{27} \gamma_{qq}^{0} \zeta_3
\Biggr\}
\nonumber\\ &&
+\textcolor{blue}{C_F T_F^2 N_F}
\Biggl\{
\frac{2 P_{32}}{729 N^4 (N+1)^4}
-\frac{55552}{729} S_1
+\frac{640}{27} S_2
-\frac{320}{81} S_3
+\frac{64}{27} S_4
\nonumber\\ &&
+\Biggl[
\frac{4 P_{10}}{27 N^2 (N+1)^2}
-\frac{160}{27} S_1
+\frac{32}{9} S_2
\Biggr]
\zeta_2
+\frac{56}{27} \gamma_{qq}^{0} \zeta_3
\Biggr\}
\nonumber\\ &&
+\textcolor{blue}{C_F C_A T_F} 
\Biggl\{
-\frac{64}{27} S_2 S_1^3
+\Biggl[
 \frac{4 P_{19}}{9 N^3 (N+1)^3}
+\frac{32}{9 N (N+1)} S_2
-\frac{80}{9} S_3
+\frac{128}{9} [S_{2,1} + S_{-2,1}]
\Biggr]
S_1^2
\nonumber\\ &&
+\Biggl[
\frac{80 (2 N+1)^2}{9 N (N+1)} S_3
+\frac{112}{9} S_2^2
+\frac{4 P_{31}}{729 N^4 (N+1)^4}
-\frac{16 (N-1) \big(2 N^3-N^2-N-2\big)}{9 N^2 (N+1)^2} S_2
\nonumber\\ &&
-\frac{208}{9} S_4
-\frac{8 \big(9 N^2+9 N+16\big)}{9 N (N+1)} S_{2,1}
+\frac{64}{3} S_{3,1}
+\frac{128 \big(10 N^2+10 N-3\big)}{27 N (N+1)} S_{-2,1}
+\frac{128}{9} S_{-2,2}
\nonumber\\ &&
-\frac{512}{9} S_{-2,1,1}
\Biggr]
S_1
-\frac{4 \big(15 N^2+15 N+14\big)}{9 N (N+1)} S_2^2
+\frac{4}{3} B_4 \gamma_{qq}^{0}
+\frac{P_{33}}{1458 N^5 (N+1)^5}
\nonumber\\ &&
+ 
\Biggl[
\frac{64 \big(10 N^2+10 N+3\big)}{27 N (N+1)}
-\frac{128}{9} S_1
\Biggr]
S_{-4}
+
\Biggl[
-\frac{64}{9} S_1^2
+\frac{64 \big(10 N^2+10 N+3\big)}{27 N (N+1)} S_1
\nonumber\\ &&
-\frac{32 \big(112 N^3+224 N^2+169 N+39\big)}{81 N (N+1)^2}
-\frac{64}{9} S_2
\Biggr] S_{-3}
-\frac{8 P_{21}}{81 N^2 (N+1)^2} S_3
\nonumber\\ &&
+\frac{4 \big(311 N^2+311 N+78\big)}{27 N (N+1)} S_4
-\frac{224}{9} S_5
-\frac{128}{9} S_{-5}
-\frac{4 \big(2 N^3-35 N^2-37 N-24\big)}{9 N^3 (N+1)^2} [S_1^2 +S_2]
\nonumber\\ &&
-\frac{8 P_{20}}{9 N^2 (N+1)^2} S_{2,1}
+
\Biggl[
-\frac{128}{27} S_1^3
+\frac{64}{9 N (N+1)} S_1^2
-\frac{64 P_{13}}{81 N^2 (N+1)^2} S_1
+\frac{32 P_{14}}{81 N^3 (N+1)^3}
\nonumber\\ &&
+\frac{640}{27} S_2
-\frac{256}{27} S_3
+\frac{256}{9} S_{2,1}
\Biggr] S_{-2}
-\frac{128}{3} S_{2,3}
+\frac{256}{9} S_{2,-3}
-\frac{8 (13 N+4) (13 N+9)}{27 N (N+1)} S_{3,1}
\nonumber\\ &&
+\frac{256}{9} S_{4,1}
+
\Biggl[
-\frac{4 P_{26}}{81 N^3 (N+1)^3}
+\frac{496}{27} S_3
-\frac{64}{3} S_{2,1}
-\frac{128}{3} S_{-2,1}
\Biggr]
S_2
\nonumber\\ &&
+\frac{64 \big(112 N^3+112 N^2-39 N+18\big)}{81 N^2 (N+1)} S_{-2,1}
+\frac{64 \big(10 N^2+10 N-3\big)}{27 N (N+1)} S_{-2,2}
-\frac{256}{9} S_{-2,3}
\nonumber\\ &&
+\gamma_{qq}^{0}
\Biggl[
-12 \zeta_4
-4 S_{2,1,1}
\Biggr]
-\frac{256}{9} S_{2,1,-2}
+\frac{64}{3} S_{2,2,1}
-\frac{256}{9} S_{3,1,1}
-\frac{256 \big(10 N^2+10 N-3\big)}{27 N (N+1)} S_{-2,1,1}
\nonumber\\ &&
-\frac{256}{9} S_{-2,2,1}
+\frac{224}{9} S_{2,1,1,1}
+\frac{1024}{9} S_{-2,1,1,1}
+\Biggl[
\frac{P_{25}}{27 N^3 (N+1)^3}
+
\Biggl[
\frac{16}{3 N (N+1)}
-\frac{32}{3} S_1
\Biggr]
S_{-2}
\nonumber\\ &&
-\frac{16}{27} S_1
-\frac{88}{9} S_2
-\frac{16}{3} S_3
-\frac{16}{3} S_{-3}
+\frac{32}{3} S_{-2,1}
\Biggr] 
\zeta_2
+(-1)^N 
\Biggl[
-\frac{32 \big(2 N^2+2 N+1\big)}{9 N^3 (N+1)^3} S_1^2
\nonumber\\ &&
+\frac{16 P_{24}}{27 N^4 (N+1)^4} S_1
-\frac{8 P_{28}}{81 N^5 (N+1)^5}
-\frac{32 \big(2 N^2+2 N+1\big)}{9 N^3 (N+1)^3} S_2
-\frac{8 \big(2 N^2+2 N+1\big)}{3 N^3 (N+1)^3} \zeta_2
\Biggr]
\nonumber\\ &&
+\Biggl[
-16 S_1^2
+\frac{4 \big(637 N^2+637 N+108\big)}{27 N (N+1)} S_1
+\frac{P_{22}}{27 N^2 (N+1)^2}+16 S_2
\Biggr]
\zeta_3
\Biggr\}~.
\end{eqnarray}
Here and in the following expressions we will abbreviate some of the terms by the leading order non-singlet anomalous dimension
\begin{eqnarray}
\gamma_{qq}^{0} = 4 \left[2 S_1 - \frac{3 N^2 + 3 N +2}{2 N (N+1)}\right].
\end{eqnarray}
normalized to the color factor $C_F$. The constant $B_4$ reads
\begin{eqnarray}
B_4 = - 4 \zeta_2 \ln^2(2) + \frac{2}{3} \ln^4(2) - \frac{13}{2} \zeta_4 + 16 \Li_4\left(\frac{1}{2}\right),
\end{eqnarray}
which is related to a multiple zeta value, cf.~\cite{Blumlein:2009cf}, and the polynomials $P_i$ read
\begin{eqnarray}
P_{10} &=& 3 N^4+6 N^3+47 N^2+20 N-12,
\\
P_{11} &=& 7 N^4+14 N^3+3 N^2-4 N-4,
\\
P_{12} &=& 15 N^4+30 N^3+15 N^2-4 N-2,
\\
P_{13} &=& 112 N^4+224 N^3+121 N^2+9 N+9,
\\
P_{14} &=& 181 N^4+266 N^3+82 N^2-3 N+18,
\\
P_{15} &=& 448 N^4+896 N^3+484 N^2+54 N+45,
\\
P_{16} &=& 561 N^4+1122 N^3+767 N^2+302 N+48,
\\
P_{17} &=& 1301 N^4+2602 N^3+2177 N^2+492 N-84,
\\
P_{18} &=& 2 N^5+7 N^4+3 N^3-9 N^2-7 N+2,
\\
P_{19} &=& 3 N^5+13 N^4-23 N^3-69 N^2-54 N-16,
\\
P_{20} &=& 12 N^5+16 N^4+18 N^3-15 N^2-5 N-8,
\\
P_{21} &=& 27 N^5+533 N^4+913 N^3+821 N^2+144 N-36,
\\
P_{22} &=& 648 N^5-2235 N^4-4542 N^3-3725 N^2-770 N-432,
\\
P_{23} &=& -87 N^6-261 N^5-321 N^4-183 N^3-52 N^2-8,
\\
P_{24} &=& 3 N^6+9 N^5+70 N^4+77 N^3+39 N^2-10 N-12,
\\
P_{25} &=& 255 N^6+765 N^5+581 N^4+151 N^3+356 N^2+276 N+72,
\\
P_{26} &=& 364 N^6+1227 N^5+1191 N^4+589 N^3+621 N^2+486 N+144,
\\
P_{27} &=& 1014 N^6+3042 N^5+3757 N^4+1703 N^3+31 N^2+93 N+162,
\\
P_{28} &=& 39 N^8+138 N^7+847 N^6+1371 N^5+1283 N^4+485 N^3+101 N^2
\nonumber\\ &&
+132 N+72,
\\
P_{29} &=& 417 N^8+1668 N^7-4822 N^6-12384 N^5-6507 N^4+740 N^3+216 N^2
\nonumber\\ &&
+144 N+432,
\\
P_{30} &=& 2307 N^8+9255 N^7+13977 N^6+7915 N^5-350 N^4-1456 N^3-106 N^2
\nonumber\\ &&
-138 N-108,
\\
P_{31} &=& 6197 N^8+24788 N^7+39126 N^6+28838 N^5+9977 N^4-702 N^3-3240 N^2
\nonumber\\ &&
-3456 N-1620,
\\
P_{32} &=& 11751 N^8+47004 N^7+93754 N^6+104364 N^5+55287 N^4+6256 N^3-2448 N^2
\nonumber\\ &&
-144 N-432,
\\
P_{33} &=& -22989 N^{10}-114945 N^9-199958 N^8-99362 N^7+179919 N^6+291355 N^5
\nonumber\\ &&
+223828 N^4
           +90936 N^3+31680 N^2+23760 N+10368,
\\
P_{34} &=& -22293 N^{10}-111465 N^9-252090 N^8-310818 N^7-225241 N^6-77573 N^5
\nonumber\\ &&
-8808 N^4 -352 N^3+256 N^2+672 N-288. 
\end{eqnarray}
Here and in the following we reduce the contributing harmonic sums and harmonic polylogarithms algebraically
\cite{Blumlein:2003gb}.

Eq.~(\ref{eq:aqq3v}) can be expressed in harmonic sums only reaching the level of weight {\sf w = 5}.
It is interesting to note that a part of the rational pre-factors rise $\propto N$
\begin{eqnarray}
T_1  = \textcolor{blue}{C_F C_A T_F} \frac{8}{3} N \left[9 \zeta_3 - 3S_3(N) - 4S_{2,1}(N) \right].
\end{eqnarray}
Its Mellin inversion implies a denominator $\propto 1/(1-x)^2$. However, as can be seen form the asymptotic expansion
\begin{eqnarray}
T_1  \propto \frac{32}{3} \textcolor{blue}{C_F C_A T_F} \ln(\bar{N})~, 
\end{eqnarray}
and the leading term behaves $\propto 1/(1-x)_+$ in $x$-space. 

The flavor non-singlet OME in the on-shell scheme is given by
\begin{eqnarray}
A_{qq,Q}^{{\rm NS, \, OMS}} &=&
1
+a_s^2 \textcolor{blue}{C_F T_F}
\Biggl\{
        -\frac{{\gamma_{qq}^0}}{3} \ln^2\left(\frac{m^2}{\mu^2}\right)
        +\ln\left(\frac{m^2}{\mu^2}\right) \biggl[
                \frac{2 P_{10}}{9 N^2 (N+1)^2}
                -\frac{80}{9} S_1
                +\frac{16 }{3} S_2
        \biggr]
\nonumber \\ &&
        +\frac{P_{43}}{54 N^3 (N+1)^3}
        -\frac{224}{27} S_1
        +\frac{40 }{9} S_2
        -\frac{8}{3} S_3
\Biggr\}
\nonumber \\ &&
+a_s^3
\Biggl\{
        \textcolor{blue}{C_F^2 T_F}
        \Biggl[
                \ln^2\left(\frac{m^2}{\mu^2}\right) \biggl[
                        \frac{4 P_{39}}{3 N^3 (N+1)^3}
                        -\frac{8}{3} \frac{(N+2) \big(3 N^3+3 N+2\big)}{N^2 (N+1)^2} S_1
\nonumber \\ &&
                        +\frac{8}{3} {\gamma_{qq}^0} S_2
                        +\frac{64 }{3} S_3
                        +\biggl(
                                -\frac{64}{3 N (N+1)}
                                +\frac{128 }{3} S_1
                        \biggr) S_{-2}
                        +\frac{64}{3} S_{-3}
                        -\frac{128}{3} S_{-2,1}
                \biggr]
\nonumber \\ &&
                +\ln\left(\frac{m^2}{\mu^2}\right) \biggl[
                        \frac{(N-1) P_{47}}{9 N^4 (N+1)^4}
                        +\biggl(
                                -\frac{8 P_{40}}{9 N^3 (N+1)^3}
                                +\frac{640 }{9} S_2
                                -\frac{256}{3} S_3
\nonumber \\ &&
                                -\frac{256}{3} S_{-2,1}
                        \biggr) S_1
                        -\frac{8 P_3 }{9 N^2 (N+1)^2} S_2
                        -\frac{64}{3} S_2^2
                        +\frac{32 \big(29 N^2+29 N+12\big) }{9 N (N+1)} S_3
\nonumber \\ &&
                        -\frac{256}{3} S_4
                        +\biggl(
                                -\frac{64 \big(16 N^2+10 N-3\big)}{9 N^2 (N+1)^2}
                                +\frac{1280 }{9} S_1
                                -\frac{128}{3} S_2
                        \biggr) S_{-2}
\nonumber \\ &&
                        +\biggl(
                                \frac{64 \big(10 N^2+10 N+3\big)}{9 N (N+1)}
                                -\frac{128}{3} S_1
                        \biggr) S_{-3}
                        -\frac{128}{3} S_{-4}
                        +\frac{128}{3} S_{3,1}
                        -\frac{128}{3} S_{-2,2}
\nonumber \\ &&
                        -\frac{128 \big(10 N^2+10 N-3\big) }{9 N (N+1)} S_{-2,1}
                        +\frac{512}{3} S_{-2,1,1}
                        +8 {\gamma_{qq}^0} \zeta_3
                \biggr]
                +\frac{P_{53}}{81 N^5 (N+1)^5}
\nonumber \\ &&
                -\frac{128}{81} \frac{\big(112 N^3+112 N^2-39 N+18\big)}{N^2 (N+1)} S_{-2,1}
                +\biggl(
                        \frac{2 P_{16}}{9 N^2 (N+1)^2}
                        -\frac{1208}{9} S_1
\nonumber \\ &&
                        -\frac{64}{3} S_2
                \biggr) \zeta_3
                +(-1)^N \biggl[
                        \ln^2\left(\frac{m^2}{\mu^2}\right) \frac{32}{3} \frac{\big(2 N^2+2 N+1\big)}{N^3 (N+1)^3}
                        +\frac{16 P_{28}}{81 N^5 (N+1)^5}
\nonumber \\ &&
                        +\ln\left(\frac{m^2}{\mu^2}\right) \biggl[
                                \frac{16 P_{24}}{9 N^4 (N+1)^4}
                                -\frac{64 \big(2 N^2+2 N+1\big) }{3 N^3 (N+1)^3} S_1
                        \biggr]
                        -\frac{32 P_{24} }{27 N^4 (N+1)^4} S_1
\nonumber \\ &&
                        +\frac{64 \big(2 N^2+2 N+1\big) }{9 N^3 (N+1)^3} S_1^2
                        +\frac{64 \big(2 N^2+2 N+1\big) }{9 N^3 (N+1)^3} S_2
                \biggr]
                +{\gamma_{qq}^0} \biggl(
                        -\frac{8 B_4}{3}
                        +12 \zeta_4
\nonumber \\ &&
                        +\frac{8}{3} S_{2,1,1}
                \biggr)
                +\biggl(
                        \frac{2 P_{51}}{81 N^4 (N+1)^4}
                        +\frac{16 P_{15} }{81 N^2 (N+1)^2} S_2
                        -\frac{64}{9} S_2^2
                        +\frac{128 }{9 N (N+1)} S_{2,1}
\nonumber \\ &&
                        -\frac{64 \big(40 N^2+40 N+9\big) }{27 N (N+1)} S_3
                        +\frac{704 }{9} S_4
                        -\frac{320}{9} S_{3,1}
                        -\frac{256}{9} S_{-2,2}
                        +\frac{1024}{9} S_{-2,1,1}
\nonumber \\ &&
                        -\frac{256 \big(10 N^2+10 N-3\big) }{27 N (N+1)} S_{-2,1}
                \biggr) S_1
                +\biggl(
                        -\frac{8 P_{18}}{9 N^3 (N+1)^3}
                        -\frac{64 }{9 N (N+1)} S_2
\nonumber \\ &&
                        +\frac{64 }{3} S_3
                        -\frac{128}{9} S_{2,1}
                        -\frac{256}{9} S_{-2,1}
                \biggr) S_1^2
                +\biggl(
                        -\frac{16 (2 N+1)}{27 N^2 (N+1)^2}
                        +\frac{128 }{27} S_2
                \biggr) S_1^3
\nonumber \\ &&
                +\biggl(
                        -\frac{8 P_{46}}{81 N^3 (N+1)^3}
                        +\frac{256 }{27} S_3
                        +\frac{256}{3} S_{-2,1}
                \biggr) S_2
                -\frac{16 \big(31 N^2+31 N-6\big) }{27 N (N+1)} S_2^2
\nonumber \\ &&
                +\frac{8 P_{35} }{81 N^2 (N+1)^2} S_3
                -\frac{176 \big(17 N^2+17 N+6\big) }{27 N (N+1)} S_4
                +\frac{512 }{9} S_5
                +\biggl(
                        -\frac{512}{9} S_{2,1}
\nonumber \\ &&
                        -\frac{64 P_{14}}{81 N^3 (N+1)^3}
                        +\frac{128 P_{13} }{81 N^2 (N+1)^2} S_1
                        -\frac{128 }{9 N (N+1)} S_1^2
                        +\frac{256}{27} S_1^3
                        -\frac{1280}{27} S_2
\nonumber \\ &&
                        +\frac{512 }{27} S_3
                \biggr) S_{-2}
                +\biggl(
                        \frac{64}{81} \frac{\big(112 N^3+224 N^2+169 N+39\big)}{N (N+1)^2} 
                        +\frac{128}{9} S_1^2
                        +\frac{128 }{9} S_2
\nonumber \\ &&
                        -\frac{128 \big(10 N^2+10 N+3\big) }{27 N (N+1)} S_1
                \biggr) S_{-3}
                +\biggl(
                        -\frac{128 \big(10 N^2+10 N+3\big)}{27 N (N+1)}
                        +\frac{256 }{9} S_1
                \biggr) S_{-4}
\nonumber \\ &&
                +\frac{256}{9} S_{-5}
                +\frac{16 P_{11} }{9 N^2 (N+1)^2} S_{2,1}
                +\frac{256}{9} S_{2,3}
                -\frac{512}{9} S_{2,-3}
                -\frac{512}{9} S_{4,1}
                +\frac{512}{9} S_{-2,3}
\nonumber \\ &&
                +\frac{16 \big(89 N^2+89 N+30\big) }{27 N (N+1)} S_{3,1}
                -\frac{128 \big(10 N^2+10 N-3\big) }{27 N (N+1)} S_{-2,2}
                +\frac{512}{9} S_{2,1,-2}
\nonumber \\ &&
                +\frac{256}{9} S_{3,1,1}
                +\frac{512 \big(10 N^2+10 N-3\big) }{27 N (N+1)} S_{-2,1,1}
                +\frac{512}{9} S_{-2,2,1}
                -\frac{2048}{9} S_{-2,1,1,1}
        \Biggr]
\nonumber \\ &&
        +\textcolor{blue}{C_A C_F T_F}
        \Biggl[
                \frac{22 {\gamma_{qq}^0}}{27} \ln^3\left(\frac{m^2}{\mu^2}\right)
                +\ln^2\left(\frac{m^2}{\mu^2}\right) \biggl[
                        \frac{2 P_{41}}{9 N^3 (N+1)^3}
                        -\frac{184}{9} S_1
                        -\frac{32}{3} S_3
\nonumber \\ &&
                        +\biggl(
                                \frac{32}{3 N (N+1)}
                                -\frac{64}{3} S_1
                        \biggr) S_{-2}
                        -\frac{32}{3} S_{-3}
                        +\frac{64}{3} S_{-2,1}
                \biggr]
\nonumber \\ &&
                +\ln\left(\frac{m^2}{\mu^2}\right) \biggl[
                        \frac{P_{48}}{81 N^4 (N+1)^4}
                        +\biggl(
                                -\frac{8 P_{42}}{81 N^3 (N+1)^3}
                                +32 S_3
                                +\frac{128}{3} S_{-2,1}
                        \biggr) S_1
\nonumber \\ &&
                        +\frac{1792 }{27} S_2
                        -\frac{16 \big(31 N^2+31 N+9\big) }{9 N (N+1)} S_3
                        +\frac{160 }{3} S_4
                        +\biggl(
                                \frac{32 \big(16 N^2+10 N-3\big)}{9 N^2 (N+1)^2}
\nonumber \\ &&
                                -\frac{640}{9} S_1
                                +\frac{64 }{3} S_2
                        \biggr) S_{-2}
                        +\biggl(
                                -\frac{32 \big(10 N^2+10 N+3\big)}{9 N (N+1)}
                                +\frac{64 }{3} S_1
                        \biggr) S_{-3}
                        +\frac{64}{3} S_{-4}
\nonumber \\ &&
                        -\frac{128}{3} S_{3,1}
                        +\frac{64 \big(10 N^2+10 N-3\big) }{9 N (N+1)} S_{-2,1}
                        +\frac{64}{3} S_{-2,2}
                        -\frac{256}{3} S_{-2,1,1}
                        -8 {\gamma_{qq}^0} \zeta_3
                \biggr]
\nonumber \\ &&
                +\frac{P_{54}}{729 N^5 (N+1)^5}
                +\frac{64}{81} \frac{\big(112 N^3+112 N^2-39 N+18\big)}{N^2 (N+1)} S_{-2,1}
\nonumber \\ &&
                +\biggl(
                        \frac{P_{38}}{27 N^2 (N+1)^2}
                        +\frac{4 \big(593 N^2+593 N+108\big) }{27 N (N+1)} S_1
                        -16 S_1^2
                        +16 S_2
                \biggr) \zeta_3
\nonumber \\ &&
                +(-1)^N \biggl[
                        -\frac{\big(2 N^2+2 N+1\big)}{N^3 (N+1)^3} \frac{16}{3} \ln^2\left(\frac{m^2}{\mu^2}\right)
                        +\ln\left(\frac{m^2}{\mu^2}\right) \biggl[
                                -\frac{8 P_{24}}{9 N^4 (N+1)^4}
\nonumber \\ &&
                                +\frac{32 \big(2 N^2+2 N+1\big) }{3 N^3 (N+1)^3} S_1
                        \biggr]
                        -\frac{8 P_{28}}{81 N^5 (N+1)^5}
                        +\frac{16 P_{24} }{27 N^4 (N+1)^4} S_1
\nonumber \\ &&
                        -\frac{32 \big(2 N^2+2 N+1\big) }{9 N^3 (N+1)^3} S_1^2
                        -\frac{32 \big(2 N^2+2 N+1\big) }{9 N^3 (N+1)^3} S_2
                \biggr]
                +{\gamma_{qq}^0} \biggl(
                        \frac{4 B_4}{3}
                        -12 \zeta_4
\nonumber \\ &&
                        -4 S_{2,1,1}
                \biggr)
                +\biggl(
                        -\frac{4 P_{52}}{729 N^4 (N+1)^4}
                        -\frac{16}{9} \frac{(N-1) \big(2 N^3-N^2-N-2\big)}{N^2 (N+1)^2} S_2
\nonumber \\ &&
                        +\frac{112}{9} S_2^2
                        +\frac{80 (2 N+1)^2 }{9 N (N+1)} S_3
                        -\frac{208}{9} S_4
                        -\frac{8 \big(9 N^2+9 N+16\big) }{9 N (N+1)} S_{2,1}
                        +\frac{64}{3} S_{3,1}
\nonumber \\ &&
                        +\frac{128 \big(10 N^2+10 N-3\big) }{27 N (N+1)} S_{-2,1}
                        +\frac{128}{9} S_{-2,2}
                        -\frac{512}{9} S_{-2,1,1}
                \biggr) S_1
\nonumber \\ &&
                +\biggl(
                        \frac{4 P_{36}}{9 N^3 (N+1)^3}
                        +\frac{32 }{9 N (N+1)} S_2
                        -\frac{80}{9} S_3
                        +\frac{128}{9} S_{2,1}
                        +\frac{128}{9} S_{-2,1}
                \biggr) S_1^2
\nonumber \\ &&
                +\biggl(
                        \frac{4 P_{45}}{81 N^3 (N+1)^3}
                        +\frac{496 }{27} S_3
                        -\frac{64}{3} S_{2,1}
                        -\frac{128}{3} S_{-2,1}
                \biggr) S_2
                -\frac{64}{27} S_1^3 S_2
\nonumber \\ &&
                -\frac{4 \big(15 N^2+15 N+14\big) }{9 N (N+1)} S_2^2
                +\frac{4 \big(443 N^2+443 N+78\big) }{27 N (N+1)} S_4
                -\frac{224}{9} S_5
\nonumber \\ &&
                -\frac{8 P_{37} }{81 N^2 (N+1)^2} S_3
                +\biggl(
                        \frac{32 P_{14}}{81 N^3 (N+1)^3}
                        -\frac{64 P_{13} }{81 N^2 (N+1)^2} S_1
                        +\frac{64 }{9 N (N+1)} S_1^2
\nonumber \\ &&
                        -\frac{128}{27} S_1^3
                        +\frac{640 }{27} S_2
                        -\frac{256}{27} S_3
                        +\frac{256}{9} S_{2,1}
                \biggr) S_{-2}
                +\biggl(
                        \frac{64 \big(10 N^2+10 N+3\big) }{27 N (N+1)} S_1
\nonumber \\ &&
                        -\frac{32}{81} \frac{\big(112 N^3+224 N^2+169 N+39\big)}{N (N+1)^2}
                        -\frac{64}{9} S_1^2
                        -\frac{64}{9} S_2
                \biggr) S_{-3}
                +\biggl(
                        -\frac{128}{9} S_1
\nonumber \\ &&
                        +\frac{64 \big(10 N^2+10 N+3\big)}{27 N (N+1)}
                \biggr) S_{-4}
                -\frac{128}{9} S_{-5}
                -\frac{8 P_{20} S_{2,1}}{9 N^2 (N+1)^2}
                -\frac{128}{3} S_{2,3}
\nonumber \\ &&
                +\frac{256}{9} S_{2,-3}
                -\frac{8 (13 N+4) (13 N+9) }{27 N (N+1)} S_{3,1}
                +\frac{256}{9} S_{4,1}
                -\frac{256}{9} S_{-2,3}
                -\frac{256}{9} S_{2,1,-2}
\nonumber \\ &&
                +\frac{64 \big(10 N^2+10 N-3\big) }{27 N (N+1)} S_{-2,2}
                +\frac{64}{3} S_{2,2,1}
                -\frac{256}{9} S_{3,1,1}
                -\frac{256}{9} S_{-2,2,1}
\nonumber \\ &&
                -\frac{256 \big(10 N^2+10 N-3\big) }{27 N (N+1)} S_{-2,1,1}
                +\frac{224}{9} S_{2,1,1,1}
                +\frac{1024}{9} S_{-2,1,1,1}
        \Biggr]
\nonumber \\ &&
        +\textcolor{blue}{C_F T_F^2}
        \Biggl[
                -\frac{16 {\gamma_{qq}^0}}{27} \ln^3\left(\frac{m^2}{\mu^2}\right)
                +\ln^2\left(\frac{m^2}{\mu^2}\right) \biggl[
                        \frac{8 P_{10}}{27 N^2 (N+1)^2}
                        -\frac{320}{27} S_1
                        +\frac{64 }{9} S_2
                \biggr]
\nonumber \\ &&
                -\frac{248 {\gamma_{qq}^0}}{81} \ln\left(\frac{m^2}{\mu^2}\right)
                -\frac{2 P_{50}}{729 N^4 (N+1)^4}
                +\frac{12064 }{729} S_1
                +\frac{64 }{81} S_2
                +\frac{320 }{81} S_3
                -\frac{64}{27} S_4
\nonumber \\ &&
                -\frac{112 {\gamma_{qq}^0} }{27} \zeta_3
        \Biggr]
        +\textcolor{blue}{C_F N_F T_F^2}
        \Biggl[
                -\frac{8 {\gamma_{qq}^0} \ln^3\left(\frac{m^2}{\mu^2}\right)}{27}
                +\ln\left(\frac{m^2}{\mu^2}\right) \biggl[
                        \frac{4 P_{44}}{81 N^3 (N+1)^3}
\nonumber \\ &&
                        -\frac{2176}{81} S_1
                        -\frac{320}{27} S_2
                        +\frac{64 }{9} S_3
                \biggr]
                +\frac{4 P_{49}}{729 N^4 (N+1)^4}
                -\frac{24064}{729} S_1
                +\frac{128 }{81} S_2
                +\frac{640 }{81} S_3
\nonumber \\ &&
                -\frac{128}{27} S_4
                +\frac{64 {\gamma_{qq}^0} }{27} \zeta_3
        \Biggr]
\Biggr\},
\end{eqnarray}
with the polynomials
\begin{eqnarray}
P_{35}    &=& 977 N^4+1954 N^3+1853 N^2+492 N-84, \\
P_{36} &=& 3 N^5+11 N^4+10 N^3+3 N^2+7 N+8, \\
P_{37} &=& 27 N^5+863 N^4+1573 N^3+1151 N^2+144 N-36, \\
P_{38} &=& 648 N^5-2103 N^4-4278 N^3-3505 N^2-682 N-432, \\
P_{39} &=& 6 N^6+18 N^5+21 N^4+24 N^3+7 N^2-4, \\
P_{40} &=& 15 N^6+45 N^5+45 N^4+143 N^3+120 N^2-8 N-24, \\
P_{41} &=& 51 N^6+153 N^5+223 N^4+191 N^3+118 N^2+48 N+24, \\
P_{42} &=& 155 N^6+465 N^5+465 N^4+155 N^3-108 N^2-108 N-54, \\
P_{43} &=& 219 N^6+657 N^5+1193 N^4+763 N^3-40 N^2-48 N+72, \\
P_{44} &=& 525 N^6+1575 N^5+1535 N^4+973 N^3+536 N^2+48 N-72, \\
P_{45} &=& 868 N^6+2469 N^5+2487 N^4+940 N^3+27 N^2+63 N+72, \\
P_{46} &=& 906 N^6+2718 N^5+3433 N^4+1595 N^3+31 N^2+93 N+162, \\
P_{47} &=& 9 N^7+45 N^6+279 N^5+1263 N^4+1348 N^3+752 N^2+112 N-48, \\
P_{48} &=& -4785 N^8-19140 N^7-18970 N^6+672 N^5+7683 N^4+1004 N^3+1272 N^2
\nonumber \\ &&
           +72 N-864, \\
P_{49} &=& 3549 N^8+14196 N^7+23870 N^6+25380 N^5+15165 N^4+1712 N^3-2016 N^2
\nonumber \\ &&
           +144 N+432, \\
P_{50} &=& 5487 N^8+21948 N^7+36370 N^6+28836 N^5+11943 N^4+4312 N^3+2016 N^2
\nonumber \\ &&
           -144 N-432, \\
P_{51} &=& 7131 N^8+28632 N^7+43326 N^6+23272 N^5-3497 N^4-5824 N^3-424 N^2
\nonumber \\ &&
           -552 N-432, \\
P_{52} &=& 10807 N^8+43228 N^7+62898 N^6+39178 N^5+7027 N^4+702 N^3+3240 N^2
\nonumber \\ &&
           +3456 N+1620, \\
P_{53} &=& -6219 N^{10}-31095 N^9-72513 N^8-95154 N^7-79721 N^6-32383 N^5-2307 N^4
\nonumber \\ &&
           +3280 N^3+1424 N^2+336 N-144, \\
P_{54} &=& 165 N^{10}+825 N^9+106856 N^8+321746 N^7+396657 N^6+247433 N^5
\nonumber \\ &&
           +126914 N^4+51804 N^3+6336 N^2+4752 N+5184.
\end{eqnarray}
The corresponding expressions in $x$-space is given in Appendix~\ref{sec:OMSx}.

The difference between the OMEs in the $\overline{\rm MS}$-scheme and the on-shell scheme for the heavy quark mass 
in $N$-space is obtained by
\begin{eqnarray}
A_{qq,Q}^{\overline{\rm MS}} - A_{qq,Q}^{\rm OMS} &=&
\textcolor{blue}{C_F^2 T_F} 
\Biggl\{ 4 \gamma_{qq}^0
\ln^2\left(\frac{m^2}{\mu^2}\right)
+\Biggl[
\frac{4 \big(21 N^4+42 N^3-7 N^2-4 N+12\big)}{3 N^2 (N+1)^2}
+\frac{32}{3} S_1 
\nonumber\\ &&
-32 S_2
\Biggr]
\ln\left(\frac{m^2}{\mu^2}\right)
+\frac{16 \big(3 N^4+6 N^3+47 N^2+20 N-12\big)}{9 N^2 (N+1)^2}
-\frac{640}{9} S_1
+\frac{128}{3} S_2
\Biggr\} . 
\nonumber\\ 
\end{eqnarray}
Here we used the same mass in both schemes symbolically to shorten the expression. The heavy quark masses in both schemes are 
given in Ref.~\cite{MASS}. In $x$-space it is given by
\begin{eqnarray}
\label{eq:MSOMS}
A_{qq,Q}^{\overline{\rm MS}} - A_{qq,Q}^{\rm OMS} &=&
\textcolor{blue}{C_F^2 T_F} 
\Biggl\{
\Biggl(\frac{1}{1-x}
\Biggl[ -32 \ln^2\left(\frac{m^2}{\mu^2}\right)
        -32 \ln\left(\frac{m^2}{\mu^2}\right)\left(\frac{1}{3} + H_0\right) 
\nonumber\\ &&
+ \frac{640}{9} + \frac{128}{3} H_0
\Biggr]\Biggr)_+ 
+
\Biggl[
-24 \ln^2\left(\frac{m^2}{\mu^2}\right)
+28 \ln\left(\frac{m^2}{\mu^2}\right)
+\frac{16}{3}
\Biggr] \delta(1-x)
\nonumber\\ &&
+16 \ln^2\left(\frac{m^2}{\mu^2}\right)(x+1)
+ \ln\left(\frac{m^2}{\mu^2}\right) \left(16 (x+1) H_0 + \frac{16}{3} (7 x-5)\right)
\nonumber\\ &&
-\frac{64}{3} (x+1) H_0
-\frac{64}{9} (11 x-1)
\Biggr\}~.
\end{eqnarray}
Here $H_{\vec{a}}(x)$ denote the harmonic polylogarithms \cite{Remiddi:1999ew}
\begin{eqnarray}
H_{b,\vec{a}}(x) = \int_0^x dy f_b(y) H_{\vec{a}}(y),~~~H_\emptyset = 1,~~~H_{\underbrace{\mbox{\scriptsize 0,...,0}}_k}(x) = 
\frac{1}{k!} \ln^k(x). 
\end{eqnarray}
The letters $f_i=f_0,f_{1},f_{-1}$ belong to the alphabet $\{1/x, 1/(1-x), 1/(1+x)\}$ and we use the abbrevation
$H_{\vec{a}} \equiv H_{\vec{a}}(x)$.
The $+$-prescription is defined by
\begin{eqnarray}
\int_0^1 dx [f(x)]_+ g(x) = \int_0^1 dx f(x) \left[g(x) - g(1)\right]~.
\end{eqnarray}

In the unpolarized case the leading term of the non-singlet OME in the small-$x$ limit is obtained from 
Eq.~(\ref{eq:ANSx})
\begin{eqnarray}
A_{qq,Q}^{(2)} &\propto& \textcolor{blue}{C_F T_F} \frac{2}{3} \ln^2(x), 
\\
A_{qq,Q}^{(3)} &\propto& \textcolor{blue}{C_F^2 T_F} \frac{14}{27} \ln^4(x). 
\end{eqnarray}
For comparison, the massless 3-loop Wilson coefficient \cite{Vermaseren:2005qc} behaves like 
\begin{eqnarray}
\hat{c}_{2,q,\rm NS}^{(3)} \propto \textcolor{blue}{C_F^2 T_F} \frac{91}{27} \ln^4(x)~.
\end{eqnarray}
These terms correspond to the so-called leading poles at $N = 0$ in the flavor non-singlet case \cite{SX}. 

We finally derive the leading asymptotic behaviour of the massive OME for large values of $N$, corresponding to the region 
$x \rightarrow 1$. In the case of the on-shell scheme it is given by
\begin{eqnarray}
A_{qq,Q}^{(2),\rm NS} &\propto& \textcolor{blue}{C_F T_F} \Biggl\{
\Biggl[\frac{8}{3} \ln^2\left(\frac{m^2}{\mu^2}\right)
+\frac{80}{9} \ln\left(\frac{m^2}{\mu^2}\right)
+\frac{224}{27}\Biggr] \left(\frac{1}{1-x}\right)_+
\nonumber\\ && 
+ \Biggl[2 \ln^2\left(\frac{m^2}{\mu^2}\right)
+ \frac{2}{3}\left[1 + 8 \zeta_2 \right] \ln\left(\frac{m^2}{\mu^2}\right)
+\frac{73}{18}
+\frac{40}{9} \zeta_2
-\frac{8}{3} \zeta_3\Biggr] \delta(1-x) \Biggr\},
\\
A_{qq,Q}^{(3),\rm NS} &\propto&
- \textcolor{blue}{C_F T_F} 
\Biggl\{
\Biggl\{
        \frac{176}{27} \textcolor{blue}{C_A} 
        -\Biggl[\frac{64}{27} \textcolor{blue}{N_F} + \frac{128}{27}\Biggr] \textcolor{blue}{T_F}
\Biggr\} 
\ln^3\left(\frac{m^2}{\mu^2}\right)
\nonumber\\ &&
+  
\Biggl\{
        -8 \textcolor{blue}{C_F}
                -\frac{320}{27} \textcolor{blue}{T_F}
                +\textcolor{blue}{C_A} 
\Biggl[
                        -\frac{184}{9}
                        +\frac{32}{3} \zeta_2
\Biggr]
\Biggl\} 
\ln^2\left(\frac{m^2}{\mu^2}\right)
\nonumber\\ &&
+ 
\Biggl\{
- \Biggl[
                        \frac{2176}{81} \textcolor{blue}{N_F}
+ \frac{1984}{81}
\Biggr] 
         \textcolor{blue}{T_F}
                +\textcolor{blue}{C_A} 
\Biggl[
                        -\frac{1240}{81}
                        +\frac{320}{9} \zeta_2
                        -\frac{224}{3} \zeta_3
\Biggr]
\nonumber\\ &&
        +\textcolor{blue}{C_F} 
\Biggl[
                -\frac{40}{3}
                +64 \zeta_3
\Biggr]
\Biggr\} 
\ln\left(\frac{m^2}{\mu^2}\right)
\nonumber\\ &&
+ 
\Biggl\{
            \textcolor{blue}{C_F} 
\Biggl[
        -\frac{2}{27} (288  B_4 - 2377)
        -\frac{1208}{9} \zeta_3
        +96 \zeta_4
\Biggr]
+ \textcolor{blue}{T_F} 
\Biggl[
                \frac{12064}{729}
                -\frac{896}{27} \zeta_3
\nonumber\\ &&
                +\textcolor{blue}{N_F} 
\Biggl[
                        -\frac{24064}{729}
                        +\frac{512}{27} \zeta_3
\Biggr]
\Biggr]
        +\textcolor{blue}{C_A} 
\Biggl[
                \frac{4}{729} (1944 B_4-8863)
                +\frac{3296}{81} \zeta_2
\nonumber\\ &&
                +60 \zeta_3
                -\frac{352}{3} \zeta_4
\Biggr]
\Biggr\}
\Biggr\} \left(\frac{1}{1-x}\right)_+
\nonumber\\ && 
+ \textcolor{blue}{C_F T_F} 
\Biggl\{
\Biggl[
        -\frac{44}{9} \textcolor{blue}{C_A}
        +\Biggl[
                \frac{16}{9} \textcolor{blue}{N_F}
                +\frac{32}{9}\Biggr] \textcolor{blue}{T_F}
\Biggr]
\ln^3\left(\frac{m^2}{\mu^2}\right)
\nonumber\\ && 
+
\Bigg\{
        \textcolor{blue}{C_F} 
\Biggl[
                8
                -16 \zeta_2
                +32 \zeta_3
\Biggr]
+                \textcolor{blue}{T_F} 
\Biggl[
                        \frac{8}{9}
                        +\frac{64}{9} \zeta_2
\Biggr]
                +\textcolor{blue}{C_A} 
\Biggl[
                        \frac{34}{3}
                        -16 \zeta_3
\Biggr]
\Biggr\}
\ln^2\left(\frac{m^2}{\mu^2}\right)
\nonumber\\ && 
+\Biggl\{
                \textcolor{blue}{T_F} 
\Biggl[
                        \frac{496}{27}
                        +\textcolor{blue}{N_F} 
\Biggl[
                                \frac{700}{27}
                                -\frac{320}{27} \zeta_2
                                +\frac{64}{9} \zeta_3
\Biggr]
\Biggr]
                +\textcolor{blue}{C_A} 
\Biggl[
                        -\frac{1595}{27}
                        +\frac{1792}{27} \zeta_2
\nonumber\\ && 
                        -\frac{224}{9} \zeta_3
                        -4 \zeta_4
\Biggr]
        +\textcolor{blue}{C_F}
\Biggl[
                1
                -\frac{40}{3} \zeta_2
                +\frac{272}{3} \zeta_3
                -\frac{232}{3} \zeta_4
\Biggr]
\Biggr\} \ln\left(\frac{m^2}{\mu^2}\right)
\nonumber\\ && 
+
        \textcolor{blue}{T_F} 
\Biggl[
                -
                \frac{3658}{243}
                +\frac{64}{81} \zeta_2
                +\frac{2336}{81} \zeta_3
                -\frac{64}{27} \zeta_4
                +\textcolor{blue}{N_F} 
\Biggl[
                        \frac{4732}{243}
                        +\frac{128}{81} \zeta_2
                        -\frac{512}{81} \zeta_3
                        -\frac{128}{27} \zeta_4
\Biggr]
\Biggr]
\nonumber\\ && 
        +\textcolor{blue}{C_A}
\Biggl[
                \frac{1}{243} (2647-1944 B_4)
                +
\Biggl[
                        \frac{3472}{81}
                        +\frac{112}{3} \zeta_3
\Biggr]
\zeta_2
                -\frac{17309}{81} \zeta_3
                +\frac{3400}{27} \zeta_4
                -\frac{112}{3} \zeta_5
\Biggr]
\nonumber\\ && 
+\textcolor{blue}{C_F}
\Biggl[
        \frac{1}{9} (144 B_4 - 691)
- \Biggl[
                 \frac{2416}{27}
                + \frac{160}{9} \zeta_3
\Biggr]
\zeta_2
        +\frac{7838}{27} \zeta_3
        -\frac{5452}{27} \zeta_4
\nonumber\\ &&
        +\frac{608}{9} \zeta_5
\Biggr]
\Biggr\}
\delta(1-x).
\end{eqnarray}
The corresponding expressions in the $\overline{\rm MS}$ scheme are easily obtained using Eq.~(\ref{eq:MSOMS}).
\section{The Massive Flavor Non-Singlet Wilson Coefficient in the Asymptotic Region}
\label{sec:6}

\vspace{1mm}
\noindent
The heavy flavor unpolarized non-singlet Wilson coefficient in the the region $Q^2 \gg m^2$, cf.~Eq.~(\ref{eq:LQ2}), is given 
by 
\begin{eqnarray}
\label{WIL:NS}
L_{q,2}^{\rm NS}(N) &=& \tfrac{1}{2}\left[1 + (-1)^N\right]
\nonumber\\ && \times
\Biggl\{
a_s^2 \textcolor{blue}{C_F T_F} \biggl\{
-\frac{8}{9} S_1^3
-\frac{2 \big(29 N^2+29 N-6\big)}{9 N (N+1)} S_1^2
+\biggl(
\frac{8}{3} S_2
-\frac{2 P_{63}}{27 N^2 (N+1)^2}
\biggr) S_1
\nonumber \\ &&
-\frac{1}{3} {\gamma_{qq}^{0}} \big(
L_M^2
+L_Q^2
\big)
+\frac{P_{78}}{27 N^3 (N+1)^3}
+\biggl(
\frac{8}{3} S_1^2
+\frac{4 \big(29 N^2+29 N-6\big)}{9 N (N+1)} S_1
\nonumber \\ &&
-8 S_2
-\frac{2 P_{60}}{9 N^2 (N+1)^2}
\biggr) L_Q
+\biggl(
\frac{2 P_{10}}{9 N^2 (N+1)^2}
-\frac{80}{9} S_1
+\frac{16}{3} S_2
\biggr) L_M
\nonumber \\ &&
-\frac{112}{9} S_3
+\frac{16}{3} S_{2,1}
+\frac{2 \big(35 N^2+35 N-2\big)}{3 N (N+1)} S_2
\biggr\}
\nonumber \\ &&
+a_s^3 \Biggl\{
\textcolor{blue}{C_F^2 T_F} \biggl\{
\biggl(
\frac{128}{27} S_2
-\frac{16 P_{57}}{27 N^2 (N+1)^2}
\biggr) S_1^3
+\biggl(
\frac{16 \big(5 N^2+5 N-4\big)}{9 N (N+1)} S_2
+16 S_3
\nonumber \\ &&
-\frac{128}{9} S_{2,1}
+\frac{P_{71}}{9 N^3 (N+1)^3}
-\frac{256}{9} S_{-2,1}
\biggr) S_1^2
+\biggl(
-\frac{64}{9} S_2^2
+\frac{8 P_{67}}{81 N^2 (N+1)^2} S_2
\nonumber \\ &&
+\frac{704}{9} S_4
+\frac{P_{83}}{162 N^4 (N+1)^4}
-\frac{8 \big(347 N^2+347 N+54\big)}{27 N (N+1)} S_3
+\frac{128}{9 N (N+1)} S_{2,1}
\nonumber \\ &&
-\frac{320}{9} S_{3,1}
-\frac{256 \big(10 N^2+10 N-3\big)}{27 N (N+1)} S_{-2,1}
-\frac{256}{9} S_{-2,2}
+\frac{1024}{9} S_{-2,1,1}
\biggr) S_1
\nonumber \\ &&
-\frac{32 \big(23 N^2+23 N-3\big)}{27 N (N+1)} S_2^2
+\frac{1}{6} {\gamma_{qq}^{0}}^2 \big(L_Q^3+L_M^2 L_Q\big)
+\frac{P_{84}}{162 N^5 (N+1)^5}
\nonumber \\ &&
-\frac{16 \big(2 N^3+2 N^2+2 N+1\big)}{3 N^3 (N+1)^3} \zeta_2
+\biggl(
\frac{256}{9} S_1
-\frac{128 \big(10 N^2+10 N+3\big)}{27 N (N+1)}
\biggr) S_{-4}
\nonumber \\ &&
+L_M L_Q \biggl(
\frac{320}{9} S_1^2
-\frac{80 \big(3 N^2+3 N+2\big)}{9 N (N+1)} S_1
\biggr)
+\biggl(
-\frac{128 \big(10 N^2+10 N+3\big)}{27 N (N+1)} S_1
\nonumber \\ &&
+\frac{128}{9} S_1^2
+\frac{64 \big(112 N^3+224 N^2+169 N+39\big)}{81 N (N+1)^2}
+\frac{128}{9} S_2
\biggr) S_{-3} 
\nonumber \\ &&
+\frac{(-1)^N}{(N+1)^3} \biggl[
-\frac{64}{3} L_M^2
+\biggl(
\frac{128}{3} S_1
-\frac{256 (4 N+1)}{9 (N+1)}
\biggr) L_M
+\frac{64 \big(2 N^2+2 N+1\big)}{9 N^3} S_1^2
\nonumber \\ &&
+\frac{16 P_{28}}{81 N^5 (N+1)^2}
+\frac{64 P_{85}}{45 (N-2) (N-1)^2 N^2 (N+1) (N+2)^2 (N+3)^3} L_Q 
\nonumber \\ &&
+\frac{16}{3} \big(
2 N^3+2 N^2+2 N+1
\big) \frac{\zeta_2}{N^3}
-\frac{32 P_{24}}{27 N^4 (N+1)} S_1
+\frac{64 \big(2 N^2+2 N+1\big)}{9 N^3} S_2
\nonumber \\ &&
-\frac{64}{3} L_Q^2
\biggr]
+\frac{8 P_{68}}{81 N^2 (N+1)^2} S_3
-\frac{176 \big(17 N^2+17 N+6\big)}{27 N (N+1)} S_4
+\frac{512}{9} S_5
+\frac{256}{9} S_{-5}
\nonumber \\ &&
+\biggl(
\frac{256}{27} S_1^3
-\frac{128}{9 N (N+1)} S_1^2
+\frac{128 P_{13}}{81 N^2 (N+1)^2} S_1
-\frac{64 P_{14}}{81 N^3 (N+1)^3}
-\frac{1280}{27} S_2
\nonumber \\ &&
+\frac{512}{27} S_3
-\frac{512}{9} S_{2,1}
\biggr) S_{-2}
+\frac{16 P_{11}}{9 N^2 (N+1)^2} S_{2,1}
+\frac{16 \big(89 N^2+89 N+30\big)}{27 N (N+1)} S_{3,1}
\nonumber \\ &&
+\frac{256}{9} S_{2,3}
-\frac{512}{9} S_{2,-3}
-\frac{512}{9} S_{4,1}
+\biggl[
-16 S_1^3
-\frac{4 \big(107 N^2+107 N-54\big)}{9 N (N+1)} S_1^2
\nonumber \\ &&
+\frac{2 P_{64}}{9 N^2 (N+1)^2} S_1
-\frac{2 P_{75}}{9 N^3 (N+1)^3}
+\biggl(
\frac{128}{3} S_1
-\frac{64}{3 N (N+1)}
\biggr) S_{-2}
+\frac{64}{3} S_3
\nonumber \\ &&
+\frac{64}{3} S_{-3}
-\frac{128}{3} S_{-2,1}
\biggr] L_Q^2
+\biggl[
-\frac{16}{3} S_1^3
-\frac{4 (N-1) (N+2)}{N (N+1)} S_1^2
+\frac{2 P_{58}}{3 N^2 (N+1)^2} S_1
\nonumber \\ &&
-\frac{2 P_{73}}{3 N^3 (N+1)^3}
+S_{-2} \biggl(
\frac{128}{3} S_1
-\frac{64}{3 N (N+1)}
\biggr)
+\frac{64}{3} \big(
S_3
+S_{-3}
-2 S_{-2,1}\big)
\biggr] L_M^2
\nonumber \\ &&
-\frac{128 \big(112 N^3+112 N^2-39 N+18\big)}{81 N^2 (N+1)} S_{-2,1}
+\biggl(
\frac{P_{70}}{81 N^3 (N+1)^3}
+\frac{400 S_3}{27}
\nonumber \\ &&
+\frac{256}{3} S_{-2,1}
\biggr) S_2
-\frac{128 \big(10 N^2+10 N-3\big) S_{-2,2}}{27 N (N+1)}
+\frac{512}{9} S_{-2,3}
+\frac{512}{9} S_{2,1,-2}
\nonumber \\ &&
+\frac{256}{9} S_{3,1,1}
+\frac{512 \big(10 N^2+10 N-3\big)}{27 N (N+1)} S_{-2,1,1}
+\biggl[
\biggl(
\frac{32}{3} S_2
-\frac{4 P_{56}}{3 N^2 (N+1)^2}
\biggr) S_1^2
\nonumber \\ &&
-\frac{160}{9} S_1^3
+\biggl(
\frac{2 P_{76}}{9 N^3 (N+1)^3}
+\frac{16 \big(59 N^2+59 N-6\big)}{9 N (N+1)} S_2
-\frac{256}{3} S_3
-\frac{256}{3} S_{-2,1}
\biggr) S_1
\nonumber \\ &&
-32 S_2^2
+\frac{P_{82}}{9 N^4 (N+1)^4}
+S_{-3} \biggl(
\frac{64 \big(10 N^2+10 N+3\big)}{9 N (N+1)}
-\frac{128}{3} S_1
\biggr)
+\frac{128}{3} S_{3,1}
\nonumber \\ &&
+S_{-2} \biggl(
-\frac{64 \big(16 N^2+10 N-3\big)}{9 N^2 (N+1)^2}
+\frac{1280}{9} S_1
-\frac{128}{3} S_2
\biggr)
-\frac{4 P_{61}}{9 N^2 (N+1)^2} S_2
\nonumber \\ &&
+\frac{32 \big(29 N^2+29 N+12\big)}{9 N (N+1)} S_3
-\frac{256}{3} S_4
-\frac{128}{3} S_{-4}
-\frac{128 \big(10 N^2+10 N-3\big)}{9 N (N+1)} S_{-2,1}
\nonumber \\ &&
-\frac{128}{3} S_{-2,2}
+\frac{512}{3} S_{-2,1,1}
\biggr] L_M
+\frac{512}{9} S_{-2,2,1}
-\frac{2048}{9} S_{-2,1,1,1}
+\biggl(
\frac{2 P_{16}}{9 N^2 (N+1)^2}
\nonumber \\ &&
-\frac{1208}{9} S_1
-\frac{64}{3} S_2
\biggr) \zeta_3
+{\gamma_{qq}^{0}} \biggl[
\frac{10}{3} S_2 L_M^2
+\biggl(
\frac{P_{55}}{9 N^2 (N+1)^2}
-\frac{8}{3} S_2
\biggr) L_Q L_M
\nonumber \\ &&
+8 \zeta_3 L_M
-\frac{8}{3} B_4
+12 \zeta_4
+\frac{22}{3} L_Q^2 S_2
+\frac{8}{3} S_{2,1,1}
\biggr]
+\biggl[
\frac{32 (4 N-1) (4 N+5)}{9 N (N+1)} S_1^3
\nonumber \\ &&
+\frac{80}{9} S_1^4
+\biggl(
\frac{2 P_{66}}{27 N^2 (N+1)^2}
-\frac{224}{3} S_2
\biggr) S_1^2
+\biggl(
-\frac{32 \big(67 N^2+67 N-21\big)}{9 N (N+1)} S_2
\nonumber \\ &&
-\frac{4 P_{79}}{27 N^3 (N+1)^3}
+\frac{640}{9} S_3
+\frac{64}{3} S_{2,1}
+\frac{512}{3} S_{-2,1}
\biggr) S_1
+48 S_2^2
+64 S_{-2}^2
\nonumber \\ &&
+\frac{4 P_{87}}{135 (N-1)^2 N^4 (N+1)^4 (N+2)^2 (N+3)^3}
-\frac{32 \big(53 N^2+77 N+4\big)}{9 N (N+1)} S_3
\nonumber \\ &&
+S_{-3} \biggl(
\frac{256}{3} S_1
-\frac{64 \big(10 N^2+22 N+3\big)}{9 N (N+1)}
\biggr)
+\frac{2 P_{65}}{9 N^2 (N+1)^2} S_2
+\biggl(
\frac{256}{3} \big(S_2-S_1^2\big)
\nonumber \\ &&
-\frac{128 \big(10 N^2+22 N-9\big)}{9 N (N+1)} S_1
-\frac{64 P_{72}}{9 (N-2) N^2 (N+1)^2 (N+3)}
\biggr) S_{-2}
+\frac{352}{3} S_4
\nonumber \\ &&
+\frac{448}{3} S_{-4}
+\frac{16 \big(9 N^2+9 N-2\big)}{3 N (N+1)} S_{2,1}
+64 S_{3,1}
+\frac{128 \big(10 N^2+22 N-9\big)}{9 N (N+1)} S_{-2,1}
\nonumber \\ &&
-\frac{256}{3} S_{-3,1}
-64 S_{2,1,1}
-\frac{512}{3} S_{-2,1,1}
+\biggl(
64 S_1
-\frac{16 \big(9 N^2-7 N+6\big)}{N (N+1)}
\biggr) \zeta_3
\biggl] L_Q 
\biggr\}
\nonumber \\ &&
+\textcolor{blue}{C_F T_F^2} \biggl\{
\biggl(
\frac{8 P_{10}}{27 N^2 (N+1)^2}
-\frac{320}{27} S_1
+\frac{64}{9} S_2
\biggr) L_M^2
-\frac{2 P_{50}}{729 N^4 (N+1)^4}
+\frac{64}{81} S_2
\nonumber \\ &&
+\frac{12064}{729} S_1
+\biggl(
\frac{32}{9} S_1^2
+\frac{16 \big(29 N^2+29 N-6\big)}{27 N (N+1)} S_1
-\frac{8 P_{60}}{27 N^2 (N+1)^2}
-\frac{32}{3} S_2
\biggr) L_Q^2
\nonumber \\ &&
+\frac{320}{81} S_3
-\frac{64}{27} S_4
+\biggl[
-\frac{896}{27} S_3
-\frac{16 \big(29 N^2+29 N-6\big)}{27 N (N+1)} S_1^2
+\frac{8 P_{77}}{81 N^3 (N+1)^3}
\nonumber \\ &&
-\frac{16 P_{62}}{81 N^2 (N+1)^2} S_1
+\frac{16 \big(35 N^2+35 N-2\big)}{9 N (N+1)} S_2
+\frac{64}{27} \big(6 S_{2,1}-S_1^3+3 S_1 S_2 \big)
\biggr] L_Q
%
%
%
\nonumber \\ &&
+{\gamma_{qq}^{0}} \biggl(
-\frac{16}{27} L_M^3
-\frac{248}{81} L_M
-\frac{8}{27} L_Q^3
-\frac{112}{27} \zeta_3
\biggr)
\biggr\}
+\textcolor{blue}{N_F C_F T_F^2} \biggl\{
\biggl(
\frac{64}{9} S_1^2
-\frac{64}{3} S_2
\nonumber \\ &&
-\frac{16 P_{60}}{27 N^2 (N+1)^2}
+\frac{32 \big(29 N^2+29 N-6\big)}{27 N (N+1)} S_1
\biggr) L_Q^2
+\biggl[
\frac{16 P_{77}}{81 N^3 (N+1)^3}
-\frac{128}{27} S_1^3
\nonumber \\ &&
-\frac{32 \big(29 N^2+29 N-6\big)}{27 N (N+1)} S_1^2
+\biggl(
\frac{128}{9} S_2
-\frac{32 P_{62}}{81 N^2 (N+1)^2}
\biggr) S_1
-\frac{1792}{27} S_3
\nonumber \\ &&
+\frac{256}{9} S_{2,1}
+\frac{32 \big(35 N^2+35 N-2\big)}{9 N (N+1)} S_2
\biggr] L_Q
+\frac{4 P_{49}}{729 N^4 (N+1)^4}
-\frac{24064}{729} S_1
\nonumber \\ &&
+\frac{128}{81} S_2
+\frac{640}{81} S_3
+\biggl(
\frac{4 P_{44}}{81 N^3 (N+1)^3}
-\frac{2176}{81} S_1
-\frac{320}{27} S_2
+\frac{64}{9} S_3
\biggr) L_M
\nonumber \\ &&
-\frac{128 S_4}{27}
+{\gamma_{qq}^{0}} \biggl(
-\frac{8}{27} L_M^3
-\frac{16}{27} L_Q^3
+\frac{64}{27} \zeta_3
\biggr)
\biggr\}
+\textcolor{blue}{C_A C_F T_F} \biggl\{
\biggl(
-\frac{80}{9} S_3
+\frac{128}{9} S_{2,1}
\nonumber \\ &&
+\frac{32}{9 N (N+1)} S_2
+\frac{4 P_{36}}{9 N^3 (N+1)^3}
+\frac{128}{9} S_{-2,1}
\biggr) S_1^2
+\biggl[
\frac{80 S_3 (2 N+1)^2}{9 N (N+1)}
+\frac{112}{9} S_2^2
\nonumber \\ &&
-\frac{4 P_{52}}{729 N^4 (N+1)^4}
-\frac{16 (N-1) \big(2 N^3-N^2-N-2\big)}{9 N^2 (N+1)^2} S_2
-\frac{208}{9} S_4
-\frac{512}{9} S_{-2,1,1}
\nonumber \\ &&
-\frac{8 \big(9 N^2+9 N+16\big)}{9 N (N+1)} S_{2,1}
+\frac{128 \big(10 N^2+10 N-3\big)}{27 N (N+1)} S_{-2,1}
+\frac{128}{9} S_{-2,2}
\nonumber \\ &&
+\frac{64}{3} S_{3,1}
\biggr] S_1
-\frac{64}{27} S_2 S_1^3
-\frac{4 \big(15 N^2+15 N+14\big)}{9 N (N+1)} S_2^2
+\frac{P_{54}}{729 N^5 (N+1)^5}
\nonumber \\ &&
+\frac{8 \big(2 N^3+2 N^2+2 N+1\big)}{3 N^3 (N+1)^3} \zeta_2
+S_{-4} \biggl(
\frac{64 \big(10 N^2+10 N+3\big)}{27 N (N+1)}
-\frac{128}{9} S_1
\biggr)
\nonumber \\ &&
+\biggl(
-\frac{64}{9} S_1^2
+\frac{64 \big(10 N^2+10 N+3\big)}{27 N (N+1)} S_1
-\frac{32 \big(112 N^3+224 N^2+169 N+39\big)}{81 N (N+1)^2}
\nonumber \\ &&
-\frac{64}{9} S_2
\biggr) S_{-3}
+(-1)^N \biggl[
\frac{32}{3 (N+1)^3} L_M^2
+\biggl(
\frac{128 (4 N+1)}{9 (N+1)^4}
-\frac{64}{3 (N+1)^3} S_1
\biggr) L_M
\nonumber \\ &&
+\frac{32}{3 (N+1)^3} L_Q^2
-\frac{32 P_{85}}{45 (N-2) (N-1)^2 N^2 (N+1)^4 (N+2)^2 (N+3)^3} L_Q
\nonumber \\ &&
-\frac{8 \big(2 N^3+2 N^2+2 N+1\big)}{3 N^3 (N+1)^3} \zeta_2
+\frac{16 P_{24}}{27 N^4 (N+1)^4} S_1
-\frac{32 \big(2 N^2+2 N+1\big)}{9 N^3 (N+1)^3} S_2
\nonumber \\ &&
-\frac{32 \big(2 N^2+2 N+1\big)}{9 N^3 (N+1)^3} S_1^2
-\frac{8 P_{28}}{81 N^5 (N+1)^5}
\biggr]
-\frac{8 P_{37}}{81 N^2 (N+1)^2} S_3
-\frac{128}{9} S_{-5}
\nonumber \\ &&
-\frac{224}{9} S_5
+\frac{4 \big(443 N^2+443 N+78\big)}{27 N (N+1)} S_4
-\frac{8 P_{20}}{9 N^2 (N+1)^2} S_{2,1}
+\biggl[
-\frac{256}{27} S_3
\nonumber \\ &&
+\frac{256}{9} S_{2,1}
-\frac{128}{27} S_1^3
+\frac{64}{9 N (N+1)} S_1^2
-\frac{64 P_{13}}{81 N^2 (N+1)^2} S_1
+\frac{32 P_{14}}{81 N^3 (N+1)^3}
\nonumber \\ &&
+\frac{640}{27} S_2
\biggr] S_{-2}
-\frac{128}{3} S_{2,3}
+\frac{256}{9} S_{2,-3}
-\frac{8 (13 N+4) (13 N+9)}{27 N (N+1)} S_{3,1}
+\frac{256}{9} S_{4,1}
\nonumber \\ &&
+S_2 \biggl(
\frac{4 P_{45}}{81 N^3 (N+1)^3}
+\frac{496}{27} S_3
-\frac{64}{3} S_{2,1}
-\frac{128}{3} S_{-2,1}
\biggr)
+\biggl[
-\frac{32}{3} S_3
-\frac{32}{3} S_{-3}
\nonumber \\ &&
+\frac{64}{3} S_{-2,1}
+\frac{2 P_{59}}{9 N (N+1)^3}
+S_{-2} \biggl(
\frac{32}{3 N (N+1)}
-\frac{64}{3} S_1
\biggr)
-\frac{184}{9} S_1
\biggr] L_M^2
\nonumber \\ &&
+\frac{64 \big(112 N^3+112 N^2-39 N+18\big)}{81 N^2 (N+1)} S_{-2,1}
+\biggl[
-\frac{16 \big(194 N^2+194 N-33\big)}{27 N (N+1)} S_1
\nonumber \\ &&
-\frac{176}{9} S_1^2
+\frac{2 P_{69}}{27 N^2 (N+1)^3}
+S_{-2} \biggl(
\frac{32}{3 N (N+1)}
-\frac{64}{3} S_1
\biggr)
+\frac{176}{3} S_2
-\frac{32}{3} S_3
\nonumber \\ &&
-\frac{32}{3} S_{-3}
+\frac{64}{3} S_{-2,1}
\biggr] L_Q^2
+\frac{64 \big(10 N^2+10 N-3\big)}{27 N (N+1)} S_{-2,2}
-\frac{256}{9} S_{-2,3}
-\frac{256}{9} S_{2,1,-2}
\nonumber \\ &&
+\frac{64}{3} S_{2,2,1}
-\frac{256}{9} S_{3,1,1}
+\biggl[
S_{-2} \biggl(
\frac{32 \big(16 N^2+10 N-3\big)}{9 N^2 (N+1)^2}
-\frac{640}{9} S_1
+\frac{64}{3} S_2
\biggr)
\nonumber \\ &&
+\frac{P_{81}}{81 N^3 (N+1)^4}
+S_{-3} \biggl(
\frac{64}{3} S_1
-\frac{32 \big(10 N^2+10 N+3\big)}{9 N (N+1)}
\biggr)
+\frac{1792}{27} S_2
+\frac{160}{3} S_4
\nonumber \\ &&
-\frac{16 \big(31 N^2+31 N+9\big)}{9 N (N+1)} S_3
+\frac{64}{3} S_{-4}
-\frac{128}{3} S_{3,1}
+\frac{64 \big(10 N^2+10 N-3\big)}{9 N (N+1)} S_{-2,1}
\nonumber \\ &&
+S_1 \biggl(
-\frac{8 P_{74}}{81 N^3 (N+1)^3}
+32 S_3
+\frac{128}{3} S_{-2,1}
\biggr)
+\frac{64}{3} S_{-2,2}
-\frac{256}{3} S_{-2,1,1}
\biggr] L_M
\nonumber \\ &&
-\frac{256 \big(10 N^2+10 N-3\big)}{27 N (N+1)} S_{-2,1,1}
-\frac{256}{9} S_{-2,2,1}
+\frac{224}{9} S_{2,1,1,1}
+\frac{1024}{9} S_{-2,1,1,1}
\nonumber \\ &&
+\biggl(
-16 S_1^2
+\frac{4 \big(593 N^2+593 N+108\big)}{27 N (N+1)} S_1
+\frac{P_{38}}{27 N^2 (N+1)^2}
+16 S_2
\biggr) \zeta_3
\nonumber \\ &&
+{\gamma_{qq}^{0}} \biggl(
\frac{22}{27} L_M^3
-8 \zeta_3 L_M
+\frac{44}{27} L_Q^3
+\frac{4}{3} B_4
-12 \zeta_4
-4 S_{2,1,1}
\biggr)
+\biggl[
\frac{352}{27} S_1^3
-\frac{32}{3} S_2^2
\nonumber \\ &&
+\biggl(
\frac{16 \big(194 N^2+194 N-33\big)}{27 N (N+1)}
+\frac{32}{3} S_2
\biggr) S_1^2
+\biggl(
-\frac{32 \big(11 N^2+11 N+3\big)}{9 N (N+1)} S_2
\nonumber \\ &&
+\frac{4 P_{80}}{81 N^3 (N+1)^3}
+32 S_3
-\frac{128}{3} S_{2,1}
-\frac{256}{3} S_{-2,1}
\biggr) S_1
-32 S_{-2}^2
+S_{-3} \biggl(
-\frac{128}{3} S_1
\nonumber \\ &&
+\frac{32 \big(10 N^2+22 N+3\big)}{9 N (N+1)}
\biggr)
-\frac{4 P_{86}}{405 (N-1)^2 N^3 (N+1)^4 (N+2)^2 (N+3)^3}
\nonumber \\ &&
+S_{-2} \biggl(
\frac{128}{3} S_1^2
+\frac{64 \big(10 N^2+22 N-9\big)}{9 N (N+1)} S_1
+\frac{32 P_{72}}{9 (N-2) N^2 (N+1)^2 (N+3)}
\nonumber \\ &&
-\frac{128}{3} S_2
\biggr)
-\frac{16 \big(230 N^3+460 N^2+213 N-11\big)}{9 N (N+1)^2} S_2
-\frac{64 \big(11 N^2+11 N-3\big)}{9 N (N+1)} S_{2,1}
\nonumber \\ &&
+\frac{16 \big(368 N^2+440 N-45\big)}{27 N (N+1)} S_3
-\frac{224}{3} \big(S_4+S_{-4}\big)
-\frac{64 \big(10 N^2+22 N-9\big)}{9 N (N+1)} S_{-2,1}
\nonumber \\ &&
+\frac{128}{3} S_{-3,1}
+64 S_{2,1,1}
+\frac{256}{3} S_{-2,1,1}
+\biggl(
\frac{32 \big(3 N^2-N+2\big)}{N (N+1)}
-64 S_1
\biggr) \zeta_3
\nonumber \\ &&
-\frac{64}{3} S_{3,1}
\biggr] L_Q
\biggr\}
+\hat{C}_{q,2}^{(3), \rm NS}(N_F)
\Biggr\} \Biggr\}. 
\end{eqnarray}
The massless Wilson coefficient ${C}_{q,2}^{(3), \rm NS}(N_F)$ has been given in \cite{Vermaseren:2005qc} and the polynomials 
$P_i$ read
\begin{eqnarray}
P_{55} &=& -3 N^4-6 N^3-47 N^2-20 N+12, \\
P_{56} &=& 19 N^4+38 N^3-9 N^2-20 N+4, \\
P_{57} &=& 28 N^4+56 N^3+28 N^2+2 N+1, \\
P_{58} &=& 33 N^4+38 N^3-15 N^2-60 N-28, \\
P_{59} &=& 51 N^4+153 N^3+223 N^2+143 N+70 \\
P_{60} &=& 57 N^4+72 N^3+29 N^2-22 N-24, \\
P_{61} &=& 141 N^4+198 N^3+169 N^2-32 N-84, \\
P_{62} &=& 235 N^4+596 N^3+319 N^2+66 N+72, \\
P_{63} &=& 359 N^4+844 N^3+443 N^2+66 N+72, \\
P_{64} &=& 501 N^4+750 N^3+325 N^2-188 N-204 \\
P_{65} &=& 1131 N^4+1926 N^3+1019 N^2-64 N-276, \\
P_{66} &=& 1139 N^4+3286 N^3+1499 N^2+504 N+828, \\
P_{67} &=& 1199 N^4+2398 N^3+1181 N^2+18 N+90, \\
P_{68} &=& 1220 N^4+2251 N^3+1772 N^2+303 N-138, \\
P_{69} &=& 1407 N^5+3297 N^4+2891 N^3+583 N^2-802 N-528, \\
P_{70} &=& -11145 N^6-30915 N^5-33923 N^4-11449 N^3+3112 N^2+120 N-1512, \\
P_{71} &=& -151 N^6-469 N^5-181 N^4+305 N^3+208 N^2+40 N+8, \\
P_{72} &=& 6 N^6-6 N^5-25 N^4+52 N^3-46 N^2-39 N-162, \\
P_{73} &=& 15 N^6+24 N^5-88 N^3-79 N^2-52 N-12, \\
P_{74} &=& 155 N^6+465 N^5+465 N^4-61 N^3-324 N^2-324 N-162 \\
P_{75} &=& 216 N^6+459 N^5+417 N^4-99 N^3-317 N^2-272 N-84, \\
P_{76} &=& 309 N^6+647 N^5+293 N^4-975 N^3-1102 N^2-316 N+24, \\
P_{77} &=& 609 N^6+1029 N^5+613 N^4-37 N^3-74 N^2+300 N+216, \\
P_{78} &=& 795 N^6+1587 N^5+1295 N^4+397 N^3+50 N^2+300 N+216, \\
P_{79} &=& 1770 N^6+4731 N^5+4483 N^4+749 N^3+55 N^2+1440 N+756, \\
P_{80} &=& 7531 N^6+26121 N^5+27447 N^4+8815 N^3+1110 N^2+936 N-324, \\
P_{81} &=& -4785 N^7-19140 N^6-19186 N^5-4584 N^4+1491 N^3-4540 N^2
\nonumber \\ && 
-1536 N+792 \\
P_{82} &=& -45 N^8-138 N^7-678 N^6+836 N^5+1615 N^4+1702 N^3+380 N^2
\nonumber \\ && 
-408 N-192, \\
P_{83} &=& 42591 N^8+161388 N^7+226272 N^6+104062 N^5-40175 N^4-43450 N^3-3928 N^2
\nonumber \\ && 
-1272 N-2160, \\
P_{84} &=& -18351 N^{10}-87156 N^9-196947 N^8-239766 N^7-157693 N^6-26288 N^5
\nonumber \\ && 
+17847 N^4+7490 N^3+2248 N^2+1896 N+144, \\
P_{85} &=& 101 N^{11}+1268 N^{10}+4423 N^9+908 N^8-20681 N^7-19546 N^6+52505 N^5
\nonumber \\ && 
+83160 N^4-4668 N^3-38934 N^2-2592 N-648, \\
P_{86} &=& 41370 N^{14}+571305 N^{13}+3141790 N^{12}+8395028 N^{11}+9302220 N^{10}-4510326 N^9
\nonumber \\ && 
-22388388 N^8-17101704 N^7+7895114 N^6+18219253 N^5+4736406 N^4
\nonumber \\ && 
-5978772 N^3-1986336 N^2+2361312 N+1283040, \\
P_{87} &=& 4140 N^{15}+54540 N^{14}+277575 N^{13}+634467 N^{12}+354380 N^{11}-1199584 N^{10}
\nonumber \\ && 
-2051492 N^9+733454 N^8+4802206 N^7+3686432 N^6-1882531 N^5-3693633 N^4
\nonumber \\ && 
-1066014 N^3+869508 N^2+897480 N+233280. 
\end{eqnarray}
The Wilson cofficient depends on the logarithms
\begin{eqnarray}
L_Q = \ln\left(\frac{Q^2}{\mu^2}\right)~~~\text{and}~~~L_M = \ln\left(\frac{m^2}{\mu^2}\right),
\end{eqnarray}
where the renormalization and factorization scales $\mu_F = \mu_R \equiv \mu$ have been set equal.
\begin{figure}[H]
\centering
\includegraphics[width=0.5\textwidth]{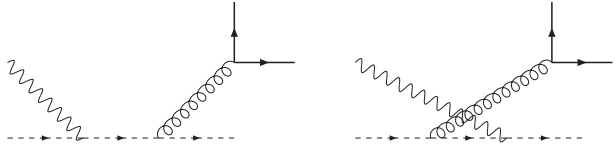}
\caption{\sf \small The diagrams contributing to Eq.~(\ref{eq:tag}) for $x < 1$ at $O(a_s^2)$.The dashed (full) lines denote 
massless (massive) quarks, respectively.} 
\label{QCDCOMP}
\end{figure}

\noindent
The corresponding expression in $x$-space is given in Appendix~\ref{LqqNSx} in terms of harmonic polylogarithms.
The analytic continuation can also be performed from the expressions in $N$-space directly, referring to their asymptotic
representation and recurrence relations, cf.~\cite{ANCONT}. Let us comment on the pole-structure of Eq.~(\ref{WIL:NS}).
Because (\ref{WIL:NS}) is a flavor non-singlet quantity, one expects the rightmost pole to be situated at $N = 0$.
Performing expansions around $N = 2$ and $N = 1$ one finds that the Wilson coefficient is finite there and the evanescent poles 
cancel.

\begin{figure}[H]
\centering
\includegraphics[width=0.22\textwidth]{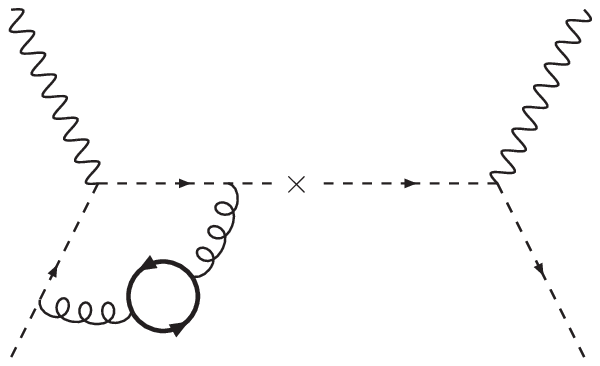} \hspace*{5mm}
\includegraphics[width=0.1\textwidth]{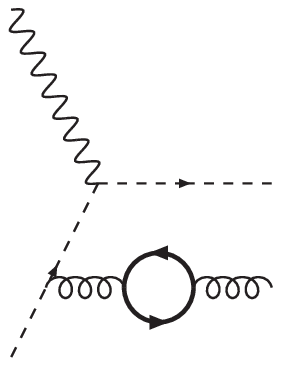} \hspace*{5mm}
\includegraphics[width=0.1\textwidth]{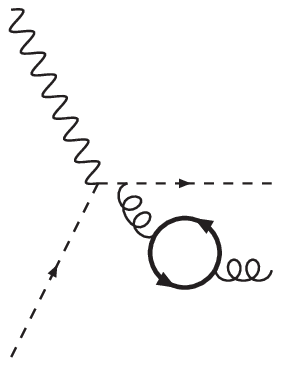}
\caption{\sf \small The diagrams of the virtual heavy flavor corrections at 2-loop order, Eq.~(\ref{eq:diff2}).
The self-energy corrections, which are not shown here, have to be added.}
\label{ONELB0Q}
\end{figure}

To compare with results in the foregoing literature, we would like to consider the results of Ref.~\cite{Buza:1995ie} to 
$O(a_s^2)$. Here the non-singlet heavy flavor Wilson coefficient was calculated in the case
of tagged heavy quarks, given by the QCD-Compton process shown in Figure~\ref{QCDCOMP}. 
This scattering process
is non-singular and can be calculated analytically for any ratio $m^2/Q^2$. The authors of \cite{Buza:1995ie}
have given the following representation for $Q^2 \gg m^2$ in terms of the massive OME and massless Wilson 
coefficients
\begin{eqnarray}
\label{eq:tag}
L_{qq,Q}^{(2), \rm NS, tagged} &=& 
a_s^2\Biggl\{\frac{1}{4} \beta_{0,Q} \gamma_{qq}^0 \ln^2\left(\frac{m^2}{\mu^2}\right) + \frac{1}{2} 
\hat{\gamma}_{qq}^{(1),\rm NS} \ln\left(\frac{m^2}{\mu^2}\right) 
\nonumber\\ && 
+ a_{qq,Q}^{(2),\rm NS} - \frac{1}{4} \beta_{0,Q} \zeta_2 \gamma_{qq}^0
\nonumber\\ &&
+ \frac{1}{4} \beta_{0,Q} \gamma_{qq}^0 \ln^2\left(\frac{Q^2}{\mu^2}\right) 
- \left[\frac{1}{2} \hat{\gamma}_{qq}^{(1),\rm NS} + \beta_{0,Q}
  c_{2,q}^{(1)} 
\right] 
\ln\left(\frac{Q^2}{\mu^2}\right) + \hat{c}_{2,q}^{(2),\rm NS}
\nonumber\\ &&
+\beta_{0,Q} \left[- \frac{1}{2} \gamma_{qq}^0 \ln\left(\frac{Q^2}{\mu^2}\right) +c_{2,q}^{(1)}\right] 
\ln\left(\frac{m^2}{\mu^2}\right) \Biggr\},
\nonumber\\ &=&
a_s^2 \Biggl\{\frac{1}{4} \beta_{0,Q} \gamma_{qq}^0 
\ln^2\left(\frac{Q^2}{m^2}\right) - \left[
\frac{1}{2} \hat{\gamma}_{qq}^{(1), \rm NS}  + \beta_{0,Q} c_{2,q}^{(1)} 
\right] \ln\left(\frac{Q^2}{m^2}\right)
\nonumber\\ &&
+ \hat{c}_{2,q}^{(2), \rm NS} + a_{qq,Q}^{(2),\rm NS} - \frac{1}{4} \beta_{0,Q} \zeta_2 \gamma_{qq}^0\Biggr\}.    
\end{eqnarray}
Note that in Eq.~(\ref{WIL:NS}) the complete heavy flavor contributions were given, i.e. including also those
due to virtual effects at 2-loop order. The difference term stems from the graphs shown in Figure~\ref{ONELB0Q}.

In these graphs the final state contains massless partons only. 
However, the heavy flavor contributions to the deep-inelastic structure functions can only be defined consistently as the

\begin{figure}[H]
\centering
\includegraphics[width=0.8\textwidth]{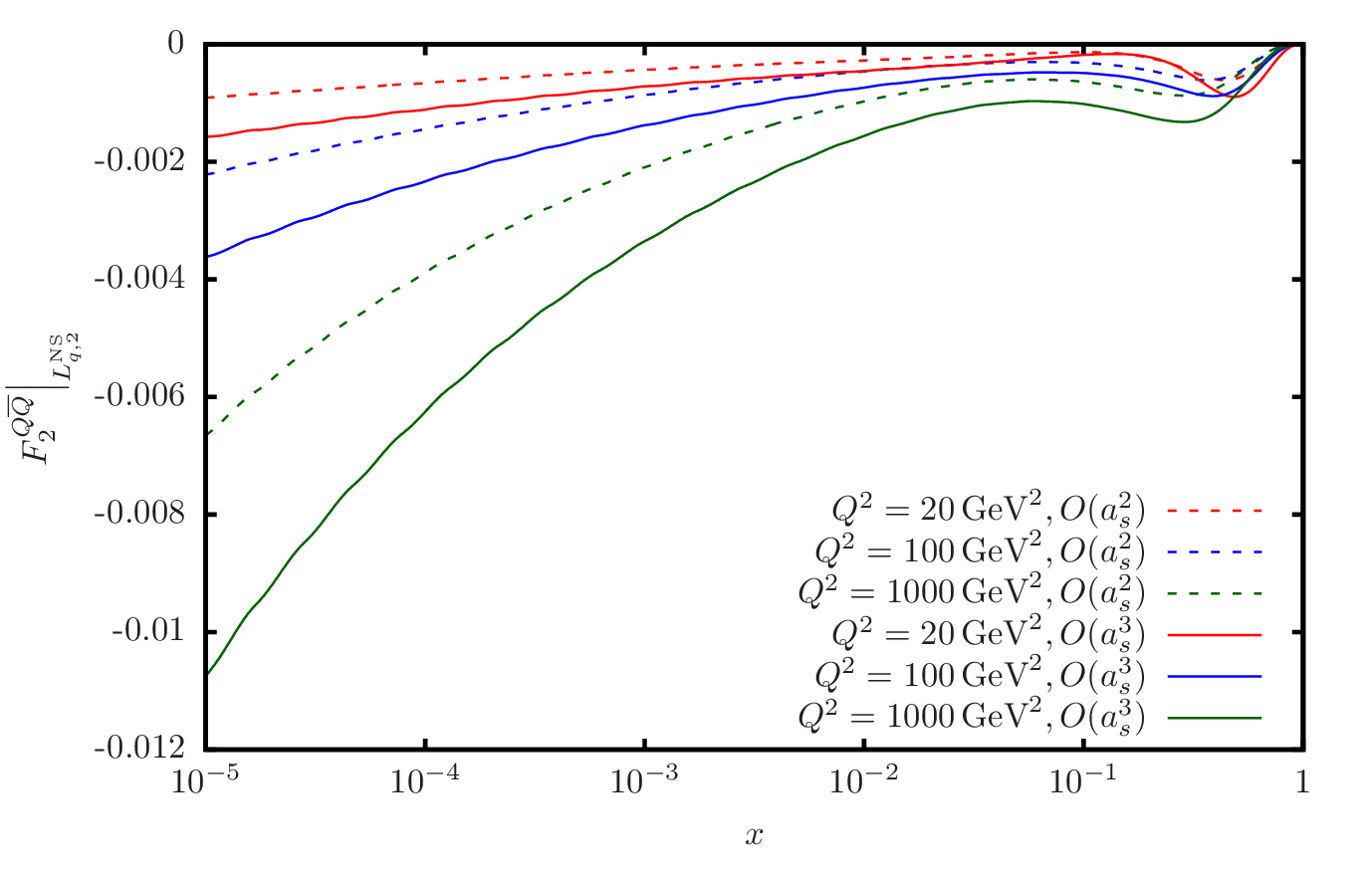}
\caption{\sf \small The flavor non-singlet contribution of the Wilson coefficient $L_{q,2}^{\rm NS}$ to the structure function
$F_2(x,Q^2)$ for the  2- and up to the 3-loop order using the NNLO parton distribution functions of Ref.~\cite{Alekhin:2013nda} 
in the on-shell 
scheme for $m_c = 1.59~\GeV$.  Here and in Figures~\ref{Fig:WILS2} and \ref{Fig:WILS3} we do not display the $O(a_s^0)$ 
terms.}
\label{Fig:WILS1}
\end{figure}
\begin{figure}[H]
\centering
\includegraphics[width=0.8\textwidth]{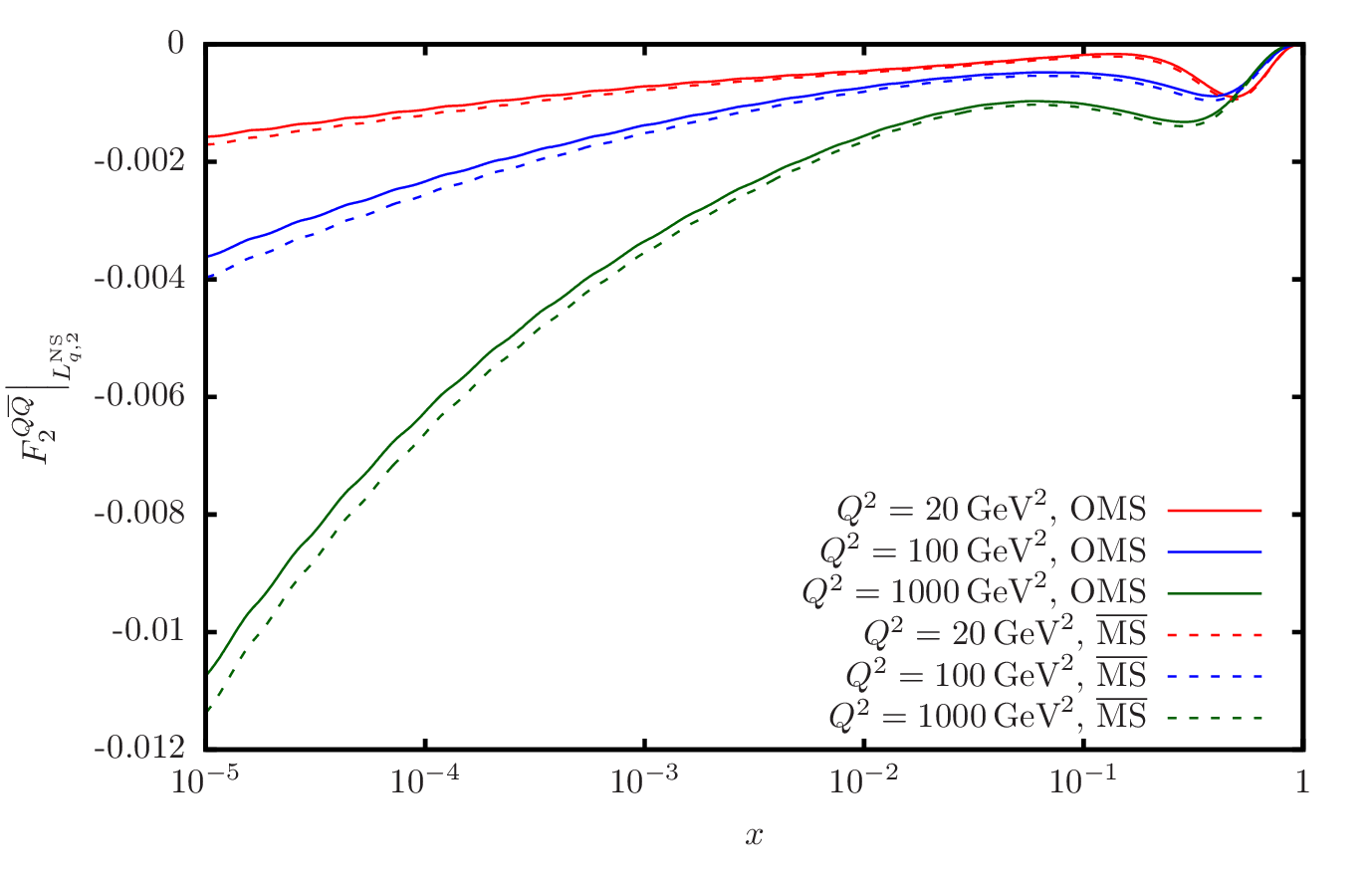}
\caption{\sf \small The flavor non-singlet contribution of the Wilson coefficient $L_{q,2}^{\rm NS}$ to the structure function
$F_2(x,Q^2)$ at 3-loop order comparing the prediction in the on-shell scheme for $m_c = 1.59~\GeV$ (dashed lines)
to the $\overline{\rm MS}$ scheme for $m_c = 1.24~\GeV$, cf. \cite{Alekhin:2012vu}, using the NNLO parton distribution 
functions of Ref.~\cite{Alekhin:2013nda}.} 
\label{Fig:WILS2}
\end{figure}

\noindent
difference between the structure function involving all light and heavy quark contributions and an expression for the structure
function considering only light flavors. The diagrams of Figure~\ref{ONELB0Q}
belong to the heavy flavor contributions at $O(a_s^2)$. In the asymptotic case $\mu^2 \gg m^2$
their contribution is given by 
\begin{eqnarray}
- a_s^2 \beta_{0,Q} \ln\left(\frac{m^2}{\mu^2}\right) 
\left[- \frac{1}{2} \gamma_{qq}^{(0)} \ln\left(\frac{Q^2}{\mu^2}\right) + c_{2,q}^{(1)}\right].
\label{eq:diff2}
\end{eqnarray}
These terms exhibit a collinear singularity, being reflected in the $\mu$-dependence of this contribution. Adding 
both terms yields Eq.~(\ref{eq:LQ2}) at 2-loop order.

In Figure~\ref{Fig:WILS1} we illustrate the contribution of the heavy flavor  non-singlet Wilson coefficient Eq.~(\ref{WIL:NS}) 
to 
the structure function $F_2(x,Q^2)$ accounting for the charm quark contributions at 2- and 3-loop order. The numerical 
calculation has been performed using the $x$-space 
representations given in Appendix~\ref{LqqNSx} and the numerical representation of the harmonic polylogarithms given in 
\cite{Gehrmann:2001pz}\footnote{For other implementations see \cite{Vollinga:2004sn,Maitre:2007kp,Buehler:2011ev}.}.

The contributions to $O(a_s^2)$ are significantly enhanced adding the $O(a_s^3)$ terms due to the gluonic and sea-quark 
contributions, despite the smallness of the additional factor $a_s$. Overall, the NNLO corrections are below $1\%$. Note,
however, that the present experimental precision for the structure function $F_2(x,Q^2)$ reaches $O(1\%)$. It will be even 
improved at high-luminosity facilities like the EIC \cite{Boer:2011fh} in the future.

In Figure~\ref{Fig:WILS2} we compare the predictions for the charm mass definition in the on-shell scheme
and in the $\overline{\rm MS}$ scheme. The corrections in the $\overline{\rm MS}$ scheme lead to slightly larger 
absolute values than those in the on-shell scheme with very similar $x$-shapes.

\begin{figure}[H]
\centering
\includegraphics[width=0.8\textwidth]{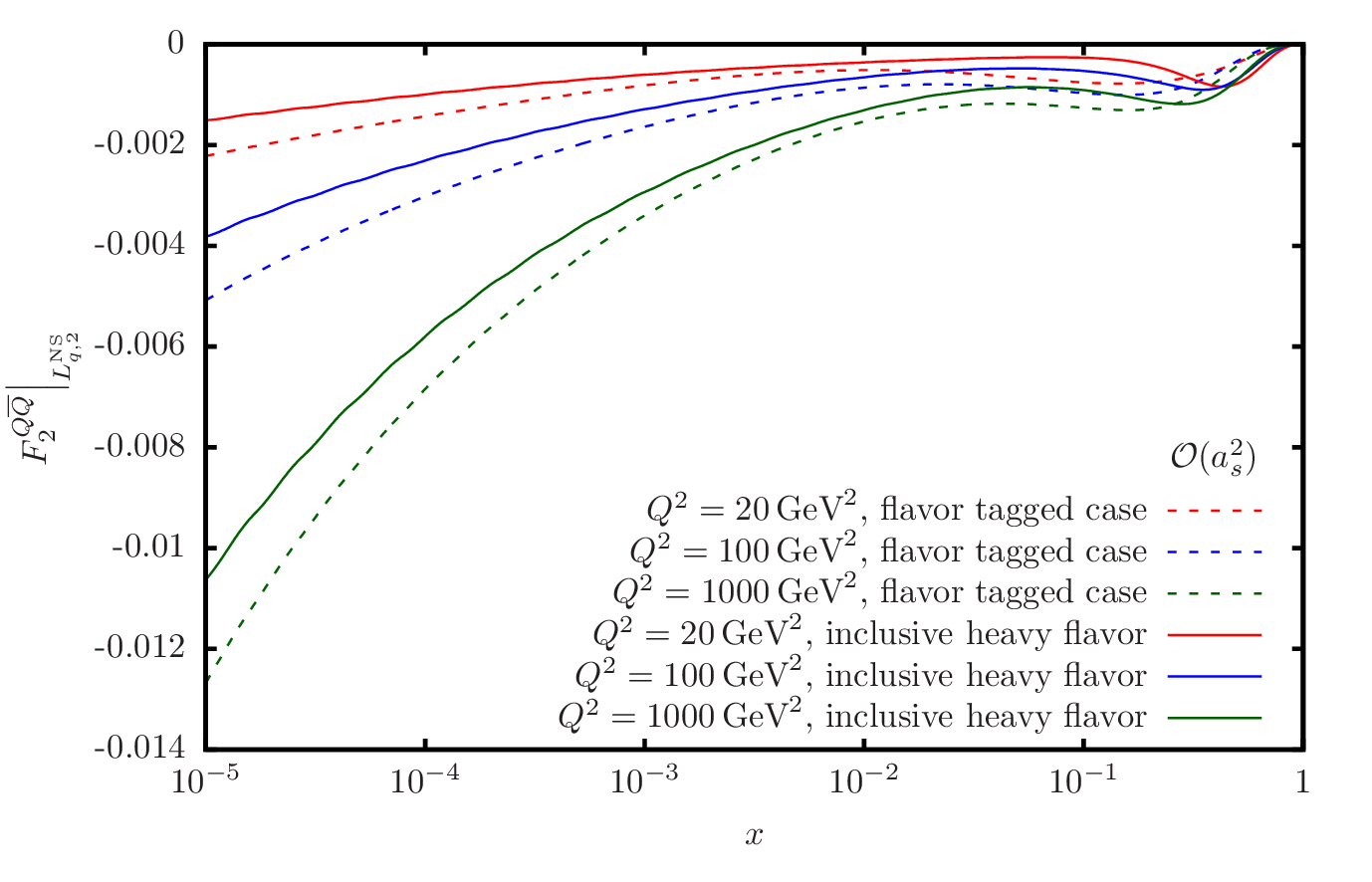}
\caption{\sf \small Comparison of the prediction of the flavor non-singlet charm contribution to $F_2(x,Q^2)$ at 2-loop order
in case of tagged charm (dashed lines) and the inclusive charm contribution (full lines) using the NLO parton distribution 
functions of Ref.~\cite{Alekhin:2012ig} in the $\overline{\rm MS}$ scheme with $m_c = 1.15~\GeV$.} 
\label{Fig:WILS3}
\end{figure}

The effect of retaining only the tagged heavy flavor contributions Eq.~(\ref{eq:tag}) is compared to the complete 
charm quark contributions to 2-loop order in Figure~\ref{Fig:WILS3}. Here we set the factorization and renormalization  
scales to $\mu^2 =  Q^2$. The absolute value of the correction is larger in the tagged case, compared to the inclusive
corrections. We note that a separation of this kind at 3-loop order is in general not possible, without defining additional 
cuts, referring to the structure function approach as outlined in Ref.~\cite{Bierenbaum:2009mv}.
\section{The Variable Flavor Number Scheme}
\label{sec:7}

\vspace{1mm}
\noindent
The transition relation from $N_F \rightarrow N_F + 1$ massless flavors of the flavor non-singlet distribution in the variable 
flavor number scheme is described by \cite{Bierenbaum:2009mv}
\begin{eqnarray}
f_k(N_F + 1, \mu^2) + \bar{f}_k(N_F + 1, \mu^2) 
&=&
A_{qq,Q}^{\rm NS} \otimes \left[f_k(N_F, \mu^2) + \bar{f}_k(N_F, \mu^2)\right]
\nonumber\\ &&
+ \frac{1}{N_F} \left\{A_{qq,Q}^{\rm PS} \otimes \Sigma(N_F,\mu^2) 
+ A_{qg,Q}^{\rm S} \otimes G(N_F,\mu^2)\right\}, 
\label{eq:VFNS}
\nonumber\\ 
\end{eqnarray}
\begin{figure}[H]
\centering
\includegraphics[width=0.8\textwidth]{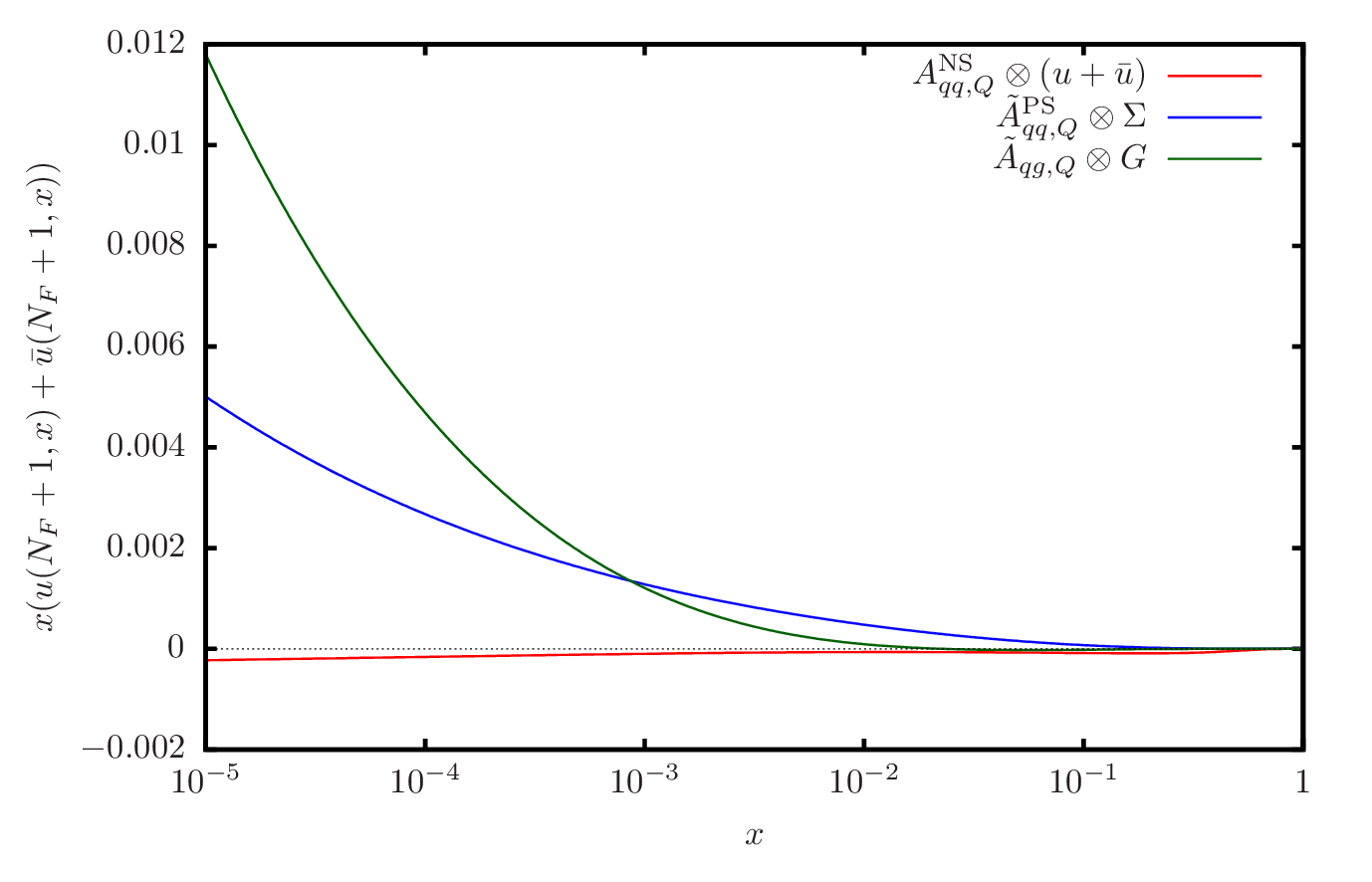}
\caption{\sf \small The contributions to the distribution $x(u+\bar{u})$ at 3-loop order for four flavors in the variable 
flavor number scheme 
matched at the scale $\mu^2 = 20~\GeV^2$ using the parton distribution functions of Ref.~\cite{Alekhin:2013nda} and the 
on-mass-shell definition of the charm quark mass $m_c = 1.59~\GeV$. The contributions due to 
the non-singlet, singlet and gluon distributions are shown individually. Here only the $O(a_s^3)$ terms are 
shown.}
\label{Fig:VFNS1}
\end{figure}

\noindent
where $\mu^2$ denotes the matching scale. All the OMEs contributing to Eq.~(\ref{eq:VFNS}) are now known to 
3-loop order, cf. also 
\cite{Ablinger:2010ty,Behring:2014eya}, which enables us to study its quantitative effects. The choice of the transition scale
is process-dependent and usually significantly larger than the mass of the quark becoming light are requested to match the 
corresponding scattering cross sections, cf. \cite{Blumlein:1998sh}.
This is obvious, because a heavy quark near production threshold is non-relativistic.

In Figure~\ref{Fig:VFNS1}, we illustrate the flavor non-singlet, singlet and gluonic contributions to 3-loop order to the 
4-flavor distribution $x(u(x,\mu^2) + \bar{u}(x,\mu^2))$ as a function of $x$ for $\mu^2 = 20~\GeV^2$. The flavor non-singlet 
contributions are very small and show a weak $x$-dependence only, while the universal singlet and gluon 
distributions grow towards small
values of $x$ and are much larger. Due to this the distributions for the down and strange quarks are nearly the same. 

\begin{figure}[H]
\centering
\includegraphics[width=0.8\textwidth]{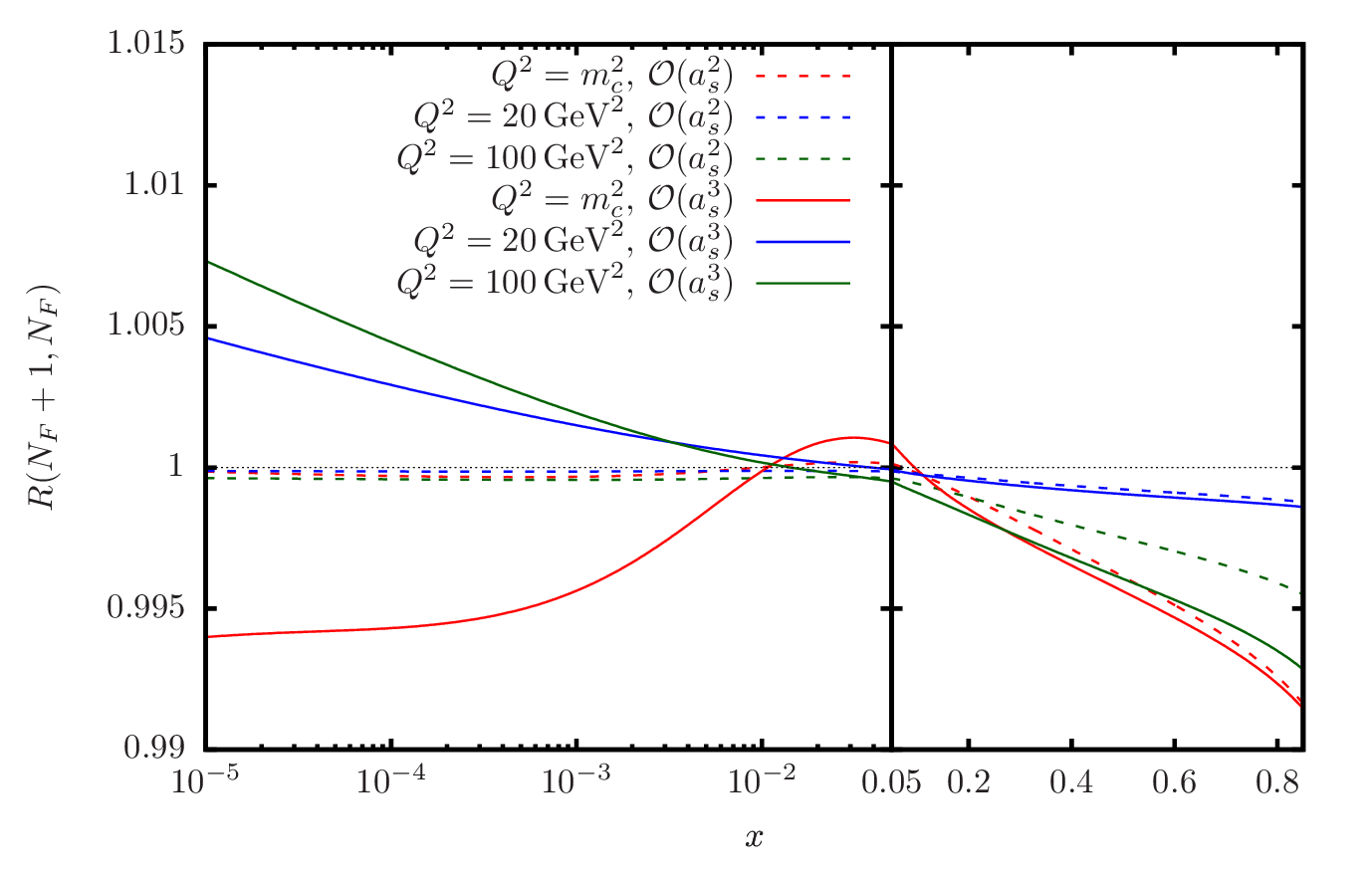}
\caption{\sf \small The ratio of the 
distribution $x(u+\bar{u})$ for four and three quark flavors at 2- and 3-loop order in the variable flavor 
number scheme matched at different scales of $Q^2$ as a function of $x$ using the parton distribution functions of 
Ref.~\cite{Alekhin:2013nda} and the on-mass-shell 
definition of the charm quark mass $m_c = 1.59~\GeV$.} 
\label{Fig:VFNS2}
\end{figure}

In Figure~\ref{Fig:VFNS2} the ratio of the momentum distribution function $x(u + \bar{u})$ is shown comparing the 4- and 
3-flavor scheme at NNLO for a variety of matching scales in the VFNS to $O(a_s^2)$ and $O(a_s^3)$. 
The ratio $R(N_F+1, N_F)$ is defined by
\begin{eqnarray}
R(N_F+1, N_F) = 
\frac{f_k(N_F+1,\mu^2) + \bar{f}_k(N_F+1,\mu^2)}{f_k(N_F,\mu^2) + \bar{f}_k(N_F,\mu^2)},
\label{eq:RNF}
\end{eqnarray}
where the numerator is given by Eq.~(\ref{eq:VFNS}).
At $O(a_s^3)$ the effects 
become larger, in particular 

\begin{figure}[H]
\centering
\includegraphics[width=0.8\textwidth]{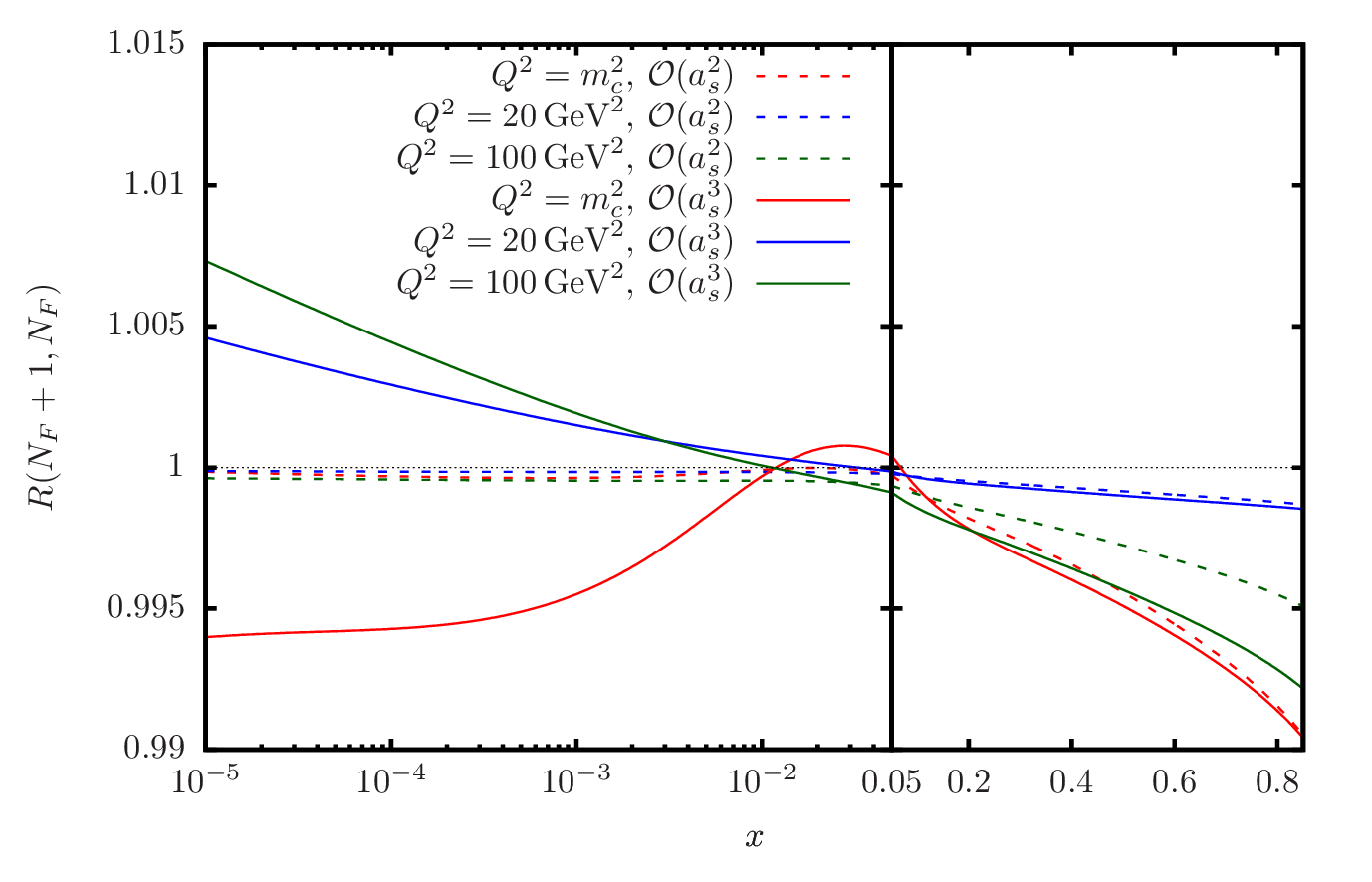}
\caption{\sf \small The ratio of the 
distribution $x(d+\bar{d})$ for four and three quark flavors at 2- and 3-loop order in the variable flavor 
number scheme matched at different scales of $Q^2$ as a function of $x$ using the parton distribution functions of 
Ref.~\cite{Alekhin:2013nda} and the on-mass-shell 
definition of the charm quark mass $m_c = 1.59~\GeV$.} 
\label{Fig:VFNS3}
\end{figure}

\noindent
at low matching scales and low values of $x$. In the region of larger $x$  the 
predictions in both orders are closer. The overall effect is of $O(\pm 0.5 \%)$. Given the fact, that the structure 
function $F_2(x,Q^2)$ can be measured at the per-cent level, these effects are only somewhat below 
present experimental precision.

\begin{figure}[H]
\centering
\includegraphics[width=0.8\textwidth]{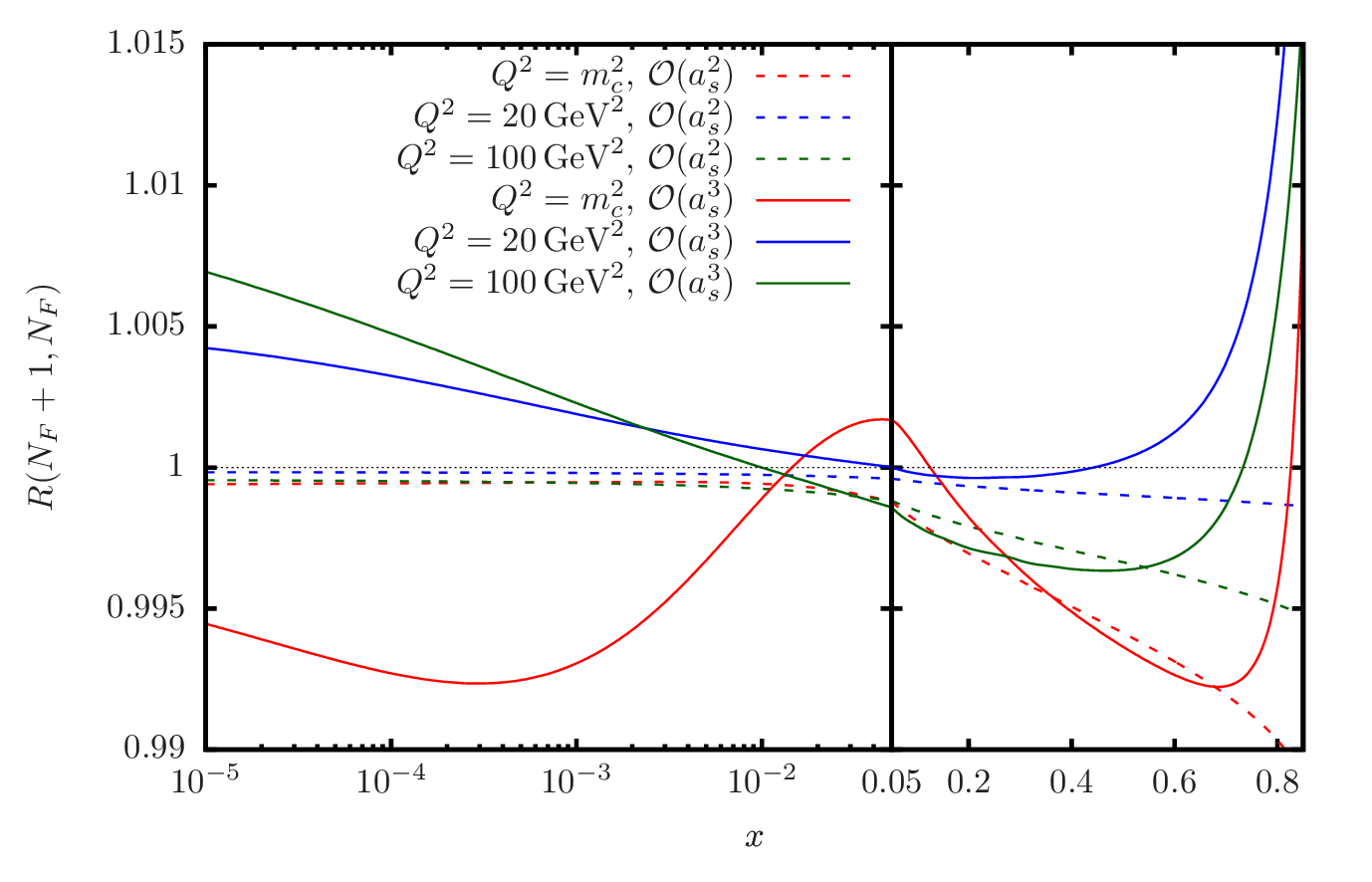}
\caption{\sf \small The ratio of the 
distribution $x(s+\bar{s})$ for four and three quark flavors at 2- and 3-loop order in the variable flavor 
number scheme matched at different scales of $Q^2$ as a function of $x$ using the parton distribution functions of 
Ref.~\cite{Alekhin:2013nda} and the on-mass-shell 
definition of the charm quark mass $m_c = 1.59~\GeV$.} 
\label{Fig:VFNS4}
\end{figure}

The corresponding ratio for the distribution $x(d + \bar{d})$ is shown in Figure~\ref{Fig:VFNS3}. The results are very similar 
to those in Figure~\ref{Fig:VFNS2}, with a slightly larger effect at higher values of $x$. In comparison, the 
results for the distribution $x(s + \bar{s})$ in Figure~\ref{Fig:VFNS4}
shows a more pronounced behaviour giving larger effects and positive corrections at $O(a_s^3)$ in the region of larger values 
of 
$x$. Overall the corrections are of $O(\pm 1\%)$ and largest at low matching scales of $\mu \sim m_c$.

\section{The Flavor Non-Singlet Massive OME: Transversity}
\label{sec:8}

\vspace*{1mm}
\noindent
In a similar way to the non-singlet OME in the vector case, the OME for transversity is obtained referring to the
operator Eq.~(\ref{eq:OP2}). The contribution to the un-renormalized OME at $O(\ep^0)$ reads 
\begin{eqnarray}
a_{qq,Q}^{(3), \rm NS, TR} &=&
\textcolor{blue}{C_F^2 T_F} \Biggl\{
\frac{128}{27} S_2 S_1^3
+\biggl[
-\frac{16 \big(N+2 (-1)^N+1\big)}{9 N (N+1)}
+\frac{64}{3} S_3
-\frac{128}{9} S_{2,1}
-\frac{256}{9} S_{-2,1}
\biggr] S_1^2
\nonumber \\ &&
+\biggl[
-\frac{64}{9} S_2^2
+\frac{7168}{81} S_2
+\frac{8 P_{92}}{27 N^2 (N+1)^2}
-\frac{2560}{27} S_3
+\frac{704}{9} S_4
-\frac{320}{9} S_{3,1}
-\frac{2560}{27} S_{-2,1}
\nonumber \\ &&
-\frac{256}{9} S_{-2,2}
+\frac{64}{3} S_{2,1,1}
+\frac{1024}{9} S_{-2,1,1}
\biggr] S_1
-\frac{496}{27} S_2^2
+\frac{32 (-1)^N \big(2 N^2+2 N+1\big)}{9 N^3 (N+1)^3}
\nonumber \\ &&
-\frac{2 P_{95}}{81 N^3 (N+1)^3}
+B_4 \biggl(16-\frac{64 S_1}{3}\biggr)
+2 \biggl(
\frac{128}{9} S_1
-\frac{640}{27}
\biggr) S_{-4}
+ \bigl(96 S_1-72\bigr) \zeta_4
\nonumber \\ &&
+2 \biggl[
\frac{64}{9} S_1^2
-\frac{640}{27} S_1
+\frac{64}{9} S_2
+\frac{3584}{81}
\biggr] S_{-3}
+\frac{10408}{81} S_3
-\frac{2992}{27} S_4
+\frac{256}{9} \bigl(S_{-5}+2 S_5\bigr)
%
\nonumber \\ &&
+\biggl[
\frac{256}{27} S_1^3
+\frac{14336}{81} S_1
-\frac{1280}{27} S_2
+\frac{512}{27} S_3
-\frac{512}{9} S_{2,1}
-\frac{64}{9 N (N+1)}
\biggr] S_{-2}
+\frac{112}{9} S_{2,1}
\nonumber \\ &&
+\frac{256}{9} S_{2,3}
-\frac{512}{9} S_{2,-3}
+\frac{1424}{27} S_{3,1}
-\frac{512}{9} S_{4,1}
-\frac{14336}{81} S_{-2,1}
+\biggl[
\frac{256}{27} S_3
+\frac{256}{3} S_{-2,1}
\nonumber \\ &&
-\frac{16 \big(169 N^2+169 N+6 (-1)^N+6\big)}{27 N (N+1)}
\biggr] S_2
-\frac{1280}{27} S_{-2,2}
+\frac{512}{9} S_{-2,3}
-16 S_{2,1,1}
\nonumber \\ &&
+\frac{512}{9} S_{2,1,-2}
+\frac{256}{9} S_{3,1,1}
+\frac{5120}{27} S_{-2,1,1}
+\frac{512}{9} S_{-2,2,1}
-\frac{2048}{9} S_{-2,1,1,1}
\nonumber \\ &&
+\biggl[
-\frac{2 \big(45 N^2+45 N+4 (-1)^N-4\big)}{3 N (N+1)}
+\frac{64}{3} S_{-2} S_1
-8 S_2
+\biggl(
\frac{32}{3} S_2+40
\biggr) S_1
+\frac{32}{3} S_3
\nonumber \\ &&
+\frac{32}{3} S_{-3}
-\frac{64}{3} S_{-2,1}
+ 1
\biggr] \zeta_2
+\biggl(-\frac{1208}{9} S_1-\frac{64}{3} S_2+\frac{350}{3}+8 
\biggr) \zeta_3
+\frac{335}{18}
\Biggr\} 
%
\nonumber \\ &&
+\textcolor{blue}{C_F T_F^2} \Biggl\{
\textcolor{blue}{N_F} \biggl[
\frac{32 P_{90}}{243 N^2 (N+1)^2}
-\frac{55552}{729} S_1
+\frac{640}{27} S_2
-\frac{320}{81} S_3
+\frac{64}{27} S_4
\nonumber \\ &&
+\biggl(
-\frac{160}{27} S_1
+\frac{32}{9} S_2
+\frac{4}{9}
\biggr) \zeta_2
+\biggl(
\frac{448}{27} S_1
-\frac{112}{9}
\biggr) \zeta_3
-\frac{674}{81}
\biggr]
+\frac{8 P_{89}}{243 N^2 (N+1)^2}
\nonumber \\ &&
-\frac{19424}{729} S_1
+\frac{1856}{81} S_2
-\frac{640}{81} S_3
+\frac{128}{27} S_4
+\left[-\frac{320}{27}S_1
+\frac{64}{9} S_2
+\frac{8}{9} \right] \zeta_2
+\left[-\frac{1024}{27} S_1 \right.
\nonumber \\ && \left.
+\frac{256}{9} \right] \zeta_3
-\frac{604}{81}
\Biggr\}
+\textcolor{blue}{C_A C_F T_F} \Biggl\{
-\frac{64}{27} S_2 S_1^3
+\biggl[
\frac{4 P_{88}}{9 N (N+1)^2 (N+2)}
-\frac{80}{9} S_3
\nonumber \\ &&
+\frac{128}{9} \bigl(S_{2,1}+S_{-2,1}\bigr)
\biggr] S_1^2
+\biggl[
\frac{112}{9} S_2^2
-\frac{16 (N-2) (2 N+3) S_2}{9 (N+1) (N+2)}
-\frac{208}{9} S_4
-8 S_{2,1}
\nonumber \\ &&
+\frac{1280}{27} S_{-2,1}
+\frac{4 P_{93}}{729 N (N+1)^2 (N+2)}
+\frac{320}{9} S_3
+\frac{64}{3} S_{3,1}
+\frac{128}{9} S_{-2,2}
-32 S_{2,1,1}
\nonumber \\ &&
-\frac{512}{9} S_{-2,1,1}
\biggr] S_1
-\frac{20}{3} S_2^2
-\frac{16 (-1)^N \big(2 N^2+2 N+1\big)}{9 N^3 (N+1)^3}
+\frac{P_{94}}{243 N^3 (N+1)^3}
\nonumber \\ &&
+\big(72-96 S_1\big) \zeta_4
+2 \biggl(
\frac{320}{27}
-\frac{64}{9} S_1
\biggr) S_{-4}
+B_4 \biggl(\frac{32}{3} S_1-8\biggr)
+2 \biggl[
-\frac{32}{9} S_1^2
+\frac{320}{27} S_1
\nonumber \\ &&
-\frac{32}{9} S_2
-\frac{1792}{81}
\biggr] S_{-3}
-\frac{8 \big(27 N^3+560 N^2+1365 N+778\big)}{81 (N+1) (N+2)} S_3
+\frac{1244}{27} S_4
-\frac{224}{9} S_5
\nonumber \\ &&
-\frac{128}{9} S_{-5}
+\frac{8 \big(2 N^2-35 N-54\big)}{9 N (N+1)^2 (N+2)} S_{1,1}
-\frac{32 \big(3 N^3+7 N^2+7 N+6\big)}{9 (N+1) (N+2)} S_{2,1}
\nonumber \\ &&
+\biggl[
-\frac{128}{27} S_1^3
-\frac{7168}{81} S_1
+\frac{640}{27} S_2
-\frac{256}{27} S_3
+\frac{256}{9} S_{2,1}
+\frac{32}{9 N (N+1)}
\biggr] S_{-2}
\nonumber \\ &&
-\frac{128}{3} S_{2,3}
+\frac{256}{9} S_{2,-3}
-\frac{1352}{27} S_{3,1}
+\frac{256}{9} S_{4,1}
+\biggl[
-\frac{4 P_{91}}{81 N (N+1)^2 (N+2)}
\nonumber \\ &&
+\frac{496}{27} S_3
-\frac{64}{3} S_{2,1}
-\frac{128}{3} S_{-2,1}
\biggr] S_2
+\frac{7168}{81} S_{-2,1}
+\frac{640}{27} S_{-2,2}
-\frac{256}{9} S_{-2,3}
\nonumber \\ &&
+24 S_{2,1,1}
-\frac{256}{9} \bigl(S_{2,1,-2}+S_{3,1,1}+S_{-2,2,1}\bigr)
+\frac{64}{3} S_{2,2,1}
-\frac{2560}{27} S_{-2,1,1}
+\frac{224}{9} S_{2,1,1,1}
\nonumber \\ &&
+\frac{1024}{9} S_{-2,1,1,1}
+\biggl[
\frac{2 \big(35 N^2+35 N+6 (-1)^N-6\big)}{9 N (N+1)}
-\frac{32}{3} S_{-2} S_1
-\frac{16}{27} S_1
-\frac{88}{9} S_2
\nonumber \\ &&
+\frac{16}{3} \bigl(2 S_{-2,1}-S_3-S_{-3}\bigr)
+\frac{5}{3} 
\biggr] \zeta_2
+\biggl[
\frac{2 \big(108 N^3-239 N^2-1137 N-646\big)}{9 (N+1) (N+2)}
\nonumber \\ &&
-16 S_1^2
+\frac{2548}{27} S_1
+16 S_2
-\frac{17}{3} 
\biggr] \zeta_3
-\frac{1879}{162}
\Biggr\}, 
\end{eqnarray}
with the polynomials
\begin{eqnarray}
P_{88} &=& 3 N^3+9 N^2+47 N+58 + 4 (-1)^N (N+1) (N+2), \\
P_{89} &=& 157 N^4+314 N^3+277 N^2-24 N-72, \\
P_{90} &=& 308 N^4+616 N^3+323 N^2-3 N-9, \\
P_{91} &=& 364 N^4+1591 N^3+2117 N^2+593 N-450 -36 (-1)^N (N+1) (N+2), \\
P_{92} &=& 769 N^4+1547 N^3+787 N^2-15 N-12 + 4 (-1)^N N (13 N+7), \\
P_{93} &=& (N+1) \big(6197 N^3+18591 N^2+15850 N+4320\big) - 108 (-1)^N (N+2) (13 N+7), \\
P_{94} &=& -1013 N^6-3039 N^5-5751 N^4-2981 N^3+1752 N^2+1872 N+432
        \nonumber \\ &&
        + 24 (-1)^N N \big(133 N^3+188 N^2+82 N-9\big), \\
P_{95} &=& 6327 N^6+18981 N^5+18457 N^4+5687 N^3-260 N^2+144 N+144
        \nonumber \\ &&
        + 8 (-1)^N N \big(133 N^3+188 N^2+82 N-9\big).
\end{eqnarray}

Let us define the leading order anomalous dimension for transversity without the color factor by
\begin{eqnarray}
\tilde{\gamma}_{qq, \rm NS, TR}^0 = 2 \left[4 S_1 - 3 \right].
\end{eqnarray}

The 2- and 3-loop contributions to the massive OME for transversity Eqs.~(\ref{eq:Aqq2}) and (\ref{eq:Aqq3}) are then 
given in the on-mass shell scheme by
\begin{eqnarray}
\lefteqn{A_{qq,Q, \rm NS, TR}^{(2)} =}  \nonumber\\ &&
\textcolor{blue}{C_F T_F} 
\Biggl\{
-\frac{1}{3} \tilde{\gamma}_{qq, \rm NS, TR}^0 \ln^2\left(\frac{m^2}{\mu^2}\right) 
+\Biggl[
        -\frac{2}{9} \left(
                -3
                -24 S_2
        \right)
        -\frac{80}{9} S_1
\Biggr] \ln\left(\frac{m^2}{\mu^2}\right)
\nonumber\\ &&
+
        \frac{73 N^2+73 N+24}{18 N (N+1)}
        -\frac{224}{27} S_1
        +\frac{40}{9} S_2
        -\frac{8}{3} S_3
\Biggr\},
\\ 
\lefteqn{A_{qq,Q, \rm NS, TR}^{(3)} =} \nonumber\\ &&
\textcolor{blue}{C_F T_F} 
\Biggl\{ - \frac{8}{27} 
        \textcolor{blue}{T_F} (2 + \textcolor{blue}{N_F}) + \frac{22}{27} \textcolor{blue}{C_A} \Biggr\} \tilde{\gamma}_{qq,\rm NS, TR}^0
        \ln^3\left(\frac{m^2}{\mu^2}\right) 
\nonumber\\ &&
+ \textcolor{blue}{C_F} \Biggl\{
                \textcolor{blue}{T_F^2} \Biggl[
                        \frac{8}{27} \big(
                                3
                                + 24 S_2
                        \big)
                        -\frac{320}{27} S_1
                \Biggl]
                +\textcolor{blue}{C_A T_F} \Biggl[
                        \frac{34}{3}
                        -\frac{184}{9} S_1
\nonumber\\ && 
                       -\frac{64}{3} S_{-2} S_1
                        -\frac{32}{3} S_3
                        -\frac{32}{3} S_{-3}
                        +\frac{64}{3} S_{-2,1}
        \Biggr]
        +\textcolor{blue}{C_F T_F} \Biggl[
                8
                +\left(
                        -8
                        +\frac{64}{3} S_2
                \right) S_1
\nonumber\\ && 
                +\frac{128}{3} S_{-2} S_1
                -16 S_2
                +\frac{64}{3} [S_3 + S_{-3}]
                -\frac{128}{3} S_{-2,1}
        \Biggr]
\Biggr\}
\ln^2\left(\frac{m^2}{\mu^2}\right)
\nonumber\\ && 
+
\Biggl\{\textcolor{blue}{C_F} \Biggl\{
                \textcolor{blue}{T_F^2} \Biggl[
                        \frac{496}{27}
                        +\textcolor{blue}{N_F} \Biggl[
                                \frac{4 \big(
                                        175 N^2+175 N-24\big)}{27 N (N+1)}
                                -\frac{2176}{81} S_1
                                -\frac{320}{27} S_2
\nonumber\\ &&                
                 +\frac{64}{9} S_3
                        \Biggr]
                        -
                        \frac{1984}{81} S_1
                \Biggr]
                +\textcolor{blue}{C_A T_F} \Biggl[
                        -\frac{11 \big(
                                145 N^2+145 N-24\big)}{27 N (N+1)}
 + \frac{1}{2} \Biggl[
                                -\frac{640}{9}
                                +\frac{128}{3} S_1
                        \Biggr] S_{-3}
\nonumber\\ &&
                        +\Biggl[
                                -\frac{8 \big(
                                        155 N^2+155 N+27\big)}{81 N (N+1)}
                                +32 S_3
                                +\frac{128}{3} S_{-2,1}
                        \Biggr] S_1
                        +\frac{1792}{27} S_2
                        -\frac{496}{9} S_3
                        +\frac{160}{3} S_4
\nonumber\\ && 
                        +\Biggl[
                                -\frac{640}{9} S_1
                                +\frac{64}{3} S_2
                        \Biggr] S_{-2}
                        +\frac{64}{3} S_{-4}
                        -\frac{128}{3} S_{3,1}
                        +\frac{640}{9} S_{-2,1}
                        +\frac{64}{3} S_{-2,2}
                        -\frac{256}{3} S_{-2,1,1}
\nonumber\\ &&                
         +\frac{1}{2} \Biggl[
                                96
                                -128 S_1
                        \Biggr] \zeta_3
                \Biggr]
        +\textcolor{blue}{C_F T_F} \Biggl[
                1
                + \frac{1}{2} \Biggl[
                        \frac{1280}{9}
                        -\frac{256}{3} S_1
                \Biggr] S_{-3}
                +\Biggl[
                        -\frac{8 \big(
                                5 N^2+5 N-4\big)}{3 N (N+1)}
\nonumber\\ &&                
         +\frac{640}{9} S_2
                        -\frac{256}{3} S_3
                        -\frac{256}{3} S_{-2,1}
                \Biggr] S_1
                -\frac{40}{3} S_2
                -\frac{64}{3} S_2^2
                +\frac{928}{9} S_3
                -\frac{256}{3} S_4
                +\Biggl[
                        \frac{1280}{9} S_1
\nonumber\\ &&                
         -\frac{128}{3} S_2
                \Biggr] S_{-2}
                -\frac{128}{3} S_{-4}
                +
                \frac{128}{3} S_{3,1}
                -\frac{1280}{9} S_{-2,1}
                -\frac{128}{3} S_{-2,2}
                +\frac{512}{3} S_{-2,1,1}
\nonumber\\ &&               
 +\frac{1}{2} \Biggl[
                        -96
                        +128 S_1
                \Biggr] \zeta_3
        \Biggr]
\Biggr\} 
        \Biggr\}
\ln\left(\frac{m^2}{\mu^2}\right)
\nonumber\\ &&
+\textcolor{blue}{C_F} \Biggl\{
        \textcolor{blue}{C_A T_F} \Biggl[
                -\frac{16 (-1)^N \big(
                        2 N^2+2 N+1\big)}{9 N^3 (N+1)^3}
                -12 \zeta_4
                +\frac{P_{102}}{243 N^3 (N+1)^3}
\nonumber\\ &&
                +\frac{4(2 N^2-35N-54)}{9 N (N+1)^2 (N+2)} [S_1^2 + S_2]
                -\frac{32 (3 N^3+7 N^2+7 N+6)}{9(N+1) (N+2)}  S_{2,1}
\nonumber\\ &&
                -\frac{8(27 N^3+890 N^2+2355 N+1438)}{81(N+1) (N+2)}  S_3 
                + 2 \Biggl[
                        \frac{320}{27}
                        -\frac{64}{9} S_1
                \Biggr] S_{-4}  
\nonumber\\ &&
              +\Biggl[
                        \frac{(216 N^3-485 N^2-2295 N-1306)}{9(N+1) (N+2)} 
                        +\frac{2372}{27} S_1
                        -16 [S_1^2 - S_2]
                \Biggl] \zeta_3
\nonumber\\ &&  
                + 
                 \Biggl[
                        84
                        -96 S_1
                \Biggr] \zeta_4
                + \Biggl[
                        -8
                        +
                        \frac{32}{3} S_1
                \Biggr] B_4
                +\Biggl[
                        -\frac{4 P_{100}}{729 N (N+1)^2 (N+2)}
\nonumber\\ && 
                        -\frac{16 (N-2) (2 N+3)}{9 (N+1) (N+2)} S_2
                        +\frac{112}{9} S_2^2
                        +\frac{320}{9} S_3
                        -\frac{208}{9} S_4
                        -8 S_{2,1}
                        +\frac{64}{3} S_{3,1}
                        +\frac{1280}{27} S_{-2,1}
\nonumber\\ && 
                        +\frac{128}{9} S_{-2,2}
                        -32 S_{2,1,1}
                        -\frac{512}{9} S_{-2,1,1}
                \Biggr] S_1
                +\Biggl[
                        \frac{4(3 N^3+13 N^2+59 N+66)}{9 N (N+1)^2 (N+2)} 
                        -\frac{80}{9} S_3
\nonumber\\ && 
                        +\frac{128}{9} [S_{2,1} + S_{-2,1}]
                \Biggr] S_1^2
                +\Biggl[
                        \frac{4 P_{96}}{81 N (N+1)^2 (N+2)}
                        +\frac{496}{27} S_3
                        -\frac{64}{3} S_{2,1}
                        -\frac{128}{3} S_{-2,1}
                \Biggr] S_2
\nonumber\\ &&                
 -\frac{64}{27} S_1^3 S_2
                -\frac{20}{3} S_2^2
                +\frac{1772}{27} S_4
                -\frac{224}{9} S_5
                +\Biggl[
                        \frac{32}{9 N (N+1)}
                        -\frac{7168}{81} S_1
                        -\frac{128}{27} S_1^3
\nonumber\\ &&                
         +\frac{640}{27} S_2
                        -\frac{256}{27} S_3
                        +\frac{256}{9} S_{2,1}
                \Biggr] S_{-2}
                +\Biggl[
                        2 \Biggl[
                                -\frac{1792}{81}
                                -\frac{32}{9} S_2
                        \Biggr]
                        +\frac{640}{27} S_1
                        -\frac{64}{9} S_1^2
                \Biggr] S_{-3}
\nonumber\\ &&                
 -\frac{128}
                {9} S_{-5}
                -
                \frac{128}{3} S_{2,3}
                +\frac{256}{9} S_{2,-3}
                -\frac{1352}{27} S_{3,1}
                +\frac{256}{9} S_{4,1}
                +\frac{7168}{81} S_{-2,1}
                +\frac{640}{27} S_{-2,2}
\nonumber\\ &&                
 -\frac{256}{9} S_{-2,3}
                +24 S_{2,1,1}
                -\frac{256}{9} S_{2,1,-2}
                +\frac{64}{3} S_{2,2,1}
                -\frac{256}{9} S_{3,1,1}
                -\frac{2560}{27} S_{-2,1,1}
\nonumber\\ &&
                -\frac{256}{9} S_{-2,2,1}
                +\frac{224}{9} S_{2,1,1,1}
                +\frac{1024}{9} S_{-2,1,1,1}
        \Biggr]
\nonumber\\ &&
        +\textcolor{blue}{T_F^2} \Biggl[
                -\frac{2 P_{98}}{243 N^2 (N+1)^2}
                +\textcolor{blue}{N_F} \Biggl[
                        \frac{4 P_{97}}{243 N^2 (N+1)^2}
                        -\frac{24064}{729} S_1
                        +\frac{128}{81} S_2
                        +\frac{640}{81} S_3
\nonumber\\ && 
                        -\frac{128}{27} S_4
                        +\Biggl[
                                -\frac{128}{9}
                                +\frac{512}{27} S_1
                        \Biggr] \zeta_3
                \Biggr]
                +\frac{12064}{729} S_1
                +\frac{64}{81} S_2
                +\frac{320}{81} S_3
                -\frac{64}{27} S_4
\nonumber\\ &&                
 +\Biggl[
                        \frac{224}{9}
                        -\frac{896}{27} S_1
                \Biggr] \zeta_3
        \Biggr]
\nonumber 
\end{eqnarray}
\begin{eqnarray}
&& 
+\textcolor{blue}{C_F T_F}  \Biggl[
        \frac{32 (-1)^N \big(
                2 N^2+2 N+1\big)}{9 N^3 (N+1)^3}
        +\frac{P_{101}}{9 N^3 (N+1)^3}
        + 2 \Biggl[
                -\frac{640}{27}
                +\frac{128}{9} S_1
        \Biggr] S_{-4}
\nonumber\\ &&
        + \Biggl[
                16
                -\frac{64}{3} S_1
        \Biggr] B_4
        +       \Biggl[
                -72
                +96 S_1
        \Biggr] \zeta_4
        +\Biggl[
                \frac{2 P_{99}}{27 N^2 (N+1)^2}
\nonumber\\ && 
                +\frac{7168}{81} S_2
                -\frac{64}{9} S_2^2
                -\frac{2560}{27} S_3
                +\frac{704}{9} S_4
                -\frac{320}{9} S_{3,1}
                -\frac{2560}{27} S_{-2,1}
                -\frac{256}{9} S_{-2,2}
\nonumber\\ &&               
 +\frac{64}{3} S_{2,1,1}
                +\frac{1024}{9} S_{-2,1,1}
        \Biggr] S_1
        +\Biggl[
                -\frac{16 (N+3)}{9 N (N+1)}
                +\frac{64}{3} S_3
                -\frac{128}{9} S_{2,1}
                -\frac{256}{9} S_{-2,1}
        \Biggr] S_1^2
\nonumber\\ && 
        +\Biggl[
                -\frac{16 \big(
                        151 N^2+151 N+12\big)}{27 N (N+1)}
                +\frac{256}{27} S_3
                +\frac{256}{3} S_{-2,1}
        \Biggr] S_2
        +\frac{128}{27} S_1^3 S_2
        -\frac{496}{27} S_2^2
\nonumber
\\ 
&& 
        +\frac{7816}{81} S_3
        -\frac{2992}{27} S_4
        +\frac{512}{9} S_5
        +\Biggl[
                -\frac{64}{9 N (N+1)}
                +\frac{14336}{81} S_1
                +\frac{256}{27} S_1^3
                -\frac{1280}{27} S_2
\nonumber\\ && 
                +\frac{512}{27} S_3
                -\frac{512}{9} S_{2,1}
        \Biggr] S_{-2}
        +\Biggl[
                2 \Biggl[
                        \frac{3584}{81}
                        +\frac{64}{9} S_2
                \Biggr]
                -\frac{1280}{27} S_1
                +\frac{128}{9} S_1^2
        \Biggr] S_{-3}
\nonumber\\ && 
        +\frac{256}{9} S_{-5}
        +\frac{112}{9} S_{2,1}
        +\frac{256}{9} S_{2,3}
        -\frac{512}{9} S_{2,-3}
        +\frac{1424}{27} S_{3,1}
        -\frac{512}{9} S_{4,1}
\nonumber\\ && 
        -\frac{14336}{81} S_{-2,1}
        -\frac{1280}{27} S_{-2,2}
        +\frac{512}{9} S_{-2,3}
        -16 S_{2,1,1}
        +\frac{512}{9} S_{2,1,-2}
        +\frac{256}{9} S_{3,1,1}
\nonumber\\ && 
        +\frac{5120}{27} S_{-2,1,1}
        +\frac{512}{9} S_{-2,2,1}
        -\frac{2048}{9} S_{-2,1,1,1}
        +\Biggl[
                \frac{374}{3}
                -\frac{1208}{9} S_1
                -\frac{64}{3} S_2
        \Biggr] \zeta_3
\Biggr]
\Biggr\}
\nonumber\\
&& - \tfrac{1}{2}\left[1 - (-1)^N\right]  \textcolor{blue}{C_F T_F \left(\frac{C_A}{2} - C_F\right)} 
\Biggr\{
\frac{1}{N (N+1)} \frac{32}{3} \ln^2\left(\frac{m^2}{\mu^2}\right)
+ 
\Biggl[\frac{32 (13 N+7)}{9 N (N+1)^2}
\nonumber\\ &&
        -\frac{64}{3 N (N+1)} S_1
\Biggr] \ln\left(\frac{m^2}{\mu^2}\right)
+
\frac{32(133 N^3+188 N^2+82 N-9)}{81 N^2 (N+1)^3} 
        -\frac{64 (13 N+7)}{27 N (N+1)^2} S_1
\nonumber\\ &&
        +\frac{64}{9 N (N+1)} [S_2 + S_1^2]
\Biggr\},
\end{eqnarray}
with the polynomials
\begin{eqnarray}
P_{96} &=& 868 N^4+3337 N^3+4079 N^2+1979 N+522,
\\
P_{97} &=& 1183 N^4+2366 N^3+943 N^2+48 N+144,
\\
P_{98} &=& 1829 N^4+3658 N^3+2069 N^2-48 N-144,
\\
P_{99} &=& 2377 N^4+4790 N^3+2657 N^2+52 N-48,
\\
P_{100} &=& 10807 N^4+43228 N^3+51983 N^2+17402 N-2808,
\\
P_{101} &=& -691 N^6-2073 N^5-2049 N^4-595 N^3+56 N^2-16 N-32,
\\
P_{102} &=&55 N^6+165 N^5+4605 N^4+5767 N^3+552 N^2-720 N+432.
\end{eqnarray}

The transformation to the $\overline{\rm MS}$ scheme is given by 
\begin{eqnarray}
A_{qq,Q, \rm NS, TR}^{\overline{\rm MS}} - A_{qq,Q, \rm NS, TR}^{\rm OMS} &=&
\textcolor{blue}{C_F^2 T_F} 
\Biggl\{
4 \tilde{\gamma}_{qq,\rm NS, TR}^{0} \ln^2\left(\frac{m^2}{\mu^2}\right)
+ \ln\left(\frac{m^2}{\mu^2}\right)
\left[\frac{32}{3} S_1-32 S_2+28\right]
\nonumber\\ &&
+\frac{16}{3}
-\frac{640}{9} S_1 + \frac{128}{3} S_2 
\Biggr\},
\end{eqnarray}
in $N$-space. Again we have represented the mass in both schemes by the same symbol $m$.
The corresponding expressions in $x$-space are given in Appendix~\ref{sec:OMSx}.
Unfortunately the massless Wilson coefficients for transversity have not been calculated yet. Therefore the corresponding
massive Wilson coefficients can only be given in the future, based on the present results.
\section{Conclusions}
\label{sec:9}

\vspace*{1mm}
\noindent
The heavy flavor non-singlet corrections to 3-loop order in the asymptotic region $Q^2 \gg m^2$ have been calculated 
in the vector case and for transversity as inclusive corrections in case of a single heavy quark at the time. We 
have presented the massive Wilson coefficient $L_{2,q}^{\rm NS}(x,Q^2)$ and the transition matrix elements $A_{qq,Q,\rm NS(TR)}$
occurring in the variable flavor scheme. They can be expressed in terms of harmonic sums in $N$-space and harmonic 
polylogarithms in $x$-space. At the technical side of the calculation 3-loop topologies up to massive
Benz-type graphs had to be calculated. We used the integration-by-parts reduction for Feynman diagrams which carry local
operator insertions, encoded in the package {\tt Reduze\,2}. The master integrals have been calculated using 
different methods like generalized hypergeometric 
functions, Mellin-Barnes techniques and differential equations, mapped onto summation problems. The latter ones were solved
applying modern summation technologies encoded in the packages {\tt Sigma, EvaluateMultiSums, HarmonicSums} and {\tt SumProduction}.

As a by-product of the present calculation we obtained the complete non-singlet anomalous dimensions at 2-loop order and 
their contributions $\propto N_F$ at 3-loop order, both in the vector and transversity cases. In the latter, we corrected 
a typo in the 15$^{\rm th}$ moment in \cite{Velizhanin:2012nm} and otherwise confirm earlier results in the literature. In case 
of transversity, the present result has been obtained in a first calculation ab initio.

The $O(a_s^3)$ corrections to $F_2(x,Q^2)$ amount to $\sim O(0.5\%)$ in the lower $x$ range in the 3-flavor scheme, with a 
small variance defining the charm mass in the on-shell or $\overline{\rm MS}$ scheme. The present experimental accuracy 
for the structure function $F_2(x,Q^2)$ is about 1\% in the small $x$ region.  We also discussed, at 2-loop order, 
differences in considering tagged heavy flavor vs. the inclusive heavy flavor corrections to $F_2(x,Q^2)$. The effect is 
smaller in the inclusive case. Finally, we considered the matching ratios in the VFNS going from 3 to 4 flavors, which now are
available at $O(a_s^3)$ accuracy. In the ratio $R(N_F+1, N_F)$ Eq.~(\ref{eq:RNF}) the 3-loop effects are much larger than 
those at 2 loops in the small $x$ region. At larger 
values of $x$ the variation is broad and bounded by about $\sim \pm 1\%$. This applies in particular to low matching scales 
as applied in many phenomenological analyses. 
\appendix
\section{Master Integrals}
\label{sec:MIres}

\vspace*{1mm}
\noindent
In the following we give the results for the master integrals as functions of the Mellin variable $N$. The integrals were 
calculated starting
from $N=0$, in other words, the shifts $N-1 \rightarrow N$ and $N-2 \rightarrow N$ were performed in the case of a fermion line insertion 
and a 3-point insertion, respectively. In some cases, the values at $N=0$ are given separately, since the general $N$ result
diverges for this value of $N$. Each integral has been expanded up to the order in $\varepsilon$ required for the present calculation. For
simplicity, we have set the mass $m$ and $\Delta.p$ to 1, and the overall factor $i S_{\varepsilon}$ has been omitted, where 
$S_{\varepsilon}$
is the spherical factor
\begin{eqnarray}
S_\ep = \exp\left[\frac{\ep}{2}\left(\gamma_E - \ln(4\pi)\right)\right]~,
\end{eqnarray}
which emerges once at each loop order.
At the end of the calculation in the $\overline{\rm MS}$ renormalization scheme $S_\ep$ is set to one.
The different master integrals read
\begin{eqnarray}
J_1 &=&
\frac{1}{(N+1)^2} \Biggl\{
-\frac{8}{3 \varepsilon ^3}
-\frac{8}{3 \varepsilon^2} S_1 
-\frac{1}{\varepsilon} \Biggl[
\frac{2}{3} S_1^2+\frac{2}{3} S_2+\frac{8}{3 (N+1)^2}+\zeta_2
\Biggr]
-\frac{1}{9} S_1^3
-\frac{8}{9} S_3
\nonumber \\ &&  
-S_1 \left(S_2+\frac{8}{3 (N+1)^2}+\zeta_2\right)
-\frac{2}{3} S_{2,1}
-\frac{5 \zeta_3}{3}
+\varepsilon \Biggl[
-S_1^2 \left(\frac{1}{4} S_2+\frac{2}{3 (N+1)^2}+\frac{\zeta_2}{4}\right)
\nonumber \\ &&  
-\left(\frac{2}{3 (N+1)^2}+\frac{\zeta_2}{4}\right) S_2
+S_1 \left(\frac{S_2}{N+1}+\frac{2}{9} S_3-S_{2,1}-\frac{5 \zeta_3}{3}\right)
-\frac{1}{72} S_1^4
-\frac{3}{8} S_2^2
\nonumber \\ &&  
+\frac{S_3}{N+1}
+\frac{1}{12} S_4
-\frac{S_{2,1}}{N+1}
-\frac{5}{3} S_{3,1}
+\frac{5}{3} S_{2,1,1}
-\frac{\zeta_2}{(N+1)^2}
+\frac{2 \zeta_3}{N+1}
-\frac{8}{3 (N+1)^4}
\nonumber \\ &&  
-\frac{23 \zeta_2^2}{16}
\Biggr]
\Biggr\},
\\
%
J_2 &=&
\frac{1}{N+1} \Biggl\{
\frac{4}{3 \varepsilon ^3}
\left(S_1+\frac{3 N+4}{N+1}\right)
+\frac{1}{\varepsilon^2} \Biggl[
\frac{1}{3} S_1^2-\frac{7}{3} S_2-\frac{2 \left(15 N^2+34 N+22\right)}{3 (N+1)^2}
  \nonumber \\ &&  
-\frac{2 (4 N+3) S_1}{3 (N+1)}
\Biggr]
+\frac{1}{\varepsilon} \Biggl[
S_1 \left(\frac{3}{2} S_2+\frac{14 N^2+30 N+11}{3 (N+1)^2}+\frac{\zeta_2}{2}\right)-\frac{(4 N+3) S_1^2}{6 (N+1)}+\frac{1}{18} S_1^3
  \nonumber \\ &&  
+\frac{(4 N+13) S_2}{6 (N+1)}+\frac{37}{9} S_3-\frac{10}{3} S_{2,1}+\frac{57 N^3+185 N^2+205 N+84}{3 (N+1)^3}+\frac{(3 N+4) \zeta_2}{2 (N+1)}
\Biggr]
 \nonumber \\ &&  
+S_1^2 \left(\frac{3}{8} S_2+\frac{14 N^2+30 N+11}{12 (N+1)^2}+\frac{\zeta_2}{8}\right)+\left(\frac{22 N^2+38 N+3}{12 (N+1)^2}-\frac{7
   \zeta_2}{8}\right) S_2
+\frac{1}{144} S_1^4
 \nonumber \\ &&  
+S_1 \biggl(-\frac{(4 N+1) S_2}{4 (N+1)}-\frac{11}{18} S_3+S_{2,1}-\frac{46 N^3+154 N^2+168 N+47}{6 (N+1)^3}-\frac{(4
   N+3) \zeta_2}{4 (N+1)}
  \nonumber \\ &&  
+\frac{17 \zeta_3}{6}\biggr)
-\frac{(4 N+3) S_1^3}{36 (N+1)}+\frac{3}{16} S_2^2-\frac{(40 N+51)
   S_3}{18 (N+1)}-\frac{83}{24} S_4+\frac{(2 N+5) S_{2,1}}{3 (N+1)}+\frac{11}{3} S_{3,1}
 \nonumber \\ &&  
-\frac{11}{3} S_{2,1,1}-\frac{\left(15 N^2+34 N+22\right)
   \zeta_2}{4 (N+1)^2}-\frac{195 N^4+826 N^3+1320 N^2+952 N+278}{6 (N+1)^4}
 \nonumber \\ &&  
+\frac{(3 N+20) \zeta_3}{6 (N+1)}
\Biggr\},
\label{J2N}
\\
J_3 &=&
\frac{1}{(N+1)^2} \Biggl\{
\frac{8}{3 \varepsilon ^3}
-\frac{4 (N+2)}{\varepsilon^2 (N+1)}
+\frac{1}{\varepsilon} \Biggl[
\frac{2 \left(9 N^2+27 N+25\right)}{3 (N+1)^2}+\zeta_2
\Biggr]
-\frac{3 (N+2) \zeta_2}{2 (N+1)}
\nonumber \\ &&  
-\frac{7 \zeta_3}{3}
-\frac{9 N^3+36 N^2+52 N+30}{(N+1)^3}
+\varepsilon \Biggl[
\frac{\left(9 N^2+27 N+25\right) \zeta_2}{4 (N+1)^2}
+\frac{7 (N+2) \zeta_3}{2 (N+1)}
\nonumber \\ &&   
-\frac{173 \zeta_2^2}{80}
+\frac{81 N^4+405 N^3+792 N^2+738 N+301}{6 (N+1)^4}
\Biggr]
\Biggr\},
\\
J_4 &=&
\frac{4}{3 \varepsilon ^3}
+\frac{2}{3 \varepsilon^2} \Biggl[
\frac{4 N+3}{N+1}-S_1
\Biggr]
+\frac{1}{\varepsilon} \Biggl[
\frac{1}{6} S_1^2-\frac{(4 N+5) S_1}{3 (N+1)}+\frac{5}{6} S_2+\frac{16 N^2+28 N+13}{3 (N+1)^2}+\frac{\zeta_2}{2}
\Biggr]
\nonumber \\ &&  
-S_1 \left(\frac{5}{12} S_2+\frac{16 N^2+36 N+19}{6 (N+1)^2}+\frac{\zeta_2}{4}\right)
-\frac{1}{36} S_1^3
+\frac{(4 N+5) S_1^2}{12 (N+1)}
+\frac{5 (4 N+5) S_2}{12 (N+1)}
\nonumber \\ &&  
-\frac{19}{18} S_3
+\frac{1}{3} S_{2,1}
+\frac{(4 N+3) \left(16 N^2+32 N+17\right)}{6 (N+1)^3}
+\frac{(4 N+3) \zeta_2}{4 (N+1)}
-\frac{7 \zeta_3}{6}
\nonumber \\ &&  
+\varepsilon \Biggl[
S_1^2 \left(\frac{5}{48} S_2+\frac{16 N^2+36 N+19}{24 (N+1)^2}+\frac{\zeta_2}{16}\right)
+\left(\frac{5 \left(16 N^2+36 N+19\right)}{24 (N+1)^2}+\frac{5 \zeta_2}{16}\right) S_2
\nonumber \\ &&  
+S_1 \biggl(-\frac{5 (4 N+5) S_2}{24 (N+1)}+\frac{19}{36} S_3-\frac{1}{6} S_{2,1}-\frac{64 N^3+208
   N^2+220 N+77}{12 (N+1)^3}-\frac{(4 N+5) \zeta_2}{8 (N+1)}
\nonumber \\ &&    
+\frac{7 \zeta_3}{12}\biggr)
+\frac{1}{288} S_1^4-\frac{(4 N+5) S_1^3}{72
   (N+1)}+\frac{1}{96} S_2^2-\frac{19 (4 N+5) S_3}{36 (N+1)}+\frac{53}{48} S_4+\frac{(4 N+5) S_{2,1}}{6 (N+1)}
\nonumber \\ &&  
+\frac{1}{6} S_{2,1,1}
+\frac{\left(16 N^2+28 N+13\right) \zeta_2}{8 (N+1)^2}
+\frac{256 N^4+960 N^3+1360 N^2+860 N+205}{12 (N+1)^4}
\nonumber \\ &&  
-\frac{1}{2} S_{3,1}
-\frac{7 (4 N+3) \zeta_3}{12 (N+1)}
-\frac{173 \zeta_2^2}{160}
\Biggr]
+\varepsilon^2 \Biggl[
-S_1^3 \left(\frac{5}{288} S_2+\frac{16 N^2+36 N+19}{144 (N+1)^2}+\frac{\zeta_2}{96}\right)
\nonumber \\ &&  
-\left(\frac{19 \left(16 N^2+36 N+19\right)}{72 (N+1)^2}+\frac{19 \zeta_2}{48}\right) S_3
+\left(\frac{16 N^2+36 N+19}{12 (N+1)^2}+\frac{\zeta_2}{8}\right) S_{2,1}
\nonumber \\ &&  
+S_1^2 \biggl(\frac{5 (4 N+5) S_2}{96 (N+1)}-\frac{19}{144} S_3+\frac{1}{24} S_{2,1}+\frac{64 N^3+208 N^2+220 N+77}{48 (N+1)^3}-\frac{7 \zeta_3}{48}
\nonumber \\ &&   
+\frac{(4 N+5) \zeta_2}{32 (N+1)}\biggr)
+S_2 \biggl(\frac{13}{144} S_3-\frac{1}{24} S_{2,1}+\frac{5 \left(64 N^3+208 N^2+220 N+77\right)}{48 (N+1)^3}
\nonumber \\ &&  
+\frac{5 (4 N+5) \zeta_2}{32 (N+1)}-\frac{35 \zeta_3}{48}\biggr)
+S_1 \biggl(\left(-\frac{5 \left(16 N^2+36 N+19\right)}{48 (N+1)^2}-\frac{5 \zeta_2}{32}\right)
   S_2-\frac{1}{192} S_2^2
\nonumber \\ &&  
+\frac{19 (4 N+5) S_3}{72 (N+1)}-\frac{53}{96} S_4-\frac{(4 N+5) S_{2,1}}{12 (N+1)}+\frac{1}{4} S_{3,1}-\frac{1}{12}
   S_{2,1,1}-\frac{\left(16 N^2+36 N+19\right) \zeta_2}{16 (N+1)^2}
\nonumber \\ &&  
-\frac{256 N^4+1088 N^3+1712 N^2+1188 N+307}{24 (N+1)^4}+\frac{7 (4 N+5) \zeta_3}{24
   (N+1)}+\frac{173 \zeta_2^2}{320}\biggr)-\frac{S_1^5}{2880}
\nonumber \\ &&  
+\frac{(4 N+5) S_1^4}{576 (N+1)}+\frac{(4 N+5) S_2^2}{192 (N+1)}+\frac{53 (4 N+5)
   S_4}{96 (N+1)}-\frac{121}{120} S_5-\frac{1}{4} S_{2,3}-\frac{(4 N+5) S_{3,1}}{4 (N+1)}
\nonumber \\ &&  
+\frac{1}{2} S_{4,1}+\frac{(4 N+5) S_{2,1,1}}{12 (N+1)}
+\frac{1}{12} S_{2,2,1}-\frac{1}{4} S_{3,1,1}+\frac{1}{12} S_{2,1,1,1}
-\frac{173 (4 N+3) \zeta_2^2}{320 (N+1)}
\nonumber \\ &&  
+\zeta_2 \left(\frac{(4 N+3) \left(16 N^2+32 N+17\right)}{16 (N+1)^3}-\frac{7 \zeta_3}{16}\right)
-\frac{7 \left(16 N^2+28 N+13\right) \zeta_3}{24 (N+1)^2}
-\frac{239 \zeta_5}{40}
\nonumber \\ &&  
+\frac{(4 N+3) \left(16 N^2+28 N+13\right) \left(16 N^2+36 N+21\right)}{24 (N+1)^5}
\Biggr],
\\
J_5 &=&
\frac{1}{N+1} \Biggl\{
\frac{8}{3 \varepsilon^3} \left(S_1+\frac{1}{N+1}\right)
-\frac{1}{\varepsilon^2} \Biggl[
\frac{2}{3} S_1^2+\frac{4 (3 N+4) S_1}{3 (N+1)}+\frac{10}{3} S_2+\frac{4 (N+2)}{(N+1)^2}
\Biggr]
\nonumber \\ &&  
+\frac{1}{\varepsilon} \Biggl[
S_1 \left(\frac{5}{3} S_2+\frac{2 \left(9 N^2+21 N+13\right)}{3 (N+1)^2}+\zeta_2\right)
+\frac{1}{9} S_1^3+\frac{(3 N+4) S_1^2}{3 (N+1)}
+\frac{5 (3 N+4) S_2}{3 (N+1)}
\nonumber \\ &&  
+\frac{38}{9} S_3-\frac{4}{3} S_{2,1}+\frac{2 \left(9 N^2+27 N+25\right)}{3 (N+1)^3}+\frac{\zeta_2}{N+1}
\Biggr]
+S_1 \biggl(-\frac{5 (3 N+4) S_2}{6 (N+1)}-\frac{19}{9} S_3
\nonumber \\ &&  
+\frac{2}{3} S_{2,1}-\frac{(3 N+4) \left(9 N^2+18 N+10\right)}{3 (N+1)^3}-\frac{(3 N+4)
   \zeta_2}{2 (N+1)}-\frac{7 \zeta_3}{3}\biggr)
-\frac{1}{72} S_1^4
-\frac{53}{12} S_4
\nonumber \\ &&  
+S_1^2 \left(-\frac{5}{12} S_2-\frac{9 N^2+21 N+13}{6 (N+1)^2}-\frac{\zeta_2}{4}\right)
+\left(-\frac{5 \left(9 N^2+21 N+13\right)}{6 (N+1)^2}-\frac{5 \zeta_2}{4}\right) S_2
\nonumber \\ &&  
-\frac{(3 N+4) S_1^3}{18 (N+1)}
-\frac{1}{24} S_2^2
-\frac{19 (3 N+4) S_3}{9 (N+1)}
+\frac{2 (3 N+4) S_{2,1}}{3 (N+1)}
+2 S_{3,1}-\frac{2}{3} S_{2,1,1}
\nonumber \\ &&  
-\frac{9 N^3+36 N^2+52 N+30}{(N+1)^4}
-\frac{3 (N+2) \zeta_2}{2 (N+1)^2}-\frac{7 \zeta_3}{3 (N+1)}
\Biggr\},
\\
J_6 &=&
\frac{1}{N+1} \Biggl\{
-\frac{8}{3 \varepsilon^3} \left(S_1+\frac{1}{N+1}\right)
+\frac{1}{\varepsilon^2} \Biggl[
-\frac{2}{3} S_1^2+\frac{4 (3 N+2) S_1}{3 (N+1)}+2 S_2+\frac{4 (3 N+4)}{3 (N+1)^2}
\Biggr]
\nonumber \\ &&  
+\frac{1}{\varepsilon} \Biggl[
S_1 \left(S_2-\frac{2 \left(9 N^2+15 N+5\right)}{3 (N+1)^2}-\zeta_2\right)-\frac{1}{9} S_1^3+\frac{(3 N+2) S_1^2}{3 (N+1)}
-\frac{(3 N+2) S_2}{N+1}
\nonumber \\ &&  
-\frac{14}{9} S_3-\frac{2 \left(9 N^2+21 N+13\right)}{3 (N+1)^3}-\frac{\zeta_2}{N+1}
\Biggr]
+S_1^2 \left(\frac{1}{4} S_2-\frac{9 N^2+15 N+5}{6 (N+1)^2}-\frac{\zeta_2}{4}\right)
\nonumber \\ &&  
+\left(\frac{9 N^2+15 N+5}{2 (N+1)^2}
+\frac{3 \zeta_2}{4}\right) S_2
-\frac{1}{72} S_1^4
+\frac{(3 N+2) S_1^3}{18 (N+1)}
-\frac{3}{8} S_2^2
+\frac{7 (3 N+2) S_3}{9 (N+1)}
\nonumber \\ &&  
+S_1 \left(-\frac{(3 N+2) S_2}{2 (N+1)}-\frac{7}{9} S_3+\frac{27 N^3+72 N^2+60 N+14}{3 (N+1)^3}+\frac{(3 N+2) \zeta_2}{2 (N+1)}+\frac{7 \zeta_3}{3}\right)
\nonumber \\ &&  
+\frac{5}{4} S_4
+\frac{(3 N+4) \left(9 N^2+18 N+10\right)}{3 (N+1)^4}
+\frac{(3 N+4) \zeta_2}{2 (N+1)^2}
+\frac{7 \zeta_3}{3 (N+1)}
\Biggr\},
\\ 
J_7 &=&
\frac{1}{N+1} \Biggl\{
\frac{8}{3 \varepsilon ^3}
-\frac{4 (5 N+6)}{3 \varepsilon^2 (N+1)}
+\frac{1}{\varepsilon} \Biggl[
\frac{38 N^2+86 N+50}{3 (N+1)^2}+\zeta_2
\Biggr]
-\frac{(5 N+6) \zeta_2}{2 (N+1)}-\frac{7 \zeta_3}{3}
\nonumber \\ &&  
-\frac{65 N^3+214 N^2+238 N+90}{3 (N+1)^3}
+\varepsilon \Biggl[
\frac{\left(19 N^2+43 N+25\right) \zeta_2}{4 (N+1)^2}
+\frac{7 (5 N+6) \zeta_3}{6 (N+1)}
\nonumber \\ &&  
+\frac{211 N^4+909 N^3+1480 N^2+1082 N+301}{6 (N+1)^4}
-\frac{173 \zeta_2^2}{80}
\Biggr]
\Biggr\},
\\
J_8 &=&
\frac{1}{N+1} \Biggl\{
%
-\frac{4}{3 N \varepsilon^2}
-\frac{2}{3 N \varepsilon} \Biggl[
S_1+\frac{N^2-N-1}{N (N+1)}
\Biggr]
+\frac{\left(2 N^2+4 N+1\right) S_1}{3 N^2 (N+1)}-\frac{S_1^2}{6 N}-\frac{S_2}{6 N}
\nonumber \\ &&  
+S_{2,1}-\frac{4 N^4+6 N^3+4 N^2+2 N+1}{3 N^3
   (N+1)^2}-\frac{\zeta_2}{2 N}-2 \zeta_3
+\varepsilon \Biggl[
\frac{\left(2 N^2+4 N+1\right) S_1^2}{12 N^2 (N+1)}
\nonumber \\ &&  
+\frac{\left(2 N^2+4 N+1\right) S_2}{12 N^2 (N+1)}
-\frac{\left(2 N^2+4 N+1\right) S_{2,1}}{2 N (N+1)}
+\frac{\left(12 N^2+19 N+1\right) \zeta_3}{6 N (N+1)}
\nonumber \\ &&  
+S_1 \left(-\frac{S_2}{12 N}-S_{2,1}-\frac{4 N^4+12 N^3+13 N^2+5 N+1}{6 N^3 (N+1)^2}-\frac{\zeta_2}{4 N}+2 \zeta_3\right)
-\frac{S_1^3}{36 N}
\nonumber \\ &&  
-\frac{4 N^6+4 N^5-5 N^4-10 N^3-6 N^2-3 N-1}{6 N^4 (N+1)^3}
-\frac{1}{2} S_2^2
-\frac{S_3}{18 N}
-\frac{1}{2} S_4-S_{3,1}
\nonumber \\ &&  
+\frac{5}{2} S_{2,1,1}
-\frac{\left(N^2-N-1\right) \zeta_2}{4 N^2 (N+1)}
-\frac{9 \zeta_2^2}{5}
\Biggr]
+\varepsilon^2 \Biggl[
\frac{2 N^2+4 N+1}{36 N^2 (N+1)} \left(S_3+\frac{1}{2} S_1^3\right)
\nonumber \\ &&  
+\frac{\left(48 N^2+71 N-1\right) S_2^2}{96 N (N+1)}
+\frac{\left(24 N^2+35 N-1\right) S_4}{48 N (N+1)}-\frac{\left(10 N^2+16 N+1\right) S_{2,1,1}}{4 N (N+1)}
\nonumber \\ &&  
+S_1^2 \left(-\frac{S_2}{48 N}+\frac{1}{2} S_{2,1}-\frac{4 N^4+12 N^3+13 N^2+5 N+1}{24 N^3 (N+1)^2}-\frac{\zeta_2}{16 N}-\zeta_3\right)
\nonumber \\ &&  
+S_2 \left(S_3-2 S_{2,1}-\frac{4 N^4+12 N^3+13 N^2+5 N+1}{24 N^3 (N+1)^2}-\frac{\zeta_2}{16 N}-\zeta_3\right)
+\frac{3 \zeta_2}{8} S_{2,1}
\nonumber \\ &&  
+\frac{4 N^4+12 N^3+13 N^2+5 N+1}{4 N^2 (N+1)^2} S_{2,1}
+S_1 \biggl(\frac{\left(2 N^2+4 N+1\right) S_2}{24 N^2 (N+1)}+\frac{1}{2} S_2^2-\frac{S_3}{36 N}
\nonumber \\ &&  
+\frac{1}{2} S_4+\frac{(2 N+3) S_{2,1}}{2 (N+1)}
+S_{3,1}
+\frac{\left(2 N^2+4 N+1\right) \zeta_2}{8 N^2 (N+1)}
-\frac{\left(24 N^2+41 N+5\right) \zeta_3}{12 N (N+1)}
\nonumber \\ &&  
-\frac{5}{2} S_{2,1,1}
+\frac{8 N^6+32 N^5+50 N^4+40 N^3+18 N^2+6 N+1}{12 N^4 (N+1)^3}+\frac{9 \zeta_2^2}{5}\biggr)-\frac{S_1^4}{288 N}
\nonumber \\ &&   
+\frac{\left(288 N^2+461 N+29\right) \zeta_2^2}{160 N (N+1)}
+\zeta_2 \left(-\frac{4 N^4+6 N^3+4 N^2+2 N+1}{8 N^3 (N+1)^2}-\frac{3 \zeta_3}{4}\right)
\nonumber \\ &&  
+3 S_{2,2,1}-\frac{5}{2} S_{3,1,1}+\frac{19}{4} S_{2,1,1,1}
-\frac{\left(24 N^4+65 N^3+54 N^2+8 N+1\right) \zeta_3}{12 N^2 (N+1)^2}
\nonumber \\ &&   
-\frac{16 N^8+56 N^7+80 N^6+60 N^5+31 N^4+16 N^3+9 N^2+4 N+1}{12 N^5 (N+1)^4}
\nonumber \\ &&    
+S_5-\frac{5}{2} S_{2,3}+\frac{(2 N+3) S_{3,1}}{2 (N+1)}-\frac{3}{2} S_{4,1}
-9 \zeta_5
\Biggr]
\Biggr\}.
\end{eqnarray}
 
The last expression diverges at $N=0$. The correct result at this value of $N$ is
\begin{eqnarray}
J_8^{N=0} &=&
-\frac{8}{3 \varepsilon^3}
-\frac{1}{\varepsilon} \left(\zeta_2+\frac{8}{3}\right)
-\frac{5}{3} \zeta_3
+\varepsilon \left(-\frac{23}{16} \zeta_2^2
                 -\zeta_2
                 +2 \zeta_3
                 -\frac{8}{3}
          \right)
\nonumber \\ &&
+\varepsilon^2 \left(\frac{9}{5} \zeta_2^2
                  -\frac{5}{8} \zeta_2 \zeta_3
                  -\frac{8}{3} \zeta_3
                  -\frac{121}{20} \zeta_5
            \right),
\\
J_9 &=&
\frac{1}{(N+1)^2} \Biggl\{
\frac{4}{3 \varepsilon^3} (3 N+4)
+\frac{2}{3 \varepsilon^2} \Biggl[
(3 N+1) S_1-\frac{15 N^2+34 N+22}{N+1}
\Biggr]
\nonumber \\ &&  
+\frac{1}{\varepsilon} \Biggl[
\frac{3 N+1}{6} \left(S_1^2+S_2\right)
-\frac{(3 N+5) (5 N+2) S_1}{3 (N+1)}
+\frac{57 N^3+185 N^2+205 N+84}{3 (N+1)^2}
\nonumber \\ &&  
+\frac{1}{2} (3 N+4) \zeta_2
\Biggr]
+S_1 \Biggl[\frac{1}{4} (3 N+1) \left(\frac{1}{3} S_2+\zeta_2\right)
           +\frac{57 N^3+188 N^2+202 N+57}{6 (N+1)^2}
\Biggr]
\nonumber \\ &&  
+\frac{1}{36} (3 N+1) S_1^3
-\frac{(3 N+5) (5 N+2) S_1^2}{12 (N+1)}-\frac{(3 N+5) (5 N+2) S_2}{12 (N+1)}+\frac{1}{18} (3 N+1) S_3
\nonumber \\ &&  
+(N-1) S_{2,1}
-\frac{\left(15 N^2+34 N+22\right) \zeta_2}{4 (N+1)}-\frac{195 N^4+826 N^3+1320 N^2+952 N}{6 (N+1)^3}
\nonumber \\ &&  
-\frac{278}{6 (N+1)^3}
+\frac{1}{6} (3 N+20) \zeta_3
+\varepsilon \Biggl[
S_1^2 \biggl(\frac{57 N^3+188 N^2+202 N+57}{24 (N+1)^2}
\nonumber \\ &&  
+\frac{1}{16} (3 N+1) \left(\frac{1}{3} S_2+\zeta_2\right)\biggr)
+\left(\frac{57 N^3+188 N^2+202 N+57}{24 (N+1)^2}+\frac{3 N+1}{16} \zeta_2\right) S_2
\nonumber \\ &&  
+S_1 \biggl(-\frac{(3 N+5) (5 N+2) S_2}{24 (N+1)}+\frac{1}{36} (3 N+1) S_3
+\frac{1}{2} (N+3) S_{2,1}+\frac{1}{12} (3 N-31) \zeta_3
\nonumber \\ &&   
-\frac{195 N^4+859 N^3+1428 N^2+1048 N+254}{12 (N+1)^3}-\frac{(3 N+5) (5 N+2) \zeta_2}{8 (N+1)}\biggr)
\nonumber \\ &&   
+\frac{1}{288} (3 N+1) S_1^4
-\frac{(3 N+5) (5 N+2) S_1^3}{72 (N+1)}
+\frac{1}{96} (27 N+73) S_2^2
+\frac{1}{48} (15 N+37) S_4
\nonumber \\ &&  
-\frac{(3 N+5) (5 N+2) S_3}{36 (N+1)}
-\frac{(N+2) (5 N-1) S_{2,1}}{2 (N+1)}
+\frac{1}{2} (N+3) S_{3,1}
-\frac{1}{2} (N+7) S_{2,1,1}
\nonumber \\ &&  
-\frac{\left(15 N^2+50 N+86\right) \zeta_3}{12 (N+1)}
+\frac{\left(57 N^3+185 N^2+205 N+84\right) \zeta_2}{8 (N+1)^2}
+\frac{57 N+460}{160} \zeta_2^2
\nonumber \\ &&  
+\frac{633 N^5+3311 N^4+6938 N^3+7286 N^2+3855 N+860}{12 (N+1)^4}
\Biggr]
\Biggr\},
\\
J_{10} &=&
\frac{1}{N+1} \Biggl\{
\frac{4}{3 \varepsilon^3}
\left(S_1+\frac{3 N^2+6 N+4}{N+1}\right)
+\frac{1}{\varepsilon^2} \Biggl[
\frac{2 \left(3 N^2+5 N-1\right) S_1}{3 (N+1)}-S_1^2-S_2
  \nonumber \\ &&  
-\frac{2 \left(21 N^3+60 N^2+58 N+22\right)}{3 (N+1)^2}
\Biggr]
+\frac{1}{\varepsilon} \Biggl[
\frac{3 N^2+9 N+7}{6 (N+1)} \left(S_1^2+S_2\right)
+\frac{1}{18} S_1^3
  \nonumber \\ &&  
+S_1 \left(\frac{1}{6} S_2-\frac{21 N^3+67 N^2+68 N+15}{3 (N+1)^2}+\frac{\zeta_2}{2}\right)
+\frac{1}{9} S_3
+\frac{\left(3 N^2+6 N+4\right) \zeta_2}{2 (N+1)}
  \nonumber \\ &&  
+2 S_{2,1}
+\frac{99 N^4+375 N^3+527 N^2+328 N+84}{3 (N+1)^3}
\Biggr]
+\frac{\left(3 N^2+5 N-1\right) S_1^3}{36 (N+1)}
 \nonumber \\ &&  
+\frac{\left(3 N^2+5 N-1\right) S_3}{18 (N+1)}
+\frac{\left(N^2+N-3\right) S_{2,1}}{N+1}
-\frac{1}{48} S_1^4
-S_1^2 \biggl(\frac{1}{8} S_2+\frac{3 \zeta_2}{8}
  \nonumber \\ &&  
+\frac{21 N^3+63 N^2+64 N+19}{12 (N+1)^2}\biggr)
+\left(-\frac{21 N^3+63 N^2+64 N+19}{12 (N+1)^2}-\frac{3 \zeta_2}{8}\right) S_2
 \nonumber \\ &&  
+S_1 \biggl(\frac{\left(3 N^2+5 N-1\right) S_2}{12 (N+1)}
           -\frac{1}{6} S_3-3 S_{2,1}
           +\frac{\left(3 N^2+5 N-1\right) \zeta_2}{4 (N+1)}
           +\frac{17 \zeta_3}{6}
  \nonumber \\ &&  
           +\frac{99 N^4+428 N^3+696 N^2+497 N+115}{6 (N+1)^3}
     \biggr)
-\frac{25}{16} S_2^2
-\frac{13}{8} S_4
-3 S_{3,1}
+7 S_{2,1,1}
 \nonumber \\ &&  
+\frac{\left(3 N^2+6 N+20\right) \zeta_3}{6 (N+1)}
-\frac{393 N^5+1866 N^4+3502 N^3+3234 N^2+1468 N+278}{6 (N+1)^4}
 \nonumber \\ &&  
-\frac{\left(21 N^3+60 N^2+58 N+22\right) \zeta_2}{4 (N+1)^2}
+\varepsilon \Biggl[
\frac{\left(3 N^2+9 N+7\right) S_1^4}{288 (N+1)}
+\frac{\left(N^2+5 N+9\right) S_{3,1}}{2 (N+1)}
  \nonumber \\ &&  
+\frac{\left(27 N^2+129 N+223\right) S_2^2}{96 (N+1)}
+\frac{\left(15 N^2+69 N+115\right) S_4}{48 (N+1)}
-\frac{\left(N^2+9 N+21\right) S_{2,1,1}}{2 (N+1)}
  \nonumber \\ &&  
+S_1^3 \left(\frac{1}{144} S_2
             -\frac{21 N^3+67 N^2+68 N+15}{72 (N+1)^2}
             +\frac{\zeta_2}{48}\right)
+\frac{S_1^5}{1440}
+\frac{151}{60} S_5
-\frac{13}{2} S_{2,3}
  \nonumber \\ &&  
+\left(\frac{\zeta_2}{24}-\frac{21 N^3+67 N^2+68 N+15}{36 (N+1)^2}\right) S_3
+\left(\frac{3 \zeta_2}{4}-\frac{7 N^3+23 N^2+22 N-1}{2 (N+1)^2}\right) S_{2,1}
  \nonumber \\ &&  
+S_1^2 \biggl(\frac{\left(3 N^2+9 N+7\right) S_2}{48 (N+1)}
             +\frac{99 N^4+384 N^3+560 N^2+361 N+75}{24 (N+1)^3}
             +\frac{1}{72} S_3
   \nonumber \\ &&   
             +\frac{5}{4} S_{2,1}
             +\frac{\left(3 N^2+9 N+7\right) \zeta_2}{16 (N+1)}
             -\frac{17 \zeta_3}{8}\biggr)
+S_2 \biggl(\frac{181}{72} S_3
           +\frac{\left(3 N^2+9 N+7\right) \zeta_2}{16 (N+1)}
   \nonumber \\ &&   
           -\frac{21}{4} S_{2,1}
           +\frac{99 N^4+384 N^3+560 N^2+361 N+75}{24 (N+1)^3}
           -\frac{17 \zeta_3}{8}\biggr)
-4 S_{4,1}
+\frac{15}{2} S_{2,2,1}
  \nonumber \\ &&  
+S_1 \biggl(\frac{\left(3 N^2+9 N+7\right) S_3}{36 (N+1)}
           +\frac{\left(N^2+5 N+9\right) S_{2,1}}{2 (N+1)}
           +\frac{121}{96} S_2^2+\frac{61}{48} S_4+\frac{5}{2} S_{3,1}
   \nonumber \\ &&   
           +\left(\frac{\zeta_2}{16}-\frac{21 N^3+67 N^2+68 N+15}{24 (N+1)^2}\right) S_2
           -\frac{13}{2} S_{2,1,1}
           +\frac{\left(3 N^2-11 N-65\right) \zeta_3}{12 (N+1)}
   \nonumber \\ &&   
           -\frac{393 N^5+2113 N^4+4576 N^3+4990 N^2+2724 N+559}{12 (N+1)^4}
           +\frac{403 \zeta_2^2}{160}
   \nonumber \\ &&   
           -\frac{\left(21 N^3+67 N^2+68 N+15\right) \zeta_2}{8 (N+1)^2}\biggr)
-\frac{\left(21 N^3+60 N^2+74 N+86\right) \zeta_3}{12 (N+1)^2}
-\frac{13}{2} S_{3,1,1}
  \nonumber \\ &&  
+\frac{25}{2} S_{2,1,1,1}
+\frac{\left(57 N^2+114 N+460\right) \zeta_2^2}{160 (N+1)}
+\frac{\left(99 N^4+375 N^3+527 N^2+328 N\right) \zeta_2}{8 (N+1)^3}
  \nonumber \\ &&  
+\frac{1419 N^6+8121 N^5+19172 N^4+23804 N^3+16295 N^2+5790 N+860}{12 (N+1)^5}
  \nonumber \\ &&  
+\frac{21}{2 (N+1)^3} \zeta_2
\Biggr]
\Biggr\},
\label{J10N}
\\ 
J_{11} &=&
\frac{1}{(N+1)^2} \Biggl\{
-\frac{8}{3 \varepsilon ^3}
+\frac{4 (3 N+4)}{3 \varepsilon^2 (N+1)}
-\frac{1}{\varepsilon} \Biggl[
\frac{2 \left(9 N^2+21 N+13\right)}{3 (N+1)^2}+\zeta_2
\Biggr]
 \nonumber \\ &&    
+\frac{(3 N+4) \left(9 N^2+18 N+10\right)}{3 (N+1)^3}+\frac{(3 N+4) \zeta_2}{2 (N+1)}+\frac{7 \zeta_3}{3}
\Biggr\},
\\
J_{12} &=&
\frac{1}{N+1} \Biggl\{
-\frac{8}{3 \varepsilon ^3} \left(S_1+\frac{1}{N+1}\right)
+\frac{1}{\varepsilon^2} \Biggl[
\frac{2}{3} S_1^2+\frac{4 (3 N+4) S_1}{3 (N+1)}+\frac{2}{3} S_2+\frac{4 (3 N+4)}{3 (N+1)^2}
\Biggr]
 \nonumber \\ &&  
+\frac{1}{\varepsilon} \Biggl[
-S_1 \left(\frac{1}{3} S_2+\frac{2 \left(9 N^2+21 N+13\right)}{3 (N+1)^2}+\zeta_2\right)
-\frac{1}{9} S_1^3
-\frac{(3 N+4) S_1^2}{3 (N+1)}
-\frac{2}{9} S_3
  \nonumber \\ &&  
-\frac{(3 N+4) S_2}{3 (N+1)}
-\frac{2 \left(9 N^2+21 N+13\right)}{3 (N+1)^3}
-\frac{\zeta_2}{N+1}
\Biggr]
+S_1 \biggl(\frac{(3 N+4) S_2}{6 (N+1)}
           +\frac{1}{9} S_3
  \nonumber \\ &&  
           +\frac{(3 N+4) \left(9 N^2+18 N+10\right)}{3 (N+1)^3}
           +\frac{(3 N+4) \zeta_2}{2 (N+1)}
           +\frac{7 \zeta_3}{3}\biggr)
+\frac{(3 N+4) S_1^3}{18 (N+1)}
+\frac{1}{72} S_1^4
 \nonumber \\ && 
+S_1^2 \left(\frac{1}{12} S_2
             +\frac{9 N^2+21 N+13}{6 (N+1)^2}
             +\frac{\zeta_2}{4}\right)
+\left(\frac{9 N^2+21 N+13}{6 (N+1)^2}+\frac{\zeta_2}{4}\right) S_2
+\frac{1}{24} S_2^2
 \nonumber \\ && 
+\frac{(3 N+4) S_3}{9 (N+1)}
+\frac{1}{12} S_4
+\frac{(3 N+4) \left(9 N^2+18 N+10\right)}{3 (N+1)^4}
+\frac{(3 N+4) \zeta_2}{2 (N+1)^2}
 \nonumber \\ && 
+\frac{7 \zeta_3}{3 (N+1)}
\Biggr\},
\\
J_{13} &=&
\frac{8}{3 \varepsilon ^3}
+\frac{2}{3 \varepsilon ^2} \left(2 S_1-\frac{9 N+7}{N+1}\right)
+\frac{1}{\varepsilon} \biggl(\frac{1}{3} S_1^2-\frac{(9 N+7) S_1}{3 (N+1)}-S_2+\frac{47 N^2+76 N+25}{6 (N+1)^2}
 \nonumber \\ && 
+\zeta_2\biggr)
+S_1 \left(-\frac{1}{2} S_2+\frac{47 N^2+76 N+25}{12 (N+1)^2}+\frac{\zeta_2}{2}\right)
+\frac{1}{18} S_1^3
-\frac{(9 N+7) S_1^2}{12 (N+1)}
+\frac{7}{9} S_3
\nonumber \\ && 
+\frac{(9 N+7) S_2}{4 (N+1)}
-\frac{133 N^3+305 N^2+175 N-5}{24 (N+1)^3}
-\frac{(9 N+7) \zeta_2}{4 (N+1)}
-\frac{7 \zeta_3}{3}
\nonumber \\ && 
+\varepsilon  \Biggl[S_1^2 \left(-\frac{1}{8} S_2+\frac{47 N^2+76 N+25}{48 (N+1)^2}+\frac{\zeta_2}{8}\right)
-\left(\frac{47 N^2+76 N+25}{16 (N+1)^2}+\frac{3 \zeta_2}{8}\right) S_2
 \nonumber \\ && 
+S_1 \left(\frac{(9 N+7) S_2}{8 (N+1)}+\frac{7}{18} S_3-\frac{133 N^3+305 N^2+175 N-5}{48 (N+1)^3}
-\frac{(9 N+7) \zeta_2}{8 (N+1)}-\frac{7 \zeta_3}{6}\right)
 \nonumber \\ && 
+\frac{1}{144} S_1^4-\frac{(9 N+7) S_1^3}{72 (N+1)}+\frac{3}{16} S_2^2
-\frac{7 (9 N+7) S_3}{36 (N+1)}-\frac{5}{8} S_4+\frac{\left(47 N^2+76 N+25\right) \zeta_2}{16 (N+1)^2}
 \nonumber \\ && 
-\frac{417 N^4+1934 N^3+3488 N^2+2914 N+959}{96 (N+1)^4}
+\frac{7 (9 N+7) \zeta_3}{12 (N+1)}-\frac{173 \zeta_2^2}{80}\Biggr]
\nonumber \\ && 
+\varepsilon ^2 \Biggl[-\frac{1}{48} S_1^3 S_2 
+\left(\frac{47 N^2+76 N+25}{144 (N+1)^2}+\frac{\zeta_2}{24}\right) \left(\frac{1}{2} S_1^3+7 S_3\right)
+S_1^2 \biggl(\frac{(9 N+7) S_2}{32 (N+1)}
  \nonumber \\ &&  
             +\frac{7}{72} S_3
             -\frac{133 N^3+305 N^2+175 N-5}{192 (N+1)^3}
             -\frac{(9 N+7) \zeta_2}{32 (N+1)}-\frac{7 \zeta_3}{24}\biggr)
+S_2 \biggl(-\frac{7}{24} S_3
  \nonumber \\ &&  
+\frac{133 N^3+305 N^2+175 N-5}{64 (N+1)^3}
+\frac{3 (9 N+7) \zeta_2}{32 (N+1)}
+\frac{7 \zeta_3}{8}\biggr)
+S_1 \biggl(-\frac{7 (9 N+7) S_3}{72 (N+1)}
  \nonumber \\ &&  
-\left(\frac{47 N^2+76 N+25}{32 (N+1)^2}+\frac{3 \zeta_2}{16}\right) S_2
+\frac{3}{32} S_2^2
-\frac{5}{16} S_4
+\frac{\left(47 N^2+76 N+25\right) \zeta_2}{32 (N+1)^2}
  \nonumber \\ &&  
-\frac{417 N^4+1934 N^3+3488 N^2+2914 N+959}{192 (N+1)^4}
+\frac{7 (9 N+7) \zeta_3}{24 (N+1)}-\frac{173 \zeta_2^2}{160}\biggr)
 \nonumber \\ && 
-\frac{(9 N+7) S_1^4}{576 (N+1)}
-\frac{3 (9 N+7) S_2^2}{64 (N+1)}
+\frac{5 (9 N+7) S_4}{32 (N+1)}
-\frac{7 \left(47 N^2+76 N+25\right) \zeta_3}{48 (N+1)^2}
 \nonumber \\ && 
+\frac{S_1^5}{1440}
+\frac{31}{60} S_5
-\zeta_2 \left(\frac{133 N^3+305 N^2+175 N-5}{64 (N+1)^3}+\frac{7 \zeta_3}{8}\right)
+\frac{173 (9 N+7) \zeta_2^2}{320 (N+1)}
 \nonumber \\ && 
+\frac{10243 N^5+50381 N^4+99626 N^3+99398 N^2+50371 N+10493}{384 (N+1)^5}
-\frac{239 \zeta_5}{20}\Biggr],
\\
J_{14} &=&
\frac{1}{N+1} \Biggl\{
-\frac{4}{\varepsilon ^3}
+\frac{2}{\varepsilon^2} \Biggl[
\frac{N}{N+1}-S_1
\Biggr]
+\frac{1}{\varepsilon} \Biggl[
\frac{N-2}{N+1} S_1-\frac{1}{2} \left(S_1^2+S_2\right)-\frac{N^2+N+3}{(N+1)^2}
  \nonumber \\ &&  
-\frac{3 \zeta_2}{2}
\Biggr]
-S_1 \left(\frac{1}{4} S_2+\frac{N^2-N+3}{2 (N+1)^2}+\frac{3 \zeta_2}{4}\right)
-\frac{1}{12} S_1^3
+\frac{(N-2) S_1^2}{4 (N+1)}
+\frac{(N-2) S_2}{4 (N+1)}
 \nonumber \\ &&  
-\frac{1}{6} S_3-S_{2,1}
+\frac{N^3+2 N^2+4 N-2}{2 (N+1)^3}
+\frac{3 N \zeta_2}{4 (N+1)}
-\frac{\zeta_3}{2}
+\varepsilon \Biggl[
S_1 \biggl(\frac{(N-2) S_2}{8 (N+1)}
   \nonumber \\ &&   
-\frac{1}{12} S_3-\frac{1}{2} S_{2,1}+\frac{N^3+2 N-8}{4 (N+1)^3}+\frac{3 (N-2) \zeta_2}{8 (N+1)}-\frac{\zeta_3}{4}\biggr)
-S_1^2 \biggl(\frac{1}{16} S_2+\frac{3 \zeta_2}{16}
   \nonumber \\ &&   
+\frac{N^2-N+3}{8 (N+1)^2}\biggr)
+\left(-\frac{N^2-N+3}{8 (N+1)^2}-\frac{3 \zeta_2}{16}\right) S_2
-\frac{1}{96} S_1^4+\frac{(N-2) S_1^3}{24 (N+1)}
-\frac{1}{2} S_{3,1}
  \nonumber \\ &&   
-\frac{9}{32} S_2^2
+\frac{(N-2) S_3}{12 (N+1)}
-\frac{5}{16} S_4
+\frac{(N-2) S_{2,1}}{2 (N+1)}
+\frac{1}{2} S_{2,1,1}
-\frac{3 \left(N^2+N+3\right) \zeta_2}{8 (N+1)^2}
  \nonumber \\ &&   
-\frac{N^4+3 N^3+6 N^2+2 N+9}{4 (N+1)^4}
+\frac{N \zeta_3}{4 (N+1)}
-\frac{57 \zeta_2^2}{160}
\Biggr]
+\varepsilon^2 \Biggl[
S_1^2 \biggl(\frac{N-2}{32 (N+1)} S_2
   \nonumber \\ &&   
-\frac{1}{48} S_3-\frac{1}{8} S_{2,1}+\frac{N^3+2 N-8}{16 (N+1)^3}
+\frac{3 (N-2) \zeta_2}{32 (N+1)}-\frac{\zeta_3}{16}\biggr)
+S_2 \biggl(-\frac{13}{48} S_3+\frac{1}{8} S_{2,1}
   \nonumber \\ &&   
+\frac{N^3+2 N-8}{16 (N+1)^3}+\frac{3 (N-2) \zeta_2}{32 (N+1)}-\frac{\zeta_3}{16}\biggr)
-\left(\frac{N^2-N+3}{24 (N+1)^2}+\frac{\zeta_2}{16}\right) \left(\frac{1}{2} S_1^3+S_3\right)
  \nonumber \\ &&   
-\frac{1}{96} S_1^3 S_2
-\left(\frac{N^2-N+3}{4 (N+1)^2}+\frac{3 \zeta_2}{8}\right) S_{2,1}
+S_1 \biggl(-\left(\frac{N^2-N+3}{16 (N+1)^2}+\frac{3 \zeta_2}{32}\right) S_2
   \nonumber \\ &&   
-\frac{9}{64} S_2^2
+\frac{N-2}{4 (N+1)} \left(\frac{1}{6} S_3+S_{2,1}+\frac{1}{2} \zeta_3\right)
-\frac{5}{32} S_4
-\frac{1}{4} S_{3,1}
+\frac{1}{4} S_{2,1,1}
-\frac{57 \zeta_2^2}{320}
   \nonumber \\ &&   
-\frac{3 \left(N^2-N+3\right) \zeta_2}{16 (N+1)^2}
-\frac{N^4+N^3+2 N^2-6 N+13}{8 (N+1)^4}
\biggr)
-\frac{1}{960} S_1^5
+\frac{57 N \zeta_2^2}{320 (N+1)}
  \nonumber \\ &&   
+\frac{N-2}{4 (N+1)}\left(\frac{1}{48} S_1^4+\frac{9}{16} S_2^2+\frac{5}{8} S_4+S_{3,1}-S_{2,1,1}\right)
-\frac{\left(N^2+N+3\right) \zeta_3}{8 (N+1)^2}
-\frac{11}{40} S_5
  \nonumber \\ &&   
+\frac{1}{4} S_{2,3}
-\frac{3}{4} S_{2,2,1}
+\frac{1}{4} S_{3,1,1}
-\frac{1}{4} S_{2,1,1,1}
+\zeta_2 \left(\frac{3 \left(N^3+2 N^2+4 N-2\right)}{16 (N+1)^3}-\frac{3 \zeta_3}{16}\right)
  \nonumber \\ &&   
+\frac{N^5+4 N^4+9 N^3+8 N^2+11 N-12}{8 (N+1)^5}
-\frac{3 \zeta_5}{40}
\Biggr]
\Biggr\},
\\
J_{15} &=&
\frac{1}{N+1} \Biggl\{
\frac{4 (N+2)}{\varepsilon ^3}
+\frac{2}{\varepsilon^2} \Biggl[
N S_1-\frac{5 N^2+12 N+8}{N+1}
\Biggr]
+\frac{1}{\varepsilon} \Biggl[
\frac{N}{2} \left(S_1^2+S_2\right)
  \nonumber \\ &&   
-\frac{\left(5 N^2+8 N+2\right) S_1}{N+1}
+\frac{17 N^3+53 N^2+57 N+22}{(N+1)^2}
+\frac{3}{2} (N+2) \zeta_2
\Biggr]
+\frac{N}{12} S_1^3
 \nonumber \\ && 
-\frac{5 N^2+8 N+2}{4 (N+1)} \left(S_1^2+S_2\right)
+S_1 \left(\frac{N}{4} S_2+\frac{17 N^3+49 N^2+45 N+12}{2 (N+1)^2}+\frac{3 N \zeta_2}{4}\right)
 \nonumber \\ && 
-\frac{3 \left(5 N^2+12 N+8\right) \zeta_2}{4
   (N+1)}-\frac{49 N^4+194 N^3+290 N^2+196 N+52}{2 (N+1)^3}+\frac{1}{2} (N+2) \zeta_3
 \nonumber \\ && 
+\frac{N}{6} S_3+N S_{2,1}
+\varepsilon \Biggl[
-\frac{5 N^2+8 N+2}{2 (N+1)} \left(\frac{1}{12} S_1^3+\frac{1}{6} S_3+S_{2,1}\right)
+S_1^2 \biggl(\frac{N}{16} S_2
   \nonumber \\ &&   
             +\frac{3 N}{16} \zeta_2
             +\frac{17 N^3+49 N^2+45 N+12}{8 (N+1)^2}
\biggr)
+S_1 \biggl(-\frac{5 N^2+8 N+2}{8 (N+1)} \left(S_2+3 \zeta_2\right)
   \nonumber \\ &&   
           +\frac{N}{12} S_3
           +\frac{N}{2} S_{2,1}
           -\frac{49 N^4+198 N^3+298 N^2+196 N+46}{4 (N+1)^3}
           +\frac{N \zeta_3}{4}\biggr)
+\frac{N}{96} S_1^4
  \nonumber \\ &&  
+\left(\frac{17 N^3+49 N^2+45 N+12}{8 (N+1)^2}+\frac{3 N \zeta_2}{16}\right) S_2
-\frac{\left(5 N^2+12 N+8\right) \zeta_3}{4 (N+1)}
+\frac{9}{32} N S_2^2
  \nonumber \\ &&  
+\frac{5 N}{16} S_4
-\frac{N}{2} S_{2,1,1}
+\frac{3 \left(17 N^3+53 N^2+57 N+22\right) \zeta_2}{8 (N+1)^2}
+\frac{57}{160} (N+2) \zeta_2^2
  \nonumber \\ &&  
+\frac{N}{2} S_{3,1}
+\frac{129 N^5+627 N^4+1216 N^3+1178 N^2+573 N+114}{4 (N+1)^4}
\Biggr]
+\varepsilon^2 \Biggl[
\frac{N}{960} S_1^5
  \nonumber \\ &&  
-\frac{\left(5 N^2+8 N+2\right) S_1^4}{192 (N+1)}
+\left(\frac{17 N^3+49 N^2+45 N+12}{48 (N+1)^2}
       +\frac{N \zeta_2}{32}+\frac{1}{96} N S_2\right) S_1^3
  \nonumber \\ &&  
+\biggl(-\frac{49 N^4+198 N^3+298 N^2+196 N+46}{16 (N+1)^3}
       -\frac{5 N^2+8 N+2}{32 (N+1)} \left(S_2+3 \zeta_2\right)
   \nonumber \\ &&   
       +\frac{N \zeta_3}{16}
       +\frac{N}{48} S_3
       +\frac{N}{8} S_{2,1}\biggr) S_1^2
+S_1 \biggl(\frac{17 N^3+49 N^2+45 N+12}{16 (N+1)^2} \left(S_2+3 \zeta_2\right)
   \nonumber \\ &&   
       +\frac{57 N \zeta_2^2}{320}
       +\frac{9}{64} N S_2^2
       +\frac{129 N^5+663 N^4+1364 N^3+1402 N^2+717 N+144}{8 (N+1)^4}
   \nonumber \\ &&   
       -\frac{5 N^2+8 N+2}{4 (N+1)} \left(\frac{\zeta_3}{2}+\frac{1}{6} S_3+S_{2,1}\right)
       +\frac{3 N \zeta_2}{32} S_2
       +\frac{5 N}{32} S_4
       +\frac{N}{4} \left(S_{3,1}-S_{2,1,1}\right)
\biggr)
  \nonumber \\ &&  
+\frac{5 N^2+8 N+2}{4 (N+1)} \left(S_{2,1,1}-S_{3,1}-\frac{3}{8} \zeta_2 S_2-\frac{5}{8} S_4-\frac{9}{16} S_2^2\right)
+\frac{3}{40} (N+2) \zeta_5
  \nonumber \\ &&  
-\frac{321 N^6+1860 N^5+4467 N^4+5686 N^3+4043 N^2+1524 N+240}{8 (N+1)^5}
+\frac{11}{40} N S_5
  \nonumber \\ &&  
-\frac{57 \left(5 N^2+12 N+8\right) \zeta_2^2}{320 (N+1)}
+\frac{\left(17 N^3+53 N^2+57 N+22\right) \zeta_3}{8 (N+1)^2}
+\frac{3}{16} (N+2) \zeta_2 \zeta_3
  \nonumber \\ &&  
-\frac{3 \left(49 N^4+194 N^3+290 N^2+196 N+52\right)}{16 (N+1)^3} \zeta_2
+\frac{17 N^3+49 N^2+45 N+12}{24 (N+1)^2} S_3
  \nonumber \\ &&  
+\frac{N \zeta_2}{16} S_3
+\left(\frac{17 N^3+49 N^2+45 N+12}{4 (N+1)^2}
       +\frac{3 N \zeta_2}{8}
\right) S_{2,1}
-\frac{N}{4} S_{2,3}
+\frac{3}{4} N S_{2,2,1}
  \nonumber \\ &&  
+S_2 \left(-\frac{49 N^4+198 N^3+298 N^2+196 N+46}{16 (N+1)^3}
           +\frac{N \zeta_3}{16}
           +\frac{13}{48} N S_3
           -\frac{1}{8} N S_{2,1}
\right)
  \nonumber \\ &&  
-\frac{N}{4} S_{3,1,1}
+\frac{N}{4} S_{2,1,1,1}
\Biggr]
\Biggr\},
\\
J_{16} &=&
\frac{1}{N+1} \Biggl\{
\frac{8}{\varepsilon ^3}
-\frac{4 (3 N+4)}{\varepsilon^2 (N+1)}
+\frac{1}{\varepsilon} \Biggl[
\frac{2 \left(7 N^2+17 N+11\right)}{(N+1)^2}+3 \zeta_2
\Biggr]
-\frac{3 (3 N+4) \zeta_2}{2 (N+1)}+\zeta_3
 \nonumber \\ && 
-\frac{15 N^3+52 N^2+62 N+26}{(N+1)^3}
+\varepsilon \Biggl[
\frac{3 \left(7 N^2+17 N+11\right) \zeta_2}{4 (N+1)^2}
-\frac{(3 N+4) \zeta_3}{2 (N+1)}
+\frac{57 \zeta_2^2}{80}
  \nonumber \\ &&  
+\frac{31 N^4+139 N^3+238 N^2+186 N+57}{2 (N+1)^4}
\Biggr]
+\varepsilon^2 \Biggl[
\frac{\left(7 N^2+17 N+11\right) \zeta_3}{4 (N+1)^2}
  \nonumber \\ &&  
+\zeta_2 \left(\frac{3 \zeta_3}{8}-\frac{3 \left(15 N^3+52 N^2+62 N+26\right)}{8 (N+1)^3}\right)
-\frac{57 (3 N+4) \zeta_2^2}{160 (N+1)}+\frac{3 \zeta_5}{20}
  \nonumber \\ &&  
-\frac{63 N^5+346 N^4+769 N^3+868 N^2+501 N+120}{4 (N+1)^5}
\Biggr]
\Biggr\},
\\
J_{17} &=&
\frac{1}{N+1} \Biggl\{
\frac{8}{3 \varepsilon ^3}
-\frac{2}{3 \varepsilon^2} \Biggl[
2 S_1+\frac{5 N+7}{N+1}
\Biggr]
+\frac{1}{\varepsilon} \Biggl[
\frac{1}{3} S_1^2+\frac{11 N^2+32 N+25}{6 (N+1)^2}+\frac{1}{3} S_2+\zeta_2
  \nonumber \\ &&  
+\frac{(5 N+7) S_1}{3 (N+1)}
\Biggr]
-S_1 \left(\frac{1}{6} S_2+\frac{11 N^2+32 N+25}{12 (N+1)^2}+\frac{\zeta_2}{2}\right)
-\frac{1}{18} S_1^3
-\frac{1}{9} S_3
-\frac{7 \zeta_3}{3}
 \nonumber \\ &&  
-\frac{5 N+7}{4 (N+1)} \left(\frac{1}{3} S_1^2+\frac{1}{3} S_2+\zeta_2\right)
+\frac{55 N^3+143 N^2+101 N+5}{24 (N+1)^3}
+\varepsilon \Biggl[
\frac{1}{24} S_2 S_1^2
 \nonumber \\ &&  
+\left(\frac{11 N^2+32 N+25}{48 (N+1)^2}+\frac{\zeta_2}{8}\right) \left(S_1^2+S_2\right)
+\frac{\left(11 N^2+32 N+25\right) \zeta_2}{16 (N+1)^2}
+\frac{1}{144} S_1^4
 \nonumber \\ &&  
+\frac{5 N+7}{12 (N+1)} \left(\frac{1}{2} S_1 S_2+\frac{3}{2} \zeta_2 S_1+\frac{1}{6} S_1^3+\frac{1}{3} S_3+7 \zeta_3\right)
+\frac{1}{48} S_2^2
+\frac{1}{24} S_4
-\frac{173 \zeta_2^2}{80}
 \nonumber \\ &&  
+S_1 \left(\frac{1}{18} S_3+\frac{7 \zeta_3}{6}-\frac{55 N^3+143 N^2+101 N+5}{48 (N+1)^3}\right)
-\frac{949 N^4+3906 N^3}{96 (N+1)^4}
 \nonumber \\ &&  
-\frac{5980 N^2+3998 N+959}{96 (N+1)^4}
\Biggr]
+\varepsilon^2 \Biggl[
-\frac{1}{144} S_1^3 S_2
-\frac{1}{72} S_3 \left(S_1^2+S_2\right)
-\frac{1}{1440} S_1^5
 \nonumber \\ &&  
-\biggl(\frac{11 N^2+32 N+25}{96 (N+1)^2}+\frac{\zeta_2}{16}\biggr) \left(\frac{1}{3} S_1^3+\frac{2}{3} S_3+S_1 S_2\right)
-\frac{5 N+7}{32 (N+1)} \biggl(\frac{1}{3} S_1^2 S_2
 \nonumber \\ &&  
                               +\zeta_2 S_1^2+\zeta_2 S_2+\frac{4}{9} S_1 S_3+\frac{28}{3} \zeta_3 S_1
                               +\frac{1}{18} S_1^4+\frac{1}{6} S_2^2+\frac{1}{3} S_4-\frac{173}{10} \zeta_2^2\biggr)
-\frac{1}{60} S_5
 \nonumber \\ &&  
+\left(\frac{55 N^3+143 N^2+101 N+5}{192 (N+1)^3}-\frac{7 \zeta_3}{24}\right) \left(S_1^2+S_2+3 \zeta_2\right)
+S_1 \biggl(-\frac{1}{96} S_2^2           
           -\frac{1}{48} S_4  
 \nonumber \\ &&  
           +\frac{949 N^4+3906 N^3+5980 N^2+3998 N+959}{192 (N+1)^4}
           +\frac{173 \zeta_2^2}{160}\biggr)
-\frac{239 \zeta_5}{20}
 \nonumber \\ &&  
-\frac{11 N^2+32 N+25}{16 (N+1)^2} \left(\frac{\zeta_2}{2} S_1+\frac{7}{3} \zeta_3\right)
+\frac{93562 N^3+97710 N^2+50871 N+10493}{384 (N+1)^5} 
 \nonumber \\ &&  
+\frac{8575 N^5+44773 N^4}{384 (N+1)^5} 
\Biggr]
\Biggr\},
\\
J_{18} &=&
\frac{1}{N+1} \Biggl\{
\frac{8}{3 \varepsilon ^3}
-\frac{2 (5 N+7)}{3 \varepsilon^2 (N+1)}
+\frac{1}{\varepsilon} \Biggl[
\frac{11 N^2+32 N+25}{6 (N+1)^2}+\zeta_2
\Biggr]
-\frac{(5 N+7) \zeta_2}{4 (N+1)}-\frac{7 \zeta_3}{3}
\nonumber \\ &&
+\frac{55 N^3+143 N^2+101 N+5}{24 (N+1)^3}
+\varepsilon \Biggl[
\frac{\left(11 N^2+32 N+25\right) \zeta_2}{16 (N+1)^2}
+\frac{7 (5 N+7) \zeta_3}{12 (N+1)}
\nonumber \\ &&
-\frac{173 \zeta_2^2}{80}
-\frac{949 N^4+3906 N^3+5980 N^2+3998 N+959}{96 (N+1)^4}
\Biggr]
\Biggr\},
\\
J_{19} &=&
\frac{1}{N+1} \Biggl\{
\frac{8}{3 \varepsilon ^3} \left(S_1+\frac{1}{N+1}\right)
-\frac{1}{\varepsilon^2} \Biggl[
2 S_1^2+\frac{4 (N+2)}{N+1} S_1+2 S_2+\frac{4 (N+2)}{(N+1)^2}
\Biggr]
 \nonumber \\ && 
+\frac{1}{\varepsilon} \Biggl[
S_1 \left(\frac{7}{3} S_2+\frac{2 \left(9 N^2+27 N+25\right)}{3 (N+1)^2}+\zeta_2\right)
+\frac{7}{9} S_1^3
+\frac{9 N+16}{3 (N+1)} \left(S_1^2+S_2\right)
+\frac{14}{9} S_3
  \nonumber \\ &&  
+\frac{2 \left(9 N^2+27 N+25\right)}{3 (N+1)^3}
+\frac{\zeta_2}{N+1}
\Biggr]
-S_1^2 \left(\frac{5}{4} S_2+\frac{9 N^2+25 N+21}{2 (N+1)^2}+\frac{3 \zeta_2}{4}\right)
-\frac{5}{4} S_4
 \nonumber \\ && 
-\left(\frac{9 N^2+25 N+21}{2 (N+1)^2}+\frac{3 \zeta_2}{4}\right) S_2
-\frac{9 N^3+36 N^2+52 N+30}{(N+1)^4}
-\frac{7 \zeta_3}{3 (N+1)}
-\frac{5}{8} S_2^2
 \nonumber \\ && 
-S_1 \left(\frac{(7 N+12) S_2}{2 (N+1)}
           +\frac{5}{3} S_3
           +\frac{9 N^3+36 N^2+52 N+30}{(N+1)^3}
           +\frac{3 (N+2) \zeta_2}{2 (N+1)}
           +\frac{7 \zeta_3}{3}\right)
 \nonumber \\ && 
-\frac{5}{24} S_1^4
-\frac{7 N+12}{3 (N+1)} \left(\frac{1}{2} S_1^3+S_3\right)
-\frac{3 (N+2) \zeta_2}{2 (N+1)^2}
+\varepsilon \Biggl[
\frac{31}{72} S_2 S_1^3
+\frac{31}{48} S_1 \left(S_2^2+2 S_4\right)
 \nonumber \\ && 
+\left(\frac{63 N^2+171 N+139}{36 (N+1)^2}+\frac{7 \zeta_2}{24}\right) \left(S_1^3+2 S_3+3 S_1 S_2\right)
+\left(S_1+\frac{1}{N+1}\right) \biggl(-\frac{173 \zeta_2^2}{80}
 \nonumber \\ && 
+\frac{\left(9 N^2+27 N+25\right) \zeta_2}{4 (N+1)^2}
                                      +\frac{81 N^4+405 N^3+792 N^2+738 N+301}{6 (N+1)^4}+\frac{7 (N+2) \zeta_3}{2 (N+1)}\biggr)
 \nonumber \\ && 
+\left(S_1^2+S_2\right) \left(\frac{31}{36} S_3+\frac{81 N^3+306 N^2+414 N+220}{12 (N+1)^3}+\frac{(9 N+16) \zeta_2}{8 (N+1)}+\frac{7 \zeta_3}{4}\right)
 \nonumber \\ && 
+\frac{45 N+76}{24 (N+1)} \left(S_2 S_1^2+\frac{4}{3} S_1 S_3+\frac{1}{6} S_1^4+\frac{1}{2} S_2^2+S_4\right)
+\frac{31}{720} S_1^5
+\frac{31}{30} S_5
\Biggr]
\Biggr\},
\\
J_{20} &=&
\frac{1}{N (N+1)} \Biggl\{
%
\frac{2}{3 \varepsilon^2} 
+\frac{1}{\varepsilon} \Biggl[
\frac{4 N^2+2 N-1}{3 N (N+1)}-\frac{2}{3} S_1
\Biggr]
-\frac{\left(4 N^2+2 N-1\right) S_1}{3 N (N+1)}+\frac{1}{3} S_1^2+\frac{1}{3} S_2
 \nonumber \\ && 
+\frac{16 N^4+24 N^3+7 N^2-N+1}{6 N^2 (N+1)^2}+\frac{\zeta_2}{4}
+\varepsilon \Biggl[
\frac{4 N^2+2 N-1}{N (N+1)} \left(\frac{1}{6} \left(S_1^2+S_2\right)+\frac{\zeta_2}{8}\right)
  \nonumber \\ &&  
+\frac{64 N^6+160 N^5+124 N^4+20 N^3-6 N^2-1}{12 N^3 (N+1)^3}
-\frac{1}{9} S_1^3
-\frac{2}{9} S_3
-\frac{7 \zeta_3}{12}
-\frac{1}{3} S_1 S_2
  \nonumber \\ &&  
-S_1 \left(\frac{16 N^4+24 N^3+7 N^2-N+1}{6 N^2 (N+1)^2}+\frac{\zeta_2}{4}\right)
\Biggr]
+\varepsilon^2 \Biggl[
\frac{1}{6} S_1^2 S_2
+\frac{1}{36} S_1^4
+\frac{1}{12} S_2^2
+\frac{1}{6} S_4
  \nonumber \\ && 
-\frac{4 N^2+2 N-1}{6 N (N+1)} \left(\frac{1}{3} S_1^3+\frac{2}{3} S_3+S_1 S_2+\frac{3}{4} \zeta_2 S_1+\frac{7}{4} \zeta_3\right)
-\frac{173 \zeta_2^2}{320}
+S_1 \biggl(\frac{2}{9} S_3+\frac{7 \zeta_3}{12}
  \nonumber \\ && 
-\frac{64 N^6+160 N^5+124 N^4+20 N^3-6 N^2-1}{12 N^3 (N+1)^3}\biggr)
+\frac{\left(16 N^4+24 N^3+7 N^2-N+1\right) \zeta_2}{16 N^2 (N+1)^2}
  \nonumber \\ && 
+\left(\frac{16 N^4+24 N^3+7 N^2-N+1}{12 N^2 (N+1)^2}+\frac{\zeta_2}{8}\right) \left(S_2+S_1^2\right)
+\frac{32 N^4+112 N^3+142 N^2}{3 (N+1)^4}
  \nonumber \\ && 
+\frac{576 N^5+61 N^4-14 N^3+6 N^2+N+1}{24 N^4 (N+1)^4}
\Biggr]
+\varepsilon^3 \Biggl[
-\frac{1}{18} S_2 S_1^3
-\frac{1}{9} S_3 \left(S_1^2+S_2\right)
  \nonumber \\ && 
+\frac{4 N^2+2 N-1}{12 N (N+1)} \biggl(\frac{1}{6} S_1^4+\frac{1}{2} S_2^2+S_4+S_2 S_1^2
                                      +\frac{3}{4} \zeta_2 \left(S_1^2+S_2\right)+\frac{7}{2} \zeta_3 S_1+\frac{4}{3} S_1 S_3
  \nonumber \\ && 
-\frac{519}{160} \zeta_2^2\biggr)
-\biggl(\frac{16 N^4+24 N^3+7 N^2-N+1}{12 N^2 (N+1)^2}+\frac{\zeta_2}{8}\biggr) \left(S_1 S_2+\frac{1}{3} S_1^3+\frac{2}{3} S_3\right) 
-\frac{1}{180} S_1^5
  \nonumber \\ && 
+\left(S_1^2+S_2+\frac{3}{4} \zeta_2\right) \biggl(\frac{64 N^6+160 N^5+124 N^4+20 N^3-6 N^2-1}{24 N^3 (N+1)^3}-\frac{7 \zeta_3}{24}\biggr)
-\frac{239 \zeta_5}{80}
  \nonumber \\ && 
-S_1 \biggl(\frac{256 N^8+896 N^7+1136 N^6+576 N^5+61 N^4-14 N^3+6 N^2+N+1}{24 N^4 (N+1)^4}
           +\frac{1}{6} S_4
  \nonumber \\ && 
           +\frac{1}{12} S_2^2
           -\frac{173 \zeta_2^2}{320}\biggr)
-\frac{2}{15} S_5
-\frac{16 N^4+24 N^3+7 N^2-N+1}{16 N^2 (N+1)^2} \left(\zeta_2 S_1+\frac{7}{3} \zeta_3\right)
  \nonumber \\ && 
+\frac{8128 N^8+6848 N^7+2548 N^6+182 N^5-47 N^4+8 N^3-7 N^2-2 N-1}{48 N^5 (N+1)^5}
  \nonumber \\ && 
+\frac{64 N^5+288 N^4}{3 (N+1)^5}
\Biggr]
\Biggr\}.
\end{eqnarray}

This result is divergent at $N=0$. Evaluating the integral separately at this value of $N$, we obtain
\begin{eqnarray} 
J_{20}^{N=0} &=&
\frac{4}{3 \varepsilon ^3}
+\frac{2}{\varepsilon ^2}
+\frac{\frac{\zeta_2}{2}+\frac{13}{3}}{\varepsilon }
+\frac{3 \zeta_2}{4}-\frac{7 \zeta_3}{6}+\frac{17}{2}
+\varepsilon \Biggl[-\frac{173 \zeta_2^2}{160}+\frac{13 \zeta_2}{8}-\frac{7 \zeta_3}{4}+\frac{205}{12}\Biggr]
\nonumber \\ &&
+\varepsilon^2 \Biggl[-\frac{519 \zeta_2^2}{320}+\zeta_2 \left(\frac{51}{16}-\frac{7 \zeta_3}{16}\right)-\frac{91 \zeta_3}{24}-\frac{239
   \zeta_5}{40}+\frac{273}{8}\Biggr]
+\varepsilon^3 \Biggl[-\frac{16117 \zeta_2^3}{5376}
 \nonumber \\ && 
-\frac{2249 \zeta_2^2}{640}+\zeta_2 \left(\frac{205}{32}-\frac{21 \zeta_3}{32}\right)+\frac{49
   \zeta_3^2}{96}-\frac{119 \zeta_3}{16}-\frac{717 \zeta_5}{80}+\frac{3277}{48}\Biggr],
\\
J_{21} &=&
\frac{1}{(N+1)^2} \Biggl\{
\frac{8}{3 \varepsilon ^3}
-\frac{4}{\varepsilon^2} \Biggl[
\frac{2}{3} S_1+\frac{N+2}{N+1}
\Biggr]
+\frac{1}{\varepsilon} \Biggl[
\frac{4}{3} S_1^2
+\frac{4}{3} S_2
+\frac{2 \left(9 N^2+27 N+25\right)}{3 (N+1)^2}
  \nonumber \\ &&  
+\frac{4 (N+2) S_1}{N+1}
+\zeta_2
\Biggr]
-S_1 \left(\frac{4}{3} S_2+\frac{2 \left(9 N^2+27 N+25\right)}{3 (N+1)^2}+\zeta_2\right)
-\frac{4}{9} S_1^3
-\frac{8}{9} S_3
 \nonumber \\ && 
-\frac{N+2}{N+1} \left(2 S_1^2+2 S_2+\frac{3}{2} \zeta_2\right)
-\frac{9 N^3+36 N^2+52 N+30}{(N+1)^3}
-\frac{7 \zeta_3}{3}
+\varepsilon \Biggl[
\frac{2}{3} S_1^2 S_2
  \nonumber \\ &&  
+\left(\frac{9 N^2+27 N+25}{3 (N+1)^2}+\frac{\zeta_2}{2}\right) \left(S_1^2+S_2\right)
+S_1 \biggl(\frac{8}{9} S_3+\frac{9 N^3+36 N^2+52 N+30}{(N+1)^3}
   \nonumber \\ &&   
+\frac{7 \zeta_3}{3}\biggr)
+\frac{N+2}{N+1} \left(\frac{2}{3} S_1^3+\frac{4}{3} S_3+\left(2 S_2+\frac{3}{2} \zeta_2\right) S_1+\frac{7}{2} \zeta_3\right)
+\frac{1}{9} S_1^4
+\frac{1}{3} S_2^2
+\frac{2}{3} S_4
  \nonumber \\ &&   
+\frac{\left(9 N^2+27 N+25\right) \zeta_2}{4 (N+1)^2}
+\frac{81 N^4+405 N^3+792 N^2+738 N+301}{6 (N+1)^4}
-\frac{173 \zeta_2^2}{80}
\Biggr]
\Biggr\},
\\
J_{22} &=&
\frac{1}{N+1} \Biggl\{
\frac{8}{\varepsilon ^3}
-\frac{4}{\varepsilon^2} \Biggl[S_1+\frac{3 N+4}{N+1}\Biggr]
+\frac{1}{\varepsilon} \Biggl[
S_1^2+\frac{2 (3 N+4) S_1}{N+1}+\frac{2 \left(7 N^2+17 N+11\right)}{(N+1)^2}
\nonumber \\ &&
+S_2+3 \zeta_2
\Biggr]
-S_1 \left(\frac{1}{2} S_2+\frac{7 N^2+17 N+11}{(N+1)^2}+\frac{3 \zeta_2}{2}\right)
-\frac{1}{6} S_1^3-\frac{(3 N+4) S_1^2}{2 (N+1)}
\nonumber \\ &&
-\frac{(3 N+4) S_2}{2 (N+1)}-\frac{1}{3} S_3-\frac{15 N^3+52 N^2+62 N+26}{(N+1)^3}-\frac{3 (3 N+4) \zeta_2}{2 (N+1)}+\zeta_3
\Biggr\}
\\
%
J_{23} &=&
\frac{1}{N+1} \Biggl\{
\frac{4}{3 \varepsilon ^3}\left(S_1+\frac{1}{N+1}\right)
+\frac{1}{\varepsilon^2} \Biggl[
-\frac{1}{3} S_1^2+\frac{2 (4 N+1) S_1}{3 (N+1)}+S_2+\frac{2 (4 N+3)}{3 (N+1)^2}
\Biggr]
\nonumber \\ &&
+\frac{1}{\varepsilon} \Biggl[
S_1 \left(\frac{5}{6} S_2+\frac{16 N^2+20 N+15}{3 (N+1)^2}+\frac{\zeta_2}{2}\right)+\frac{1}{18} S_1^3-\frac{(4 N+1) S_1^2}{6 (N+1)}+\frac{(12
   N+11) S_2}{6 (N+1)}
\nonumber \\ &&
-\frac{17}{9} S_3+\frac{2}{3} S_{2,1}+\frac{16 N^2+28 N+13}{3 (N+1)^3}+\frac{\zeta_2}{2 (N+1)}
\Biggr]
-\frac{1}{144} S_1^4
+\frac{(4 N+1) S_1^3}{36 (N+1)}
\nonumber \\ &&
+S_1^2 \left(-\frac{5}{24} S_2-\frac{16 N^2+20 N+7}{12 (N+1)^2}-\frac{\zeta_2}{8}\right)
+\left(\frac{48 N^2+92 N+29}{12 (N+1)^2}+\frac{3 \zeta_2}{8}\right) S_2
\nonumber \\ &&
+S_1 \biggl(\frac{4 N+1}{4 (N+1)}\left(\frac{5}{3} S_2+\zeta_2\right)
-\frac{19}{18} S_3+\frac{1}{3} S_{2,1}+\frac{64 N^3+144 N^2+140 N+33}{6 (N+1)^3}
\nonumber \\ &&
-\frac{7 \zeta_3}{6}\biggr)
-\frac{11}{16} S_2^2
-\frac{(68 N+53) S_3}{18 (N+1)}
+\frac{43}{24} S_4
+\frac{(4 N+3) S_{2,1}}{3 (N+1)}
-\frac{1}{3} S_{3,1}
+\frac{1}{3} S_{2,1,1}
\nonumber \\ &&
+\frac{(4 N+3) \left(16 N^2+32 N+17\right)}{6 (N+1)^4}
+\frac{(4 N+3) \zeta_2}{4 (N+1)^2}
-\frac{7 \zeta_3}{6 (N+1)}
\Biggr\},
\\
J_{24} &=&
\frac{8}{3 \varepsilon ^3}
\left(\frac{S_1}{(N+1)^2}-2 S_3+S_{2,1}-\frac{1}{(N+1)^3}\right)
+\frac{1}{\varepsilon^2} \Biggl[
-\frac{8}{3} S_2^2-\frac{8 S_2}{3 (N+1)^2}+\frac{16 S_3}{3 (N+1)}
\nonumber \\ &&
+\frac{32}{3} S_4+S_1 \left(\frac{16}{3} S_3-\frac{8}{3}
   S_{2,1}-\frac{32}{3 (N+1)^3}\right)-\frac{8 S_{2,1}}{3 (N+1)}-16 S_{3,1}+\frac{16}{3} S_{2,1,1}
\Biggr]
\nonumber \\ &&
+\frac{1}{\varepsilon} \Biggl[
S_1 \biggl(\frac{8}{3} S_2^2+\frac{S_2}{3 (N+1)^2}-\frac{16 S_3}{3 (N+1)}-\frac{32}{3} S_4+\frac{8 S_{2,1}}{3 (N+1)}+16 S_{3,1}
-\frac{16}{3} S_{2,1,1}
\nonumber \\ &&
+\frac{\zeta_2}{(N+1)^2}+\frac{16}{(N+1)^4}\biggr)+\left(-\frac{58}{9 (N+1)^2}-2 \zeta_2\right)
   S_3+\left(\frac{2}{(N+1)^2}+\zeta_2\right) S_{2,1}
\nonumber \\ &&
+\frac{S_1^3}{9 (N+1)^2}+\left(-\frac{8}{3} S_3+\frac{4}{3} S_{2,1}-\frac{5}{3
   (N+1)^3}\right) S_1^2+\frac{8 S_2^2}{3 (N+1)}-\frac{32 S_4}{3 (N+1)}-\frac{4}{3} S_5
\nonumber \\ &&
+S_2 \left(\frac{56}{3} S_3-\frac{20}{3}
   S_{2,1}+\frac{11}{3 (N+1)^3}\right)-\frac{76}{3} S_{2,3}+\frac{16 S_{3,1}}{N+1}+\frac{76}{3} S_{4,1}-\frac{16 S_{2,1,1}}{3 (N+1)}
\nonumber \\ &&
+\frac{22}{3} S_{2,2,1}-\frac{92}{3} S_{3,1,1}+\frac{26}{3} S_{2,1,1,1}-\frac{\zeta_2}{(N+1)^3}-\frac{8}{3 (N+1)^5}
\Biggr]
-\frac{100}{3 (N+1)^5} S_1
\nonumber \\ &&
+S_1 \Biggl[\left(\frac{28}{3 (N+1)^2}+2 \zeta_2\right) S_3
+\left(-\frac{11}{3 (N+1)^2}-\zeta_2\right) S_{2,1}
-\frac{8 S_2^2}{3 (N+1)}
-\frac{76}{3} S_{4,1}
\nonumber \\ &&
+\left(-\frac{56}{3} S_3+\frac{20}{3} S_{2,1}-\frac{4}{3 (N+1)^3}\right) S_2
+\frac{32 S_4}{3 (N+1)}
+\frac{4}{3} S_5+\frac{76}{3} S_{2,3}
-\frac{16 S_{3,1}}{N+1}
\nonumber \\ &&
+\frac{16 S_{2,1,1}}{3 (N+1)}
-\frac{22}{3} S_{2,2,1}
+\frac{92}{3} S_{3,1,1}
-\frac{26}{3} S_{2,1,1,1}
-\frac{4 \zeta_2}{(N+1)^3}
-\frac{7 \zeta_3}{3 (N+1)^2}
\Biggr]
\nonumber \\ &&
+S_3 \left(-\frac{14}{9} S_{2,1}+\frac{2 \zeta_2}{N+1}+\frac{100}{9 (N+1)^3}+\frac{2 \zeta_3}{3}\right)
+\left(-\frac{23}{6 (N+1)^2}-\zeta_2\right) S_2^2
\nonumber \\ &&
-S_{2,1} \left(\frac{\zeta_2}{N+1}+\frac{19}{3 (N+1)^3}+\frac{7 \zeta_3}{3}\right)
+\left(\frac{101}{6 (N+1)^2}+4 \zeta_2\right) S_4
-\frac{79}{3 (N+1)^2} S_{3,1}
\nonumber \\ &&
-6 \zeta_2 S_{3,1}
+S_2 \biggl(-\frac{56 S_3}{3 (N+1)}+\frac{49}{6} S_4+\frac{20 S_{2,1}}{3 (N+1)}+28 S_{3,1}
-\frac{26}{3} S_{2,1,1}-\frac{\zeta_2}{(N+1)^2}
\nonumber \\ &&
-\frac{13}{2 (N+1)^4}\biggr)
+\left(\frac{7}{(N+1)^2}+2 \zeta_2\right) S_{2,1,1}
+\left(\frac{8}{9} S_3-\frac{4}{9} S_{2,1}-\frac{4}{9 (N+1)^3}\right) S_1^3
\nonumber \\ &&
+\left(-\frac{4}{3} S_2^2+\frac{8 S_3}{3 (N+1)}+\frac{16}{3} S_4-\frac{4 S_{2,1}}{3 (N+1)}-8 S_{3,1}+\frac{8}{3} S_{2,1,1}+\frac{17}{6 (N+1)^4}\right) S_1^2
\nonumber \\ &&
+\frac{1}{18} S_2^3
+\frac{94}{9} S_3^2
-\frac{5}{6} S_{2,1}^2
+\frac{4 S_5}{3 (N+1)}
+\frac{385}{9} S_6
+\frac{76 S_{2,3}}{3 (N+1)}
-\frac{76 S_{4,1}}{3 (N+1)}
\nonumber \\ &&
-\frac{235}{6} S_{4,2}
-\frac{52}{3} S_{5,1}
-\frac{22 S_{2,2,1}}{3 (N+1)}
-\frac{139}{3} S_{2,3,1}
+\frac{92 S_{3,1,1}}{3 (N+1)}
-2 S_{3,2,1}
+49 S_{4,1,1}
\nonumber \\ &&
-\frac{26 S_{2,1,1,1}}{3 (N+1)}
+13 S_{2,2,1,1}
-52 S_{3,1,1,1}
+\frac{40}{3} S_{2,1,1,1,1}
-\frac{5 \zeta_3}{3 (N+1)^3},
\\
J_{25} &=&
\frac{1}{N+2} \Biggl\{
\frac{16}{3 \varepsilon ^3}
\left(S_1+\frac{1}{N+1}\right)
+\frac{1}{\varepsilon^2} \Biggl[
-\frac{4}{3} S_1^2+\frac{8 (4 N+3) S_1}{3 (N+1)}-4 S_2+\frac{16 (2 N+1)}{3 (N+1)^2}
\Biggr]
\nonumber \\ &&
+\frac{1}{\varepsilon} \Biggl[
S_1 \left(\frac{2}{3} S_2+\frac{4 \left(16 N^2+28 N+13\right)}{3 (N+1)^2}+2 \zeta_2\right)+\frac{2}{9} S_1^3-\frac{2 (4 N+3) S_1^2}{3 (N+1)}
+\frac{16}{9} S_3
\nonumber \\ &&
-\frac{2 (12 N+11) S_2}{3 (N+1)}
+\frac{8 \left(8 N^2+12 N+5\right)}{3 (N+1)^3}+\frac{2 \zeta_2}{N+1}
\Biggr]
+S_1 \biggl(\frac{(4 N+3) S_2}{3 (N+1)}
-\frac{2}{9} S_3
\nonumber \\ && 
+\frac{2 (4 N+3) \left(16 N^2+32 N+17\right)}{3 (N+1)^3}
+\frac{(4 N+3) \zeta_2}{N+1}
-\frac{14 \zeta_3}{3}\biggr)
+\frac{(4 N+3) S_1^3}{9 (N+1)}
\nonumber \\ && 
+S_1^2 \left(-\frac{1}{6} S_2-\frac{16 N^2+28 N+13}{3 (N+1)^2}-\frac{\zeta_2}{2}\right)
+\left(-\frac{48 N^2+92 N+45}{3 (N+1)^2}-\frac{3 \zeta_2}{2}\right) S_2
\nonumber \\ &&
-\frac{1}{36} S_1^4
-\frac{1}{12} S_2^2
+\frac{2 (16 N+15) S_3}{9 (N+1)}
-\frac{5}{6} S_4
+\frac{4 \left(32 N^3+80 N^2+68 N+19\right)}{3 (N+1)^4}
\nonumber \\ && 
+\frac{2 (2 N+1) \zeta_2}{(N+1)^2}
-\frac{14 \zeta_3}{3 (N+1)}
\Biggr\},
\\
J_{26} &=&
\frac{1}{\varepsilon ^3}
\left(4 S_1+\frac{4}{3} S_2+\frac{4 (3 N+4)}{3 (N+1)^2}\right)
+\frac{1}{\varepsilon^2} \Biggl[
-\frac{2 \left(15 N^2+27 N+14\right) S_1}{3 (N+1)^2}+S_1^2-\frac{11}{3} S_2
\nonumber \\ &&
-\frac{2}{3} S_3-\frac{4}{3} S_{2,1}-\frac{2 \left(15 N^2+34
   N+22\right)}{3 (N+1)^3}
\Biggr]
+\frac{1}{\varepsilon} \Biggl[
\left(\frac{43 N^2+89 N+44}{6 (N+1)^2}+\frac{\zeta_2}{2}\right) S_2
\nonumber \\ &&
-\frac{\left(15 N^2+27 N+14\right) S_1^2}{6 (N+1)^2}+S_1 \left(\frac{1}{2}
   S_2+\frac{57 N^3+156 N^2+140 N+47}{3 (N+1)^3}+\frac{3 \zeta_2}{2}\right)
\nonumber \\ &&
+\frac{1}{6} S_1^3+\frac{8}{3} S_3+\frac{1}{3} S_4-\frac{4}{3}
   S_{2,1}+\frac{8}{3} S_{3,1}-\frac{2}{3} S_{2,1,1}+\frac{57 N^3+185 N^2+205 N+84}{3 (N+1)^4}
\nonumber \\ && 
+\frac{(3 N+4) \zeta_2}{2 (N+1)^2}
\Biggr]
-\left(\frac{102 N^2+201 N+101}{18 (N+1)^2}+\frac{\zeta_2}{4}\right) S_3
-\frac{\left(15 N^2+27 N+14\right) S_1^3}{36 (N+1)^2}
\nonumber \\ &&  
+\left(\frac{16 N^2+35 N+13}{3 (N+1)^2}-\frac{\zeta_2}{2}\right) S_{2,1}
+S_2 \biggl(-\frac{149 N^3+462 N^2+478 N+159}{12 (N+1)^3}
\nonumber \\ &&
-\frac{11 \zeta_2}{8}+\frac{17 \zeta_3}{6}\biggr)
+S_1^2 \left(\frac{1}{8} S_2+\frac{57 N^3+156 N^2+140 N+47}{12 (N+1)^3}+\frac{3 \zeta_2}{8}\right)
+\frac{1}{48} S_1^4
\nonumber \\ &&  
+S_1 \biggl(-\frac{\left(15 N^2+27 N+14\right) S_2}{12 (N+1)^2}+\frac{1}{6} S_3+S_{2,1}-\frac{\left(15 N^2+27 N+14\right) \zeta_2}{4 (N+1)^2}
+\frac{\zeta_3}{2}
\nonumber \\ &&
-\frac{195 N^4+723 N^3+982 N^2+578 N+138}{6 (N+1)^4}\biggr)
+\frac{9}{16} S_2^2
-\frac{13}{24} S_4
-\frac{1}{6} S_5
-\frac{1}{3} S_{3,1}
\nonumber \\ &&  
-\frac{4}{3} S_{4,1}
-\frac{8}{3} S_{2,1,1}
-2 S_{2,2,1}
+\frac{4}{3} S_{3,1,1}
-\frac{1}{3} S_{2,1,1,1}
-\frac{\left(15 N^2+34 N+22\right) \zeta_2}{4 (N+1)^3}
\nonumber \\ &&  
-\frac{195 N^4+826 N^3+1320 N^2+952 N+278}{6 (N+1)^5}
+\frac{(3 N+20) \zeta_3}{6 (N+1)^2},
\\
J_{27} &=&
\frac{1}{N+1} \Biggl\{
-\frac{8}{3 \varepsilon ^3 (N+1)}
-\frac{4 S_1}{\varepsilon^2 (N+1)}
+\frac{1}{\varepsilon} \Biggl[
-\frac{S_1^2}{N+1}+\frac{14 S_1}{3 (N+1)^2}+\frac{S_2}{3 (N+1)}
\nonumber \\ &&
-4 S_3+2 S_{2,1}-\frac{\zeta_2}{N+1}-\frac{8}{3 (N+1)^3}
\Biggr]
+S_1 \biggl(-\frac{S_2}{2 (N+1)}+6 S_3-\frac{3 \zeta_2}{2 (N+1)}
\nonumber \\ &&
-2 S_{2,1}-\frac{43}{3 (N+1)^3}-2 \zeta_3\biggr)
-\frac{S_1^3}{6 (N+1)}+\frac{7 S_1^2}{6 (N+1)^2}-2 S_2^2-\frac{13 S_2}{6 (N+1)^2}
\nonumber \\ &&
+\frac{19 S_3}{3 (N+1)}+10 S_4-\frac{11 S_{2,1}}{3 (N+1)}-16 S_{3,1}+5
   S_{2,1,1}-\frac{5 \zeta_3}{3 (N+1)}
\Biggr\},
\\
J_{28} &=&
\frac{1}{(N+1) (N+2)} \Biggl\{
\frac{8 (3 N+4)}{3 \varepsilon ^3} \left(S_1+\frac{1}{N+1}\right)
+\frac{2}{3 \varepsilon^2} \Biggl[
(3 N+4) S_1^2-(9 N+16) S_2
\nonumber \\ &&
-\frac{2 \left(9 N^3+44 N^2+60 N+24\right) S_1}{(N+1) (N+2)}
-\frac{2 \left(9 N^3+50 N^2+82 N+44\right)}{(N+1)^2 (N+2)}
\Biggr]
\nonumber \\ &&
+\frac{1}{\varepsilon} \Biggl[
-\frac{\left(9 N^3+44 N^2+60 N+24\right) S_1^2}{3 (N+1) (N+2)}+\frac{\left(27 N^3+140 N^2+224 N+112\right) S_2}{3 (N+1) (N+2)}
\nonumber \\ &&
+S_1 \biggl(\frac{1}{3} (3 N+4) S_2+\frac{2 \left(27 N^5+213 N^4+651 N^3+912 N^2+556 N+96\right)}{3 (N+1)^2 (N+2)^2}
\nonumber \\ && 
+(3 N+4) \zeta_2\biggr)
+\frac{1}{9} (3 N+4) S_1^3+\frac{2}{9} (21 N+82) S_3-\frac{4}{3} (3 N+10) S_{2,1}
+\frac{(3 N+4) \zeta_2}{N+1}
\nonumber \\ &&
+\frac{2 (3 N+4) \left(9 N^4+65 N^3+175 N^2+196 N+84\right)}{3 (N+1)^3 (N+2)^2}
\Biggr]
+\frac{2}{9} (21 N+82) S_3
\nonumber \\ &&
-\frac{\left(9 N^3+44 N^2+60 N+24\right) S_1^2}{3 (N+1) (N+2)}+\frac{\left(27 N^3+140 N^2+224 N+112\right) S_2}{3 (N+1) (N+2)}
\nonumber \\ &&
+S_1 \Biggl[(3 N+4) \left(\frac{1}{3} S_2+\zeta_2\right)+\frac{2 \left(27 N^5+213 N^4+651 N^3+912 N^2+556 N+96\right)}{3 (N+1)^2 (N+2)^2}\Biggr]
\nonumber \\ &&
+\frac{1}{9} (3 N+4) S_1^3
+\frac{2 (3 N+4) \left(9 N^4+65 N^3+175 N^2+196 N+84\right)}{3 (N+1)^3 (N+2)^2}
+\frac{(3 N+4) \zeta_2}{N+1}
\nonumber \\ &&
-\frac{4}{3} (3 N+10) S_{2,1}
\Biggr\},
\\
J_{29} &=&
-\frac{8}{3 \varepsilon^3} \left(2 S_{-2}+ S_2-\frac{2 (-1)^N-1}{(N+1)^2}\right)
+\frac{1}{\varepsilon^2} \Biggl[
-\frac{16}{3} S_1 S_{-2}
-S_1 \left(\frac{8}{3} S_2+\frac{16 (-1)^N}{3 (N+1)^2}\right)
 \nonumber \\ && 
+\frac{4 (3 N+1)}{3 (N+1)} \left(S_2+2 S_{-2}\right)
+\frac{4}{3} S_3
+\frac{32}{3} S_{-2,1}
-\frac{4 \left(6 (-1)^N N-3 N+10 (-1)^N-4\right)}{3 (N+1)^3}
 \nonumber \\ && 
+\frac{8}{3} S_{2,1}
\Biggr]
+\frac{1}{\varepsilon} \Biggl[
\left(\frac{2 \left(-9 N^2-12 N+4 (-1)^N-3\right)}{3 (N+1)^2}-\zeta_2\right) S_2
+\frac{8}{3} S_{-2} \left(S_2-S_1^2\right)
 \nonumber \\ && 
-2 \zeta_2 S_{-2}
+\frac{3 N+1}{N+1} \left(\frac{8}{3} S_1 S_{-2}-4 S_{-2}-\frac{2}{3} S_3-\frac{16}{3} S_{-2,1}-\frac{4}{3} S_{2,1}+\frac{4}{3} S_2 S_1\right)
+\frac{4}{3} S_2^2
 \nonumber \\ && 
+\left(\frac{8 (-1)^N}{3 (N+1)^2}-\frac{4}{3} S_2\right) S_1^2
+\left(\frac{4}{3} S_3+\frac{32}{3} S_{-2,1}+\frac{8}{3} S_{2,1}+\frac{8 (-1)^N (3 N+5)}{3 (N+1)^3}\right) S_1
 \nonumber \\ && 
-\frac{2}{3} S_4
+\frac{16}{3} S_{-2,2}
-\frac{4}{3} S_{3,1}
+\frac{2 \left((-1)^N \left(18 N^2+48 N+38\right)-9 N^2-21 N-13\right)}{3 (N+1)^4}
 \nonumber \\ && 
-\frac{64}{3} S_{-2,1,1}
-\frac{8}{3} S_{2,1,1}
+\frac{\left(2 (-1)^N-1\right) \zeta_2}{(N+1)^2}
\Biggr]
-\left(\frac{4}{9} S_2+\frac{8 (-1)^N}{9 (N+1)^2}\right) S_1^3
\nonumber \\ && 
+\frac{3 N+1}{N+1} \Biggl[
\frac{2}{3} S_2 S_1^2
-\frac{2}{3} \left(S_3+8 S_{-2,1}+2 S_{2,1}\right) S_1
-\frac{2}{3} S_2^2
+\frac{1}{3} S_4
+8 S_{-2,1}
-\frac{8}{3} S_{-2,2}
 \nonumber \\ && 
+2 S_{2,1}
+\left(\frac{4}{3} S_1^2-4 S_1+6+\zeta_2-\frac{4}{3} S_2\right) S_{-2}
+\frac{2}{3} S_{3,1}
+\frac{32}{3} S_{-2,1,1}
+\frac{4}{3} S_{2,1,1}
 \nonumber \\ && 
+\frac{\zeta_2}{2} S_2
\Biggr]
+\left(-\frac{4 (-1)^N (3 N+5)}{3 (N+1)^3}       
       +\frac{2}{3} S_3
       +\frac{16}{3} S_{-2,1}
       +\frac{4}{3} S_{2,1}\right) S_1^2
-\frac{2 (-1)^N \zeta_2}{(N+1)^2} S_1
\nonumber \\ && 
+\Biggl[\frac{4}{3} S_2^2
       -\left(\frac{2 \left(9 N^2+12 N+4 (-1)^N+3\right)}{3 (N+1)^2}+\zeta_2\right) S_2
       -\frac{2}{3} S_4       
       +\frac{16}{3} S_{-2,2}       
       -\frac{4}{3} S_{3,1}
 \nonumber \\ && 
       -\frac{64}{3} S_{-2,1,1}
       -\frac{8}{3} S_{2,1,1}
       -\frac{4 (-1)^N \left(9 N^2+24 N+19\right)}{3 (N+1)^4}
\Biggr] S_1
-\frac{7 \left(-1+2 (-1)^N\right) \zeta_3}{3 (N+1)^2}
\nonumber \\ && 
-\frac{\left(6 (-1)^N N-3 N+10 (-1)^N-4\right) \zeta_2}{2 (N+1)^3}
+\left(\frac{\zeta_2}{2}+\frac{27 N^2+36 N-16 (-1)^N+9}{9 (N+1)^2}\right) S_3
\nonumber \\ && 
+S_2 \biggl(\frac{27 N^3+63 N^2-12 (-1)^N N+45 N-20 (-1)^N+9}{3 (N+1)^3}           
           +\frac{7 \zeta_3}{3}
           -\frac{14}{9} S_3
           -16 S_{-2,1}
\nonumber \\ && 
           -\frac{4}{3} S_{2,1}
\biggr)
+S_{-2} \biggl[-\frac{8}{9} S_1^3              
              +\biggl(\frac{8}{3} S_2-2 \zeta_2\biggr) S_1
              +\frac{14 \zeta_3}{3}              
              -\frac{16}{9} S_3
              +\frac{32}{3} S_{2,1}
\biggr]
+4 \zeta_2 S_{-2,1}
\nonumber \\ && 
+\frac{1}{3} S_5
-\frac{32}{3} S_{-2,3}
+\frac{32}{3} S_{2,-3}
+\zeta_2 S_{2,1}
+\frac{2}{3} S_{4,1}
-\frac{32}{3} S_{-2,2,1}
-\frac{32}{3} S_{2,1,-2}
+\frac{4}{3} S_{3,1,1}
\nonumber \\ && 
-\frac{(-1)^N \left(54 N^3+198 N^2+258 N+130\right)-27 N^3-90 N^2-102 N-40}{3 (N+1)^5}
\nonumber \\ && 
+\frac{128}{3} S_{-2,1,1,1}
+\frac{8}{3} S_{2,1,1,1},
\\
J_{30} &=&
\frac{4}{3 \varepsilon ^3}
\left(-2 S_{-2}-S_2+\frac{2 (-1)^N-1}{(N+1)^2}\right)
+\frac{2}{3 \varepsilon ^2}
\Biggl[4 S_{-3}+S_{-2} \left(4 S_1-\frac{4 (2 N+1)}{N+1}\right)
 \nonumber \\ && 
-\frac{2 (2 N+1) S_2}{N+1}+S_1 \left(2 S_2
-\frac{4 (-1)^N}{(N+1)^2}\right)+7 S_3-2 S_{2,1}+\frac{8 N (-1)^N-4 N+3}{(N+1)^3}\Biggr]
\nonumber \\ &&
+\frac{1}{\varepsilon}
\Biggl[S_{-2} \left(-\frac{4}{3} S_1^2+\frac{8 (2 N+1) S_1}{3 (N+1)}-\frac{4}{3} S_2-\frac{8 \left(4 N^2+6 N+3\right)}{3 (N+1)^2}-\zeta_2\right)
 \nonumber \\ && 
+\left(-\frac{16 N^2+24 N-4 (-1)^N+11}{3 (N+1)^2}-\frac{\zeta_2}{2}\right) S_2+\left(\frac{4 (-1)^N+1}{3
   (N+1)^2}-\frac{2}{3} S_2\right) S_1^2
 \nonumber \\ && 
+\left(\frac{4 (2 N+1) S_2}{3 (N+1)}-\frac{14}{3} S_3+\frac{4}{3} S_{2,1}-\frac{16 (-1)^N N}{3 (N+1)^3}\right) S_1
-\frac{8}{3} S_{-4}
+\frac{14 (2 N+1) S_3}{3 (N+1)}
 \nonumber \\ && 
+S_{-3} \left(\frac{8 (2 N+1)}{3 (N+1)}-\frac{8}{3} S_1\right)
-7 S_4
-\frac{4 (2 N+1) S_{2,1}}{3 (N+1)}
+4 S_{3,1}
-\frac{2}{3} S_{2,1,1}
 \nonumber \\ && 
+\frac{(-1)^N \left(32 N^2+32 N+24\right)-16 N^2-4 N-11}{3 (N+1)^4}
+\frac{\left(2 (-1)^N-1\right) \zeta_2}{2 (N+1)^2}\Biggr]
\nonumber \\ &&
+S_{-2} \Biggl[S_1 \left(\frac{4}{3} S_2
+\frac{8 \left(4 N^2+6 N+3\right)}{3 (N+1)^2}
+\zeta_2\right)
+\frac{4}{9} S_1^3
-\frac{4 (2 N+1) S_1^2}{3 (N+1)}
+\frac{8}{9} S_3
 \nonumber \\ && 
-\frac{4 (2 N+1) S_2}{3 (N+1)}
-\frac{8 (2 N+1) \left(4 N^2+8 N+5\right)}{3 (N+1)^3}
-\frac{(2 N+1) \zeta_2}{N+1}
+\frac{7 \zeta_3}{3}\Biggr]
+\frac{13}{2} S_5
\nonumber \\ &&
+S_1 \Biggl[\left(\frac{32 N^2+48 N-8 (-1)^N+21}{6 (N+1)^2}+\frac{\zeta_2}{2}\right) S_2
+\frac{2 N+1}{3 (N+1)} \left(4 S_{2,1}-14 S_3\right)
 \nonumber \\ && 
+7 S_4
-4 S_{3,1}
+\frac{2}{3} S_{2,1,1}
-\frac{8 (-1)^N \left(4 N^2+4 N+3\right)}{3 (N+1)^4}
-\frac{(-1)^N \zeta_2}{(N+1)^2}
\Biggr]
+\frac{4}{3} S_1^2 S_{-3}
\nonumber \\ &&
+\frac{2 N+1}{N+1} \left(4 S_{3,1}
                         -\frac{8}{3} S_1 S_{-3}
                         -\frac{\zeta_2}{2} S_2
                         -\frac{2}{3} S_2 S_1^2
                         -\frac{8}{3} S_{-4}
                         -7 S_4
                         -\frac{2}{3} S_{2,1,1}
\right)
+\frac{8}{3} S_{-5}
\nonumber \\ &&
+\frac{4 N^2+6 N+3}{3 (N+1)^2} \left(8 S_{-3}-4 S_{2,1}\right)
+\left(\frac{168 N^2+252 N-8 (-1)^N+123}{9 (N+1)^2}+\frac{7 \zeta_2}{4}\right) S_3
\nonumber \\ &&
+S_{-3} \left(\frac{4}{3} S_2+\zeta_2\right)
-\frac{1}{3} S_{2,1,1,1}
+\left(\frac{7}{3} S_3-\frac{2}{3} S_{2,1}+\frac{16 (-1)^N N+4 N+1}{6 (N+1)^3}\right) S_1^2
\nonumber \\ &&
-\frac{\zeta_2}{2} S_{2,1}
+\left(\frac{2}{9} S_2-\frac{8 (-1)^N+3}{18 (N+1)^2}\right) S_1^3
+\frac{\left(8 (-1)^N N-4 N+3\right) \zeta_2}{4 (N+1)^3}
+\frac{8}{3} S_1 S_{-4}
\nonumber \\ &&
+S_2 \left(-\frac{11}{9} S_3-\frac{64 N^3+160 N^2-16 (-1)^N N+140 N+39}{6 (N+1)^3}+\frac{7 \zeta_3}{6}\right)
+\frac{10}{3} S_{2,3}
\nonumber \\ &&
-\frac{7 \left(2 (-1)^N-1\right) \zeta_3}{6 (N+1)^2}
-6 S_{4,1}
-\frac{2}{3} S_{2,2,1}
+\frac{10}{3} S_{3,1,1}
-\frac{64 N^3+80 N^2+60 N-19}{6 (N+1)^5}
\nonumber \\ &&
+\frac{(-1)^N \left(64 N^3+128 N^2+112 N+16\right)}{3 (N+1)^5},
\\
J_{31} &=&
\frac{1}{N+1} \Biggl\{
\frac{16}{3 \varepsilon ^3}
-\frac{8 (5 N+6)}{3 \varepsilon^2 (N+1)}
+\frac{1}{\varepsilon} \Biggl[
\frac{4 \left(19 N^2+43 N+25\right)}{3 (N+1)^2}+2 \zeta_2
\Biggr]
-\frac{(5 N+6) \zeta_2}{N+1}
 \nonumber \\ && 
+\frac{58 \zeta_3}{3}
-\frac{2 \left(65 N^3+214 N^2+238 N+90\right)}{3 (N+1)^3}
+\varepsilon \Biggl[\frac{19 N^2+43 N+25}{2 (N+1)^2} \zeta_2+16 \zeta_2 \ln ^2(2)
  \nonumber \\ &&  
                +\frac{211 N^4+909 N^3+1480 N^2+1082 N+301}{3 (N+1)^4}
                -\frac{29 (5 N+6) \zeta_3}{3 (N+1)}
                -64 \text{Li}_4 \left(\frac{1}{2}\right)
  \nonumber \\ &&  
                +\frac{1107 \zeta_2^2}{40}
                -\frac{8 \ln ^4(2)}{3}
\Biggr]
\Biggr\}.
\label{J31N}
\end{eqnarray}

The last three integrals are independent of $N$,
\begin{eqnarray}
J_{32} &=&
\frac{8}{3 \varepsilon ^3}
-\frac{14}{3 \varepsilon ^2}
+\frac{\zeta_2+\frac{25}{6}}{\varepsilon }
-\frac{7 \zeta_2}{4}-\frac{7 \zeta_3}{3}+\frac{5}{24}
+\varepsilon \Biggl[-\frac{173 \zeta_2^2}{80}+\frac{25 \zeta_2}{16}+\frac{49 \zeta_3}{12}-\frac{959}{96}\Biggr]
\nonumber \\ &&
+\varepsilon^2 \Biggl[\frac{1211 \zeta_2^2}{320}+\zeta_2 \left(\frac{5}{64}-\frac{7 \zeta_3}{8}\right)-\frac{175 \zeta_3}{48}-\frac{239
   \zeta_5}{20}+\frac{10493}{384}\Biggr]
+\varepsilon^3 \Biggl[-\frac{16117 \zeta_2^3}{2688}
\nonumber \\ &&
-\frac{865 \zeta_2^2}{256}+\zeta_2 \left(\frac{49 \zeta_3}{32}-\frac{959}{256}\right)+\frac{49
   \zeta_3^2}{48}-\frac{35 \zeta_3}{192}+\frac{1673 \zeta_5}{80}-\frac{85175}{1536}\Biggr]
+\varepsilon^4 \Biggl[\frac{16117 \zeta_2^3}{1536}
\nonumber \\ &&
+\zeta_2^2 \left(\frac{1211 \zeta_3}{640}-\frac{173}{1024}\right)+\zeta_2 \left(-\frac{175
   \zeta_3}{128}-\frac{717 \zeta_5}{160}+\frac{10493}{1024}\right)-\frac{343 \zeta_3^2}{192}+\frac{6713 \zeta_3}{768}
\nonumber \\ &&
-\frac{1195 \zeta_5}{64}-\frac{4647 \zeta_7}{112}+\frac{610085}{6144}\Biggr],
%
\\
%
J_{33} &=&
\frac{8}{\varepsilon^3}
-\frac{12}{\varepsilon^2}
+\frac{3 \zeta_2+12}{\varepsilon}
-\frac{9 \zeta_2}{2}+\zeta_3-10
+\varepsilon \left(\frac{57 \zeta_2^2}{80}+\frac{9 \zeta_2}{2}-\frac{3 \zeta_3}{2}+\frac{15}{2}\right)
\nonumber \\ &&
+\varepsilon^2 \left(-\frac{171 \zeta_2^2}{160}+\zeta_2 \left(\frac{3 \zeta_3}{8}-\frac{15}{4}\right)+\frac{3 
\zeta_3}{2}+\frac{3 (\zeta_5-35)}{20}\right),
\\
J_{34} &=&
\frac{16}{\varepsilon ^3}
-\frac{92}{3 \varepsilon ^2}
+\frac{6 \zeta_2+35}{\varepsilon }
-\frac{23 \zeta_2}{2}+2 \zeta_3-\frac{275}{12}
+\varepsilon \left(\frac{57 \zeta_2^2}{40}+\frac{105 \zeta_2}{8}+\frac{89 \zeta_3}{6}-\frac{189}{16}\right)
\nonumber \\ &&
+\varepsilon^2 \biggl(-64 \text{Li}_4\left(\frac{1}{2}\right)+\frac{783 \zeta_2^2}{32}+\zeta_2 \left(\frac{3 \zeta_3}{4}-\frac{275}{32}+16 \ln
   ^2(2)\right)-\frac{525 \zeta_3}{8}+\frac{3 \zeta_5}{10}
 \nonumber \\ && 
+\frac{14917}{192}-\frac{8 \ln ^4(2)}{3}\biggr).
\label{J34N}
\end{eqnarray}
\section{The Operator Matrix Elements in \boldmath $x$-Space}
\label{sec:OMSx}

\vspace*{1mm}
\noindent
The analytic continuation of the OME both in the flavor non-singlet vector case and for transversity leads to different 
expressions considering the even and odd moments from 3-loop order onwards, since these sets of moments
correspond to different processes due to current crossing, cf.~\cite{CRO}.
To account for this, we define the following functions
\begin{eqnarray}
A_{qq,Q}^{\rm NS}(x) &=& A_{qq,Q}^{\rm NS,a}(x) + (-1)^N A_{qq,Q}^{\rm NS,b}(x), \\
A_{qq,Q}^{\rm NS,even}(x) &=& A_{qq,Q}^{\rm NS,a}(x) + A_{qq,Q}^{\rm NS,b}(x), \\
A_{qq,Q}^{\rm NS,odd}(x) &=& A_{qq,Q}^{\rm NS,a}(x) - A_{qq,Q}^{\rm NS,b}(x)~.
\end{eqnarray}
One obtains in the non-singlet vector case

\begin{eqnarray}
A_{qq,Q}^{\rm NS,b}(x) &=&
a_s^3
        \textcolor{blue}{\left(\frac{C_A}{2}-C_F\right) C_F T_F}
        \Biggl[
                \ln^2\left(\frac{m^2}{\mu^2}\right) \biggl[
                        -\frac{32}{3} (x+1) H_0
                        +\frac{64}{3} (x-1)
\nonumber\\&&
                        +\frac{16 \big(x^2+1\big)}{3 (x+1)} \biggl(
                                -H_0^2
                                +4 H_{-1} H_0
                                -4 H_{0,-1}
                                +2 \zeta_2
                        \biggr)
                \biggr]
\nonumber\\&&
                +\ln\left(\frac{m^2}{\mu^2}\right) \biggl[
                        \frac{16 \big(x^2+x+1\big)}{3 (x+1)} \biggl(
                                -6 H_0^2
                                +16 H_0 H_{-1}
                                -16 H_{0,-1}
                        \biggr)
\nonumber\\&&
                        -\frac{656}{9} (x+1) H_0
                        +\frac{976}{9} (x-1)
                        -\frac{128}{3} (x-1) H_1
                        +\frac{64}{3} (x+1) H_{0,1}
\nonumber\\&&
                        +\frac{16 \big(x^2+1\big)}{3 (x+1)} \biggl(
                                -\frac{2}{3} H_0^3
                                -\frac{1}{3} H_0^2
                                +2 H_{-1} H_0^2
                                +\frac{16}{3} H_0 H_{-1}
                                -\frac{16}{3} H_{0,-1}
\nonumber\\&&
                                -8 H_{-1} H_{0,1}
                                +8 H_{0,-1,1}
                                -4 H_{0,0,-1}
                                +4 H_{0,0,1}
                                +8 H_{0,1,-1}
                                -6 \zeta_3
\nonumber\\&&
                                +\biggl(
                                        8 H_{-1}
                                        -2 H_0
                                        +\frac{20}{3}
                                \biggr) \zeta_2
                        \biggr)
                \biggr]
                +\frac{16}{9} (x-1) \biggl(
                        8 H_1^2
                        -\frac{122}{3} H_1
                        +\frac{793}{9}
                \biggr)
\nonumber\\&&
                +\frac{32}{9} (x+1) \biggl(
                        -\frac{280}{9} H_0
                        -2 H_0 H_{-1}^2
                        +4 H_{-1} H_{0,-1}
                        +\frac{41}{3} H_{0,1}
                        -4 H_{0,-1,-1}
\nonumber\\&&
                        -4 H_{0,1,1}
                \biggr)
                +\frac{16 \big(x^2+x+1\big)}{3 (x+1)} \biggl(
                        -\frac{4}{3} H_0^3
                        -11 H_0^2
                        +4 H_0^2 H_{-1}
                        +\frac{232}{9} H_0 H_{-1}
\nonumber\\&&
                        -\frac{232}{9} H_{0,-1}
                        -\frac{32}{3} H_{-1} H_{0,1}
                        +\frac{32}{3} H_{0,-1,1}
                        -8 H_{0,0,-1}
                        +8 H_{0,0,1}
\nonumber\\&&
                        +\frac{32}{3} H_{0,1,-1}
                        +\biggl(
                                8 H_{-1}
                               -4 H_0
                                -\frac{16}{3}
                        \biggr) \zeta_2
                        -\frac{8}{3} \zeta_3
                \biggr)
\nonumber\\&&
                +\frac{16 \big(x^2+1\big)}{9 (x+1)} \biggl(
                        -\frac{1}{3} H_0^4
                        -\frac{2}{9} H_0^3
                        +\frac{73}{18} H_0^2
                        +\frac{8}{3} H_{-1}^3 H_0
                        +\frac{21}{2} \zeta_4
                        -2 H_0^2 H_{-1}^2
\nonumber\\&&
                        +\frac{4}{3} H_0^3 H_{-1}
                        +\frac{2}{3} H_0^2 H_{-1}
                        +\frac{100}{9} H_0 H_{-1}
                        -8 H_{-1}^2 H_{0,-1}
                        -\frac{100}{9} H_{0,-1}
\nonumber\\&&
                        -\frac{32}{3} H_{-1} H_{0,1}
                        +16 H_{-1} H_{0,-1,-1}
                        +\frac{32}{3} H_{0,-1,1}
                        +8 H_{-1} H_{0,0,-1}
\nonumber\\&&
                        -\frac{4}{3} H_{0,0,-1}
                        -8 H_{-1} H_{0,0,1}
                        +\frac{4}{3} H_{0,0,1}
                        +\frac{32}{3} H_{0,1,-1}
                        +16 H_{-1} H_{0,1,1}
\nonumber\\&&
                        -16 H_{0,-1,-1,-1}
                        -16 H_{0,-1,1,1}
                        -8 H_{0,0,-1,-1}
                        +8 H_{0,0,-1,1}
                        -8 H_{0,0,0,-1}
\nonumber\\&&
                        +8 H_{0,0,0,1}
                        +8 H_{0,0,1,-1}
                        -8 H_{0,0,1,1}
                        -16 H_{0,1,-1,1}
                        -16 H_{0,1,1,-1}
\nonumber\\&&
                        +\biggl(
                                4 H_{-1}^2
                                +\frac{44}{3} H_{-1}
                                -2 H_0^2
                                +4 H_{-1} H_0
                                -\frac{2}{3} H_0
                                +\frac{296}{9}
                        \biggr) \zeta_2
\nonumber\\&&
                        +\biggl(
                                -16 H_{-1}
                                +2 H_0
                                -16
                        \biggr) \zeta_3
                \biggr)
        \Biggr].
\end{eqnarray}
For all the $x$-space expressions we have performed numerical checks using the code {\tt HPL 2.0} \cite{Maitre:2007kp} 
comparing to the available fixed moments.

In the case of transversity the OME reads

\begin{eqnarray}
A_{qq,Q}^{\rm NS, TR,b}(x) &=&
a_s^3
\Biggl\{
        \textcolor{blue}{C_F T_F \left(\frac{C_A}{2}-C_F\right)}
        \Biggl[
                \ln^2\left(\frac{m^2}{\mu^2}\right) \biggl[
                        -\frac{16}{3} (x-1)
                        +\frac{x}{x+1} \biggl(
                                -\frac{128}{3} H_{-1} H_0
\nonumber \\ &&
                                +\frac{32}{3} H_0^2
                                +\frac{128}{3} H_{0,-1}
                                -\frac{64}{3} \zeta_2
                        \biggr)
                \biggr]
                +\ln\left(\frac{m^2}{\mu^2}\right) \biggl[
                        (1-x) \biggl(
                                \frac{112}{9}
                                -\frac{32}{3} H_1
                        \biggr)
\nonumber \\ &&
                        +\frac{x}{x+1} \biggl(
                                -\frac{1280}{9} H_{-1} H_0
                                +\biggl(
                                        \frac{320}{9}
                                        -\frac{64}{3} H_{-1}
                                \biggr) H_0^2
                                +\frac{64}{9} H_0^3
                                +\frac{1280}{9} H_{0,-1}
\nonumber \\ &&
                                +\frac{256}{3} H_{-1} H_{0,1}
                                -\frac{256}{3} H_{0,-1,1}
                                +\frac{128}{3} H_{0,0,-1}
                                -\frac{128}{3} H_{0,0,1}
                                -\frac{256}{3} H_{0,1,-1}
\nonumber \\ &&
                                +\biggl(
                                        -\frac{640}{9}
                                        -\frac{256}{3} H_{-1}
                                        +\frac{64}{3} H_0
                                \biggr) \zeta_2
                                +64 \zeta_3
                        \biggr)
                \biggr]
                +(1-x) \biggl(
			\frac{1168}{81}
                	-\frac{224}{27} H_1
\nonumber \\ &&
                	+\frac{32}{9} H_1^2
		\biggr)
                -\frac{16}{9} (x+1) H_0
                +\frac{1}{x+1} \biggl(
                        \biggl(
                                \frac{64}{81} \big(9 x^2-206 x+9\big) H_0
                                -\frac{640}{27} x H_0^2
\nonumber \\ &&
                                -\frac{128}{27} x H_0^3
                        \biggr) H_{-1}
                        -\frac{256}{27} x H_{-1}^3 H_0
                        -\frac{16}{81} \big(9 x^2-206 x+9\big) H_0^2
                        +\frac{64}{9} x H_{-1}^2 H_0^2
\nonumber \\ &&
                        +\frac{640}{81} x H_0^3
                        +\frac{32}{27} x H_0^4
                        +\biggl(
                                -\frac{64}{81} \big(9 x^2-206 x+9\big)
                                +\frac{256}{9} x H_{-1}^2
                        \biggr) H_{0,-1}
\nonumber \\ &&
                        +\frac{2560}{27} x H_{-1} H_{0,1}
                        -\frac{512}{9} x H_{-1} H_{0,-1,-1}
                        -\frac{2560}{27} x H_{0,-1,1}
                        +\biggl(
                                \frac{1280}{27} x
\nonumber \\ &&
                                -\frac{256}{9} x H_{-1}
                        \biggr) H_{0,0,-1}
                        +\biggl(
                                -\frac{1280}{27} x
                                +\frac{256}{9} x H_{-1}
                        \biggr) H_{0,0,1}
                        -\frac{2560}{27} x H_{0,1,-1}
\nonumber \\ &&
                        -\frac{512}{9} x H_{-1} H_{0,1,1}
                        +\frac{512}{9} x H_{0,-1,-1,-1}
                        +\frac{512}{9} x H_{0,-1,1,1}
                        +\frac{256}{9} x H_{0,0,-1,-1}
\nonumber \\ &&
                        -\frac{256}{9} x H_{0,0,-1,1}
                        +\frac{256}{9} x H_{0,0,0,-1}
                        -\frac{256}{9} x H_{0,0,0,1}
                        -\frac{256}{9} x H_{0,0,1,-1}
\nonumber \\ &&
                        +\frac{256}{9} x H_{0,0,1,1}
                        +\frac{512}{9} x H_{0,1,-1,1}
                        +\frac{512}{9} x H_{0,1,1,-1}
                        +\biggl(
                                \frac{32}{81} \big(9 x^2-206 x+9\big)
\nonumber \\ &&
                                -\frac{2560}{27} x H_{-1}
                                -\frac{128}{9} x H_{-1}^2
                                +\biggl(
                                        \frac{640}{27} x
                                        -\frac{128}{9} x H_{-1}
                                \biggr) H_0
                                +\frac{64}{9} x H_0^2
                        \biggr) \zeta_2
\nonumber \\ &&
                        +\biggl(
                                \frac{640}{9} x
                                +\frac{512}{9} x H_{-1}
                                -\frac{64}{9} x H_0
                        \biggr) \zeta_3
                        -\frac{112}{3} x \zeta_4
                \biggr)
        \Biggr]
\Biggr\}.
\end{eqnarray}

The corresponding expressions in the $\overline{\rm MS}$ scheme are obtained by the transformation
\begin{eqnarray}
A_{qq,Q,\rm NS, TR}^{\overline{\rm MS}} - A_{qq,Q,\rm NS, TR}^{\rm OMS} &=&
\textcolor{blue}{C_F^2 T_F} 
\Biggl\{
-\Biggl(\frac{1}{1-x} \Biggl[
  32 \ln^2\left(\frac{m^2}{\mu^2}\right)
+ \left[32 H_0+\frac{32}{3}\right] \ln\left(\frac{m^2}{\mu^2}\right)
\nonumber\\ &&
-\frac{128}{3} H_0
-\frac{640}{9}
\Biggr] \Biggr)_+
+ \Biggl[-24 \ln^2\left(\frac{m^2}{\mu^2}\right) + 28 \ln\left(\frac{m^2}{\mu^2}\right) +\frac{16}{3}\Biggr]
\nonumber\\ &&
\times \delta(1-x)
\nonumber\\ &&
+32 \ln^2\left(\frac{m^2}{\mu^2}\right)
+\ln\left(\frac{m^2}{\mu^2}\right) \left[32 H_0+\frac{32}{3}\right]
-\frac{128}{3} H_0
-\frac{640}{9}
\Biggr\}~.
\nonumber\\
\end{eqnarray}
\section{The Unpolarized Non-Singlet Wilson Coefficient in \boldmath $x$-Space}
\label{LqqNSx}

\vspace*{1mm}
\noindent
The unpolarized non-singlet heavy flavor Wilson coefficient for $Q^2 \gg m^2$ in $x$-space is given by

where $\hat{C}_{q,2}^{(3), \rm NS}$ refers to the massless Wilson coefficient, cf.~\cite{Vermaseren:2005qc}.

\vspace{5mm}
\noindent
{\bf Acknowledgment.}~
We would like to thank I. Bierenbaum, S. Klein, C.G.~Raab and A. Vogt for discussions, and
M.~Steinhauser for providing the code {\tt MATAD 3.0}. The graphs have been drawn using 
{\tt Axodraw}~\cite{Vermaseren:1994je}. This work was supported in part by DFG Sonderforschungsbereich 
Transregio 9, Computergest\"utzte Theoretische Teilchenphysik, Studienstiftung des Deutschen Volkes, 
the Austrian Science Fund (FWF) grants P20347-N18 and SFB F50 (F5009-N15), the European Commission through 
contract PITN-GA-2010-264564 ({LHCPhenoNet}) and PITN-GA-2012-316704 ({HIGGSTOOLS}), by the Research Center 
``Elementary Forces and Mathematical Foundations (EMG)'' of J. Gutenberg University Mainz and DFG, and by FP7 
ERC Starting Grant  257638 PAGAP.



\begin{thebibliography}{100}
%
\bibitem{PDF}
  S.~Alekhin, G.~Altarelli, N.~Amapane, J.~Andersen, V.~Andreev, M.~Arneodo, V.~Avati and J.~Baines {\it et al.},
  hep-ph/0601012;
  hep-ph/0601013;\\
  Z.~J.~Ajaltouni, S.~Albino, G.~Altarelli, F.~Ambroglini, J.~Anderson, G.~Antchev, M.~Arneodo and P.~Aspell {\it et al.},
  arXiv:0903.3861 [hep-ph].
%
\bibitem{Bethke:2011tr}
  S.~Bethke et al., 
  {\sf Proceedings of the Workshop on Precision Measurements of $\alpha_s$ (2011)},
  arXiv:1110.0016 [hep-ph];\\
  S.~Moch, S.~Weinzierl et al., 
  arXiv:1405.4781 [hep-ph].
%
\bibitem{Alekhin:2012vu}
  S.~Alekhin, J.~Bl\"umlein, K.~Daum, K.~Lipka and S.~Moch,
  Phys.\ Lett.\ B {\bf 720} (2013) 172
  [arXiv:1212.2355 [hep-ph]].
%
\bibitem{CHTL}
  E.~Laenen, S.~Riemersma, J.~Smith and W.~L.~van Neerven,
  Nucl.\ Phys.\ B {\bf 392} (1993) 162,
229;\\
  S.~Riemersma, J.~Smith and W.~L.~van Neerven,
  Phys.\ Lett.\ B {\bf 347} (1995) 143
  [hep-ph/9411431].
%
\bibitem{Alekhin:2003ev}
  S.~I.~Alekhin and J.~Bl\"umlein,
  Phys.\ Lett.\ B {\bf 594} (2004) 299 
  [hep-ph/0404034].
%
\bibitem{Buza:1995ie}
  M.~Buza, Y.~Matiounine, J.~Smith, R.~Migneron and W.~L.~van Neerven,
  Nucl.\ Phys.\  B {\bf 472} (1996) 611
  [arXiv:hep-ph/9601302].
%
\bibitem{Bierenbaum:2007qe}
I.~Bierenbaum, J.~Bl\"umlein and S.~Klein,
  Nucl.\ Phys.\  B {\bf 780} (2007) 40
  [arXiv:hep-ph/0703285].
%
\bibitem{Buza:1996xr}
  M.~Buza, Y.~Matiounine, J.~Smith and W.~L.~van Neerven,
  Nucl.\ Phys.\ B {\bf 485} (1997) 420
  [hep-ph/9608342].
%
\bibitem{Bierenbaum:2007pn}
  I.~Bierenbaum, J.~Bl\"umlein and S.~Klein,
  arXiv:0706.2738 [hep-ph].
%
\bibitem{Buza:1996wv}
  M.~Buza, Y.~Matiounine, J.~Smith and W.~L.~van Neerven,
  Eur.\ Phys.\ J.\ C {\bf 1} (1998) 301
  [hep-ph/9612398].
%
\bibitem{Bierenbaum:2009zt}
  I.~Bierenbaum, J.~Bl\"umlein and S.~Klein,
  Phys.\ Lett.\ B {\bf 672} (2009) 401
  [arXiv:0901.0669 [hep-ph]].
%
\bibitem{Buza:1997mg}
  M.~Buza and W.~L.~van Neerven,
  Nucl.\ Phys.\ B {\bf 500} (1997) 301
  [hep-ph/9702242].
%
\bibitem{Blumlein:2014fqa}
  J.~Bl\"umlein, A.~Hasselhuhn and T.~Pfoh,
  Nucl.\ Phys.\ B {\bf 881} (2014) 1
  [arXiv:1401.4352 [hep-ph]].
%
\bibitem{Blumlein:2006mh}
  J.~Bl\"umlein, A.~De Freitas, W.~L.~van Neerven and S.~Klein,
  Nucl.\ Phys.\  B {\bf 755} (2006) 272
  [arXiv:hep-ph/0608024].
%
\bibitem{Behring:2014eya}
  A.~Behring, I.~Bierenbaum, J.~Bl\"umlein, A.~De Freitas, S.~Klein and F.~Wi\ss{}brock,
  arXiv:1403.6356 [hep-ph].
%
\bibitem{Bierenbaum:2009mv}
  I.~Bierenbaum, J.~Bl\"umlein and S.~Klein,
  Nucl.\ Phys.\ B {\bf 820} (2009) 417
  [arXiv:0904.3563 [hep-ph]].
%
\bibitem{Blumlein:2009rg}
  J.~Bl\"umlein, S.~Klein and B.~T\"odtli,
  Phys.\ Rev.\ D {\bf 80} (2009) 094010
  [arXiv:0909.1547 [hep-ph]].
%
\bibitem{Bierenbaum:2008yu}
  I.~Bierenbaum, J.~Bl\"umlein, S.~Klein and C.~Schneider,
  Nucl.\ Phys.\  B {\bf 803} (2008) 1
  [arXiv:0803.0273 [hep-ph]].
%
\bibitem{Ablinger:2010ty}
  J.~Ablinger, J.~Bl\"umlein, S.~Klein, C.~Schneider and F.~Wi\ss{}brock,
  Nucl.\ Phys.\ B {\bf 844} (2011) 26 
  [arXiv:1008.3347 [hep-ph]].
%
\bibitem{Blumlein:2012vq}
  J.~Bl\"umlein, A.~Hasselhuhn, S.~Klein and C.~Schneider,
  Nucl.\ Phys.\ B {\bf 866} (2013) 196
  [arXiv:1205.4184 [hep-ph]].
%
\bibitem{Ablinger:2014uka}
  J.~Ablinger, J.~Bl\"umlein, A.~De Freitas, A.~Hasselhuhn, A.~von Manteuffel, M.~Round and C.~Schneider,
  arXiv:1405.4259 [hep-ph], Nucl. Phys. B {\bf 885} (2014) 280.
%
\bibitem{Ablinger:2014lka}
  J.~Ablinger, J.~Bl\"umlein, A.~De Freitas, A.~Hasselhuhn, A.~von Manteuffel, M.~Round, C.~Schneider and F.~Wi\ss{}brock,
  Nucl.\ Phys.\ B {\bf 882} (2014) 263
  [arXiv:1402.0359 [hep-ph]].
%
\bibitem{Barone:2001sp}
  V.~Barone, A.~Drago and P.~G.~Ratcliffe,
  Phys.\ Rept.\  {\bf 359} (2002) 1
  [arXiv:hep-ph/0104283].
%
\bibitem{BW1}
  J.~Ablinger, J.~Bl\"umlein, S.~Klein, C.~Schneider and F.~Wi\ss{}brock,
  arXiv:1106.5937 [hep-ph];\\
  J.~Ablinger, J.~Bl\"umlein, A.~Hasselhuhn, S.~Klein, C.~Schneider and F.~Wi\ss{}brock,
  PoS~(RADCOR2011)~031
  [arXiv:1202.2700 [hep-ph]];\\
J. Bl\"umlein and F. Wi\ss{}brock, DESY 14--019.
%
\bibitem{Vermaseren:2005qc}
  J.~A.~M.~Vermaseren, A.~Vogt and S.~Moch,
  Nucl.\ Phys.\ B {\bf 724} (2005) 3
  [hep-ph/0504242].
%
\bibitem{WT}
  J.~C.~Ward,
  Phys.\ Rev.\  {\bf 78} (1950) 182;\\
  Y.~Takahashi,
  Nuovo Cim.\  {\bf 6} (1957) 371.
%
\bibitem{Nogueira:1991ex}
  P.~Nogueira,
  J.\ Comput.\ Phys.\  {\bf 105} (1993) 279.
%
\bibitem{Ablinger:2014yaa}
  J.~Ablinger, J.~Bl\"umlein, C.~Raab, C.~Schneider and F.~Wi\ss{}brock,
  arXiv:1403.1137 [hep-ph], Nucl. Phys. B {\bf 885} (2014) 409.
%
\bibitem{Tentyukov:2007mu}
  M.~Tentyukov and J.~A.~M.~Vermaseren,
  Comput.\ Phys.\ Commun.\  {\bf 181} (2010) 1419
  [hep-ph/0702279];\\
  J.~A.~M.~Vermaseren,
  arXiv:math-ph/0010025.
%
\bibitem{IBP}
J. Lagrange, {\sf Nouvelles recherches sur la nature et la propagation
du son}, Miscellanea Taurinensis, t. II, 1760-61; Oeuvres t. I, p. 263;\\
C.F. Gauss, {Theoria attractionis corporum sphaeroidicorum ellipticorum
homogeneorum methodo novo tractate}, Commentationes societas scientiarum
Gottingensis recentiores, Vol III, 1813, Werke Bd. {\bf V} pp. 5-7;\\
G. Green, {\sf Essay on the Mathematical Theory of Electricity and
Magnetism}, Nottingham, 1828 [Green Papers, pp. 1-115];\\
M. Ostrogradski, Mem. Ac. Sci. St. Peters., {\bf 6}, (1831) 39;\\
  K.~G.~Chetyrkin, A.~L.~Kataev and F.~V.~Tkachov,
  Nucl.\ Phys.\  B {\bf 174} (1980) 345.
%
\bibitem{vonManteuffel:2012np}
  A.~von Manteuffel and C.~Studerus,
  arXiv:1201.4330 [hep-ph];\\
  C.~Studerus,
  Comput.\ Phys.\ Commun.\  {\bf 181} (2010) 1293
  [arXiv:0912.2546 [physics.comp-ph]].
%
\bibitem{FERMAT}
R.H.~Lewis, Computer Algebra System {\tt Fermat}, {\tt http://home.bway.net/lewis.}
%
\bibitem{Bauer:2000cp}
  C.~W.~Bauer, A.~Frink and R.~Kreckel,
  Symbolic Computation {\bf 33} (2002) 1,
  cs/0004015 [cs-sc].
%
\bibitem{Ablinger:2012qm}
  J.~Ablinger, J.~Bl\"umlein, A.~Hasselhuhn, S.~Klein, C.~Schneider and F.~Wi\ss{}brock,
  Nucl.\ Phys.\ B {\bf 864} (2012) 52
  [arXiv:1206.2252 [hep-ph]].
%
\bibitem{Anastasiou:2004vj}
  C.~Anastasiou and A.~Lazopoulos,
  JHEP {\bf 0407} (2004) 046
  [hep-ph/0404258].
%
\bibitem{Smirnov:2013dia}
  A.~V.~Smirnov and V.~A.~Smirnov,
  Comput.\ Phys.\ Commun.\  {\bf 184} (2013) 2820
  [arXiv:1302.5885 [hep-ph]];\\
  A.~V.~Smirnov,
  JHEP {\bf 0810} (2008) 107
  [arXiv:0807.3243 [hep-ph]].
%
\bibitem{Lee:2012cn} 
  R.~N.~Lee,
  arXiv:1212.2685 [hep-ph].
%
%
\bibitem{Laporta:1996mq}
  S.~Laporta and E.~Remiddi,
  Phys.\ Lett.\ B {\bf 379} (1996) 283
  [hep-ph/9602417].
%
\bibitem{Tkachov:1981wb}
  F.~V.~Tkachov,
  Phys.\ Lett.\ B {\bf 100} (1981) 65.
%
\bibitem{Chetyrkin:1981qh}
  K.~G.~Chetyrkin and F.~V.~Tkachov,
  Nucl.\ Phys.\ B {\bf 192} (1981) 159.
%
\bibitem{Laporta:2001dd}
  S.~Laporta,
  Int.\ J.\ Mod.\ Phys.\ A {\bf 15} (2000) 5087
  [hep-ph/0102033].
%
\bibitem{EMSSP}
  J.~Ablinger, J.~Bl\"umlein, S.~Klein and C.~Schneider,
  Nucl.\ Phys.\ Proc.\ Suppl.\  {\bf 205-206} (2010) 110
  [arXiv:1006.4797 [math-ph]];\\
  J.~Bl\"umlein, A.~Hasselhuhn and C.~Schneider,
  PoS RADCOR {\bf 2011} (2011) 032
  [arXiv:1202.4303 [math-ph]];\\
  C.~Schneider,
  arXiv:1310.0160 [cs.SC], {Proc. of ACAT 2013} in press. 
%
\bibitem{SIG1}
C.~Schneider, {S\'em.~Lothar. Combin.\/} {\bf 56} (2007) 1,
 article B56b.
%
\bibitem{SIG2}
C.~Schneider, {{Computer Algebra in Quantum Field Theory: Integration,
  Summation and Special Functions}\/} Texts and Monographs in Symbolic
  Computation eds. C.~Schneider and J.~Bl\"umlein  (Springer, Wien, 2013) 325 
  arXiv:1304.4134 [cs.SC].
%
\bibitem{Karr:81}
M.~Karr 1981 {J.~ACM\/} {\bf 28} (1981) 305.
%
\bibitem{Schneider:01}
C.~Schneider,  
{\sf Symbolic Summation in Difference Fields\/} Ph.D. Thesis
RISC, Johannes Kepler University, Linz technical report 01-17 (2001).
%
\bibitem{Schneider:05a}
C.~Schneider, {J. Differ. Equations Appl.\/} {\bf 11} (2005) 799.
%
\bibitem{Schneider:07d}
C.~Schneider, {J. Algebra Appl.\/} {\bf 6} (2007) 415.
%
\bibitem{Schneider:08c}
C.~Schneider, {J. Symbolic Comput.\/} {\bf 43} (2008) 611.
  [arXiv:0808.2543v1].
%
\bibitem{Schneider:10a}
C.~Schneider, {Appl. Algebra Engrg. Comm. Comput.\/} {\bf 21} (2010) 1.
%
\bibitem{Schneider:10b}
C.~Schneider, {\sf {Motives, Quantum Field Theory, and Pseudodifferential
  Operators}\/} ({\sf Clay Mathematics Proceedings\/} vol~12) ed Carey A,
  Ellwood D, Paycha S and Rosenberg S (Amer. Math. Soc) (2010), 285, 
  arXiv:0808.2543.
%
\bibitem{Schneider:10c}
C.~Schneider, {Ann. Comb.\/} {\bf 14} (2010)  533
[arXiv:0808.2596].
%
\bibitem{Schneider:13b}
C.~Schneider, 
in~: Lecture Notes in Computer Science (LNCS)
eds. J. Guitierrez, J. Schicho, M. Weimann, in press,
arXiv:1307.7887 [cs.SC] (2013).
%
\bibitem{HARMONICSUMS}
  J.~Ablinger,
  arXiv:1305.0687 [math-ph];\\
  J.~Ablinger, J.~Bl\"umlein and C.~Schneider,
  J.\ Math.\ Phys.\  {\bf 52} (2011) 102301
  [arXiv:1105.6063 [math-ph]];
  J.\ Math.\ Phys.\  {\bf 54} (2013) 082301
  [arXiv:1302.0378 [math-ph]].
%
\bibitem{Ablinger:2010kw}
  J.~Ablinger,
  arXiv:1011.1176 [math-ph].
%
\bibitem{GHYP}
W.N. Bailey, {\sf Generalized Hypergeometric Series}, (Cambridge University
Press,  Cambridge, 1935).
%
\bibitem{Slater}
L.J. Slater, {\sf Generalized Hypergeometric Functions}, (Cambridge University  
Press, Cambridge, 1966).
%
\bibitem{Appell}
P. Appell and J. Kamp\'{e} de F\'{e}riet, {\sf Fonctions
Hyperg\'{e}om\'{e}triques et Hypersp\'{e}riques, Polynomes D' Hermite},
(Gauthier-Villars, Paris, 1926);\\
P. Appell, {\sf Les Fonctions Hyperg\"{e}om\'{e}triques de Plusieur
Variables}, (Gauthier-Villars, Paris, 1925);\\
J. Kamp\'{e} de F\'{e}riet, {\sf La fonction
hyperg\"{e}om\'{e}trique},(Gauthier-Villars, Paris, 1937);\\
H. Exton, {\sf Multiple Hypergeometric Functions and Applications},
(Ellis Horwood, Chichester, 1976);\\
H. Exton, {\sf Handbook of Hypergeometric Integrals},
(Ellis Horwood, Chichester, 1978);\\
H.M. Srivastava and P.W. Karlsson, {\sf Multiple Gaussian Hypergeometric
Series}, (Ellis Horwood, Chicester, 1985).
%
\bibitem{Hamberg}
R.~Hamberg, {\sf Second Order Gluonic Contributions to Physical Quantities}, Ph.D. Thesis, Univ. of Leiden 
(1991).
%
\bibitem{Steinhauser:2000ry}
  M.~Steinhauser,
  Comput.\ Phys.\ Commun.\  {\bf 134} (2001) 335
  [hep-ph/0009029].
%
\bibitem{MELB}
E.W. Barnes, Proc. Lond. Math. Soc. (2) {\bf 6} (1908) 141; Quart.
Journ. Math. {\bf 41} (1910) 136;\\ 
H. Mellin,
Math. Ann. {\bf 68} (1910) 305.
%
\bibitem{Czakon:2005rk}
  M.~Czakon,
  Comput.\ Phys.\ Commun.\  {\bf 175} (2006) 559
  [hep-ph/0511200].
%
\bibitem{Smirnov:2009up}
  A.~V.~Smirnov and V.~A.~Smirnov,
  Eur.\ Phys.\ J.\ C {\bf 62} (2009) 445
  [arXiv:0901.0386 [hep-ph]].
%
\bibitem{HSUM}
  J.~A.~M.~Vermaseren,
  Int.\ J.\ Mod.\ Phys.\ A {\bf 14} (1999) 2037
  [hep-ph/9806280];\\
  J.~Bl\"umlein and S.~Kurth,
  Phys.\ Rev.\ D {\bf 60} (1999) 014018,
  [hep-ph/9810241].
%
\bibitem{ANDI2}
  E.~G.~Floratos, D.~A.~Ross and C.~T.~Sachrajda,
  Nucl.\ Phys.\ B {\bf 129} (1977) 66
   [Erratum-ibid.\ B {\bf 139} (1978) 545];\\
  A.~Gonzalez-Arroyo, C.~Lopez and F.~J.~Yndurain,
  Nucl.\ Phys.\ B {\bf 153} (1979) 161;\\
  G.~Curci, W.~Furmanski and R.~Petronzio,
  Nucl.\ Phys.\ B {\bf 175} (1980) 27;\\
  E.~G.~Floratos, C.~Kounnas and R.~Lacaze,
  Nucl.\ Phys.\ B {\bf 192} (1981) 417;\\
  S.~Moch and J.~A.~M.~Vermaseren,
  {Nucl.\ Phys.}\ B {\bf 573} (2000) 853
  [hep-ph/9912355].
%
\bibitem{ANDI3}
  S.~A.~Larin, T.~van Ritbergen and J.~A.~M.~Vermaseren,
  Nucl.\ Phys.\  B {\bf 427} (1994) 41;\\
  S.~A.~Larin, P.~Nogueira, T.~van Ritbergen and J.~A.~M.~Vermaseren,
  Nucl.\ Phys.\  B {\bf 492} (1997) 338
  [arXiv:hep-ph/9605317];\\
  A.~Retey and J.~A.~M.~Vermaseren,
  Nucl.\ Phys.\  B {\bf 604} (2001) 281
  [arXiv:hep-ph/0007294];\\
  J.~Bl\"umlein and J.~A.~M.~Vermaseren,
  Phys.\ Lett.\  B {\bf 606} (2005) 130
  [arXiv:hep-ph/0411111].
%
\bibitem{Moch:2004pa}
  S.~Moch, J.~A.~M.~Vermaseren and A.~Vogt,
  Nucl.\ Phys.\ B {\bf 688} (2004) 101
  [hep-ph/0403192].
%
\bibitem{TR2LOOP}
  S.~Kumano and M.~Miyama,
  Phys.\ Rev.\  D {\bf 56} (1997) 2504
  [arXiv:hep-ph/9706420];\\
  W.~Vogelsang,
  Phys.\ Rev.\  D {\bf 57} (1998) 1886
  [arXiv:hep-ph/9706511] and references therein;\\
  A.~Hayashigaki, Y.~Kanazawa and Y.~Koike,
  Phys.\ Rev.\ D {\bf 56} (1997) 7350
  [hep-ph/9707208].
%
\bibitem{GRAC}
  J.~A.~Gracey,
  Nucl.\ Phys.\  B {\bf 662} (2003) 247
  [arXiv:hep-ph/0304113];
  Nucl.\ Phys.\  B {\bf 667} (2003) 242
  [arXiv:hep-ph/0306163];
  JHEP {\bf 0610} (2006) 040
  [arXiv:hep-ph/0609231];
  Phys.\ Lett.\  B {\bf 643} (2006) 374
  [arXiv:hep-ph/0611071].
%
\bibitem{Bagaev:2012bw}
  A.~A.~Bagaev, A.~V.~Bednyakov, A.~F.~Pikelner and V.~N.~Velizhanin,
  Phys.\ Lett.\ B {\bf 714} (2012) 76
  [arXiv:1206.2890 [hep-ph]].
%
\bibitem{Velizhanin:2012nm}
  V.~N.~Velizhanin,
  Nucl.\ Phys.\ B {\bf 864} (2012) 113
  [arXiv:1203.1022 [hep-ph]].
%
\bibitem{Blumlein:2009cf}
  J.~Bl\"umlein, D.~J.~Broadhurst and J.~A.~M.~Vermaseren,
  Comput.\ Phys.\ Commun.\  {\bf 181} (2010) 582
  [arXiv:0907.2557 [math-ph]].
%
\bibitem{Blumlein:2003gb}
  J.~Bl\"umlein,
  Comput.\ Phys.\ Commun.\  {\bf 159} (2004) 19
  [hep-ph/0311046].
%
\bibitem{MASS}
  N.~Gray, D.~J.~Broadhurst, W.~Grafe and K.~Schilcher,
  Z.\ Phys.\ C {\bf 48} (1990) 673;\\
  K.~G.~Chetyrkin and M.~Steinhauser,
  Nucl.\ Phys.\ B {\bf 573} (2000) 617
  [hep-ph/9911434];\\
  K.~Melnikov and T.~v.~Ritbergen,
  Phys.\ Lett.\ B {\bf 482} (2000) 99
  [hep-ph/9912391];
  Nucl.\ Phys.\ B {\bf 591} (2000) 515
  [hep-ph/0005131].
%
\bibitem{Remiddi:1999ew}
  E.~Remiddi and J.~A.~M.~Vermaseren,
  Int.\ J.\ Mod.\ Phys.\ A {\bf 15} (2000) 725
  [hep-ph/9905237].
%
\bibitem{SX}
  R.~Kirschner and L.N.~Lipatov,
  Nucl.\ Phys.\ B {\bf 213} (1983) 122;\\
  J.~Bl\"umlein and A.~Vogt,
  Phys.\ Lett.\ B {\bf 370} (1996) 149
  [hep-ph/9510410].
%
\bibitem{ANCONT}
  J.~Bl\"umlein,
  Comput.\ Phys.\ Commun.\  {\bf 133} (2000) 76
  [hep-ph/0003100];
  Comput.\ Phys.\ Commun.\  {\bf 180} (2009) 2218
  [arXiv:0901.3106 [hep-ph]];
  Proceedings 
  of the Workshop ``Motives, Quantum Field
  Theory, and Pseudodifferential Operators'',
       held at the Clay Mathematics
       Institute, Boston University, June 2--13, 2008, 
       Clay Mathematics Proceedings {\bf 12} (2010) 167, 
       Eds. A.~Carey, D.~Ellwood, S.~Paycha, S. Rosenberg,
  arXiv:0901.0837 [math-ph];\\
  A.~V.~Kotikov and V.~N.~Velizhanin,
  hep-ph/0501274;\\
  J.~Bl\"umlein and S.-O.~Moch,
  Phys.\ Lett.\ B {\bf 614} (2005) 53
  [hep-ph/0503188].
%
\bibitem{Gehrmann:2001pz}
  T.~Gehrmann and E.~Remiddi,
  Comput.\ Phys.\ Commun.\  {\bf 141} (2001) 296
  [hep-ph/0107173].
%
\bibitem{Vollinga:2004sn}
  J.~Vollinga and S.~Weinzierl,
  Comput.\ Phys.\ Commun.\  {\bf 167} (2005) 177
  [hep-ph/0410259].
%
%
\bibitem{Maitre:2007kp}
  D.~Ma\^{i}tre,
  Comput.\ Phys.\ Commun.\  {\bf 183} (2012) 846
  [hep-ph/0703052 [hep-ph]];
  Comput.\ Phys.\ Commun.\  {\bf 174} (2006) 222
  [hep-ph/0507152].
%
%
\bibitem{Buehler:2011ev}
  S.~Buehler and C.~Duhr,
  arXiv:1106.5739 [hep-ph].
%
\bibitem{Alekhin:2013nda}
  S.~Alekhin, J.~Bl\"umlein and S.~Moch,
  Phys.\ Rev.\ D {\bf 89} (2014) 054028
  [arXiv:1310.3059 [hep-ph]].
%
\bibitem{Boer:2011fh}
  D.~Boer, M.~Diehl, R.~Milner, R.~Venugopalan, W.~Vogelsang, D.~Kaplan, H.~Montgomery and S.~Vigdor {\it et al.},
  arXiv:1108.1713 [nucl-th];\\
  A.~Accardi, J.~L.~Albacete, M.~Anselmino, N.~Armesto, E.~C.~Aschenauer, A.~Bacchetta, D.~Boer and W.~Brooks {\it et al.},
  arXiv:1212.1701 [nucl-ex].
%
\bibitem{Alekhin:2012ig}
  S.~Alekhin, J.~Bl\"umlein and S.~Moch,
  Phys.\ Rev.\ D {\bf 86} (2012) 054009
  [arXiv:1202.2281 [hep-ph]].
%
\bibitem{Blumlein:1998sh}
  J.~Bl\"umlein and W.~L.~van Neerven,
  Phys.\ Lett.\ B {\bf 450} (1999) 417
  [hep-ph/9811351].
%
\bibitem{CRO}
  H.~D.~Politzer,
  Phys.\ Rept.\  {\bf 14} (1974) 129;\\
  J.~Bl\"umlein and N.~Kochelev,
  Phys.\ Lett.\ B {\bf 381} (1996) 296
  [hep-ph/9603397];
  Nucl.\ Phys.\ B {\bf 498} (1997) 285
  [hep-ph/9612318].
%
\bibitem{Vermaseren:1994je}
  J.~A.~M.~Vermaseren,
  Comput.\ Phys.\ Commun.\  {\bf 83} (1994) 45.
\end{thebibliography}
\end{document}